\documentclass[aps,reprint,pre,superscriptaddress,nofootinbib]{revtex4-1}

\usepackage{xpatch}
\usepackage{graphicx}
\usepackage{tikz}
\usetikzlibrary{matrix}
\usepackage{pifont}
\usepackage{paralist}
\usepackage{color}
\usepackage{amsfonts,amssymb,amsthm,amsmath}
\usepackage{hyperref}
\usepackage{bm}
\usepackage{subfig}
\usepackage{listings}
\usepackage{verbatim}
\usepackage{lmodern}
\usepackage{threeparttable}
\usepackage{centernot}
\usepackage{float}

\counterwithin{table}{section}
\counterwithin{figure}{section}

\renewcommand{\figurename}{{\bf FIG.}}
\renewcommand{\tablename}{{\bf TABLE}}

\renewcommand{\thefigure}{{\Roman{section}.\arabic{figure}}}
\renewcommand{\thetable}{{\Roman{section}.\arabic{table}}}

\definecolor{codegreen}{rgb}{0,0.6,0}
\definecolor{codegray}{rgb}{0.5,0.5,0.5}
\definecolor{codepurple}{rgb}{0.58,0,0.82}
\definecolor{backcolour}{rgb}{0.95,0.95,0.92}

\lstdefinestyle{mystyle}{
    backgroundcolor=\color{backcolour},
    commentstyle=\color{codegreen},
    keywordstyle=\color{magenta},
    numberstyle=\tiny\color{codegray},
    stringstyle=\color{codepurple},
    basicstyle=\footnotesize,
    breakatwhitespace=false,
    breaklines=true,
    captionpos=b,
    keepspaces=true,
    numbers=left,
    numbersep=5pt,
    columns=full1flexible,
    showspaces=false,
    showstringspaces=false,
    showtabs=false,
    tabsize=2
}

\definecolor{green}{RGB}{0,255,0}
\definecolor{cyan}{RGB}{0,183,235}
\definecolor{blue}{RGB}{0,0,255}
\definecolor{violet}{RGB}{199,21,133}
\definecolor{orange}{RGB}{255,127,0}

\definecolor{red_i}{RGB}{128,0,0}
\definecolor{cyan_i}{RGB}{0,255,102}
\definecolor{blue_i}{RGB}{55,55,200}
\definecolor{orange_i}{RGB}{255,102,0}
\definecolor{brown_l}{RGB}{120,75,40}
\definecolor{magenta_l}{RGB}{128,0,128}
\definecolor{magenta_ll}{RGB}{180,0,180}
\definecolor{blue_l}{RGB}{0,146,146}
\definecolor{black_l}{RGB}{36,31,28}
\definecolor{orange_l}{RGB}{255,40,0}

\theoremstyle{definition}
\newtheorem{theorem}{Theorem}[]
\newtheorem{proposition}{Proposition}[section]

\newtheorem{definition}{Definition}[section]
\newtheorem{example}{Example}[section]

\newcommand{\proofprop}[3]{\vspace{\topsep}\noindent\textbf{Proof of Proposition~\ref{#1}:} #2\vspace{0.1cm}\vspace{-\topsep}\begin{proof}#3\end{proof}}
\newcommand{\prooftheorem}[3]{\vspace{\topsep}\noindent\textbf{Proof of Theorem~\ref{#1}:} #2\vspace{0.1cm}\vspace{-\topsep}\begin{proof}#3\end{proof}}

\newcommand{\beq}{\begin{equation}}
\newcommand{\eeq}{\end{equation}}
\newcommand{\beqa}{\begin{eqnarray}}
\newcommand{\eeqa}{\end{eqnarray}}
\newcommand{\es}{\enspace}

\newcommand{\comp}[2]{{#1\circ #2}}

\newcommand{\incomp}{\hspace{0.1cm} /\hspace{-0.1cm} /\hspace{0.1cm} }

\newcommand{\cmark}{{\color{green}${\bf x}$}}
\newcommand{\cxmark}{{\color{green}${\bf x}$}}
\newcommand{\piq}{\bm{\pi}_{\Q}}
\newcommand{\pxq}{p_{\Q}}

\newcommand{\pit}{\bm{\pi}_{\T}}

\newcommand{\pitmark}{{\color{blue}$\pit$}}

\newcommand{\pii}{\bm{\pi}_{\N}}

\newcommand{\piimark}{{\color{blue}$\pii$}}
\newcommand{\piij}{\bm{\pi}_{\L}}

\newcommand{\piijmark}{{\color{blue}$\piij$}}

\newcommand{\pxt}{{p}_{\T}}

\newcommand{\pxtmark}{{\color{blue}$\pxt$}}

\newcommand{\pxi}{{p}_{\N}}

\newcommand{\pximark}{{\color{blue}$\pxi$}}

\newcommand{\pxl}{{p}_{\L}}

\newcommand{\pxlmark}{{\color{blue}$\pxl$}}

\newcommand{\muq}{\bm{\mu}_{\Q}}

\newcommand{\mut}{\bm{\mu}_{\T}}
\newcommand{\mutmark}{{\color{violet}$\mut$}}
\newcommand{\mui}{\bm{\mu}_{\N}}
\newcommand{\muimark}{{\color{violet}$\mui$}}
\newcommand{\muij}{\bm{\mu}_{\L}}
\newcommand{\muijmark}{{\color{violet}$\muij$}}
\newcommand{\pmark}{{\color{blue}$p$}}

\newcommand{\mumark}{{\color{violet}$\mu$}}
\newcommand{\xmark}{{\color{red}$-$}}

\newcommand{\lc}{\chi_{\lambda}}
\newcommand{\lcmark}{$\lc$}

\newcommand{\sgnmark}{{\color{black}$H$}}
\renewcommand{\t}[1]{\overline{#1}}
\newcommand{\sgn}{\lN}

\newcommand{\Prob}[1]{P_{#1}}
\newcommand{\CProb}[2]{P_{#1 | #2}}
\newcommand{\model}[1]{\text{P}[#1]}
\newcommand{\partition}[1]{\G_#1}

\newcommand{\et}{(i,j,t)}

\newcommand{\V}{\mathcal{V}}
\newcommand{\G}{\mathcal{G}}
\newcommand{\C}{\mathcal{C}}
\newcommand{\E}{\mathcal{E}}
\newcommand{\Gl}{G_\L}
\newcommand{\Gt}{{\Gamma}^{t}} %^{\rm s}}
\newcommand{\Gs}{{\bf\Gamma}}

\newcommand{\Thl}{{\Theta_{\ell}}}
\newcommand{\Thij}{\Theta_{(i,j)}}
\newcommand{\Dt}{{\Delta t}}
\newcommand{\nij}{{n_{(i,j)}}}
\newcommand{\wij}{{w_{(i,j)}}}
\newcommand{\ai}{{a_i}}
\newcommand{\si}{{s_i}}
\newcommand{\Dtau}{{\Delta\tau}}
\newcommand{\Dtauf}{{\bm{\Dtau}}}
\newcommand{\tauf}{{\bm{\tau}}}
\newcommand{\alphaf}{\bm{\alpha}}
\newcommand{\Dalpha}{{\Delta\alpha}}
\newcommand{\Dalphaf}{\bm{\Delta\alpha}}
\newcommand{\tmin}{{t_{\rm min}}}
\newcommand{\tmax}{{t_{\rm max}}}
\newcommand{\Et}{{A^t}}

\newcommand{\Estat}{{L}}

\newcommand{\iso}{{\rm iso}}
\newcommand{\isof}{{\bf iso}}
\newcommand{\sgnf}{{\bm \chi}_{\mathbb{N}^+}}
\renewcommand{\ij}{{(i,j)}}

\newcommand{\N}{\mathcal{V}}

\newcommand{\ul}{t_\ij}
\newcommand{\taul}{\tau_\ij}
\newcommand{\Dtaul}{\Dtau_\ij}
\newcommand{\nl}{{n_\ij}}
\newcommand{\wl}{{w_\ij}}
\newcommand{\vi}{v_i}
\newcommand{\alphai}{\alpha_i}
\newcommand{\Dalphai}{\Dalpha_i}
\renewcommand{\P}{{\bf P}}
\newcommand{\R}{\mathcal{R}}
\renewcommand{\L}{\mathcal{L}}
\newcommand{\dit}{d_i^t}

\newcommand{\x}{{\bf x}}

\newcommand{\X}{\mathcal{X}}
\newcommand{\Y}{\mathcal{Y}}
\newcommand{\y}{{\bf y}}
\newcommand{\z}{{\bf z}}
\newcommand{\f}{{\bf f}}
\newcommand{\g}{{\bf g}}
\renewcommand{\k}{{\bf k}}
\renewcommand{\d}{{\bf d}}

\newcommand{\kstat}{{\bf k}}
\newcommand{\Gstat}{{G}^{\rm stat}}

\newcommand{\iout}{{i\rightarrow}}
\newcommand{\iin}{{i\leftarrow}}
\newcommand{\kout}[1]{k_{#1\rightarrow}}
\newcommand{\kin}[1]{k_{#1\leftarrow}}
\newcommand{\sout}[1]{s_{#1\rightarrow}}
\renewcommand{\sin}[1]{s_{#1\leftarrow}}
\newcommand{\aout}[1]{a_{#1\rightarrow}}
\newcommand{\ain}[1]{a_{#1\leftarrow}}
\newcommand{\dtout}[2]{{d}_{#1\rightarrow}^{#2}}
\newcommand{\dtin}[2]{{d}_{#1\leftarrow}^{#2}}

\newcommand{\soutf}{{\bf s}_{\rightarrow}}

\newcommand{\doutf}{{\bf d}_{\rightarrow}}

\renewcommand{\t}{{\bf t}}
\newcommand{\n}{{\bf n}}
\renewcommand{\a}{{\bf a}}
\newcommand{\w}{{\bf w}}
\newcommand{\s}{{\bf s}}
\newcommand{\Ebf}{{\bf A}}
\newcommand{\Thf}{{\bf\Theta}}

\newcommand{\perl}{{\bf per}}
\newcommand{\Phif}{{\bf\Phi}}
\newcommand{\Gf}{{\bf\Gamma}}
\newcommand{\gf}{{\bm \sigma}}
\newcommand{\Sigmaf}{{\bf \Sigma}}

\newcommand{\Ml}{{\mathcal{M}_\ij}}
\newcommand{\Mi}{{\mathcal{M}_i}}

\newcommand{\T}{\mathcal{T}}
\newcommand{\Gss}{G_\T}
\newcommand{\Q}{\mathcal{Q}}

\newcommand{\tinf}{\theta}

\newcommand{\lN}{\chi_{\mathbb{N}^+}}

\renewcommand{\checkmark}{\textcolor{blue}{\ding{51}}}
\newcommand{\crossmark}{\textcolor{red}{\ding{55}}}

\newcommand{\new}[1]{#1}

\makeatletter
\def\l@subsubsection#1#2{}
\makeatother
\newcommand{\nocontentsline}[3]{}
\newcommand{\tocless}[2]{\bgroup\let\addcontentsline=\nocontentsline#1{#2}\egroup}

\makeatletter
\patchcmd{\@ssect@ltx}
    {\addcontentsline{toc}{#1}{\protect\numberline{}#8}}
    {}
    {}
    {}
\makeatother

\begin{document}

\title{Randomized reference models for temporal networks}
\author{L. Gauvin}
\affiliation{Data Science Lab, ISI Foundation, Turin, Italy.}
\author{M. G\'enois}
\affiliation{Aix Marseille Univ., Univ. Toulon, CNRS, CPT, Marseille, France.}
\author{M. Karsai}
\affiliation{Department of Network and Data Science, Central European University, 1100 Vienna, Austria.}
\affiliation{Universit\'e de Lyon, ENS de Lyon, INRIA, CNRS, LIP-UMR 5668, IXXI, 69364, Cedex 07 Lyon, France.}
\affiliation{Alfr\'ed R\'enyi Insititute of Mathematics, H-1053 Budapest, Hungary}
\author{M. Kivel\"a}
\affiliation{Department of Computer Science, Aalto University School of Science, P.O. Box 12200, FI-00076 Aalto, Finland.}
\author{T. Takaguchi}
\affiliation{National Institute of Information and Communications Technology, 4-2-1 Nukui-Kitamachi, Koganei, Tokyo  184-8795, Japan.}
\author{E. Valdano}
\affiliation{Center for Biomedical Modeling, The Semel Institute for Neuroscience and Human Behavior, David Geffen School of Medicine, 760 Westwood Plaza, University of California Los Angeles, Los Angeles, CA 90024 USA.}
\author{C.~L. Vestergaard}
\email{cvestergaard@gmail.com}
\affiliation{Aix Marseille Univ., Univ. Toulon, CNRS, CPT, Marseille, France.}
\affiliation{Decision and Bayesian Computation, Department of Computational Biology, Department of Neuroscience, CNRS USR 3756, CNRS UMR 3571, Institut Pasteur, 25 rue du Docteur Roux, Paris, 75015, France.}

\date{\today}
%------------------------------------------------------------------------
\begin{abstract}
\noindent
Many dynamical systems can be successfully analyzed by representing them as networks. Empirically measured networks and dynamic processes that take place in these situations show heterogeneous, non-Markovian, and intrinsically correlated topologies and dynamics. This makes their analysis particularly challenging. Randomized reference models (RRMs) have emerged as a general and versatile toolbox for studying such systems. Defined as random networks with given features constrained to match those of an input (empirical) network, they may for example be used to identify important features of empirical networks and their effects on dynamical processes unfolding in the network. RRMs are typically implemented as procedures that reshuffle an empirical network, making them very generally applicable. However, the effects of most shuffling procedures on network features remain poorly understood, rendering their use non-trivial and susceptible to misinterpretation. Here we propose a unified framework for classifying and understanding microcanonical RRMs (MRRMs) that sample networks with uniform probability. Focusing on temporal networks, we survey applications of MRRMs found in literature, and we use this framework to build a taxonomy of MRRMs that proposes a canonical naming convention, classifies them, and deduces their effects on a range of important network features. We furthermore show that certain classes of MRRMs may be applied in sequential composition to generate new MRRMs from the existing ones surveyed in this article. We finally provide a tutorial showing how to apply a series of MRRMs to analyze how different network features affect a dynamic process in an empirical temporal network. Our taxonomy provides a reference for the use of MRRMs, and the theoretical foundations laid here may further serve as a base for the development of a principled and automatized way to generate and apply randomized reference models for the study of networked systems.
\end{abstract}
%------------------------------------------------------------------------
\begin{titlepage}
\maketitle
\end{titlepage}

\onecolumngrid
\clearpage
\tableofcontents
%\vspace{1cm}~
\pagebreak
\twocolumngrid

%========================================================================
\section{Introduction}
%========================================================================
\noindent
Random network models are responsible of major parts of our theoretical understanding of networked systems and practical knowledge extracted from networked data. 
Well-known examples of such models include the Erd\H{o}s-R\'enyi model~\cite{Erdos1960O} -- \new{which generates random networks with fixed numbers of nodes and links} -- and the configuration model~\cite{Newman2003S,Fosdick2018C} -- \new{which fixes} the degree sequence. 
A main application of these models is as null models for hypothesis testing, 
though their use goes beyond this. 
They may notably be used more generally to investigate the relationship between different network features~\cite{Kovanen2011T, Karsai2012U, Karsai2012C, Kovanen2013T, Orsini2015Q} and their roles in dynamic phenomena~\cite{Karsai2011S, Rocha2011S, Miritello2011D, Starnini2012R, Kivela2012M, Gauvin2013A, Karimi2013T, Takaguchi2013B, Holme2014B, Karsai2014T, Backlund2014E, Cardillo2014E, Thomas2015D, Delvenne2015D, Saramaki2015E, Genois2015C, Valdano2015I, Holme2016T, Barrat2008Dynamical, Gleeson2011High, Pastor-Satorras2001E, Jeong2000L, Sole2001C, Holme2002A, Bollobas2004R, Albert2000E, Cohen2000R, Bollobas2004R}.
To underline their general scope, we here call these type of models {\sl randomized reference models} (RRMs). 
Much of what we know of the behavior of dynamic processes in networks is based on them~\cite{Barrat2008Dynamical, Gleeson2011High}, and they stand behind many prominent results in network science such as the absence of epidemic threshold~\cite{Pastor-Satorras2001E}, the vulnerability to attacks~\cite{Jeong2000L, Sole2001C, Holme2002A, Bollobas2004R}, and the robustness to failures~\cite{Albert2000E, Cohen2000R, Bollobas2004R} in certain types of networks. 
These models are also integral parts of many methods for network data analysis, such as popular network clustering methods \cite{Newman2004Finding, Karrer2011S}, network motif analysis~\cite{Maslov2002S, Milo2002N}, and the analysis of structural correlations~\cite{Watts1998C, Onnela2007S}.
The above is just a small selection of applications, but the examples are legion.

As network science has matured there has been an increasing need to go beyond the simple graph representation for networks, and at the same time repeat the success of RRMs for these new types of networks.
An important extension to simple graphs is temporal networks, which allow the networks' topology to evolve in time.
RRMs\footnote{In the temporal network literature RRMs are also known as {\sl null models}~\cite{Karsai2011S, Bajardi2011D, Kovanen2011T, Squartini2011RI, Squartini2011RII, Holme2012T, Jurgens2012T, Gauvin2013A, Karimi2013T, Kovanen2013T, Cardillo2014E, Valdano2015I, Holme2015M, Saracco2016D, Thomas2015D, Zhang2017C}; {\sl reshuffling methods}~\cite{Genois2015C}, {\sl randomization techniques}~\cite{Posfai2014S, Holme2015M}, {\sl randomization procedures}~\cite{Bajardi2011D, Gauvin2013A, Takaguchi2013B, Posfai2014S, Holme2015M}, {\sl randomization strategies}~\cite{Gauvin2013A}, {\sl randomization schemes}~\cite{Holme2015M}, {\sl randomization methods}~\cite{Takaguchi2012I}, or simply {\sl randomizations}~\cite{Holme2005N, Holme2012T, Takaguchi2012I, Takaguchi2013B, Takaguchi2013I, Posfai2014S, Holme2015M, Takaguchi2016C}).} have also emerged as a powerful toolbox for the study of the dynamics on and of temporal networks~\cite{Holme2012T, Holme2015M, Masuda2016G} and have been applied to complex systems in a broad range of fields, including sociology, epidemiology, infrastructure, economics, and biology. 
They have been used to study how given \new{temporal} network features affect other node- or interaction-level features~\cite{Kovanen2011T, Karsai2012U, Karsai2012C, Kovanen2013T, Orsini2015Q}, and how the features affect dynamical processes unfolding in the network~\cite{Karsai2011S, Rocha2011S, Miritello2011D, Starnini2012R, Kivela2012M, Gauvin2013A, Karimi2013T, Takaguchi2013B, Holme2014B, Karsai2014T, Backlund2014E, Cardillo2014E, Thomas2015D, Delvenne2015D, Saramaki2015E, Genois2015C, Valdano2015I, Holme2016T}
as well as the network's controllability~\cite{Posfai2014S, Li2017T, Zhang2017C}.
Systems studied using temporal network RRMs include: human face-to-face interactions and physical proximity~\cite{Takaguchi2012I, Starnini2012R, Takaguchi2013I, Gauvin2013A, Takaguchi2013B, Karimi2013T, Backlund2014E, Valdano2015I, Delvenne2015D, Takaguchi2016C, Holme2016T, Tang2010S, Cardillo2014E, Thomas2015D, Holme2016T}; 
prostitution networks~\cite{Rocha2011S, Karimi2013T, Delvenne2015D, Holme2016T}; 
functional connections in the brain~\cite{Valencia2008dynamic, Tang2010S, Sun2019D}; 
human mobility~\cite{Pan2011P, Alessandretti2016E}; 
livestock transport~\cite{Bajardi2011D}; 
mobile phone calls and text messages~\cite{Karsai2011S, Miritello2011D, Kivela2012M, Kovanen2013T, Backlund2014E};
email correspondences~\cite{Holme2005N, Karsai2011S, Takaguchi2013B, Karimi2013T, Backlund2014E, Posfai2014S, Delvenne2015D, Takaguchi2016C};
online communities~\cite{Holme2005N, Tang2010S, Karimi2013T, Redmond2014I, Delvenne2015D, Sun2015C};
editing of Wikipedia pages~\cite{Jurgens2012T};
and world trade~\cite{Squartini2011RI, Squartini2011RII, Saracco2016D}.

The popularity of RRMs for the study of complex networks may be explained by the fact that they can often be defined simply as numerical procedures that generate random networks by shuffling the original data, thus avoiding the need to specify a complete generative model.
The resulting set of randomized networks typically serves as a null reference that is compared to the original temporal network, or it may be compared to a second set of networks generated by another RRM. 
For example, by comparing how a given dynamic process evolves on the original network with how it evolves on different sets of randomized networks, we may identify how various features of the network affect the dynamic process.

The algorithmic definition of RRMs as shuffling methods makes them simple to apply in very general settings and with little domain-specific tweaking needed.
However, an important downside to the algorithmic representation is that the effects of RRMs on network features are rarely investigated systematically and remain poorly understood.
This lack of systematic understanding of the methods is not only a theoretical problem but it has led to severe practical problems in the literature.
First, there are no unified naming conventions for the RRMs. 
This makes it difficult to compare the methods used in different studies and has lead to a situation where the algorithms producing equivalent RRMs are given a multitude of different names, and possibly worse, where multiple algorithms producing different RRMs are given the same names~\cite{Holme2019map} 
\new{(%we list the different names given in the literature for all the MRRMs reviewed here in 
see also Section~\ref{sec:classification})}.
%\note{CLV: Refer to Section~\ref{sec:classification} here and/or give one example of a model with many names and one of different models with the same?}.
Second, researchers are confronted with the problems of how to choose and develop randomization techniques, in which order they should be applied, and how to interpret the results.
These are crucial choices in order to be able to identify important features for each given dynamical phenomenon and for each temporal network under study (problems that are non-trivial even for the study of simple graphs~\cite{Orsini2015Q}).

We review temporal network RRMs used in the literature and find that most of them fall into a class of methods that gives a uniform probability of sampling all networks with a given set of features constrained to the same value as that of the original data \cite{Kivela2012M}.
\new{These RRMs are described} by the concept of microcanonical ensembles from statistical physics~\cite{Jaynes1957I,Presse2013P}.
\new{We will consequently call them} {\sl microcanonical} randomized reference models (MRRMs) and represent them in a formal framework where they are fully defined by the set of features they constrain. 
This principled approach has several advantages over the algorithmic representation:
As MRRMs are completely defined by the constraints that they impose, we propose an unambiguous naming convention for MRRMs of temporal networks based on these constraints.
Furthermore, \new{the theoretical} framework enables us to build a taxonomy of existing MRRMs, which lists their effects on important temporal network features and  orders them by the amount of features they constrain. 
This hierarchy allows researchers to apply MRRMs so that the fixed features of the original data are systematically reduced.
We also show how and when new MRRMs can be devised by applying previously implemented algorithms one after another.
\new{We finally illustrate how series of MRRMs may be applied
to analyze temporal network data with a walk-through example.}
%, and we review applications of MRRMs in the literature in light of the theoretical framework}. 

Reference models, which keep parts of the features of original data and shuffle the rest, are clearly widely applicable outside of temporal networks. 
For example, MRRMs are closely related to exact (permutation) tests of classical statistics~\cite{Good2005P} and to conditionally uniform graph tests (CUGTs) found in the sociology literature~\cite{Katz1957P,Snijders1991E,Holme2005N}. 
Furthermore, even though we are here mostly concerned with temporal networks, our framework of MRRMs is directly applicable to  
a far more general class of systems, which can be considered as a realization of a state in a predefined discrete state space. 
In particular, it may be applied e.g. to correlation matrices \new{and} to more general types of relational data such as multilayer networks or hypergraphs.  

It is our aim that, in addition to categorizing previous RRMs and surveying the literature, the unified framework and taxonomy we present would serve as a starting point for the development of a general and principled randomization-based approach for the characterization and analysis of networked dynamical systems. 
To this end, we provide 
%\new{two walk-through examples illustrating how series of MRRMs may be applied
%%a practical guide to researchers who want to apply MRRMs 
%to analyze temporal network data. } %, and we review applications of MRRMs in the literature}. 
%We furthermore provide
 (at \url{https://github.com/mgenois/RandTempNet}) a pure python library implementing the MRRMs presented in this article; 
furthermore, for applications to larger networks we provide (at \url{https://github.com/bolozna/Events}) a fast Python library with core functions written in C++ implementing most of the MRRMs.

%------------------------------------------------------------------------------------------------------------------
%\subsection{Road map to this article}
%\label{sec:roadmap}
%%------------------------------------------------------------------------------------------------------------------
%\note{CLV: Maybe this overview is not needed given that we have a table of contents. Also the paragraph starting with ``We review temporal network RRMs'' above basically gives the same outline, though a bit less detailed.}
\new{We begin in Section~\ref{sec:definitions} by providing theoretical definitions needed to describe temporal networks (Section~\ref{sec:temp-net}) and MRRMs (Section~\ref{sec:RRM}). 
We use these to define classes of shuffling methods that are used to generate randomized temporal networks in practice (Section~\ref{sec:implementation}).
%In Section~\ref{sec:null_model} we detail and discuss the use of MRRMs as null models. 
%\note{CLV: added section on hypothesis testing.}.
In Section~\ref{sec:applications} we survey applications of MRRMs found in the literature, using the theoretical foundations posed in the previous section to provide consistent names for the MRRMs used in each study.
In Section~\ref{sec:hierarchies} we develop the theoretical framework needed to %introduce the concepts and definitions needed to 
compare and build hierarchies of network features and of MRRMs (Section~\ref{sec:theory-comparison}), 
and we use this to build a taxonomy of MRRMs found in the literature that characterizes and orders them hierarchically %and list their effects on important temporal network features 
(Section~\ref{sec:classification}).  
In Section~\ref{sec:composition} we develop a theory for composing MRRMs by applying one algorithm after another (Section~\ref{sec:theory-comparison}). 
We exploit this to show which classes of shuffling methods can be combined to form new MRRMs from existing ones (Section~\ref{sec:shufflings-composition}), 
and we classify several MRRMs found in the literature that are compositions of two other MRRMs (Section~\ref{sec:compositions}). 
In Section~\ref{sec:other_models} we provide a brief overview of temporal network RRMs that do not fall into the class of microcanonical methods. 
In Section~\ref{sec:spreading_example} %we give a procedure for statistical analysis using MRRMs, and 
we finally provide %two different 
a walk-through example showcasing how to apply nested series of MRRMs to analyze an empirical temporal network. 
}

\new{%Two appendices complement the main text: 
% \note{CLV: move some to supplement / remove?}. 
An appendix %~\ref{app:proofs} 
provides formal definitions that are left out in the main text and gives proofs for the propositions and theorems developed in Sections~\ref{sec:theory-comparison}, \ref{sec:theory-composition}, and \ref{sec:shufflings-composition}.
% ;
% Appendix~\ref{app:tables} provides two tables: one defining general types of temporal network features such as ordered sequences or unordered multisets of node- or link-level features, and another describing the effects of MRRMs on a large selection of temporal network features.
We furthermore provide a supplemental material, which includes supplementary tables detailing features of temporal networks as well as four supplementary notes:
%Appendix~\ref{sec:features} 
% \note{CLV move to supplement/remove?}. %, and we establish their hierarchy. 
%Appendix~\ref{sec:instant-event_shuffling} 
Supplementary Note~1 discusses how we can randomize the durations of the events in a temporal network using shuffling methods for networks with instantaneous events; 
Supplementary Note~2 provides detailed definitions for a large selection of important features of temporal networks and orders them hierarchically; 
%, which is sufficient to fully describe the MRRMs found in our literature review; 
% \note{CLV move to supplement?}. 
%An appendix lists additional features for describing MRRMs for directed temporal networks (Appendix~\ref{app:directed}).
Supplementary Note~3 provides a second walk-through example describing an analysis of features of a face-to-face interaction network using series of MRRMs; 
Supplementary Note~4 finally lists the names of the algorithms in the python library used for generating the MRRMs reviewed here.}

\section{Fundamental definitions}
\label{sec:definitions}
%========================================================================
\noindent
\new{In this section we define fundamental concepts %and derive general results 
for temporal networks (Section~\ref{sec:temp-net}) and for microcanonical randomized reference models (MRRMs) (Section~\ref{sec:RRM}).
Based on this, we propose a canonical naming convention for MRRMs that unambiguously and completely defines a MRRM (Section~\ref{sec:RRM}). 
We finally describe several important classes of algorithms for sampling randomized networks and we classify them according to  what overall structures of a temporal network they randomize (Section~\ref{sec:implementation}).} %, such as the aggregated topology or the temporal order of events.}
 
We focus on microcanonical RRMs as these represent the class of maximum entropy models that can be obtained directly by randomly sampling constrained permutations of an empirically observed network (i.e.\ by shuffling the network). 
\new{As shuffling %These algorithms are often implemented as methods that reshuffle elements of a temporal network, and 
is by far the most dominant method for generating randomized temporal networks,
we shall in the following refer to any algorithm for generating random networks according to a given MRRM as a {\sl shuffling method} or simply a {\sl shuffling}.}
% We note \new{finally} that, while we here apply the formalism to the study of temporal networks\new{, %, the results derived in subsections ~\ref{sec:RRM} and ~\ref{sec:combination} apply 
% it applies} more generally to MRRMs for any system with a %finite and 
% discrete state space. 
% \footnote{To apply the formalism to other types of data it suffices to replace the state space $\G$ (Def.~\ref{def:state_space}) in the definition of a MRRM (Def.~\ref{def:MRRM}) %\ref{def:feature} and \ref{def:MRRM} 
% by the appropriate state space, e.g.\ of multilayer networks or of hypergraphs. %To describe the MRRMs one must additionally define features corresponding to the data structure as needed.
% }.  %is done in Section~\ref{sec:features} for temporal networks.}.

%Subsection~\ref{sec:temp-net} provides definitions of a temporal network and of the important notion of a temporal network {\sl feature}, defined as any function that takes a temporal network as input and which is used as the constraint imposed by a MRRM. The subsection also presents two-level representations of a temporal network which facilitate definition of temporal network features (Section \ref{sec:features}) and the implementation of many MRRMs (Section \ref{sec:implementation}).
%Subsection~\ref{sec:RRM} provides a rigorous definition of a MRRM and introduces its basic properties.

%------------------------------------------------------------------------
\subsection{Temporal network}
\label{sec:temp-net}
%------------------------------------------------------------------------
\noindent
We consider a system consisting of $N$ individual nodes engaging in intermittent dyadic interactions observed over a period of time from $t=\tmin$ to $t=\tmax$; in a social network, for example, the nodes are persons, in an ecological network nodes are species, while in a transport network they are locations.
A temporal network is our representation of such an observation.

\begin{definition}
\label{def:temporal_network}
{\sl Temporal network\footnote{Our definition of a temporal network is equivalent to what is called a link stream in~\cite{Latapy2017S}.}.}
A temporal network $G=(\N,\C)$  is defined by the set of nodes $\N=\{1,2,...,N\}$ and a set of events  
$\C=\{c_1, c_2, \ldots, c_C\}$, where each event 
\new{$c_q=(i,j,t,\tau)$ represents an interaction between nodes $i$ and $j$ during the time-interval $[t,t+\tau)$}.
%$c_q=(i_q,j_q,t_q,\tau_q)$ denotes an interaction between nodes $i_q$ and $j_q$ during the time-interval $[t_q,t_q+\tau_q)$.
\end{definition}

Definition~\ref{def:temporal_network} encompasses both temporal networks with directed interactions (e.g.\ for phone-call or instant messaging networks) and undirected interactions (e.g.\ for face-to-face or proximity networks), but does not consider possible weights of the events.
For directed networks, we may adopt the convention that the direction of an event \new{$(i, j, t, \tau)$ is from $i$ to $j$}.
For undirected networks, the presence of the event \new{$(i, j, t, \tau)$ implies the symmetric interaction $(j, i, t, \tau)$, and in practice we may impose $i<j$ to avoid double counting.} %for efficient data storage.

For simplicity, we consider only undirected temporal networks in the main text. %, and in the same spirit of simplicity we have not considered possible weights of events in the above definition.
However, all models and methods may be applied directly to temporal networks with directed and/or weighted events
%Furthermore, they may easily be extended to explicitly take into account the directionality or weights of interactions 
by defining the appropriate features (see 
\new{Supplementary Table~\ref{tab:features-directed}} for features that explicitly account for directed events).
%Appendix~\ref{app:tables} 
%\note{CLV: move this appendix subsection to supplement?} \note{CLV: check if description there is clear when referenced from here.}). 
% and Section~\ref{sec:applications}).

\begin{figure}
  \begin{tikzpicture}
    \node at(0,0){ \includegraphics{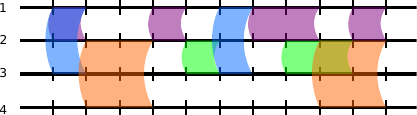} };
    \node at(-4,1){\bf (a) };
    %---------------------------------------    
    \node at(0,-2.8){ \includegraphics{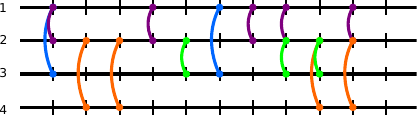} };
    \node at(-4,-1.8){\bf (b) };
  \end{tikzpicture}
  \caption{
{\bf Graphical ({\sl node-timeline}) representations of temporal network.}
  Each node is represented as a timeline and events as links between corresponding timelines at corresponding points in time. 
  (a) Temporal network with event durations  recorded at discrete-time resolution.
%   constructed by concatenating temporally adjacent instantaneous events for each pair of nodes.
  (b) Instant-event temporal network representation of the temporal network in (a).
}
  \label{fig:node-timeline}
\end{figure}

\begin{example} 
\label{ex:node-timeline}
Figure~\ref{fig:node-timeline}(a) shows a schematic temporal network consisting of four nodes. 
\end{example}

For some systems, e.g., email communications or instant messaging~\cite{Holme2005N,Karsai2011S,Takaguchi2013B,Karimi2013T,Backlund2014E,Posfai2014S,Delvenne2015D,Takaguchi2016C}, events are instantaneous; in other cases, event durations are so short compared to the time-intervals between them, the {\sl inter-event durations}, that they may be treated as instantaneous~\cite{Karsai2011S,Kivela2012M}. 
Both cases are included in the above framework by setting $\tau=0$.  
We may then reduce our representation of the sequence of events to a sequence of reduced events, leading us to the \new{following} definition.

\begin{definition}
{\sl Instant-event temporal network.}
\label{def:event_network}
An instant-event temporal network $G=(\N,\E)$  is defined by the set of nodes $\N=\{1,2,...,N\}$ and a set of \new{instantaneous} events $\E=\{e_1,e_2,\ldots,e_E\}$, where each event \new{$e_q=(i,j,t)$ describes an interaction between nodes $i$ and $j$ at time $t$}, but where the duration is implicit. 
\end{definition}

Some systems with continuous dyadic activity, notably face-to-face interaction and proximity networks~\cite{Fournet2014C,Stopczynski2014M}, are recorded with a coarse time resolution at evenly spaced points in time, $t=\tmin, \tmin+\Dt, \tmin+2\Dt, \ldots, \tmax-\Dt$.
In this case we may also represent the system as an {\sl instant-event network}, where the events mark a beginning of an activity at each measurement time and the time-resolution $\tau = \Dt$ is implicit. 
% Alternatively, and more commonly, the system is represented as a temporal network, where consecutive measurements of activity between the same pair of nodes are merged into a single event with the duration $\tau$ indicating the length of the event.
%Both representations will come in handy for defining different types of MRRMs. 

\new{
\begin{example} 
\label{ex:node-timeline-instant}
Figure~\ref{fig:node-timeline}(b) shows a schematic discrete-time instant-event temporal network corresponding to the temporal network of Example \ref{ex:node-timeline} (shown in Fig.~\ref{fig:node-timeline}(a)).
% consisting of four nodes (Def.~\ref{def:event_network}) and the temporal network (Def.~\ref{def:temporal_network}) built by concatenating adjacent instantaneous events. 
\end{example}
}

%\note{Added figure of ``node-timeline'' representation of the same network as in Figs.~\ref{fig:link-timeline} and \ref{fig:snapshot-sequence}? (i.e.\ the representation shown in Fig. 1(b) of Petter and Jari's ``Temporal networks'' review paper.) This follows one of Ref.\ 2's suggestions and provides ``non-two-level" visual representation of a temporal net.}

%------------------------------------------------------------------------------------------------------------------
\subsubsection{\new{Temporal network feature}}
%------------------------------------------------------------------------------------------------------------------
\noindent
\new{We are in general interested in comparing the values of given features of an observed temporal network (either recorded empirically or generated by a model) to their values in randomized networks generated by a MRRM that constrains certain other features. 
To formalize both the notion of a temporal network feature and that of a MRRM, we first need to define a {\sl state space} comprising all temporal networks.}

\new{\begin{definition}
{\sl State space $\G$.}
\label{def:state_space}
The {\sl state space} $\G$ is a predefined finite\footnote{Note that from a practical point of view it is enough for our purposes to consider only finite state spaces. Considering all possible temporal networks would make the state space at least countably infinite, but as all of the reference models encountered in the literature keep the system size (measured in the number of nodes \new{and the length of the observation period}) fixed, we can \new{always fix the} state space to contain only networks of fixed size. Furthermore, \new{the time can always} be considered finite as the %observation windows and 
measurement resolution is finite.}  set of temporal networks. 
\end{definition}}

\new{With the definition of the state space $\G$, representing the world of possible networks, we can now define any feature of a network as a function on $\G$.}

\begin{definition}
{\sl Temporal network feature.}
\label{def:feature}
A {\sl feature} $\x$ is any function that takes as an input any temporal network $G\in\G$. 
Formally, given a \new{state space $\G$ (Def.~\ref{def:state_space}), $\x$ is a function that has $\G$ as domain}. 
\end{definition}

Often a feature is a vector-valued function. 
However, the definition allows for more general functions, e.g. a function that returns a graph.
Furthermore, the feature does not need to be a structural feature of a network. 
It may for example quantify the outcome of a dynamic process on the network, such as the expected time needed for a contagion to infect a given proportion of the nodes.

An important temporal network feature is the {\sl static graph}, which summarizes the time-aggregated topology of a temporal network. 

\begin{definition}
{\sl Static graph.}
\label{def:static_graph}
The static graph, $\Gstat$, is a function which returns a simple (i.e.\ static, unweighted, and undirected) graph $\Gstat(G)=(\N,\L)$ with the same set of nodes $\N$ as the original temporal network $G$ and the set of {\sl links} $\L=\{ (i,j) : (i,j,t,\tau) \in \C \} $, which includes all pairs of nodes $(i,j)$ that interact at least once in $G$. 
\end{definition}

Note that by Def.~\ref{def:static_graph} the static graph is a feature of a temporal network. Conversely, we may see the temporal network as a direct generalization of the static graph to include information about the time-evolution of the system's topology~\cite{Holme2012T}. 
Note also that here we have defined the static graph as an unweighted graph, but one could also use the number of events between each pair of nodes or their cumulative duration to define link weights. % (see \new{Appendix} \ref{sec:features}).

%\note{CLV: add also a definition of weighted graphs? - a weighted graph could e.g.\ serve as a simpler example for the composition of MRRMs with a MRRM that keeps $p(\w)$ and randomizes the topology with one that keeps the topology and randomizes the weights.}

%-------------------------------------------------------------------------
\subsubsection{Two-level temporal network representations} 
\label{sec:nested}
%-------------------------------------------------------------------------
\noindent
Sometimes it is useful to separate the static structure and the temporal aspect in the definition of the temporal network as opposed to having them mixed together like in definitions \ref{def:temporal_network} and \ref{def:event_network}. 
This can be done by separating these two aspects into two levels, either by first defining the static network structure and then \new{the activation times of each link}
%how it changes in time, 
or by first defining the sequence of activation times and then the network structure at each of those times~\cite{Holme2015M}. 
We call the first of these options a \emph{link-timeline network} and the second a \emph{snapshot-graph sequence}. 
These two-level temporal network representations will be practical 
%for visualizing temporal networks, 
%defining important temporal network features  (Appendix~\ref{sec:features}) %, and 
for designing and implementing temporal network MRRMs (Sec.~\ref{sec:implementation}).

\begin{definition}
{\sl Link-timeline network.}
\label{def:link-timeline_graph}
\new{A link-timeline network represents a temporal network using an edge-valued graph  $\Gl = \left(\N,\L,\Thf\right)$. 
It uses} the static graph $\Gstat=(\N,\L)$ to indicate the pairs of nodes that interact at least once during the observation period (Def.~\ref{def:static_graph}).
To each link $\ij\in\L$ \new{it associates} a {\sl timeline} $\Thij\in\Thf$, which indicates when the corresponding nodes interact.
Each timeline is given by a sequence,
\beq
  \small \Thij = \left( \left(\ul^1,\taul^1\right), \left(\ul^2,\taul^2\right), \ldots, \left(\ul^\nl,\taul^\nl\right) \right) \es,
  \label{eq:link-timeline}
\eeq
where $\ul^m$ is the start of the $m$th event on link $\ij$, with $\taul^m$ its duration, and $\nl$ is the total number of events taking place over the link~\footnote{A link-timeline network is equivalent to the {\sl interval graph} defined in \cite{Holme2012T}. 
%it is furthermore equivalent to the link stream defined in~\cite{Latapy2017S} \note{CLV: is it the same, or is it Def.~\ref{def:temporal_network} that is?}, but where the set of possible event times was defined explicitly. 
If we set $\tau_m=0$, i.e.\ in the instantaneous event limit, we obtain what is termed {\sl contact sequences} in~\cite{Holme2012T}.}.
\end{definition}

\begin{figure}
  \begin{tikzpicture}
    \node at(-0.4,-0.2){\includegraphics[width=0.17\textwidth]{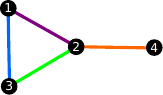}};
    \node at(-4.,1.){\bf(a)};
    \node at(0.2,-3.3){\includegraphics[width=0.4\textwidth]{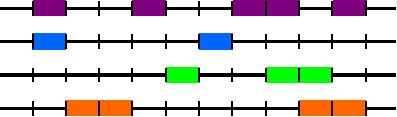}};
    \node at(-3.9,-2.4){ $\Theta_{(1,2)}$};
    \node at(-3.9,-3.025){ $\Theta_{(1,3)}$};
    \node at(-3.9,-3.65){ $\Theta_{(2,3)}$};
    \node at(-3.9,-4.275){ $\Theta_{(2,4)}$};
    \node at(-4.,-1.7){\bf(b)};
  \end{tikzpicture}
  \caption{{\bf Link-timeline network.} 
  Graphical representation of the link-timeline \new{representation $\Gl$ of the discrete-time temporal network shown in Fig.~\ref{fig:node-timeline}.}
  (a) Static graph \new{$\Gstat=(\N,\L)$} showing links between nodes. %, i.e.\ which nodes interact.   
  Links are drawn between pairs $(i,j)$ of nodes that interact at least once. 
  (b) Timelines $\Thf=\left( \Theta_{(1,2)}, \Theta_{(1,3)}, \Theta_{(2,3)}, \Theta_{(2,4)}  \right) $ of the links $\L$ in the graph, showing when each link is active.}
  \label{fig:link-timeline}
\end{figure}

\begin{example}
\label{ex:link-timeline}
%Figure~\ref{fig:link-timeline} shows an example link-timeline network of a temporal network consisting of four nodes and recorded at finite time-resolution.  
\new{Figure~\ref{fig:link-timeline} shows the link-timeline representation of the temporal network of Example~\ref{ex:node-timeline}.} % and Fig.~\ref{fig:node-timeline}}.  
%Panel (a) shows the static graph \new{$\Gstat=(\N,\L)$}, while panel (b) shows the link timelines $\Thf$. 
\end{example}

Alternative to the link timeline \new{representation} we may think of a temporal network as a time-varying sequence of instantaneous graph snapshots. This leads to the following definition:

\begin{definition}
\label{def:snapshot-sequence}
{\sl Snapshot-graph sequence.} 
A {\sl snapshot-graph sequence}, $\Gss=(\T,\Gs)$, represents a temporal network using a sequence of times, $\T=(t_1,t_2,\ldots,t_T)$, and a sequence of snapshot graphs,
\beq
  \Gs=(\Gamma^{1},\Gamma^{2},\ldots,\Gamma^{T}) \es,
  \label{eq:snapshot-sequence}
\eeq
where for each $m=1,2,\ldots,T$, $\Gamma^m\in\Gs$ is associated to $t_m\in\T$.
The {\sl snapshot graphs} are defined as graphs $\Gamma^m=(\N,\E^{t_m})$, where $\N$ is the set of nodes and $\E^t$ is the set of edges for which there is an event taking place at time $t$,
\beq
  \E^t= \left\{ (i,j) : (i,j,t)\in\E \right\} \es.
  \label{eq:snapshot}
\eeq
\end{definition}

Instantaneous-event networks can be represented as snapshot-graph sequences by constructing the sequence of times $\T$ as the times at which at least one event takes place. This is a natural representation especially for networks which are recorded with a fixed time resolution $\Dt$, as the sequence of times becomes $\T=(\Dt,2\Dt,\ldots,T)$, and if the time resolution is coarse enough so that the individual snapshot graphs do not become too sparse.
We use the shorthand $\Gt$ to refer to the snapshot graph 
associated with the time $t \in \T$ (i.e.\ $\Gamma^{m}$ for $m$ such that $t_m=t$).

\begin{figure}
  \begin{tikzpicture}
    \node at (0,0){\includegraphics[width=0.44\textwidth]{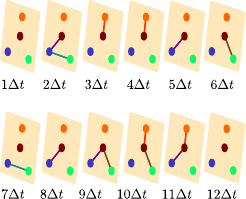}};
  \end{tikzpicture}
  \caption{{\bf Snapshot-graph sequence.} Sequence of snapshot graphs of the temporal network shown in \new{Figs.~\ref{fig:node-timeline} and \ref{fig:link-timeline}}.}
  \label{fig:snapshot-sequence}
\end{figure}

\begin{example}
\label{ex:snapshot-sequence}
Figure~\ref{fig:snapshot-sequence} shows the snapshot-graph sequence for the temporal network of Example~\ref{ex:node-timeline-instant}.  
\end{example}

Note that the two-level temporal networks do not add anything new to the temporal network structure. They are simply alternative ways of representing them because any temporal network can be uniquely represented as a link-timeline network and any instant-event temporal network can be uniquely represented as a snapshot-graph sequence. Despite this, the representations are often used for specific types of systems and they come with their own perspective on temporal networks. 

The link-timeline networks are often used for data that is sparse in time such that only a small fraction of links are active at each time instant. Furthermore, because the static network is made explicit in its definition it is easy to think \new{of the temporal network as having} a latent static network which manifests as activation events of the links. For example, for an email communication data represented as link-timeline network one might consider the static graph as an acquaintance or friendship network where each social tie is activated during the communication events. The structure also guides the generation of randomized reference models, because it is easy to either randomize the static graph and keep the link timelines, or randomize the timelines while keeping the static graph. 

The snapshot-graph sequences are more likely used for data that are dense in time such that each snapshot graph contains a reasonable number of links. Furthermore, this representation is natural for networks that change in time while the network nature of the system is still important at each separate time instance. For example, for networks in which the structure changes on the same or longer timescales as dynamics on that network, it is important to be able to look at the topology of the network at each time instance. Here, again, the structure guides the construction of randomized reference models: It is convenient to define shuffling methods where the order of the snapshot graphs change or where each snapshot is independently randomized. 

The two-level temporal network representations \new{thus} provide convenient ways to define and generate MRRMs that constrain certain overall properties. 
We will explore this in detail in \new{Subsection~\ref{sec:implementation} below} and see that many RRMs found in the literature are implemented this way in Section~\ref{sec:classification}.

%------------------------------------------------------------------------
\subsection{Microcanonical randomized reference model (MRRM)}
\label{sec:RRM}
%------------------------------------------------------------------------
\noindent
Here we give a rigorous definition of a microcanonical randomized reference model (MRRM). 
\new{We use this to propose an unambiguous {\sl canonical} naming convention for MRRMs.
We finally discuss several equivalent ways to represent a MRRM, each of which is useful for different purposes.}
%develop several related concepts and properties of MRRMs. We will use these to consistently describe and hierarchically rank MRRMs in Sections~\ref{sec:implementation} and \ref{sec:classification}.

%\subsubsection{Basic definitions}
We consider a predetermined state space $\G$ (Def.~\ref{def:state_space}), where states are here temporal networks, and  a single observation $G^*\in\G$, termed the {\sl input network}.

Many procedures leading to a sampling from a conditional probability distribution $P(G | G^*)$ defined on $\G$ could be considered to be {\sl randomized reference models} (RRMs). 
In order for such models to be useful for testing hypothesis and finding effect sizes they need to retain some of the properties of the original network $G^*$ and randomize others in a controlled way. 
In the context of \new{simple} graphs, the most popular choices of RRMs include Erd\H{o}s-R\'enyi (ER) models \cite{Erdos1960O,Newman2003S}, configuration models \cite{Fosdick2018C,Newman2003S}, and exponential random graph models \cite{Newman2003S}. 

Here we will focus on models that exactly preserve certain features but are otherwise maximally random.
\new{The principle of maximum entropy formalizes this notion, %of maximizing randomness under given constraints, 
and it provides theoretical justification for using such models as they are the least possible biased w.r.t.\ all other degrees of freedom~\cite{Presse2013P, Jaynes1957I}.
Maximum entropy models that impose exact (i.e.\ {\sl hard}) constraints are called
microcanonical models, and we will thus refer to RRMs that exactly constrain given feature values as {\sl microcanonical randomized reference models} (MRRMs). }

%In the context of static graphs the variant of the ER model that returns uniformly at random a graph with $N$ nodes and $L$ edges, and the variant of the configuration model that returns uniformly a randomly selected graph with degree sequence $\kstat=(k_i )_{i\in\N}$, also known as the {\sl Maslov-Sneppen} model~\cite{Maslov2002S}, are MRRMs.

%We will next formally define MRRMs and show that they have several attractive properties. We shall also see in Section~\ref{sec:classification} that most RRMs for temporal networks generated by shuffling an input network are of this type.

\begin{definition}
\label{def:MRRM}
{\sl Microcanonical randomized reference model (MRRM).}
Consider any function $\x$ that has the set $\G$ as domain  (i.e.\ a temporal network feature, Def.~\ref{def:feature}). A MRRM %, denoted by P[$\x$], 
is then a model which given $G^* \in \G$ returns $G \in \G$ with probability: 
\beq
  \Prob{\x}(G|G^*) = \frac{\delta_{\x(G),\x(G^*)}}{\Omega_{\x}(G^*)} \es,
  \label{eq:p(G|x)}
\eeq
where $\delta$ is the Kronecker delta function, 
and $\Omega_{\x}(G^*)=\sum_{G\in\G}\delta_{\x(G),\x(G^*)}$ is a normalization constant.
\end{definition}

We will sometimes use the shorthand notation $\x^*=\x(G^*)$, and $\Omega_{\x^*}=\Omega_{\x}(G^*)$. 
Furthermore, because the conditional probability depends only on the value of $\x$ in $G^*$ we can define the notation $\Prob{\x}(G|\x^*)=\Prob{\x}(G|G^*)$. 

In the above definition the feature function $\x$ defines the features of  $G^*$ that are retained in the randomized reference model.  
In statistical physics terms $\Omega_{\x^*}$ is the {\sl microcanonical partition function}. 
% and is equal to the {\sl multiplicity} of states. % \new{$G\in\G_{x^*}$}. 

Note that restricting ourselves to a single feature entails no loss in generality since any number of distinct features may be combined into one tuple-valued feature, e.g., for two distinct features $\x$ and $\y$, we may simply define a third tuple-valued feature $\z=(\x,\y)$.
%(see Def.~\ref{def:intersection} below). 
\new{This defines what we shall call the {\sl intersection} of two MRRMs:}

\begin{definition}
\label{def:intersection}
{\sl Intersection of randomized reference models.}
The \emph{intersection} of two features $\x$ and $\y$ is the tuple (\x,\y), and for the associated MRRM we write $\model{\x, \y} = \model{(\x, \y)}$
\end{definition}

%Note that the intersection by definition gives another MRRM.

%\new{
%\begin{example}
%In the context of static graphs the variant of the ER model that returns uniformly at random a graph with $N$ nodes and $L$ edges, and the variant of the configuration model that returns uniformly a randomly selected graph with degree sequence $\kstat=(k_i )_{i\in\N}$, also known as the {\sl Maslov-Sneppen} model~\cite{Maslov2002S}, are MRRMs.
%\end{example}
%}

%------------------------------------------------------------------------------------------------------------------
\subsubsection{Naming convention}
\label{sec:naming}
%------------------------------------------------------------------------------------------------------------------
\noindent
\new{A MRRM is completely defined by the feature(s) it constrains (Def.~\ref{def:MRRM}). 
This lets us propose a rigorous naming convention that specifies a MRRM by listing the corresponding features}. % it constrains}.

\begin{definition}
\label{def:naming-MRRMs}
{\sl Naming convention for MRRMs.} 
\new{A MRRM that constrains the individual temporal network features $\x_1,\x_2,\ldots,\x_Q$ is named P[$\x_1,\x_2,\ldots,\x_Q$].}
\end{definition}

Note that our naming convention is not unique \new{as the list of features is not required to be non-overlapping, so we may always} devise different ways to name the same MRRM (for a practical example see the description \new{of the MRRM P[$\w,\t$] in Section~\ref{sec:intersections-TS-Snaps}%, which constrains the static topology, link weights, and the number of events at each time
}). 
\new{It is however} unambiguous as a set of features uniquely defines a single MRRM (Def.~\ref{def:MRRM}). 
This means that a name always uniquely defines a MRRM.

\begin{example}
\new{In the context of simple graphs, the variant of the ER model~\cite{Erdos1960O} that returns uniformly at random a graph with $N$ nodes and $L$ edges, and the variant of the configuration model that returns uniformly a randomly selected simple graph with degree sequence $\kstat=(k_i )_{i\in\N}$, also known as the {\sl Maslov-Sneppen} model~\cite{Maslov2002S}, are MRRMs.}
%The Maslov-Sneppen model~\cite{Maslov2002S} is an example of a MRRM for static graphs. 
In the space of simple graphs with $N$ nodes, the ER model is defined as P[$L$].
It maps an input graph $G^*$ to a microcanonical ensemble of graphs that all have the same \new{number of links $L$ as the input graph $G^*$}, but are otherwise uniformly random.
\new{The Maslov-Sneppen model is defined as P[$\kstat$], 
%where $\k$ is the sequence of node degrees, 
and it maps $G^*$ to a microcanonical ensemble of graphs that all have the same sequence $\k$ of node degrees as $G^*$.}
\end{example}

\new{Note that MRRMs are always defined relative to a state space $\G$ (Def.~\ref{def:state_space}), which should also be specified. 
In the context of reference models for temporal networks $\G$ contains all networks with the same set of nodes $\N$ and the same temporal duration $\tmax-\tmin$ as the original (input) network. 
%in order to keep the randomized networks the same size as the original. 
%, and the number of instantaneous events $E$, we 
We do thus not need to include these features in the names of temporal network MRRMs as they are always constrained. } 
%, and we will exclude them in the following in order to avoid clutter.}

\begin{example}
\new{A popular temporal network MRRM is the model which randomizes the time stamps of the instantaneous events completely inside each timeline without changing the aggregated topology of the network, leading the events to follow a Poisson process on each timeline.
In the space of instant event temporal networks with a fixed set of nodes and observation interval it is named $\model{\w}$. 
Here $\w=(\wij)_{\ij\in\L}$ is the sequence of link weights, which retains the number of instantaneous events on each link and the links' placement in the static graph.
Several different names has been used in the literature to designate this MRRM: {\sl random time(s)}~\cite{Holme2005N,Holme2012T,Posfai2014S,Holme2015M}, {\sl uniformly random times}~\cite{Kivela2012M}, {\sl temporal mixed edges}~\cite{Bajardi2011D}, {\sl Poissonized inter-event intervals}~\cite{Takaguchi2012I}, and {\sl SRan}~\cite{Starnini2012R}.}
\end{example}

%-----------------------------------------------------------------
\subsubsection{MRRM representations}
\label{sec:representations}
%-----------------------------------------------------------------
\noindent
While the definition of MRRMs is written as a conditional probability it is often useful to use alternative representations of MRRMs.
\new{Namely, as a {\sl shuffling method} that uniformly samples randomized networks, as a partition of the state space, and as a transition matrix between states.
All of these representations are equivalent in a sense that they completely and uniquely specify a MRRM. 
%(in the case of 
%\new{conditional probabilities}
%shuffling methods 
%this is by definition). For example, each $\model{\x}$ defines exactly one partition and each partition defines exactly one $\model{\x}$.
%Note that two \new{different} feature functions $\x$ and $\y$ might correspond to the same partition, transition matrix, or shuffling method, but in this case the two functions also give the same conditional probabilities in Def.~\ref{def:MRRM} and we say that $\model{\x} = \model{\y}$. 
The power of the equivalence between the different %\new{theoretical} 
representations is that any result proven for one representation automatically carries over to the others.
We will in the following switch between the representations to use the one that is most convenient in each context\new{: 
the definition as a conditional probability notably provided a consistent naming convention that fully characterizes any MRRM (Def.~\ref{def:naming-MRRMs}), 
shuffling methods are how MRRMs are implemented in practice 
(we will explore this in Subsection~\ref{sec:implementation} below),
while the partition and matrix pictures will provide theoretical underpinnings for building hierarchies of MRRMs (Section~\ref{sec:hierarchies}) and for generating new MRRMs from existing ones (Section~\ref{sec:composition}).}
}
%\new{%We here present three:
%The first of these links the theoretical definition of a MRRM (Def.~\ref{def:MRRM}) to the way it is generated in practice, i.e.\ by sampling randomized networks, while the two others provide alternative theoretical descriptions which highlight the deep connections of MRRMs to set partitions and to linear algebra.}

\begin{definition} \label{def:mrrm_representations}
{\sl MRRM representations:}
 %\note{CLV: I changed the order in which the representations are presented.}
\begin{enumerate}
   \item {\sl Shuffling method.}
     An algorithm that transforms $G^*$ into $G$ according to Def.~\ref{def:MRRM}. These algorithms often shuffle some elements of $G^*$. Note that multiple algorithms or shuffling procedures might correspond to the same MRRM and in this case these are considered here to be the same shuffling method.
   \item {\sl Partition of the state space.} The feature function $\x$ (Def.~\ref{def:feature}) defines an equivalence relation and thus partitions the state space $\G$ (Def.~\ref{def:state_space})~\cite{Hrbacek1999I}: 
   Given $\x$, one can construct a partition of the state space $\{ \G_i \}$ (i.e., a set of subsets of $\G$ where each element of $\G$ is in exactly one subset) such that $G,G'\in\G_i$ if $\x(G) = \x(G')$.
   The set which $G^* \in \G$ belongs to in this partition is the {\sl $\x$-equivalence class} of $G^*$ and is denoted by $\G_{\x^*} = \{G\in\G : \x(G) = \x^*\}$. Note that the partition function which normalizes the conditional probability (Def.~\ref{def:MRRM}) is the cardinality of this set, $\Omega_{\x^*} = |\G_{\x^*}|$.
   \item {\sl Transition matrix.} 
     A MRRM is a symmetric linear stochastic operator mapping the state space $\G$ to itself.
     For a given indexing of the state space $\G$, we can represent a MRRM by a transition matrix $\P^\x$ with elements
     \beq
     \P^\x_{ij} = \Prob{\x}(G_j|G_i) \es.
     \label{eq:transition_matrix}
     \eeq
     $\P^\x$ is always a block diagonal matrix where inside each block the elements have the same value. 
 \end{enumerate}
\end{definition}

\begin{figure}
  \begin{tikzpicture}
    \node at(-1.6,3.7){ \includegraphics[width=0.33in]{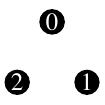} }; 
    \node at(-0.5,3.7){ \includegraphics[width=0.33in]{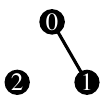} }; 
    \node at(0.6,3.7){ \includegraphics[width=0.33in]{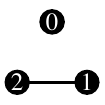} }; 
    \node at(1.7,3.7){ \includegraphics[width=0.33in]{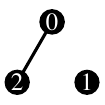} }; 
    \node at(2.8,3.7){ \includegraphics[width=0.33in]{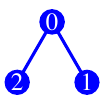} }; 
    \node at(3.9,3.7){ \includegraphics[width=0.33in]{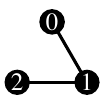} }; 
    \node at(5.0,3.7){ \includegraphics[width=0.33in]{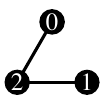} }; 
    \node at(6.1,3.7){ \includegraphics[width=0.33in]{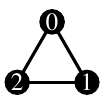} }; 
    \node at(-1.6,3.05){ $G_1$ };
    \node at(-0.5,3.05){ $G_2$ };
    \node at(0.6,3.05){ $G_3$ };
    \node at(1.7,3.05){ $G_4$ };
    \node at(2.8,3.05){ \color{blue}{$G_5$} };
    \node at(3.9,3.05){ $G_6$ };
    \node at(5.0,3.05){ $G_7$ };
    \node at(6.1,3.05){ $G_8$ };
    \node at(-2.2,2.45){ $L$ };
    \node at(-1.6,2.45){ 0 };
    \node at(-0.5,2.45){ 1 };
    \node at(0.6,2.45){ 1 };
    \node at(1.7,2.45){ 1 };
    \node at(2.8,2.45){ \color{blue}{2} };
    \node at(3.9,2.45){ 2 };
    \node at(5.0,2.45){ 2 };
    \node at(6.1,2.45){ 3 };
    \node at(-2.2,4.1){\bf (a)  };
    \draw[blue] (2.8,3.2) ellipse (0.6cm and 1.2cm);
    
    \node at(-0.2,-0.5){ \includegraphics[width=1.6in]{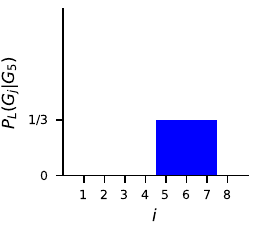} };
    \node at(-2.2,1.3){\bf (b)  };
    
    \node at(4.3,-0.2){
    \begin{minipage}{1.65in}\flushleft\footnotesize
    L = \text{number\_of\_links($G_5$)}\\
    while $|E| < L$: \\
    \hspace{5mm}  $i$ = rand($0$,$N$)\\
    \hspace{5mm}  $j$ = rand($0$,$N$)\\
    \hspace{5mm}  if $(i,j)$ and $(j,i)$ $\notin$  $E$:\\
    \hspace{10mm}     $E \leftarrow (i,j)$
    \end{minipage}
    };
     \node at(2.4,1.3){\bf (c)  };

    \node at(0.,-4.2){ \includegraphics[width=1.65in]{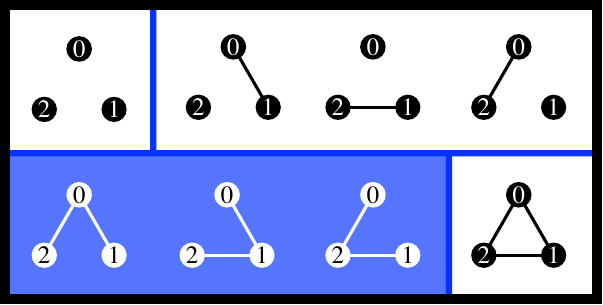} };
    \node at(-2.2,-2.6){\bf (d)  };

 	\matrix at(4.5,-4) [matrix of math nodes,left delimiter=(,right delimiter=)]
 	{
    	\hspace{-0.3cm} \includegraphics[width=1.2in]{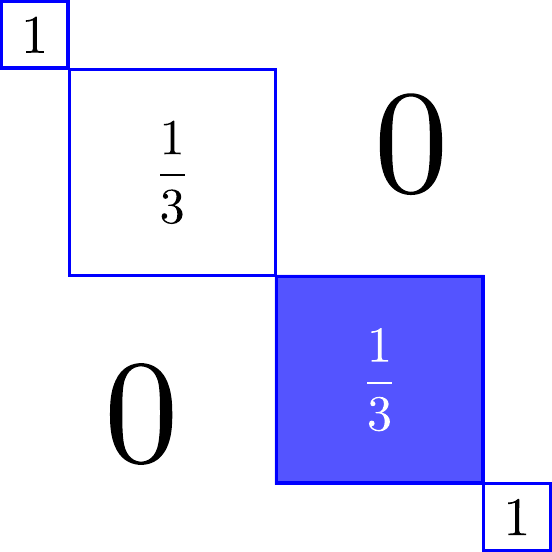}  \hspace{-0.3cm}\\
    };   
    \node at(2.4,-2.6){\bf (e)  };
  \end{tikzpicture}
  \caption{{
  {\bf Equivalent MRRM representations.} 
  We consider the state space of all simple graphs with three nodes and the MRRM P[$L$] which constrains the number of edges $L$ in a graph. The state space $\G$ contains eight states (graphs) and
  is ordered as shown in (a). The input network is taken to be $G^*=G_5$ (highlighted in blue) \new{for which $L^*=L(G_5)=2$}. 
Panels (b)-(e) illustrate graphically the four equivalent representations 
%(Defs.~\ref{def:MRRM} and \ref{def:mrrm_representations}) 
of P[$L$] when applied to $G_5$:
(b) The conditional probability $P_L(G_j|G_5)$ gives the probability to generate each state $G_j$ using P[$L$] with $G_5$ as the input state. 
(c) A  shuffling method corresponding to P[$L$] samples graphs $G_j$ from the \new{subset of networks with $L(G_5)=2$ links}, $\G_{L(G_5)}\subset\G$ according to $P_L(G_j|G_5)$.
(d) The partition of $\G$ induced by P[$L$] consists of the four distinct sets: $\G_0 = \{G_1\}$, $\G_1 = \{G_2, G_3, G_4\}$, $\G_2 = \{G_5, G_6, G_7\}$ (marked in blue), and $\G_3=\{G_8\}$.  
(e) The block-diagonal stochastic matrix $\P^L$ gives the probability to generate any output state $G_j$ from any given input state $G_i$ (the block corresponding to the input state $G^*=G_5$ is marked in blue). }
}
  \label{fig:MRRM_representations}
\end{figure}

{
\begin{example}
\label{ex:MRRM_representations}
To illustrate the different MRRM representations, we consider the state space $\G$ %=\{G_1,G_2,G_3,G_4,G_5,G_6,G_7,G_8\}$ 
of all static graphs with 3 nodes and the MRRM P[$L$], defined by the feature $L$ which returns the number of edges in the network, corresponding to the Erd\H{o}s-R\'enyi random graph model ${\rm ER}(3,L)$.
We number the 8 graphs in $\G$ %such that $G_1$ is the graph with 0 links, $G_2$ , $G_3$ , and $G_4$ are the graphs with 1 link, $G_5$ , $G_6$ , and $G_7$ are the graphs with 2 links and $G_8$ is the graph with 3 links 
such that $L(G_1)=0$, $L(G_i)=1$ for $i=2,3,4$, $L(G_i)=2$ for $i=5,6,7$, and $L(G_8)=3$ [Fig.~\ref{fig:MRRM_representations}(a)], and we take as input state the graph $G^*=G_5$.
Figure~\ref{fig:MRRM_representations} illustrates the four different representations of %the MRRM 
P[$L$] %introduced above 
for the state space $\G$ and the input graph $G^*=G_5$.
Note that in a real application the number of states is typically much much larger than in this example. 
This in particular means that almost all states are sampled at most once in practice.
\end{example}
}

% All of these representations are equivalent in a sense that they completely and uniquely specify a MRRM (in the case of 
% \new{conditional probabilities}
% %shuffling methods 
% this is by definition). For example, each $\model{\x}$ defines exactly one partition and each partition defines exactly one $\model{\x}$.
% Note that two \new{different} feature functions $\x$ and $\y$ might correspond to the same partition, transition matrix, or shuffling method, but in this case the two functions also give the same conditional probabilities in Def.~\ref{def:MRRM} and we say that $\model{\x} = \model{\y}$. 
% The power of the equivalence between the different %\new{theoretical} 
% representations is that any result proven for one representation automatically carries over to the others.
% We will in the following text switch between the representations to use the one that is most convenient in each context\new{: 
% the definition as a conditional probability notably provided a simple naming convention that fully characterizes any MRRM (Def.~\ref{def:naming-MRRMs}), 
% shuffling methods are how MRRMs are implemented in practice (we will explore this in the following subsection),
% while the partition and matrix pictures will provide theoretical underpinnings for building hierarchies of MRRMs and for generating new MRRMs from existing ones (Sections~\ref{sec:hierarchies} and \ref{sec:composition}).}

%------------------------------------------------------------------------
\subsection{Shuffling methods and classes of temporal network MRRMs} 
\label{sec:implementation}
%------------------------------------------------------------------------
% \note{CLV: Make this section less formal, and move formal definitions to Appendix.}

\noindent
\new{We describe here} several important classes of shuffling methods \new{that} are used to formulate and generate MRRMs in practice.
The classes are formulated depending on which parts of a temporal network they randomize. 

\new{All the MRRMs we will encounter are implemented by methods that
shuffle the positions of the events in a temporal network, or for instant-event networks, the instantaneous events. 
We shall call the former type of shuffling method an {\sl event shuffling} and the latter an {\sl instant-event shuffling}.
The shufflings are generally implemented by randomizing any or all of the indices $i$, $j$, and $t$ in the events $(i,j,t,\tau)\in\C$ or in the instantaneous events $(i,j,t)\in\E$.}

These methods are practical for generating reference models as they all conserve the nodes $\V$, the temporal duration $\tmax-\tmin$ and the number of events, $C$ (or instantaneous events, $E$). 
%for temporal networks with or without event durations). 
In addition to these features, event shufflings also conserve the events' durations, i.e.\ the multiset $p(\tauf)=[\tau]_{(i,j,t,\tau)\in\C}$, which contains the durations of all events in $\C$ including duplicate values. 
%, as it is only their positions that are randomized.
Different shuffling methods additionally constrain other network features, but they all conserve at least the above features.

\begin{example}
The most random event shuffling possible, $\model{p(\tauf)}$, is the one that conserves only the events' durations and otherwise redistributes them completely at random, i.e.\ it draws the triplets $(i,j,t)$ at random without replacement.
\end{example}

\begin{example}
The most random instant-event shuffling is $\model{E}$. 
It draws the triplets $(i,j,t)$ at random without replacement and conserves only the number of instantaneous events $E$.
\end{example}

We furthermore define several more restricted classes of shuffling methods that randomize specific temporal or topological aspects of a network using the two level representations introduced in Section~\ref{sec:nested} above.

%-----------------------------------------------------------------------
\subsubsection{Link and timeline shufflings}
%------------------------------------------------------------------------
\noindent
Based on the link-timeline representation (Def.~\ref{def:link-timeline_graph}), we define the following two classes of shuffling methods.
%{\sl link shufflings}, which randomize the static graph of a network but not the individual timelines, and {\sl timeline shufflings}, which randomize the timelines but not the static topology.

{\sl Link shufflings} conserve the content of the timelines, i.e.\ the multiset $\pxl(\Thf)=[\Thij]_{\ij\in\L}$, but randomizes their placement.
%as well as any constraints on the static topology (i.e.\ on the configuration $\L$ of links in $\Gstat$). 
In practice, they are implemented by randomizing the links $\L$ in the static graph, using any shuffling method for static graphs, and redistributing the timelines $\Thij\in\Thf$ on the new links without replacement. 
Note that link shufflings do not necessarily randomize the static topology of the network completely since the static graph shuffling may constrain any feature of $\Gstat$, e.g.\ the nodes' degrees $\k$.

\begin{figure}
  \begin{tikzpicture}
    \node at(-0.4,0){ \includegraphics[width=0.28\textwidth]{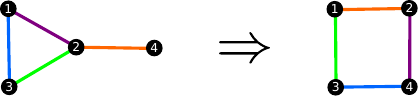} };
    \node at(-4.,0.8){ (a) };
    \node at(0,-2){ \includegraphics[width=0.45\textwidth]{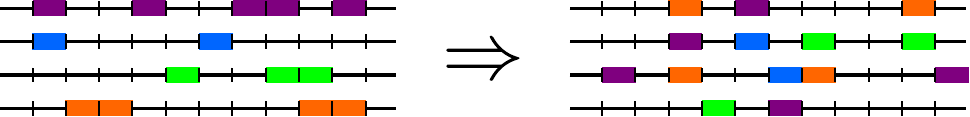}  };
    \node at(-4.,-1.1){ (b) };
  \end{tikzpicture}
  \caption{
  %\note{CLV: change the illustrations to match Figure~\ref{fig:link-timeline}.}
  \new{{\bf Illustration of the most random link and timeline shufflings.}
  (a) The most random link shuffling, P[$\pxl(\Thf)$], completely randomizes the static graph but conserves the content of the individual timelines.
  (b) The most random timeline shuffling, P[$\L,E$], redistributes the instantaneous events between all timelines at random while conserving the static topology.}}
  \label{fig:ex-link-timeline_shuffling}
\end{figure}

\begin{example}
Using the Erd\H{o}s-R\'enyi  model for randomizing the static graph $\Gstat$ leads to the most random link shuffling possible, P[$\pxl(\Thf)$] [Fig.~\ref{fig:ex-link-timeline_shuffling}(a)], while randomizing the $\Gstat$  using the configuration model leads to the more constrained link shuffling P[$\k$,$\pxl(\Thf)$], which constrains the nodes' degrees in $\Gstat$.
\end{example}

{\sl Timeline shufflings}, on the other hand, constrain the network's static topology, $\Gstat=(\N,\L)$ and randomizes the content of the timelines $\Thij\in\Thf$.
% , as well as any constraints on the content of the individual timelines $\Thij\in\Thf$. 
In practice they are implemented by redistributing the (instantaneous) events in or between the timelines.
Similarly to link shufflings, the timelines are not necessarily completely randomized as timeline shufflings may additionally constrain any feature of $\Thf$.

\begin{example}
The most random timeline shuffling, P[$\L,E$], is obtained by redistributing the instantaneous events in an instant-event network at random between the timelines [Fig.~\ref{fig:ex-link-timeline_shuffling}(b)].
\end{example}

%-----------------------------------------------------------------------
\subsubsection{Sequence and snapshot shufflings}
%-----------------------------------------------------------------------
\noindent
Based on the snapshot-sequence representation (Def.~\ref{def:snapshot-sequence}), 
we define the following two classes of shuffling methods.
%{\sl sequence shufflings}, which randomize the order of the snapshots but not the individual snapshot graphs, and {\sl  snapshot shufflings}, which randomize individual snapshot graphs but not their order.

\new{Sequence shufflings constrain the content of instantaneous snapshot graphs, i.e. the multiset $\pxt(\Gf)=[\Gt]_{t\in\T}$, and randomize the order of the snapshots.
They are implemented by shuffling the order of the snapshots.
}

\begin{figure}
  \begin{tikzpicture}
    \node at(0,0){ \includegraphics[height=0.055\textheight]{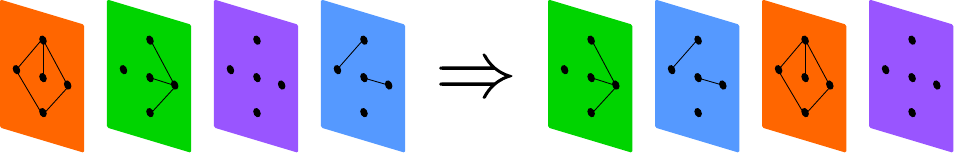} };
    \node at(-4.,0.8){ (a) };
    \node at(0,-2){ \includegraphics[height=0.055\textheight]{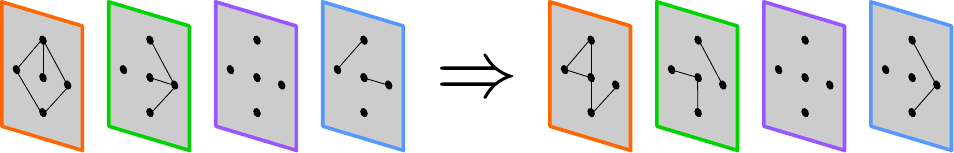}  }; 
    \node at(-4.,-1.1){ (b) };
  \end{tikzpicture}
  \caption{\new{{\bf Illustration of the most random sequence and snapshot shufflings.}
  (a) The most random sequence shuffling, P[$\pxt(\Gs)$], conserves all the individual snapshot graphs but randomizes their temporal order.
  (b) The most random sequence shuffling, P[$\t$], completely randomizes each snapshot graph while conserving the temporal ordering of the snapshots.}}
  \label{fig:ex-sequence-snapshot_shuffling}
\end{figure}

\begin{example}
Shuffling the temporal order of the individual snapshots completely at random leads to the most random sequence shuffling, P[$\pxt(\Gs)$] [Fig.~\ref{fig:ex-sequence-snapshot_shuffling}(a)]. \end{example}

Snapshot shufflings constrain the time of each event, i.e.\ $\t=(t)_{(i,j,t)\in\E}$ and randomzes the individual snapshot graphs $\Gt\in\Gf$.
They are typically implemented by randomizing the snapshot graphs individually and independently using any shuffling method for static graphs.

\begin{example}
Using the ER model to randomize each individual snapshot graph leads to the snapshot shuffling P[$\t$], which is the most random snapshot shuffling [Fig.~\ref{fig:ex-sequence-snapshot_shuffling}(b)]. 
\end{example}

%-----------------------------------------------------------------------
\subsubsection{Intersections of shuffling methods}
%-----------------------------------------------------------------------
\noindent
\new{As we shall see in the following, several MRRMs exist which constrain both the content of individual timelines, i.e.\ $\pxl(\Thf)$, and the static topology, i.e.\ $\Gstat = (\N, \L)$. This makes them intersections (Def.~\ref{def:intersection}) of link and timeline shufflings.
They are typically implemented in a manner similar to link shufflings by redistributing the timelines between the links in $\Gstat$, but without changing $\Gstat$.
}

\begin{figure}
  \begin{tikzpicture}
    \node at(0,0){ \includegraphics[height=0.05\textheight]{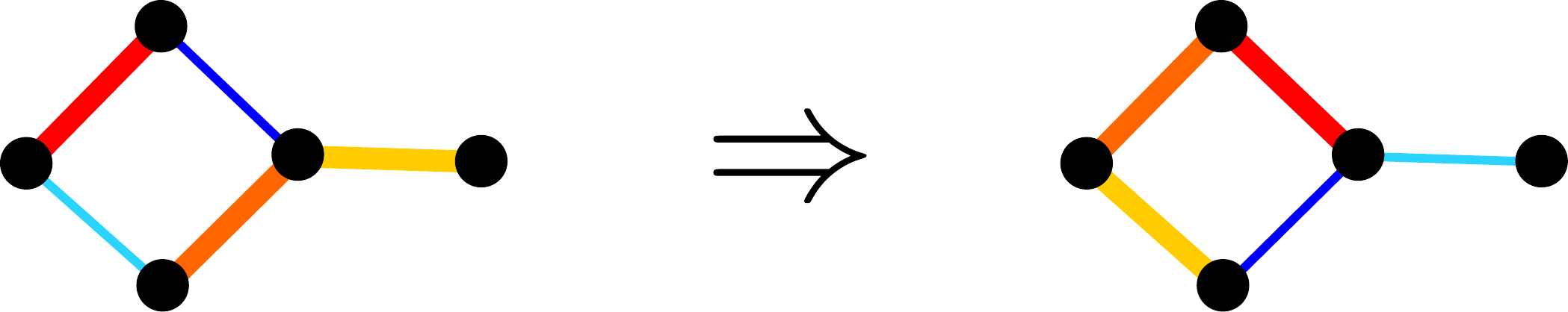} };
    \node at(-4.,0.8){ (a) };
    \node at(0,-2){ \includegraphics[height=0.045\textheight]{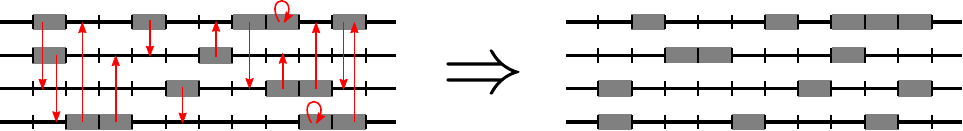}  }; 
    \node at(-4.,-1.1){ (b) };
  \end{tikzpicture}
  \caption{\new{{\bf Illustration of intersections between shuffling methods.}
  (a) The most random link-timeline intersection, $\model{\L,\pxt(\Gs)}$, constrains the static topology redistributes the individual timelines on the links at random.
  (b) The most random timeline-snapshot intersection, $\model{\L,\t}$, conserves the timestamp of each instantaneous event and redistributes them at random between the existing links.}}
  \label{fig:ex-intersections}
\end{figure}

\new{
\begin{example}
The intersection between the most random link shuffling, $\model{\pxl(\Thf)}$ and the most random timeline shuffling, $\model{\L,E}$, defines the most random link-timeline intersection: $\model{\L,\pxl(\Thf)}$ [Fig.~\ref{fig:ex-intersections}(a)].
This model constrains both the static topology and all temporal correlations on individual links, but destroys correlations between network topology and dynamics.
\end{example}
}

\new{Other MRRMs constraint both the static topology, i.e.\ $\L$, and the timestamps of the events, i.e.\ $\t$. 
These are thus intersections of timeline and snapshot shufflings.
They are typically implemented by exchanging the timestamps of the events inside each timeline, or alternatively by redistributing events between existing links while keeping their timestamps unchanged.}

\new{
\begin{example}
The intersection between the most random timeline shuffling, $\model{\L,E}$, and the most random snapshot shuffling, $\model{\t}$,
defines the most random timeline-snapshot intersection: $\model{\L,\t}$ [Fig.~\ref{fig:ex-intersections}(b)].
\end{example}
}

%-----------------------------------------------------------------------
\subsubsection{Compositions of shuffling methods}
%-----------------------------------------------------------------------
\noindent
\new{The final classes of shuffling methods that we will encounter are methods that generate randomized networks by applying a pair of different shuffling methods in {\sl composition}, i.e.\ by applying the second shuffling to the randomized networks generated by the first.}

\new{Not all compositions generate a microcanonical RRM however. 
They are e.g.\ not guaranteed to sample the randomized networks uniformly.
But as we will show in Section~\ref{sec:composition}, compositions between link shufflings and timeline shufflings and between sequence shufflings and snapshot shufflings always result in a MRRM.
Several such compositions have been used in the literature to produce MRRMs that randomize both topological and temporal aspects of a network at the same time (we describe and characterize them in Section~\ref{sec:compositions}).}

\new{
\begin{example}
The composition of the link shuffling $\model{\pxl(\Thf)}$ with the timeline shuffling $\model{\L,E}$ results in the MRRM $\model{L,E}$ which randomizes both the static topology and the temporal order of events while conserving the number of links $L=|\L|$ in the static graph.
\end{example}

\begin{example}
The composition of the sequence shuffling $\model{\pxt(\Gs)}$ with the snapshot shuffling $\model{\t}$ results in the MRRM $\model{p(\Ebf)}$ which randomizes both the topology of snapshots and their temporal order while conserving the multiset of the number of events in each snapshot, $p(\Ebf) = [|\E^t|]_{t\in\T}$. 
\end{example}
}

% %========================================================================
% \section{Microcanonical randomized reference models as null models}
% \label{sec:null_model}
% %========================================================================
% %\note{Add this short section on using MRRMs as null models, e.g.\ for hypothesis testing?}
% %\note{This section would basically contain the first part of Section~\ref{sec:testing}, before the two walkthrough examples. With little tweaking it could present first the procedure for null-model based testing.
% %It could then move on to state that using a single null model is appropriate when the choice model can be clearly justified based on theoretical grounds. 
% %However, for temporal networks we are most often interested in comparing to a series of reference models, either because we do not know how to choose our null model, or simply because we are interested in exploratory analysis of a network's features and their effects.}
% %\note{Section~\ref{sec:testing} was renamed to ``Analyzing temporal networks using series of microcanonical randomized reference models'' and its intro shortened accordingly.}
% \input{Sec-Null_model}

%========================================================================
\section{Survey of applications of randomized reference models}
\label{sec:applications}
%========================================================================
\noindent
The applications of MRRMs for temporal networks are manifold, but all follow two main directions: (i) studying how the network and ongoing dynamical processes are controlled by the effects of temporal and structural correlations that characterize empirical temporal networks, (ii) highlighting statistically significant features in temporal networks.

(i) Dynamical processes have been studied by using data-driven models, where temporal networks are obtained from real data, while the ongoing dynamical process is modeled by using any conventional process definition \cite{Holme2012T,Pastor-Satorras2015E} and typically simulated numerically on the empirical and randomized temporal networks~\cite{Pastor-Satorras2015E,Vestergaard2015T}. 
One common assumption in all these models is that information can flow between interacting entities only during their interactions. This way the direction, temporal, and structural position, duration, and the order of interactions become utmost important from the point of view of the dynamical process.
MRRMs provide a way to systematically eliminate the effects of these features and to study their influence on the ongoing dynamical process.
This methodology has recently shown to be successful in indicating the importance of temporality, bursty dynamics, community structure, weight-topology correlations, and higher-order temporal correlations on the evolution of dynamical processes, just to mention a few examples.

(ii) MRRMs have commonly been used as null models to find statistically significant features in temporal networks (often termed interaction motifs) or correlations between network dynamics and node attributes. 
This approach is conceptually the same as using the configuration model to detect overrepresented subgraphs (termed {\sl motifs}) in static networks \cite{Alon2007N,ShenOrr2002N,Milo2002N}. The difference here is that the studied networks vary in time, which induces further challenges\new{, and in particular drastically increases the number of possible null models}.

\new{We here review the main research directions and a selection of main results obtained using MRRMs to study temporal networks.
We use the naming convention developed in the preceding section to provide consistent names for the shuffling methods applied in the different studies, and we classify them according to which aspects of a temporal network they randomize.}

\new{As we shall see in this section, several studies apply a single MRRM as null model analogous to standard hypothesis testing.
However, in many cases %several reference models are applied to study the same network. This may either be because 
we may not know how to choose the right null model, in which case it is problematic to choose an arbitrary model since results may crucially depend on this choice~\cite{Beber2012A, Orsini2015Q, Fosdick2018C}.
%The results of null model testing crucially depend on the choice of null model~\cite{Beber2012A, Orsini2015Q, Fosdick2018C}.
%Using a single null model is thus appropriate when the choice of the model can be clearly justified based on theoretical grounds.
%In many cases however we may not know how to choose the right null model~\cite{Orsini2015Q}.
In other cases we are not interested in performing null hypothesis testing at all, but rather in investigating how a range of different features of a network affect each other or how they affect a given dynamical phenomenon.
Instead of basing our analysis on a single model, we want in these cases to apply and compare a series of related MRRMs to understand how the various features and their combinations change the results. 
Sections \ref{sec:hierarchies} and \ref{sec:composition} develop the theoretical machinery needed to compare and order network features and MRRMs, and they provide a taxonomy of the MRRMs found in the literature which fully describes them, orders them, and characterize their effects on temporal network features.
We refer to this taxonomy for detailed descriptions of each MRRM encountered in this section.}

In the first three subsections of this section (Subsections~\ref{sec:contagion}--\ref{sec:evol_games}), we review studies applying MRRMs to study various dynamical processes in empirical temporal networks. 
In the fourth subsection (Subsection~\ref{sec:motifs}) applications to inferring statistically significant motifs and correlations in network dynamics will be discussed.
Finally, in the last subsection (Subsection~\ref{sec:controllability}) we discuss a pair of recent papers that have applied MRRMs to study temporal network controllability.
We will in the following include reference models that are not MRRMs 
(such models are briefly discussed in Section~\ref{sec:other_models}). 
%While they do not come with the same formal statistical guarantees as maximum entropy RRMs do, they have nevertheless been useful in identifying important features in temporal networks.

%------------------------------------------------------------------------------------------------------------------
\subsection{Contagion processes}
\label{sec:contagion}
%------------------------------------------------------------------------------------------------------------------
\noindent
Contagion phenomena is the family of dynamical processes that has been studied the most using MRRMs. Since epidemics, information, or influence are all transmitted by person-to-person interactions to a large extent, the approximation provided by contact-data-driven simulations are indeed closer to reality than other conventional methods based solely on analytical models. MRRMs became important in this case to help understand which temporal or structural features of real temporal networks control the speed, size, or the critical threshold of the outbreak of any kind of contagion process. In the following we will address various types of contagion dynamics ranging from simple to complex spreading processes, focusing on findings that are due to MRRMs. For detailed definitions and discussion of the different contagion processes we refer readers to the recent review by Pastor-Satorras et al.~\cite{Pastor-Satorras2015E}.

\subsubsection{SI process}
\label{sec:SI}
\noindent
The susceptible-infected (SI) process is the simplest possible contagion model. Here nodes can be in two mutually exclusive states: susceptible (S) or infectious (I). Susceptible nodes (initially everyone except an initial seed node) become infected with rate $\beta$ when in contact with an infected node. %This process does not display a phase transition since each node (belonging to the seeded component of the network) becomes infected in the end (almost surely), i.e.\ the fraction of infected nodes $\langle I(t) \rangle /N$ tends to $1$. 
The single parameter $\beta$ controls the speed of saturation, thus by considering the limit $\beta\to\infty$ one can simulate the fastest possible contagion dynamics on a given network. In this case the infection times correspond to the temporal distances between the seed and the nodes that get infected. This can be seen as a ``light-cone" defining the horizon of propagation in the temporal network \cite{Holme2005N}.

Early motivation to use RRMs of temporal networks was to understand why models of information diffusion unfold extremely slowly in various communication networks even when modeled by the fastest possible spreading model, i.e.\ an SI process with $\beta\to\infty$ \cite{Karsai2011S}. %In the study two mobile communication networks and an email network were considered to study this phenomenon\footnote{Since the available datasets were not long enough to simulate full penetration of the dynamical process, periodic temporal boundary conditions were applied letting the process to continue from the beginning of the event sequence once it reached the last event. Note that this condition introduces some biases, by generating non-existing time-respecting paths \cite{Pan2011P}, but with a minor effect which does not qualitatively change the results.}.
The study introduced four MRRMs in order to quantify the contributions of topology and various temporal features of communication data to the spreading speed: %and measured the average fraction of infected nodes to study the early and late time behavior of the spreading process. 
%As compared to the diffusion on the original sequence %(which takes about $700$ days for full penetration) 
%the fastest null model was the 
(1) a model termed {\sl configuration model} (corresponding to the composition of the link shuffling P[$\kstat$,$\lc$,$\pxl(\Thf)$] and the timeline-snapshot intersection P[$\w$,$\t$]---Sec.~\ref{sec:compositions}), removing all structural and temporal correlations while keeping only the empirical heterogeneities in the node degrees, $\k$, in the distribution of link weights, $p(\w)$, and in the cumulative activity over time, $\Ebf$.
%At the same time, the largest contribution to the overall acceleration effect appeared once applying 
(2) a model termed {\sl time shuffled} (the timeline-snapshot intersection P[$\w$,$\t$]---Sec.~\ref{sec:intersections-TS-Snaps}), (3) the {\sl link-sequence shuffled} model P[$\L$,$\pxl(\Thf)$]---Sec.~\ref{sec:intersections-LS-TS}), and  (4) the {\sl equal-weight link-sequence shuffled} model (the link-timeline intersection P[$\w$,$\pxl(\Thf)$]---Sec.~\ref{sec:intersections-LS-TS}), which eliminates all causal correlations between events taking place on adjacent links but conserves the weighted network structure and temporal correlations in individual timelines. The conclusions was that shuffling more in general makes the spreading faster, and 
%This leads to the conclusions that
that he bursty interaction dynamics and the Granovetterian weight-topology correlations \cite{Granovetter1973S} are dominantly responsible for the slow spread of information in these systems. 
%On the other hand the \new{{\sl equal-weight link-sequence shuffled} model (the link-timeline intersection P[$\w$,$\pxl(\Thf)$]---}Sec.~\ref{sec:intersections-LS-TS}), which eliminates all causal correlations between events taking place on adjacent links but conserves the weighted network structure and temporal correlations in individual timelines, was observed to slightly slow down the process during the early phase, while accelerating it in the long run. 
%As an alternative possible explanation, 

Effects of circadian fluctuations were studied in Ref.~\cite{Karsai2011S} via two canonical RRMs (Sec.~\ref{sec:generative}), where interaction times were generated by either a homogeneous or an inhomogeneous Poisson process. %(with the rate of creation of new events set equal to either the mean (time-averaged) or the instantaneous rate of event creation, respectively).
The first model thus conserved the average link weights, 
%$\langle\wij\rangle$ 
while the second additionally conserved %and the average link weights as well as t
tyhe average activity at each point in time. 
% $(\langle\wij\rangle,\langle\Et\rangle)$, 
%respectively. 
%, i.e. they kept the original static structure unchanged, but induced synthetic interaction sequences.
%These generative models demonstrated that although circadian fluctuations may cause short term fluctuations in the overall speed of the spreading process, on the longer temporal scale they have negligible effects on the spreading dynamics.
The effect of circadian fluctuations could also be studied with MRRMs, as was done in a follow-up study by Kivel\"a et al. \cite{Kivela2012M}, who in addition applied 
%Three other null models were introduced by Kivel\"a et al. \cite{Kivela2012M} to study the same process on the same empirical networks. They applied \new{
a model termed {\sl uniformly random times} (the timeline shuffling P[$\w$]---Sec.~\ref{sec:timeline-shufflings}) to randomize all temporal correlations, including the circadian patterns, while conserving the aggregated structure.
In order to clarify the role of network topology, they also introduced  new models termed {\sl configuration model} and {\sl random network} (the link shufflings P[$\k,\lc,\pxl(\Thf)$] and P[$\lc,\pxl(\Thf)$], respectively---Sec.~\ref{sec:link-shufflings}) to randomize the static network topology while conserving all temporal correlations in individual timelines. 
%They concluded that, while temporal correlations have much stronger effects on the dynamics, the heterogeneous degree distribution of the underlying social structure initially accelerates the spreading while slowing it down on the long run. Additionally, their main conclusion was that the slow diffusion can be partially explained by the timings of individual call sequences. The spreading is strongly constrained by the frequency of interactions, i.e.\ the high variance in the inter-event durations causes the residual waiting (relay times) times to be long, and makes the spreading slower than the Poissonian case.

Another study by Gauvin et al.~\cite{Gauvin2013A} analyzed face-to-face interaction networks and employed MRRMs to identify the effective dynamical features, responsible for driving the diffusion of epidemics in local settings like schools, hospitals, or scientific conferences. To understand the dominant temporal factors driving the epidemics in these cases, they took both a bottom-up approach by using generative network models, and a top-down approach by employing two shuffling methods and a bootstrap method. They shuffled event and inter-event durations on individual links using \new{the model termed {\sl interval shuffling} (the timeline shuffling P[$\piij(\tauf$),$\piij(\Dtauf$)]---}Sec.~\ref{sec:timeline-shufflings}), they shuffled the timelines between existing links using \new{the {\sl link shuffling} model (the link-timeline intersection shuffling P[$\L$,$\pxl(\Thf)$]---}Sec.~\ref{sec:intersections-LS-TS}), and they finally bootstrapped the global distribution of event durations $p(\tauf)$ while keeping the number of events $\n$ on each link fixed (Sec.~\ref{sec:bootstrapping}). 
%In this study, time was not taken as a global measure but interpreted to be node specific. Each node was assigned with an {\sl activity clock} measuring the time that a node spent in interaction with others. This way, for an SI process, which was initiated from a seed node $i$ at time $t$ and which reached node $j$ at time $t_j$, the arrival time of node $j$ was not defined as $t_j-t$ but the cumulative duration of all of $j$'s events during the period from $t$ to $t_j$. Simulating the SI process this way, they measured the distribution of arrival {\sl activity} times, defined as the time it takes for the infection, starting from a random seed node, to reach a node in the network  (measured using the nodes activity clock). To compare different null models they calculated Kullback-Leibler divergences between the corresponding arrival-time distributions. From these measurements they concluded that the bursty nature of interaction dynamics has the strongest effect on the speed of spreading, while the heterogeneity in the number of events per link $\n$ and the synchronized contact patterns (typical in a school during breaks) also have a strong effect on the contagion dynamics.

Perotti et al. \cite{Perotti2014T} studied the effect of temporal sparsity, an entropy-based measure quantifying temporal heterogeneities on the empirical scale of average inter-event durations $\langle \Dtau_\ij^m \rangle$. As a reference model the authors used \new{the timeline shuffling} P[$\w$] (Sec.~\ref{sec:timeline-shufflings}). 
They showed via the numerical analysis of several temporal datasets and using analytical calculations that there is a linear correspondence between the temporal sparsity of a temporal network and the slowing down of a simulated SI process.

A unique temporal interaction dataset was studied by Rocha et al.~\cite{Rocha2011S}, which recorded the interaction events of sex sellers and buyers in Brazil.
The system is a temporal bipartite network where connections only exist between sellers and buyers. Using this dataset the authors studied, among other questions, the effects of temporal and structural correlations on simulated SI (and SIR) processes. They introduced three different MRRMs imposing a bipartite network structure.
%(obtained using the metadata-dependent blockmodel MRRMs discussed in Sec.~\ref{sec:metadata} with two groups and a perfectly anti-diagonal contact matrix between the groups.
Their first model, \new{{\sl random topological} (the metadata link shuffling P[$\k$,$\pxl(\Thf)$,$\gf$,$\Sigma_\L$]---Sec.~\ref{sec:metadata})}, was used to destroy any structural correlations in the bipartite structure while keeping temporal heterogeneities in the individual timelines unchanged. Conversely, their second null model, termed \new{{\sl random dynamic} (the timeline-snapshot intersection P[$\w$,$\t$]---}Sec.~\ref{sec:intersections-TS-Snaps}), destroyed all the temporal structure except global activity patterns, but kept the weighted (bipartite) network structure unchanged. 
Their third model, \new{{\sl random dynamic topological} 
%P[$\kstat$,$p(\w)$,$\t$,$\gf$,$\Sigma_\L$]  
was generated as the composition of the two others (Sec.~\ref{sec:compositions})}. 
Interestingly, they observed that bursty patterns accelerate the spreading dynamics, contrary to other studies~\cite{Karsai2011S,Kivela2012M,Gauvin2013A,Perotti2014T,Miritello2011D}. At the same time they showed that structural correlations slow down the dynamics in the long run, and by applying the two reference models at the same time, that bursty temporal patterns and structural correlations together slows spreading initially and speeds it up for later times.
The authors arrived at the same conclusion using SIR model dynamics. %(for definition see Section \ref{sec:SIR} below), with the additional observation that temporal effects cause relatively high epidemic thresholds as compared to degree-heterogeneous static networks, where thresholds are vanishing \cite{Pastor-Satorras2015E}. 
Note that the accelerating effect of burstiness in this case was explained later by the non-stationarity of the temporal network \cite{Holme2014B,Holme2016T}

\subsubsection{SIR and SIS processes}
\label{sec:SIR}
\noindent
The Susceptible-Infected-Recovered (SIR) and Susceptible-Infected-Susceptible (SIS) processes are two other dynamical processes that have been widely studied on temporal networks using MRRMs.
In addition to the S$\to$I transition of the SI process, in the SIR (SIS) process infected nodes transition spontaneously to a recovered, R (susceptible, S), state with rate $\gamma$ (or after a fixed time $\tinf$), after which they cannot (can) be re-infected. These processes are characterized by the basic reproduction number $R_0 =\beta/\gamma$  and display a phase-transition between a non-endemic and an endemic phase. An analogy with information diffusion can easily be drawn, where the infection is associated to the exposure to a given information, while spontaneous recovery mimics that the agent later forgets the given information.

One of the first studies addressing SIR dynamics using MRRMs was published by Miritello et al.~\cite{Miritello2011D} and investigated mobile phone communication networks. 
%In their model they used $\beta$ as the control parameter while letting the recovery time $\tinf$ for each node be constant. 
They used two reference models. 
The first was \new{the timeline-snapshot intersection P[$\w$,$\t$] (Sec.~\ref{sec:intersections-TS-Snaps})}, used to study the effects of bursty interaction dynamics on global information spreading. Their second null model applied a local shuffling scheme that cannot evidently be interpreted as a MRRM for networks since it considers only local information and not the whole network: 
%It investigates the effects of group conversations on local information spreading. 
%In this case t
% They considered an event $(i\rightarrow j)$ and its preceding one $(*\rightarrow i)$ reaching the node $i$. To eliminate local causal correlations between the two events they randomized relay times by selecting randomly a time for the $(*\rightarrow i)$ event from the times of all events observed in the dataset.  
Both reference models preserve the link weights $\w$, the duration of interactions, and also the circadian rhythms of human communications. As their first conclusion, they realized that relay times depend on two competing properties of communication. While burstiness induces large transmission times, thus hindering any possible infection, causual interaction patterns translate into an abundance of short relay times, favoring the probability of propagation. 
%They also showed that the outbreak size of the simulated SIR process depends counterintuitively on $\beta$. More precisely, if $\beta$ is small the spreading is faster and reach a larger fraction of nodes in the original temporal network than in the P[$\w$,$\t$] model; while on the contrary, if $\beta$ is large the process evolves slower and unfolds in smaller cascades on the original data relative to the P[$\w$,$\t$] model. If $\beta$ is large, the information propagates easily but it is affected strongly by large inter-event durations and local correlations, while if $\beta$ is small the propagation is more successful at the local scale thus reaching a larger fraction of nodes in the original temporal network even if temporal correlations are present. To quantitatively explain these effects the authors introduce the {\sl dynamical tie strength} to represent the network as static and show that the phenomena can be explained by the competition of heterogeneous interaction patterns and local causal correlations.

G\'enois et al. studied the effects of sampling of face-to-face interaction data on data-driven simulations of SIR and SIS processes \cite{Genois2015C}, and proposed an algorithm for compensating for the sampling effect by reconstructing surrogate versions of the missing contacts from the incomplete data, taking into account the network group structure and heterogeneous distributions of $n_\ij$, $\tau_\ij^m$, and $\Dtau_\ij^m$. Using the reconstructed data instead of the sampled data allowed to trade in a large underestimation of the epidemic risk by a small overestimation; here the epidemic risk was quantified by the fraction of recovered (susceptible) nodes in the stationary state and the probability that the epidemics reached at least $20\%$ of the population. They used MRRMs to investigate and explain the reasons for the small overestimation of the epidemic risk when using the reconstructed networks.
They applied following reference models:  \new{a method termed {\sl link shuffling} (the link-timeline intersection P[$\L,\pxl(\Thf)$]---Sec.~\ref{sec:intersections-LS-TS});  
their {\sl CM-shuffling} (the blockmodel link shuffling P[$\pxl(\Thf)$,$\gf$,$\Sigma_\L$]---Sec.~\ref{sec:metadata}); 
a bootstrap method, resampling $p(\n)$, $p(\tauf)$, $p(\Dtauf)$, and $p(\t^1)$ (Sec.~\ref{sec:bootstrapping}}); 
and they finally applied P[$\pxl(\Thf)$,$\gf$,$\Sigma_\L$] in composition with the bootstrap method.
This allowed them to conclude that the overestimation was due to higher order temporal and structural correlations in the empirical temporal networks, which however are notoriously hard to quantify and to model.

The effect of \new{the timing of the first and last activations 
%birth and death 
of the links in a network} on epidemic spreading was demonstrated by Holme and Liljeros~\cite{Holme2014B} using twelve empirical temporal networks. They investigated an ongoing link picture where the lifetime of social ties is irrelevant as links are assumed to be created and end before and after the observation period; and a link turnover picture where social links are assigned with a lifetime being created and dissolved during the observation. To understand which case is more relevant for modeling epidemic spreading, they defined three deterministic {\sl poor man's reference models}~\cite{Holme2015M}  (see Sec.~\ref{sec:non-MaxEnt}). Their first reference model conserved the timings of the first and last events on each link, $\t^1$, $\t^w$, respectively, as well as the links' weights $\w$, and equalized all inter-event durations in the timelines, eliminating the effects of heterogeneous inter-event durations. Their second and third models aimed to neutralize the effects of the beginning and ending times of active intervals, thus they shifted the active periods of each link either to the beginning or to the end of the observation period, i.e.\ they set $t^1_\ij=\tmin$ or $t^w_\ij=\tmax$ for all $\ij\in\L$, respectively, while keeping the original sequence $\Dtauf$ of inter-event durations on the links. The authors presented an exhaustive analysis by simulating SIR and SIS processes on each dataset using the original event sequences, and each reference model.
%They explored the entire phase space in each case. They concluded that for both processes, while shifting activity periods (either way) induce large differences in the final fraction of infected nodes, equalizing the inter-event durations while keeping the times of the first and last events on each link only marginally changes the outcome. This indicates that it is enough to consider only the observed lifetime of links while their fine-grained dynamics is less relevant in terms of modeling the spreading process.

Valdano et al.~\cite{Valdano2015I} proposed an infection propagator approach to compute the epidemic threshold of discrete time SIS (and SIRS) processes on temporal networks. Their aim was to account for more realistic effects, namely a varying force of infection per contact, the possibility of waning immunity, and limited time resolution of the temporal network. To better understand the effects of temporal aggregation and correlations on the estimation of the epidemic threshold in face-to-face interaction datasets recorded in school settings,
%they used three different MRRMs, as well as a heuristic reference model, and applied them on face-to-face interaction datasets recorded in school settings. 
\new{they employed three MRRMs: 
{\sl reshuffle} (the snapshot shuffling P[$\pxt(\Gs)$]---Sec.~\ref{sec:sequence_shufflings}), 
{\sl reconfigure} (the timeline-snapshot intersection P[$\w$,$\t$]---Sec.~\ref{sec:intersections-TS-Snaps}), 
and {\sl anonymize} (the snapshot shuffling P[$\isof(\Gs)$]---}Sec.~\ref{sec:snapshot_shufflings}).
They measured, for different recovery rates, how the epidemic threshold changed as a function of the aggregation time window relative to the case with the highest temporal resolution. They considered two different aggregation strategies: where the link weights (i) were or (ii) were not considered. %They showed that the obtained thresholds were mostly independent of the cumulative activity of the network, and more related to specific time-evolving topological structures. 
Finally, they considered a fourth, heuristic, reference model, which shuffled the snapshot order, but only within a given number of slices, this way keeping control on the length of temporal correlations it destroyed (see description in Sec.~\ref{sec:non-MaxEnt}). %They showed that long range temporal correlations, which in turn lead to repeated interactions and strong weight-topology correlations, must be considered to provide a good approximation of the epidemic threshold on short temporal scales and for slow epidemics.

Finally, there has been a single study using MRRMs with rumor spreading dynamics \cite{Karsai2014T}. It considered the Daley-Kendall model, which is very similar to the SIR model with the exception that nodes do not recover spontaneously but via interactions with other infected or recovered (stifler) nodes. The aim of this study was to understand the effects of memory processes, inducing repeated interactions between people, on the global mitigation of rumors in large social networks. Using a mobile phone communication dataset they utilized a specific directed temporal network snapshot shuffling, P[$\doutf$], which constrained the instantaneous out-degree $\dtout{i}{m}$ of each node in each snapshot (see Supplementary Table~\ref{tab:features-directed}). 
In practice this amounted to randomizing the called person for each event in order to eliminate the effects of repeated interactions over the same link. This MRRM randomized the topological and temporal correlations in the network, destroyed link weights, and increased the static node degrees considerably.
Results were confronted with corresponding model simulations, which verified that memory effects play the same role in data-driven models as was observed in the case of synthetic model processes, namely they keep rumors local due to repeated interactions over strong ties.

\subsubsection{Threshold models}

\noindent
A third family of spreading processes are {\sl complex contagion} processes, which are often used to model social contagion.
These models capture the effects of social influence, which is considered via a non-linear mechanism for contaminating neighboring nodes (typically a threshold mechanism). In the conventional definition of threshold models \cite{Watts2002S} nodes can be either of two mutually exclusive states, non-adopter (i.e.\ susceptible) -- initially all but one node -- and adopter (i.e.\ infectious) -- initially a randomly selected {\sl seed} node -- and each node $i$ is assigned a threshold $\phi_i$ defining the number $k^I_i$ or fraction $k^I_i/k_i$ of adopter neighbors necessary to make the node (with total degree $k_i$) adopt. We refer to the first variant as the Watts threshold model with {\sl absolute} thresholds, and the second as the Watts threshold model with {\sl relative} thresholds.
The central question here is the condition needed to induce a large adoption cascade that spreads all around the network. These models 
are highly constrained by the network structure and dynamics as the distribution of individual thresholds determine the conditions for global cascades. 
This is fundamentally different from the SIR type of dynamics (called \textit{simple contagion processes}) which are highly stochastic, driven by random infection and recovery. 
%In the latter case, transmission of infection is not fully determined by structural properties but possible even via a single stimuli coming from an infected neighbor.
The conventional threshold model introduced by Watts \cite{Watts2002S}, and other related dynamical processes have been thoroughly studied on static networks, however their behavior on temporal networks has been addressed only recently by studies using RRMs.

Karimi and Holme~\cite{Karimi2013T} studied two different threshold models on six empirical datasets of time-resolved human interactions. %They studied both the relative threshold and the absolute threshold Watts models~\cite{Watts2002S}, where the numbers of adopted neighbors $k^I_i(\delta)$ and all neighbors $k_i(\delta)$ were calculated over a given {\sl memory} time window $\delta$. Their main goal was to identify the effect of temporal and structural correlations on the size of the emerging cascade as function of the threshold $\phi$ (chosen to be equal for every node) and the memory time window size. For the model with fractional thresholds they observed that the cascade size decreased with $\phi$ and with the time window length, while for the absolute threshold model it increased with longer time windows. 
They employed two MRRMs: 
one called \new{{\sl time reshuffle} (the timeline-snapshot intersection P[$\w$,$\t$]---Sec.~\ref{sec:intersections-TS-Snaps}) 
and anoter termed {\sl Erd\H{o}s-R\'enyi} (the link shuffling P[$\pxl(\Thf)$]---}Sec.~\ref{sec:link-shufflings}). 
Application of P[$\w$,$\t$] allowed them to conclude that burstiness plays an important role on how large cascades can appear in complex contagions.
%They found that in the fractional threshold model temporal correlations (burstiness) allowed smaller cascades to evolve in most of the cases, while in the absolute threshold model the effect was the contrary. An exception they found was a conference setting, where temporal correlations increased the cascade size, while structural correlations slightly decreased it. As they explained, this may be due to specific constraints in this setting as bursty interaction patterns appeared synchronously during the conference breaks where also a large number of simultaneous interactions appeared between people discussing in groups.
Backlund et al.~\cite{Backlund2014E} also studied the effects of temporal correlations on cascades in slightly different threshold models on temporal networks. %In their study, they introduced a stochastic and a deterministic threshold model. Their stochastic model is a linear threshold model where the probability of adoption increases linearly with the fraction of adopting neighbors observed in a finite time window prior to the actual interaction. Note that in this case, rapidly repeated interactions with an adopted neighbor does not increase the adoption probability per interaction. However, since adoption potentially occurs after every interaction, bursty interaction patterns evidently affects the adoption process. Conversely, in their deterministic model they employed the conventional deterministic threshold rule, thus assigning a relative threshold to a node (the same for each of them, as in Ref.~\cite{Karimi2013T}), which then certainly adopts after this fraction of adopted neighbors has been reached within a finite observation window. Note that in each model when calculating the actual threshold of a node they considered the static degree $k_i$, aggregated over the whole observation period, in the \new{denominator}, and not the degree $k_i(\delta)$, aggregated over the time window $\delta$ only as in \cite{Karimi2013T}. 
They applied two MRRMs to four different temporal interaction datasets. They used the P[$\w$,$\t$] (Sec.~\ref{sec:intersections-TS-Snaps}) model to destroy all temporal correlations while keeping circadian fluctuations, and introduced another model, P[$\perl(\Thf)$] (Sec.~\ref{sec:timeline-shufflings}), that randomly shifts each individual timeline using periodic boundary conditions to keep all temporal correlations inside each timeline and destroy correlations between events on adjacent links as well as circadian fluctuations.
%After simulating both models, 
They found that %increasing the memory length (time window size) facilitates spreading, and so does 
the removal of temporal correlations using P[$\w$,$\t$] facilitates spreading. This way they concluded that burstiness negatively affects the size of the emerging cascades. At the same time, they found that higher order temporal-structural correlations, removed by P[$\perl(\Thf)$], facilitate the emergence of large cascades. 
%In addition, they observed that for the deterministic model, high degree nodes tend to block the spreading process, contrary to the case of simple contagion. For complex contagion, hubs are unlikely to interact with enough adopters to reach their adoption threshold.

A somewhat different picture was proposed by Takaguchi et al.~\cite{Takaguchi2013B}, where the authors used a threshold model denoted {\sl history dependent contagion}. This model is an extension of an SI process with a threshold mechanism. Here each node has an internal variable measuring the concentration of pathogen and is increased by unity after a stimuli arrived via temporal interactions with infected neighbors. However, this concentration decays exponentially as function of time in the absence of interaction with infected nodes. A node becomes infected if its actual concentration reaches a given threshold, after which it remains in the infected state. They simulated this model on two different temporal interaction networks and measured the fraction of adopters as function of time. In order to identify the effects of bursty interaction patterns they used \new{a model called {\sl randomly permuted times} (the timeline-snapshot intersection P[$\w$,$\t$]---}Sec.~\ref{sec:intersections-TS-Snaps}), which led to slower spreading dynamics. From this they argued that burstiness increases the speed of spreading in both datasets. Furthermore, they showed through the analysis of single link dynamics, that this acceleration was mostly due to the bursty patterns on separate links and not due to correlations between bursty events on adjacent links or to the overall structure of the network.

%------------------------------------------------------------------------------------------------------------------
\subsection{Random walks}
\label{sec:RW}
%------------------------------------------------------------------------------------------------------------------
\noindent
Random walks are some of the simplest and most studied dynamical processes on networks. On a temporal network, a random walk is defined by a walker, which is located at a node at time $t$, and can be re-located to one of the node's current neighbors in each timestep. The walker chooses the neighbor to which it jumps either at random or with a probability proportional some link weight. 
%A central measure is here the \textit{mean-first passage time} (MFPT), defined as the average time taken by the random walker to arrive for the first time at a given node, starting from some initial position in the network. Another important measure is the \textit{coverage} defined as the number of different vertices that have been visited by the walker up to time $t$.

Starnini et al.~\cite{Starnini2012R} studied stationary properties of random walks on temporal networks, and used reference models to define ways to synthetically extend their temporal face-to-face interactions datasets with a limited observation length. They assumed periodic temporal boundary conditions on their empirical temporal network (their first model), with weak induced biases as discussed in an earlier paper~\cite{Pan2011P}.
Their second model, \new{{\sl SRan} (the timeline shuffling P[$\w$]---}Sec.~\ref{sec:timeline-shufflings}), kept all weighted features of the aggregated network, but destroyed all temporal correlations and induced Poissonian interaction dynamics.
Finally, they introduced a third heuristic reference model in which they impose a delta function constraint on the number of events starting at each time step (Sec~\ref{sec:non-MaxEnt}), randomly drawing the pairs of nodes that interact in order to approximately conserve $\n$ and finally bootstrap the event durations from $p(\tauf)$.
This approximately conserves certain important statistical properties of the empirical event sequence, namely $p(\n)$ and $p(\tauf)$, but not $\Ebf$ and $p(\Dtauf)$. 
%After providing a mean field solution, t
They measured the \textit{mean-first passage time} (MFPT), defined as the average time taken by the random walker to arrive for the first time at a given node starting from some initial position in the network, and the \textit{coverage}, defined as the number of different vertices that have been visited by the walker up to time $t$, on both the original temporal network and synthetic sequences.
They found that the results for empirical sequences deviated systematically from the mean field prediction and from the results for the reference models, inducing a slowdown in coverage and MFPT. They concluded that this slowdown is  not due to the heterogeneity of the durations of conversations, but uniquely due to what they term {\sl temporal correlations} (which, given the reference models they tested, encompasses the time-varying cumulative activity, the broad distribution of inter-event durations, and higher-order temporal correlations between different events).

Delvenne et al.~\cite{Delvenne2015D} also addressed random walks on temporal networks. They used MRRMs in order to understand which factor is dominant in determining the relaxation time of linear dynamical processes to their stationary state. They introduced a general formalism for linear dynamics on temporal networks, and showed that the asymptotic dynamics is determined by the competition between three factors: a structural factor (i.e.\ community structure) associated with the spectral properties of the Laplacian of the static network, and two temporal factors associated to the shape of the waiting-time distribution, namely its burstiness coefficient (defined in~\cite{Goh2008B}) and the decay rate of its tail. They demonstrated their methodology on six empirical temporal interaction networks and used two RRMs. 
\new{A MRRM termed {\sl randomized structure} (the link shuffling P[$\k,\pxl(\Thf)$]---}Sec.~\ref{sec:link-shufflings}) aimed to remove the effects of the structural correlations. %In this case they found that in sparse networks, structure remains the dominant determinant of the dynamics as sparsity results in the inevitable creation of bottlenecks for diffusion even in a random network. On the other hand, in denser structures the removal of communities leads to the dominance of temporal features. 
The other null model, a generative reference model using a homogeneous Poisson process to generate events and constraining only $\Gstat$ and the mean number of events $\langle E\rangle$ (Sec.~\ref{sec:generative}), destroyed all temporal and weight correlations while conserving the static network structure, leading to the evident dominance of the network structure in regulating the convergence to stationarity.

A greedy random walk process and a non-backtracking random walk process were studied by Saram\"aki and Holme on eight different human interaction datasets in Ref.~\cite{Saramaki2015E}.
A greedy random walker always moves from the occupied node to one of its neighbors whenever possible. Thus its dynamics is more sensitive to local temporal correlations in the network.
A non-backtracking greedy random walker is additionally forbidden to return to its previous position. Thus, it is forced to move to a new neighbor or wait until the next event which moves it to a new neighbor. The authors studied what types of temporal correlations are determinant during these dynamics by using the \new{{\sl time-stamp shuffling} (the timeline-snapshot shuffling P[$\w$,$\t$]---}Sec.~\ref{sec:intersections-TS-Snaps}) and measuring the coverage of the walker after a fixed number of moves. 
They found that after removing temporal correlations using P[$\w$,$\t$], the walker reached considerably more nodes. 
%They explained this observation by the dominant effects of bursty (ping-pong) event trains on single links which trap the walker for a long time going back and forth between two nodes. In addition, they also indicated somewhat weaker effects of larger temporal x such as triangles. 
They finally traced the entropy of the greedy walkers and concluded that, on average, the entropy production rates measured in the original event sequences were lower than for randomized data, indicating more predictable node sequences of visited nodes in the empirical case.

%-------------------------------------------------------------------------
\subsection{Evolutionary games}
\label{sec:evol_games}
%-------------------------------------------------------------------------
\noindent
Evolutionary games~\cite{Nowak2006E} define another set of dynamical processes which have historically been studied on networks. They are analogous to several social dilemmas where the balance of local and global payoffs drive the decision of interacting agents.
Any agent may choose between two strategies (Cooperation or Defection) and can receive four different payoffs (Reward, Punishment, Sucker, or Temptation). The relative values of Temptation and Sucker determines the game, where players update their strategy depending on the state of their neighbors with a given frequency and tend to find an optimal strategy to maximize their benefits.

Cardillo et al.~\cite{Cardillo2014E} studied various evolutionary games on temporal networks and asked two questions: ``Does the interplay between the time scale associated with graph evolution and that corresponding to strategy updates affect the classical results about the enhancement of cooperation driven by network reciprocity?'' and ``what is the role of the temporal correlations of network dynamics in the evolution of cooperation?''. They analyzed two human interaction sequences, and for comparison they applied a shuffling termed \new{{\sl random ordered} (the sequence shuffling P[$\pxt(\Gs)$]---}Sec.~\ref{sec:sequence_shufflings}), and the activity-driven model \cite{Perra2012A}. As a parameter to control the time-scale of the network, they varied the size of the integration time window defining a single snapshot of the temporal networks and measured the fraction of cooperators after the simulated dynamics reached equilibrium. They showed for all social dilemmas studied that cooperation is seriously hindered when agent strategies are updated too frequently relative to the typical time scale of interactions, and that temporal correlations between links are present and lead to relatively small giant components of the graphs obtained at small aggregation intervals. 
However, when one uses randomized or synthetic time-varying networks that preserve the original activity potentials but destroys temporal correlations, the structural patterns change dramatically. 
Effects of the temporal resolution on cooperation are smoothed out, and due to the lack of temporal and structural correlations, cooperation may persistently evolve even for moderately small time periods.

%------------------------------------------------------------------------------------------------------------------
\subsection{Temporal motifs and networks with attributes}
\label{sec:motifs}
%------------------------------------------------------------------------------------------------------------------
\noindent
Another direction of application of RRMs is to highlight significant temporal correlations or motifs in interaction signals or when the interaction sequences may correlate with additional node attributes.

For directed temporal networks, one simple application of MRRMs was introduced by Karsai et al.~\cite{Karsai2012U}, who analyzed the correlated activity patterns of individuals, which induced bursty event trains. They found that the number of consecutive events arriving in clusters are distributed as a power-law. To identify the reason behind this observation they used a MRRM that shuffled the inter-event durations between  consecutive event pairs, P[$\soutf,p(\Dtauf)$] (see Supplementary Table~\ref{tab:features-directed}). 
%for definitions of features of directed temporal networks). 
They found that in the shuffled signal, bursty event trains were exponentially distributed, which evidently indicated that bursty trains were induced by intrinsic correlations in the original system and were not simply due to the broad distribution of inter-event durations.

In another study, Karsai et al.~\cite{Karsai2012C} also applied this framework to identify whether correlated bursty trains of individuals is a property of nodes or links. Using a large mobile phone call interaction dataset, the observation was made that bursty train size distributions were almost the same for nodes and links. This suggests that such correlated event trains were mostly induced by conversations by single peers rather than by group conversations. To further verify this picture, the fraction of bursty trains of a given size emerging between a varying number of individuals were calculated in the empirical event sequence and in shuffled networks generated using a MRRM ] 
where the receivers of calls were shuffled between calls of the actual caller (P[$\w$,$p(\Dtauf)$---Supplementary Table~\ref{tab:features-directed}). 
This reference model leaves the timing of each event unchanged, thus leading to the observation of the same bursty trains, and it keeps the instantaneous and static out-degrees of individuals. However, since the receivers are shuffled, potential correlations that induce bursty trains on single links are eliminated. Results showed that the fraction of single link bursty trains drops from $\sim 80\%$ to $\sim 20\%$ after shuffling in call and SMS sequences. This supports the hypothesis that single link bursty trains are significantly more frequent than one would expect from the null hypothesis, which is then rejected.

%However, r
Real temporal networks commonly reveal more complicated temporal motifs, whose detection was first addressed by Kovanen et al.~\cite{Kovanen2011T}. They proposed a method to identify  mesoscale causal temporal patterns in interaction sequences where events of nodes do not overlap in time. This framework can be used to identify overrepresented patterns, called temporal motifs which are not only similar topologically but also in the temporal order of the events. 
RRMs are crucial in this framework for
%They propose different RRMs to 
quantifying the significance of different temporal motifs. 
They used \new{{\sl time-shuffling} (the timeline-snapshot intersection P[$\w$,$\t$]---}Sec.~\ref{sec:intersections-TS-Snaps}), and they introduced a non-maximum-entropy reference model which biases the sampling of the %ensemble of 
temporal networks defined by P[$\w$,$\t$] in order to keep some temporal correlations in the sequence (see Sec.~\ref{sec:non-MaxEnt}). To do so, they selected randomly for each event in a motif $m$ other events from the sequence and chose the one which was the closest in time to the original event in focus. If $m=1$ the model is identical to P[$\w$,$\t$], while the larger $m$ is the more candidate events there are, thus the more likely it is to find one close to the original event. They furthermore suggested that to remove causal correlations from the sequence, one may simply reverse the interaction sequence and repeat the motif detection procedure (see Sec.~\ref{sec:non-MaxEnt}). They used these reference models in the same spirit as the configuration model is typically used to identify motifs in static networks~\cite{Alon2007N,ShenOrr2002N}. Here, applying P[$\w$,$\t$] and its biased version as null models to detect motifs consisting of three events, they found that motifs between two nodes, i.e.\ bursty link trains, are the most frequent, and motifs which consist of potentially casually correlated events are more common than non-causal ones.

In another study by the same authors \cite{Kovanen2013T}, the same methodology was used to identify motifs in temporal networks where nodes (individuals) were assigned with metadata attributes like gender, age, and mobile subscription types. Beyond the P[$\w$,$\t$] model (Sec.~\ref{sec:intersections-TS-Snaps}), the authors introduced the metadata MRRM termed \new{{\sl node type shuffled data} (the metadata shuffling P[$G,p(\gf),\Sigma_\L$]---}Sec.~\ref{sec:metadata}), which shuffles single attributes between nodes. %(or equivalently sampled the static network structure among graphs isometric to $\Gstat$). 
In addition, they applied the biased version of P[$\w$,$\t$] introduced in \cite{Kovanen2011T} (see Sec.~\ref{sec:non-MaxEnt}), which accounts for the frequency of motif emergence in the corresponding static weighted network without considering node attributes. 
Using this non-maximum-entropy reference model and the two MRRMs they found gender-related differences in communication patterns and showed the existence of temporal homophily, i.e.\ the tendency of similar individuals to participate in communication patterns beyond what would be expected on the basis of their average interaction frequencies.

The dynamics of egocentric network evolution was studied by Kikas et al.~\cite{Kikas2013B}, where they used a large evolving online social network to analyze bursty link creation patterns. First of all they realized that link creation dynamics evolve through correlated bursty trains. They verified this observation by comparing the distribution of inter-event durations (measured between consecutive link creation events) to those generated by the directed-network MRRM P[$\soutf,p(\Dtauf)$] (see Supplementary Table~\ref{tab:features-directed}), where inter-event durations were randomly shuffled. In addition, they classified users based on their link creation activity signals (where activity was measured as the number of new links added within a given month). They showed that bursty periods of link creation are likely to appear shortly after the creation of a user account, or when a user actively use free or paid services provided by the online social service. In order to verify these correlations they used a reference model 
where they shuffled link creation activity values between the active months of a given user and found considerably weaker correlations between the randomized link creation activity signals and service usage activity signals of people.

Finally, in a different framework, a special kind of metadata reference model was also used by Karsai et al.\ \cite{Karsai2016L} to demonstrate whether the effect of social influence or homophily is dominating during the adoption dynamics of online services on static networks.
This reference model did not consider randomizing the temporal networks, but rather node attributes linked to the dynamics of the game (i.e.\ a purely metadata MRRM -- Sec.~\ref{sec:metadata} -- coupled with a dynamical process on the network); we  include it in this survey to demonstrate the scope of maximum entropy shuffling methods beyond randomizing structural network features.
%In case of real adoption cascades, these two mechanisms may lead to similar collective adoption patters at the system level. In reality, influence-driven adoption of an ego can happen once one or more of its neighbors have adopted, since their actions may then influence the decision of the central ego. Consequently, the time-ordering of adoptions of the ego and its neighbors matters in this case. Homophily-driven adoption is, however, different. Homophily drives social tie formation such that similar people tend to be connected in the social structure. In this case connected people may adopt because they have similar interests, but the time ordering of their adoptions would not matter. Therefore, it is valid to assume that adoption could evolve in clusters due to homophily, but these adoptions would appear in a random order. In order to demonstrate these differences, 
The authors used a reference model where they shuffled all adoption times between adopted nodes and confronted the emerging adoption rates of innovator, vulnerable, and stable adopters (for definitions see \cite{Watts2002S,Karsai2016L}) to the adoption rates observed in the empirical system.
They found that after shuffling the rate of innovators considerably increased, while the rate of influence driven (vulnerable and stable) adoptions dropped. This verified that adoption times matters during real adoption dynamics, thus the social spreading process was predominantly driven by social influence. Note that in this case the network was static and shuffling was applied on the observed dynamical process. %We mention this example here to demonstrate the potential of MRRMs in other settings.

%------------------------------------------------------------------------------------------------------------------
\subsection{Network controllability}
\label{sec:controllability}
%------------------------------------------------------------------------------------------------------------------
\noindent
We finally mention two recent studies of the controllability of temporal networks that have leveraged MRRMs.
Control of a dynamical system aims at guiding a system to a desired state by designing the inputs to the system~\cite{Kirk1971O}.
Although control theory has a long history as a branch of engineering applied to diverse subjects, it was only recently that we saw a general theory of the controllability of the systems in which elements interact in a networked manner~\cite{Liu2011C}.
%The key finding was that sparse networks require more driver nodes (i.e.\ the nodes receiving the designed input) than dense networks, and that the driver nodes are not necessarily hubs in general~\cite{Liu2011C}.
%An algorithm to approximate the minimal set of driver nodes was also proposed in \cite{Liu2011C}, based on finding the maximum matching in the network.

It is natural to think of extending the theory for static networks to temporal networks.
P\'{o}sfai and H\"{o}vel made the first study in this direction, in which they considered a discrete-time linear dynamical system with time-varying interactions~\cite{Posfai2014S}.
Their focal measure of controllability is the size of the structural controllable subspace.
The structural controllable subspace is defined by the subset of nodes satisfying that any of their final states at time $t$ is realizable from any initial state in at most a number $\tau$ of time-steps by appropriately tuning the non-zero elements of the adjacency and input matrices as well as the input signals.
First, they proved a theorem stating that a node subset is a structural controllable subspace if and only if any node in the subset are connected to disjoint time-respecting paths from the nodes receiving the input signals.
This theorem implies that, keeping the same average instantaneous degree, the temporal network with a heavy-tailed distribution of instantaneous node degrees, $\pit(\d)$, is more difficult to control than a network with a homogeneous $\pit(\d)$ because the presence of hubs in snapshots decreases the number of disjoint time-respecting paths.
They examine this theoretical argument by comparing the structural controllable subspace for an empirical temporal network to the \new{ones} produced by \new{five MRRMs: 
one termed {\sl random time} (the timeline shuffling P[$\w$]---Sec.~\ref{sec:timeline-shufflings}), 
one termed {\sl random network} (P[$\t$]---Sec.~\ref{sec:snapshot_shufflings}),
one termed {\sl degree-preserved network} (P[$\d$]---Sec.~\ref{sec:snapshot_shufflings}), 
and two MRRMs both termed {\sl shuffled time} (the sequence shufflings P[$\pxt(\Gs)$] and P[$\pxt(\Gs)$,$\sgnf(\Ebf)$]---}Sec.~\ref{sec:sequence_shufflings}).
The sizes of the maximum structural controllable subspace for P[$\w$,$\t$] and P[$\t$] were generally larger than that for the original network.
This result suggests that the homogenization of $\pit(\d)$ and thus the elimination of hubs in snapshots increases the controllability of networks, which is consistent with the theoretical argument.
For the other two MRRMs, the controllability of networks generated by P[$\d$], which conserves the instantanteous degrees, is almost the same as the original network, and networks generated by P[$\pxt(\Gs)$,$\sgnf(\Ebf)$], which randomizes the temporal order of the snapshots, has a slightly lower controllability than the original network.
These results imply that the higher-order structural correlations in snapshots have little effect on network controllability and that the temporal correlations over successive snapshot present in the original network contribute to enhance the controllability to some extent.

Recently, Li et al.~\cite{Li2017T} showed that temporal networks have a fundamental advantage in controllability compared to their static network counterparts.
They compared the time and energy required to achieve full structural controllability of the network, when driving nodes in the sequence of snapshots or the single aggregated network. 
The numerical experiments on multiple empirical networks demonstrated that temporal networks can be fully controlled more efficiently in terms of both time and energy than their static counterparts.
They argued that this advantage comes from temporality itself, but not from particular temporal features, by showing that a set of different reference models \new{achieve more efficient controllability than their aggregated counterparts.
The models employed were:
{\sl randomly permuted times} (P[$\w,\t$]---Sec.~\ref{sec:intersections-TS-Snaps}), 
{\sl randomized edges} (P[$\k,\pxl(\Thf)$]---Sec.~\ref{sec:link-shufflings}), 
and {\sl randomized edges with randomly permuted times} (the composition of the two%, P[$\kstat,p(\w),\t$]
---Sec.~\ref{sec:compositions}), 
as well as time reversal (Sec.~\ref{sec:non-MaxEnt}).}

%========================================================================
\section{Characterizing and ordering microcanonical randomized reference models}
\label{sec:hierarchies}
%========================================================================
\noindent
Some MRRMs \new{randomize more (i.e.\ conserve less structure)} than others. 
We \new{here} formalize this notion which allows us to compare MRRMs{, which will let us build hierarchies between them and between different network features (Section~\ref{sec:theory-comparison})}. 
Such hierarchies turn out to be useful for classification of MRRMs \new{ and for understanding how they affect different network features. 
We will use this in Section \ref{sec:classification} to build a taxonomy of MRRMs found in the literature which orders them and characterizes their effects on temporal network features}. 
%Furthermore, sequences of MRRMs where each one shuffles slightly more structure than the previous one are often used in practice~\cite{Kivela2012M,Orsini2015Q} \new{(we employ this in two walkthrough examples in Section \ref{sec:testing})}.

%\new{We formalize in this section the notion that one MRRM randomizes more than another (Section~\ref{sec:theory-comparison})}, and the partial order of MRRMs induced by this notion of {\sl comparability} allows us to build a \new{taxonomy of of MRRMs that fully characterizes them and orders them hierarchically} (Section \ref{sec:classification}). 
%These hierarchies are also useful in the practical employment of MRRMs (Section \ref{sec:testing}).

%------------------------------------------------------------------------
\subsection{Theory: Hierarchies of MRRMs}
\label{sec:theory-comparison}
%------------------------------------------------------------------------
\noindent
\new{We here develop the theory needed to compare and order MRRMs.
%While the present review is focused on temporal networks, this theoretical framework is applicable to microcanonical models for states representing any discrete structure.
%To underline this generality we will in this subsection simply refer to individual temporal networks as states of a given state space, which is understood to be composed of temporal networks for the purpose of this review.
To keep the presentation as accessible as possible, we have relegated proofs to Appendix~\ref{app:proofs}.}

\new{The central concept for building hierarchies of MRRMs is that of {\sl comparability}}.

\begin{definition}
{\sl Comparability.}
\label{def:comparability}
%\new{We will write $\model{\x}\leq\model{\y}$ if the ensemble of states (e.g.\ temporal networks) generated by $\model{\x}$ is always contained in the ensemble generated by $\model{\y}$, that is if $\G_{\x(G)}\subseteq\G_{\y(G)}$ for all $G\in\G$.}
\new{We will write $\model{\x}\leq\model{\y}$ if the set of states (e.g.\ temporal networks) that can be reached by applying $\model{\x}$ to any state is always contained in the set reached by $\model{\y}$, that is if $\G_{\x(G)}\subseteq\G_{\y(G)}$ for all $G\in\G$.}
We say that $\model{\x}$ and $\model{\y}$ are {\sl comparable} if $\model{\x} \leq \model{\y}$ or $\model{\x} \geq \model{\y}$.
\end{definition}

\begin{figure}
  \begin{tikzpicture}
    \node at(-2.2,0){\includegraphics[width=1.2in]{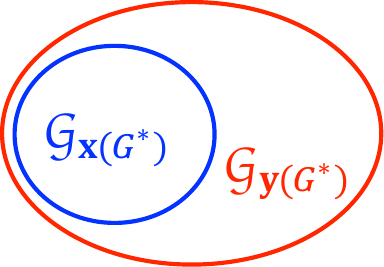}};
    \node at(1.6,0){\includegraphics[width=1.2in]{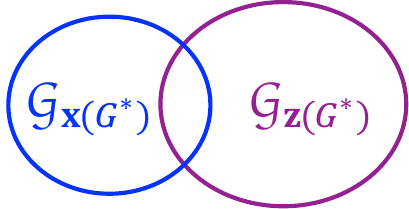}};
%    \node at(0,-2.8){\includegraphics[width=3.375in]{Comparability.pdf}};
    \node at(-1.6,-3.1){\includegraphics[width=2.in]{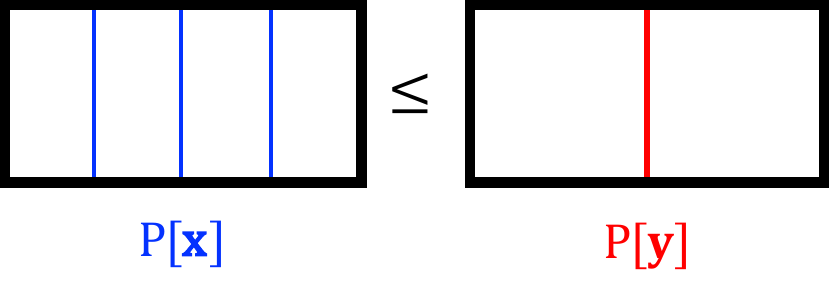}};
    \node at(-1.6,-5.1){\includegraphics[width=2.in]{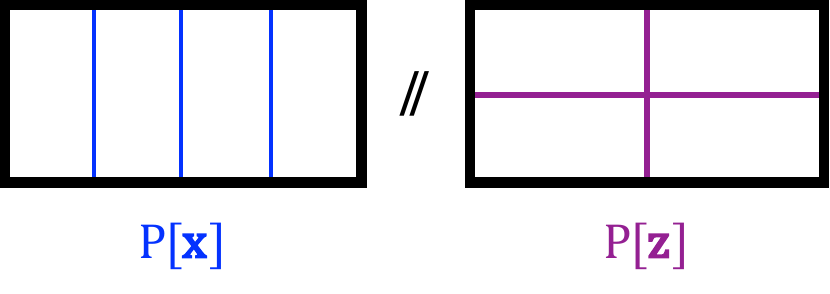}};
    \node at(2.6,-3.5){\includegraphics[width=0.5in]{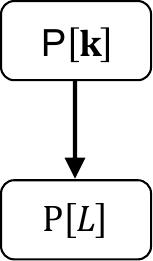}};
    \node at(-4.1,1){\bf(a)};
    \node at(-4.2,-1.8){\bf(b)};
    \node at(1.6,-1.8){\bf(c)};
  \end{tikzpicture}  
  \caption{{\bf Comparability.}
  \new{(a) Two MRRMs satisfy $\model{\x}\leq\model{\y}$ if the ensemble $\G_{\x(G^*)}$ generated by $\model{\x}$ is contained in the ensemble $\G_{\y(G^*)}$ generated by $\model{\y}$ for all $G^*\in\G$.
  Conversely, if the ensembles generated by two MRRMs $\model{\x}$ and $\model{\z}$ only overlap partially, the MRRMs are not comparable, $\model{\x} \incomp
 \model{\z}$.
  (b) In terms of partitions, $\model{\x}\leq\model{\y}$ means that P[$\x$] is finer than P[$\y$]. Conversely, if $\model{\x}\incomp\model{\z}$, neither partition is a refinement of the other.
  (c) Hasse diagram depicting the hierarchy of the two MRRMs P[$\k$] and P[$L$]: the arrow indicates that P[$\k$] is finer than P[$L$].}  
%  \note{+ third model that is not comparable with P[$\k$] (e.g.\ P[$\cf$])?.} 
  }
  \label{fig:comparability}
\end{figure}

%The intuition behind this notation is that when $\model{\x} \leq \model{\y}$, then $\model{\y}$ \new{always randomizes given network structures more than $\model{\x}$ 
\new{The definition of comparability gives a precision notion that one MRRM randomizes more than another [Fig.~\ref{fig:comparability}(a)]}. 
%\new{The precise meaning of this notion being that the ensembles of networks generated by $\model{\x}$ are subset of those generated by $\model{\y}$ [Fig.~\ref{fig:comparability}(a)].}
%Due to the correspondence between a MRRM and its feature function, we shall likewise use the notation $\x\leq\y$ when referring to the feature functions \note{CLV: do we actually do this in the main text now?}. 
\new{In practice 
%the ensembles generated by a MRRM are typically very large, so 
it is often difficult to show directly that the ensembles of states generated by one model are always subsets of those generated by another.
%In practice it is often difficult to compare directly the ensembles of states generated by two  model as they depend not only on the models but also on the particular input network to which they are applied.
To compare MRRMs, we shall instead use that if the features that one MRRM constrains can be written as a function of the features constrained by another then 
%this means that 
the two MRRMs are comparable.}

\new{
\begin{proposition}
\label{prop:comparability-functional_relation}
{\sl Equivalence between comparability of MRRMs and a functional relation between their features.}
Two MRRMs $\model{\x}$ and $\model{\y}$ are comparable and  $\model{\x} \leq \model{\y}$ if and only if there exists a function $\f$ for which $\y(G)=\f(\x(G))$ for all states $G\in\G$. 
\end{proposition}
}

\new{The following simple example illustrates how this proposition can be applied to order MRRMs.}

\begin{example}
\label{ex:partial_order}
In the space of all static graphs with $N$ nodes, one can define the MRRMs corresponding to the Erd\H{o}s-R\'enyi random graph model~\cite{Erdos1960O}, P[$L$], and the Maslov-Sneppen model~\cite{Maslov2002S}, P[$\kstat$]. We have $L=\sum_i k_i/2$, so $L$ is a function of $\k$ and ${\rm P}[L]\geq {\rm P}[\k]$ \new{by Proposition~\ref{prop:comparability-functional_relation}}.
Conversely, $\kstat$ is not a function of $L$ as networks with different degree sequences can have the same number of links\new{, so ${\rm P}[L] \nleq {\rm P}[\kstat]$ which means that $\model{L}$ and $\model{\k}$ are not equivalent (i.e.\ $\model{L}\neq\model{\k}$).}
%\note{CLV: add third MRRM that is comparable with P[$L$] but not P[$\k$]?}
%The relationships between the MRRMs can be illustrated using a Hasse diagram as shown in Fig.~\ref{fig:ex-hierarchy}.
\end{example}

Representing MRRMs as partitions (see Def.~\ref{def:mrrm_representations}) provides a useful and intuitive way of thinking about comparisons of MRRMs, because the MRRM comparison relation is exactly the natural comparison relation of the partitions.

\begin{proposition}
\label{the:partition_refinements}
{\sl Equivalence with partition refinements.}
$\model{\x} \leq \model{\y}$ if and only if the partition $\partition{\x}$ is finer than $\partition{\y}$.
\end{proposition}

%\new{
%\begin{corollary}
%\label{the:partition_refinements}
%{\sl Equivalence with partition refinements.}
%$\model{\x} \leq \model{\y}$ if and only if the partition $\partition{\x}$ is finer than $\partition{\y}$.
%\end{corollary}
%}

Borrowing the terminology from the theory of set partitions, we say for $\model{\x} \leq \model{\y}$ that $\model{\x}$ is {\sl finer} than $\model{\y}$ and equivalently that $\model{\y}$ is {\sl coarser} than $\model{\x}$. We will also refer to $\model{\x}$ as a {\sl refinement} of $\model{\y}$ and to $\model{\y}$ as a {\sl coarsening} of $\model{\x}$. 
{Figure~\ref{fig:comparability}(b) illustrates the concept of comparability in terms of partitions.}

The partition representation is especially useful here as the properties of refinements of set partitions are inherited to the comparison relation of the MRRMs. 
For example, we can now see that the use of the notation \new{$\leq$} is appropriate as the relation it denotes is indeed a partial order.
%
%\begin{corollary}
%\label{the:partial_order}
%{\sl Comparability induces a partial order.}
%The relation \new{$\leq$} is a partial order over the %state space.
%space of MRRMs.
%\end{corollary}
%The above corollary 
This follows immediately from the fact that partition refinement relations give partial orders~\cite{Hrbacek1999I}. As with any partially ordered set, one can draw Hasse diagrams to display the relationships between different MRRMs, and this turns out to be a convenient way of visually organizing the various MRRMs found in the literature \new{[Fig.~\ref{fig:comparability}(c); see also Section~\ref{sec:classification}]}. 

\subsubsection{\new{The space of MRRMs}}
\label{sec:MRRM_space}
\noindent
The set partitions always have uniquely defined minimum and maximum partitions, and these are meaningful in the case of MRRMs. We call them the zero and unity elements.

\begin{definition}
\label{def:zero_unity_elements}
{\sl Zero and unity elements.}
The \emph{zero element}, $\model{0}=\model{G}$, is the MRRM which shuffles nothing, i.e.\ the one that always returns the input network and where the feature returns the entire temporal network. The \emph{unity element}, $\model{1}$, is the MRRM that shuffles everything, i.e.\ the one that returns all networks in the state space with equal probability and where the feature is constant and does not depend on the input.
\end{definition}

The zero element corresponds to the partition where each network is in its own set and the unity element to the partition where there is only a single set. The zero and unity elements are always in the top and bottom of any hierarchy of MRRMs, respectively: $\model{0} \leq \model{\x} \leq \model{1}$ for any $\x$. 

\begin{figure}
  \begin{tikzpicture}
    \node at( 1, 0){ \includegraphics[width=0.33in]{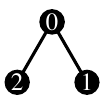} }; 
    \node at( 1,-1){ \includegraphics[width=0.33in]{{g6}.pdf} }; 
    \node at( 1,-2){ \includegraphics[width=0.33in]{{g7}.pdf} }; 
    \node at( 0.2, 0.7){ $\G$ };
    \node at( 0.2, 0){ $G_1$ };
    \node at( 0.2,-1){ $G_2$ };
    \node at( 0.2,-2){ $G_3$ };
    \node at(-0.2, 1){\bf (a) };
%-------------------------------------------
    \node at( 2.2, 0.7){ $L$ };
    \node at( 2.2, 0){ 2 };
    \node at( 2.2,-1){ 2 };
    \node at( 2.2,-2){ 2 };
    \node at( 3.2, 0.7){ $\k$ };
    \node at( 3.2, 0){ (2,1,1) };
    \node at( 3.2,-1){ (1,2,1) };
    \node at( 3.2,-2){ (1,1,2) };
    \node at( 1.8, 1){\bf (b) };
%-------------------------------------------
    \node at( 5.8, 0){ \includegraphics[width=1.in]{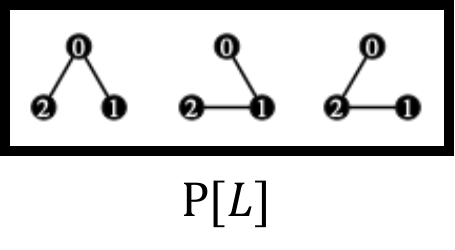} };
    \node at( 4.4, 1){\bf (c) };
    \node at( 5.8,-1.8){ \includegraphics[width=1.in]{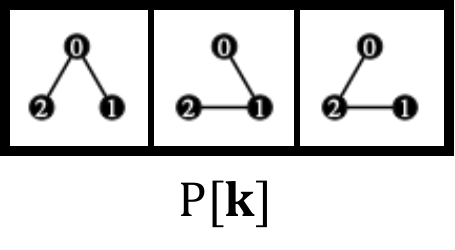} };
    \node at( 4.4,-0.8){\bf (d) };
  \end{tikzpicture}
  \caption{\new{{\bf Example of zero and unity elements in a space of MRRMs.}
   (a) State space $\G$ consisting of all simple graphs with three nodes and two edges.
   (b) Values of the features constrained by the two MRRMs $\model{L}$ and $\model{\k}$ for each state $\G_i\in\G$. 
   (c) Partition of $\G$ induced by $\model{L}$.
   (d) Partition of $\G$ induced by $\model{\k}$.}}
  \label{fig:ex:partial_order2}
\end{figure}

\begin{example}
\label{ex:partial_order2}
We continue from Example \ref{ex:partial_order}, limiting the state space to the set of simple graphs consisting of 3 nodes, $\N=\{1,2,3\}$, and 2 links. 
There are three such graphs: $\{G_1, G_2, G_3\}$ \new{[Fig.~\ref{fig:ex:partial_order2}(a)]}. %, with $\L(G_1)=((1,2); (1,3))$, $\L(G_2)=((1,2); (2,3))$, $\L(G_2)=((1,3); (2,3))$. 
Since the number of links is the same in all graphs \new{[Fig.~\ref{fig:ex:partial_order2}(b)]}, the partition of the ER model contains only one set $\partition{L} = \{ \{G_1, G_2, G_3 \} \}$ \new{[Fig.~\ref{fig:ex:partial_order2}(c)]}. 
However, the degree sequences of the networks differ \new{[Fig.~\ref{fig:ex:partial_order2}(b)]}, 
% , $\kstat(G_1)=(2,1,1)$,  $\kstat(G_2)=(1,2,1)$, and $\kstat(G_2)=(1,1,2)$, 
so the partition related to the Maslov-Sneppen model separates all networks $\partition{\kstat}=\{ \{ G_1 \},  \{ G_2 \}, \{ G_3 \} \}$ \new{[Fig.~\ref{fig:ex:partial_order2}(d)]}. 
%The partition $\partition{\kstat}$ is a refinement of $\partition{L}$ and thus $\model{\kstat} \leq \model{L} $. 
\new{For this state space the Maslov-Sneppen model is thus} the zero element $\model{\kstat}=\model{0}$ and the ER model is the unity element $\model{L}=\model{1}$.
\end{example}

As we shall see, rich hierarchical structure can be found between the zero and unity elements. 
%, and as we will see, the structure can be very rich. 
Again, the set partition representation gives us a glimpse of the theoretical understanding of this structure.
The total number of possible MRRMs for a given state space $\G$ is the same as the number of possible partitions of the space. This is given by the Bell number $B_{\Omega}$~\cite{Stanley2011E}, which grows faster than exponentially with the state space size $\Omega=|\G|$~\cite{Berend2010I}.
We also know that even though the number of MRRMs in the hierarchy can be large, it can only be relatively flat as compared to this number: The largest possible number of MRRMs all satisfying a total order (i.e.\ for which ${\rm P}[\x_1]\geq{\rm P}[\x_2]\geq{\rm P}[\x_3]\ldots$) is the maximum chain length in the set of partitions of the state space, which is equal to $\Omega+1$.
Thus, if we \new{restrict ourselves to} selecting a collection of MRRMs that is totally ordered~\cite{Orsini2015Q}, 
we can at most include an exponentially vanishing part of the possible MRRMs.
In practice these theoretical limitations are not of much concern as the number of possible networks, $\Omega$, typically is extremely large and it is not possible in practice to explore even a small fraction of all possible MRRMs for a given state space.
\new{Of more practical concern is the fact} that we are often interested in studying a set of MRRMs that do not satisfy a linear order. 
In the context of temporal networks, this is notably the case for MRRMs that randomize temporal features and MRRMs that randomize topological features of a network~\cite{Kivela2012M}. 
% Since such MRRMs are generally not comparable, %(see Section~\ref{sec:classification}), 
% we find ourselves constrained to use and study sets of MRRMs for temporal networks that do not satisfy a linear order~\cite{Kivela2012M}.

%\subsubsection{Intersection of MRRMs}
%\noindent
%\new{For building hierarchies of MRRMs, it is finally interesting to} define 

% For any two given features $\x$ and $\y$ and associated MRRMs P[$\x$] and P[$\y$], we define the {\sl intersection} of the MRRMs, P[$\x,\y$], as the MRRM that constrains both features simultaneously. We can write the following definition:\note{MKi: we already had intersections earlier when we defined MRRMs. Should we move this definition there?}

% \begin{definition}
% \label{def:intersection}
% {\sl Intersection of randomized reference models.}
% The \emph{intersection} of two features $\x$ and $\y$ is (\x,\y), and for the associated MRRM we write $\model{\x, \y} = \model{(\x, \y)}$
% \end{definition}

% Note that the intersection by definition gives another MRRM.
% In terms of conditional probabilities the intersection becomes
% \beq
%   P_{\inter{\x}{\y}}(G|G^*) = \frac{\delta_{\x(G),\x^*}\delta_{\y(G),\y^*}}{\Omega_{(\x^*,\y^*)}} \es,
% \eeq
% where $G^*$ is the input network, $\x^*=\x(G^*)$, and $\y^*=\y(G^*)$.
% \note{MKi: Should we move this formula above to appendix, since we (probably) only use it for proofs?}

\begin{figure}
  \includegraphics[width=3.375in]{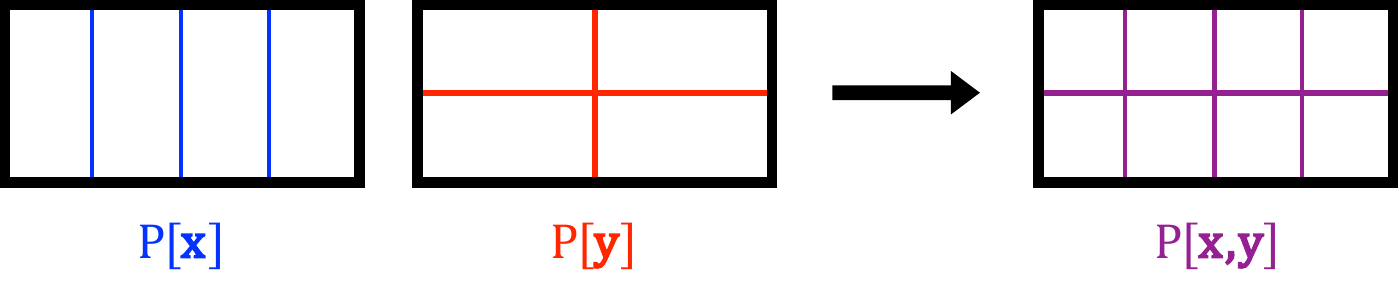}
  \caption{{{\bf Intersection of MRRMs.} The intersection P[$\x,\y$] of two MRRMs, P[$\x$] and P[$\y$], shuffles less than either of the two. In terms of partitions, P[$\x,\y$] produces the greatest lower bound of P[$\x$] and P[$\y$].}
  }
  \label{fig:intersection}
\end{figure}

\new{{\sl Intersections} of MRRMs (Def.~\ref{def:intersection}) play an important role in hierarchies of MRRMs:
for two MRRMs, $\model{\x}$ and $\model{\y}$, their intersection $\model{\x,\y}$ defines the maximally random MRRM that shuffles less than both of them.
In terms of set partitions,} the partition of $\G$ induced by $\model{\x,\y}$ is trivially given by the set of pairwise intersections between the $\x$-equivalence classes and the $\y$-equivalence classes, i.e. $\G_{(\x,\y)}(G) = \G_{\x}(G) \cap \G_{\y}(G)$ for all $G\in\G$, and $\Omega_{(\x^*,\y^*)} = |\G_{\x^*} \cap \G_{\y^*}| = \sum_{G\in\G}\delta_{\x(G),\x^*}\delta_{\y(G),\y^*}$ (Fig.~\ref{fig:intersection}).
%That is, the partition $\G_{(\x,\y)}$ is finer than both $\G_{\x}$ and $ \G_{\y}$, and the MRRM P[$\x$,$\y$] shuffles less %(or equally) 
%than P[$\x$] and P[$\y$] {(Fig.~\ref{fig:intersection})}. 

The effects of intersection with the zero and unity elements are also easy to see. The unity is a neutral element that has no effect on the intersection $\model{\x, 1}=\model{\x}$, because adding a constant to the feature function output does not affect the partitioning of the networks at all. The zero is an absorbing element $\model{\x, 0} = \model{0}$, because adding extra information to the feature function that already contains the full network doesn't change anything. In fact, from set partitions we know that the intersection gives the greatest lower bound of the two partitions~\cite{Hrbacek1999I}.
%------------------------------------------------------------------------
\subsection{Taxonomy of temporal network MRRMs found in the literature}
\label{sec:classification}
%------------------------------------------------------------------------
\noindent
\new{Using the theory developed above, we now
%In this \new{subsection we 
describe and classify MRRMs found in the literature.}
% based on the theory developed in the preceding subsection.} % Sections~\ref{sec:theory} and \ref{sec:implementation} and the features defined in Section~\ref{sec:features}.
We also introduce several new MRRMs %, \new{some which} are event-shuffling versions of existing instant-event shufflings and vice-versa\new{, and some 
\new{which we will use in the walk-through example in Section~\ref{sec:spreading_example}}.
We \new{use our canonical naming convention (Def.~\ref{def:naming-MRRMs}) to} provide unambiguous names for the MRRMs\new{, %(see Def.~\ref{def:naming-MRRMs} below) 
and we order them hierarchically and describe which temporal network features they conserve.} % listed in Table~\ref{tab:features}.

\new{We separate the descriptions of event shufflings which shuffle %described here fall in two general classes depending on whether they are applied to 
the events in temporal networks while conserving their durations, 
 % (Def.~\ref{def:event_shuffling}), 
%(Def.~\ref{def:temporal_network}) 
and instant-event shufflings which shuffle instant-event temporal networks  
%(Def.~\ref{def:event_network}): shuffle events in temporal networks while conserving event durations; 
 %(Def.~\ref{def:instant-event_shuffling})}.
(Sec.~\ref{sec:implementation}).}
%shuffle instantaneous events in instant-event temporal networks. 
\new{(Note though} that it is possible to randomize the event durations of a temporal network by first discretizing its events and then shuffling this instant-event network using an instant-event shuffling\new{---see Supplementary Note 1}).
%Appendix~\ref{sec:instant-event_shuffling} \note{CLV: move this appendix to supplement?})}.)

%The MRRMs are described in detail below and their effects on network features are summarized in Table~\ref{tab:effects-contact}.
%Three figures additionally provide hierarchies of the MRRMs ordered by whether they randomize more or less (using Def.~\ref{def:comparability} and Proposition~\ref{the:partial_order}):
%Fig.~\ref{fig:RRMs-LS} shows the hierarchy of link shufflings (Def.~\ref{def:link_shuffling}); 
%Fig.~\ref{fig:RRMs-TS} shows the hierarchy of timeline shufflings (Def.~\ref{def:timeline_shuffling}), including sequence shufflings (Def.~\ref{def:sequence_shuffling}); 
%%Fig.~\ref{fig:RRMs-SeqS} shows the hierarchy of sequence shufflings (Def.~\ref{def:sequence_shuffling}); 
%Fig.~\ref{fig:RRMs-SnapS} shows the hierarchy of snapshot shufflings (Def.~\ref{def:snapshot_shuffling}). 

%This section is organized as follows.
%Subsection~\ref{sec:naming} first defines the naming convention we will use for MRRMs for temporal networks.
\new{The following subsections list and describe the different MRRMs.
For each MRRM we provide a canonical name as well as an informal name that may be easier to retain and use outside of the formal definition. 
We furthermore describe the features that the MRRM constrains, %, and we list other features that are conserved as a consequence. 
%(Table~\ref{tab:features} and Table~\ref{tab:constraint_levels} in Appendix~\ref{app:tables} provide detailed definitions of these) 
give references to the literature, and provide details on how the MRRM is implemented algorithmically.
A comprehensive table (Table~\ref{tab:features}) provides detailed definitions of temporal network features of interest, and a pair of tables at the end of this section (Tables~\ref{tab:effects-contact} and \ref{tab:effects-metadata}) show how each MRRM affects them.}
%Subsection~\ref{sec:instant-event_shuffling} describes how to apply instant-event shufflings to generate MRRMs that randomize event durations of temporal networks. 

\new{Subsection~\ref{sec:features_short} introduces temporal network features needed to describe and characterize the MRRMs described here.}
Subsection~\ref{sec:contact_shuffling} presents the \new{coarsest}, i.e. the maximally random, instant-event and event shufflings. 
Subsections~\ref{sec:link-shufflings}--\ref{sec:snapshot_shufflings} present the four restricted classes of shufflings defined in Section~\ref{sec:implementation}, namely link shufflings (\ref{sec:link-shufflings}), timeline shufflings (\ref{sec:timeline-shufflings}), sequence shufflings (\ref{sec:sequence_shufflings}), and snapshot shufflings (\ref{sec:snapshot_shufflings}). 
%Event and instant-event shufflings are presented in separate lists in each subsection. %, and the possibility of application of  an instant-event shuffling to temporal networks with contact durations is noted individually.
\new{Subsection~\ref{sec:intersections-LS-TS} describes MRRMs that are intersections of link and timeline shufflings and Subsection~\ref{sec:intersections-TS-Snaps} describes intersections of timeline and snapshot shufflings.}
%Subsection~\ref{sec:intersections} describes several MRRMs that can be classified as intersections of link and timeline shufflings or of timeline and snapshot shufflings.
%Subsection~\ref{sec:compositions} describes compositions of MRRMs found in the literature.
Subsection~\ref{sec:metadata} introduces and describes MRRMs that use additional metadata on nodes.
%Subsection~\ref{sec:other_models} finally surveys reference models that are not MRRMs. These are namely {\sl canonical} RRMs and reference models that do not maximize entropy. Of the latter we discuss in particular models based on bootstrapping. %, and finally reference models that directly randomize summary statistics (i.e.\ features) but do not take into account the networked structure of the system.

% \new{Tables~\ref{tab:effects-contact} and \ref{tab:effects-metadata} 
% %in Appendix~\ref{sec:effects-tables} 
% list the effe the MRRMs described here on a large selection of temporal network features,
% and Table~\ref{tab:features} 
% %and \ref{tab:constraint_levels} 
% provides detailed definitions of these features (see also Supplementary Supplementary Note 2 for more details on how they are constructed and ordered). }
% all these features, and Appendix \ref{sec:features} details how they are constructed and ordered \note{CLV: move appendix to supplement / remove?}. 
%  in-depth description and classification of these temporal network featu
%------------------------------------------------------------------------
\subsubsection{\new{Important temporal network features}}
\label{sec:features_short}
%------------------------------------------------------------------------
\noindent
% A typical goal when employing MRRMs is to investigate how given predefined features of a temporal network constrain other features of the network, or alternatively how they affect a dynamical process unfolding on the network.
Many MRRMs constrain features that can be described as an ordered sequence of lower-dimensional features, e.g., the degree sequence of a static graph, $\kstat$, is given by the sequence of the individual node degrees $k_i$.
\new{Other features of interest are those defined as functions of such a sequence. 
The most important in practice are empirical distributions of feature values, e.g.\ the degree distribution $p(\k)$,
and moments, e.g.\ the mean degree $\mu(\k)$. }
%the distribution individual features' values and their mean.} 
%Before getting into the detailed description of the different MRRMs, we first define three types of features that are often constrained by MRRMs in practice (Subsection~\ref{sec:features_short}). These are an ordered sequence of features, e.g.\ the sequence $\k$ of node degrees in the static network; 

%\new{
%Two tables found in Appendix~\ref{app:tables} together describe all features needed to define and characterize the MRRMs encountered in this review.
%Table~\ref{tab:features} lists and defines  elementary temporal network features 
%and Table~\ref{tab:constraint_levels} describes different distributions and moments that can be constructed as functions of a, possibly nested, sequence of such features.
%We here briefly describe how sequences, distributions, and means of features are constructed.}
%Appendix~\ref{sec:features} 
%\note{CLV: move appendix to supplement / remove?}.}

%\new{We define here such a sequence of features as well as two other statistics of a sequence of features that are often constrained by a MRRM: (1) the distribution of feature values, and (2) their mean.
%Appendix~\ref{sec:features} provides an in-depth discussion and classification of temporal network features and the tables found in Appendix~\ref{sec:feature-tables} provide complete descriptions of all features needed to characterize the MRRMs found in this review.}

We begin by introducing the ordered sequence of a collection of features. 
It retains both the values of the individual features and what they designate in the network.
A MRRM that constrains such a sequence thus produces reference networks with exactly the same values and configuration of these features as in the input network.
In order to make the notation simpler, and without loss of generality, we will assume that features returning values of multiple named entities, such as nodes or
links, return them as sequences that have an arbitrary but fixed order. 

\begin{definition}
{\sl Sequence of features.} 
\label{def:sequence}
A {\sl sequence} of features is a tuple $\x=(x_q)_{q\in\Q}$ of individual features ordered according to an arbitrary but fixed index $q\in\Q$. 
\end{definition}

The individual features $x_q$ may be any functions, e.g.\ scalar functions, sequences of other features (i.e.\ vectors), or graphs. 
When the individual features are vectors we will generally use boldface to indicate this, i.e.\ $\x_q = (x_q^r)_{r\in\R_q}$.
In this case we will refer to $\x=(\x_q)_{q\in\Q}$ as a {\sl sequence of sequences}.
%\new{We will refer to sequec}

Typically, each $x_q$ depends on a different part of the temporal network such as a node $i$, a link $\ij$, or a time $t$.  
We shall in the following use a subscript to index individual topological features 
(e.g.\ $x_i$ or $x_\ij$ for a feature of a single node or link, respectively)
and superscript for temporal ones (e.g.\ $x^t$ for a feature of a single snapshot). 
Such features are generally scalar and are assembled into a sequence that runs over all values of the index, i.e.\ over all nodes $i\in\N$, all links $\ij\in\L$, or all times $t\in\T$.

\begin{example}
The sequence of static degrees $\k=(k_i)_{i\in\N}$ is a paradigmatic example of a sequence of scalar features.
\end{example}

Individual features that depend both on topology and time are given both a subscript and a superscript index (e.g.\ $x_i^m$ or $x_\ij^m$, where $m$ refers to a \new{given} temporal ordering). 
Such features are generally assembled into a sequence of sequence that runs over both indices.

\begin{example}
An important example of a sequence of features that are themselves sequences of scalar features is the sequence of instantaneous node degrees 
% at each point in time. 
%It is given by 
% $\d = ((\dit)_{t\in\T})_{i\in\N} = ((\dit)_{i\in\N})_{t\in\T}$, 
$\d = (\d^t)_{t \in \T}$, where $\d^t = (\dit)_{i\in\N}$ is the degree sequence of the snapshot graph at time $t$, and 
where each instantaneous degree $\dit$ is the number of events that node $i$ partakes in at time $t$. %, i.e.\ in the snapshot graph , $\Gt=(\N,\E^t)$. 
Note that since the ranges of $i$ and $t$ do not depend on each other, we can reverse their order, $\d = (\d_i)_{i \in \N}$ with $\d_i = (\dit)_{t\in\T}$, which is formally the same feature since it imposes the same constraints.
\end{example}

\new{Table~\ref{tab:features} lists and defines a selection of elementary temporal network features. It serves as reference when reading the description of MRRMs in the taxonomy below.}

%-----------------------------------------------------------------------
\begin{table*}[p]
  \caption{{\bf \new{Elementary features of a temporal network}.} 
  Below, ``$(\cdot)$'' denotes a sequence, ``$\{\cdot\}$'' denotes a set, ``$|\cdot|$'' denotes the cardinality of a set, and ``$:$'' means {\sl for which} or {\sl such that}.} %, and ``$\exists$'' is short for {\sl there exists}.}
  \label{tab:features}
  \begin{threeparttable}
  \begin{tabular}{lll}
    \hline
    \hline
    {\sl Symbol} & {\sl Meaning of symbol} & {\sl Definition}\\
    \hline
%    \multicolumn{3}{l}{{\bf\em General definitions}}\\
    $[\tmin,\tmax]$ & Period of observation. \\
    $G$ & (Instant-event) temporal network. & $G=(\N,\C)$  (Def.~\ref{def:temporal_network})$^{\rm a}$ / $G=(\N,\E)$   (Def.~\ref{def:event_network})$^{\rm b}$ \\
    $\N$ & Set of nodes in $G$. & $\N = \{1,2,\ldots,N\}$\\
    $\C$ / $\E$ & Set of (instantaneous) events in $G$. & $\C = \{c_1,c_2,\ldots,c_C\}$ $^{\rm a}$ / $\E = \{e_1,e_2,\ldots,e_E\}$ $^{\rm b}$\\
    $c_q$ / $e_q$ & $q$th (instantaneous) event. & $c_q = (i_q,j_q,t_q,\tau_q)$  $^{\rm a}$ / $e_q = (i_q,j_q,t_q)$ $^{\rm b}$\\
%    $i, j$ & Node indices.\\
    $i_q, j_q$ & Indices for nodes partaking in the $q$th event.\\
    $t_q$ & Start time of the $q$th event.\\
    $\tau_q$ & Duration of the $q$th event.$^{\rm c}$\\
    $N$ & Number of nodes in $G$. & $N=|\N|$\\
    $C$ / $E$ & Number of  events in $G$. & $C=|\C|$ $^{\rm a}$ / $E=|\E|$ $^{\rm b}$\\
    \multicolumn{2}{l}{{\bf\em Link-timeline representation}}\\
    $\Gl$ & Link-timeline  network. & $\Gl = \left(\Gstat,\Thf\right)$ (Def.~\ref{def:link-timeline_graph}) \\
    $\Gstat$ & Static graph. & $\Gstat=(\N,\L)$ \\
    $\L$ & Links in $\Gstat$. & $\L=\{(i,j) : (i,j,t,\tau)\in\C\}$ $^{\rm a}$ / $\L=\{(i,j) : (i,j,t)\in\E\}$ $^{\rm b}$\\
    $L$ & Number of links in $\Gstat$. & $L=|\L|$\\
    $\N_i$ & Neighborhood of node $i$. & $\{j: (i,j)\in\L\}$\\
    $\Thf$ & Sequence of timelines. & $\Thf=(\Thij)_{\ij\in\L}$ \\
    $\Thij$ & Link timeline. & $\Thij=\left(\left(\ul^1,\taul^1\right), \left(\ul^2,\taul^2\right), \ldots, \left(\ul^\nl,\taul^\nl\right)\right)$ $^{\rm a}$  / %[Eq.~(\ref{eq:link-timeline})]
    \\
    && $\Thij=\left(\ul^1, \ul^2, \ldots, \ul^\wl\right)$ $^{\rm b}$   
    \\
    \multicolumn{2}{l}{{\bf\em Snapshot-sequence representation} $^{\rm d}$}\\
    $\Gss$ & Snapshot-graph sequence & $\Gss=(\T,\Gs)$  (Def.~\ref{def:snapshot-sequence})$^{\rm d}$\\
    $\T$ & Sequence of snapshot times. & $\T=(t_m)_{m=1}^T$ $^{\rm d}$ \\
    $\Gs$ & Sequence of snapshot graphs. & $\Gs = (\Gt)_{t\in\T}$ $^{\rm d}$ %[Eq.~(\ref{eq:snapshot-sequence})]
    \\
    $\Gt$ & Snapshot graph at time $t$. & $\Gt=(\N,\E^t)$ $^{\rm d}$\\
    $\E^t$ & Instantaneous events at time $t$. & $\E^t= \left\{(i,j) : (i,j,t)\in\E \right\}$ $^{\rm d}$  %[Eq.~(\ref{eq:snapshot})]
    \\
    \multicolumn{2}{l}{{\bf\em Topological-temporal (two-level) features}}\\
    $\ul^m$ & Event start time. % of $m$th event on timeline $\ij$. 
    & Start time of $m$th event in timeline $\Thij$ (Def.~\ref{def:link-timeline_graph}) %See definition of $\Thij$ above.
    \\
    $\taul^m$  & Event duration. & Duration of $m$th event in timeline $\Thij$ (Def.~\ref{def:link-timeline_graph})  $^{\rm c}$ %See definition of $\Thij$ above.
    \\
    $\Dtaul^m$ & Inter-event duration. & $\Dtaul^m=\ul^{m+1}-(\ul^{m}+\taul^m)$ $^{\rm a}$ / $\Dtaul^m=\ul^{m+1}-\ul^{m}$ $^{\rm b}$ \\
    $\ul^w$ & End time of last event on timeline. & $\ul^w=\ul^\nl+\taul^\nl$ $^{\rm a}$ / $\ul^w=\ul^\wl$ $^{\rm b}$\\
    $\dit$ & Instantaneous degree at time $t$. & $\dit=|\{j : (i,j,t',\tau)\in\C\ {\rm and}\ t'\leq t<t'+\tau\}|$ $^{\rm a}$ / $\dit=|\{j : (i,j)\in\E^t\}|$ $^{\rm b}$\\
    $\vi^m$ & Activity start time. &  Start time of $m$th interval of consecutive activity of node $i$. %(Example~\ref{ex:node-timeline}) %See definition of $\Thij$ above 
    \\
    {$\alphai^m$} & Activity duration. & Duration of $m$th interval of consecutive activity of node $i$.$^{\rm c}$ % (Example~\ref{ex:node-timeline}) %See definition of $\Phi_i$ above. 
    \\
    $\Dalphai^m$ & Inactivity duration. & $\Dalphai^m=\vi^{m+1}-(\vi^m+\alphai^m)$ $^{\rm a}$ / $\Dalphai^m=\vi^{m+1}-\vi^m$ $^{\rm b}$\\
    \multicolumn{3}{l}{{\bf\em Aggregated (one-level) features}}\\
    $\nl$ & Link event frequency. & $\nl=|\Thij|$ $^{\rm c}$\\
    $\wl$ & Link weight. & $\wl=\sum_{m=1}^\nl \taul^m$ $^{\rm a}$ / $\wl=|\Thij|$ $^{\rm b}$ 
    \\
    $\ai$ & Node activity. & $\ai=\sum_{j\in\N_i} \nl$  $^{\rm c}$\\
    $\si$ & Node strength. & $\si=\sum_{j\in\N_i} \wl$  \\
    $k_i$ & Node degree. & $k_i=|\N_i|$\\
%    $\Vstat$ & Number of active nodes. & $\Vstat = |\{i : \exists j\in\N : (i,j)\in\L\}|$\\
    $\Et$ & Cumulative activity at time $t$. & $\Et=|\E^t|$ \\ %=\sum_{i\in\N}\dit/2$\\
%    $E$ & Total number of events. & $E=\sum_{m=1}^T E^{t_m}$\\
    \multicolumn{2}{l}{{\bf\em Special features}}\\
    % $\Phi_i$ & Node timeline. & $\Phi_i=\left((\vi^1,\alphai^1),(\vi^2,\alphai^2),\ldots,(\vi^{n^a_i},\alphai^{n^a_i})\right)$ $^{\rm a}$ / % (see definitions of $\vi^m$ and $\alphai^m$ above) 
%    $\Phi_i=\left(\vi^1,\vi^2,\ldots,\vi^{n^a_i}\right)$ $^{\rm b}$ \\
    $\lc$ & Indicator of connectedness of $\Gstat$. & $\lc=1$ if $\Gstat$ is connected, $\lc=0$ elsewise. % (Example~\ref{ex:connectedness}).
    \\
    $\iso(\Gt)$ & Isomorphism class of $\Gt$. & Set of graphs obtained by all permutations of node indices in $\Gt$. %(Example~\ref{ex:isomorphism}).
    \\
   \new{$\lN(A^t)$} & \new{Indicator of activity at $t$.} & Indicator function for $A^t \in \mathbb{N}^+$, returning 0 if $A^t=0$ and 1 if $A^t\geq1$.
    \\
    \new{${\rm per}(\Thij)$} & \new{All periodic shifts of timeline $\Thij$.} &  \new{$[\Thij^{\Delta T}]_{\Delta T\in\T}$, where each $(t_\ij^m)' =  t_\ij^m + \Delta T \mod (\tmax - \tmin)$.} %  for all $m$.}
    \\
%    $\Vt$ & Global node activity. & $\Vt = |\{i : \exists j\in\N : (i,j,t)\in\E^t\}|$\\
%    \multicolumn{2}{l}{{\bf\em Features using meta-data}}\\
%    $\CMt$ & Group contact tensor.\\
%    $\gt$ & (Time-resolved) group appartenance of nodes.\\
    \hline
    \hline
  \end{tabular}
  \begin{tablenotes}
    \item[{\rm a}]  Definition for a temporal network with event durations. \\
    \item[{\rm b}]  Definition for an instant-event temporal network. \\
    \item[{\rm c}]  Only defined for a temporal network with event durations. \\
    \item[{\rm d}]  Only defined for an instant-event temporal network. \\
  \end{tablenotes}
  \end{threeparttable}
\end{table*}
%-----------------------------------------------------------------------

Instead of constraining an ordered sequence itself, many MRRMs constrain marginal distributions %or moments 
of a sequence.
A distribution of feature values %, denoted $p_{\Q'}(\x)$,
returns the number of times each possible value of individual features $x_q\in\x$ %or of $x_q^r\in\x_q$ 
appears in a measured sequence.
%$\x^*=\x(G^*)$.}
We formally define a distribution as a multiset, and we will in the following use the two terms interchangeably.

\begin{definition}
\label{def:unordered_set}
{\sl \new{Distribution} of feature values.} 
Given a sequence of features, $\x$, we can define a distribution for it.
%, where the individual features $\x_q$ may be scalar, sequences of scalars, or more general functions. 
% The distribution of feature values %, denoted $p_{\Q'}(\x)$, 
% returns the number of times each possible value of $\x_q$ %or of $x_q^r\in\x_q$ 
% appears in a measured sequence $\x^*=\x(G^*)$. 
A distribution is defined as the multiset \new{ $[x_q]_{q\in\Q}$} containing the values of all elements $x_q\in\x$ including duplicate values.
% \note{CLV: Should we change the description of the multiset to the following?:
% A distribution is defined as the multiset  $[\x_q]_{q\in\Q} = \{(\bm\xi, m(\bm\xi): \bm\xi\in\x\}$ containing all the different values $\bm\xi$ of the elements in $\x$ and their multiplicities $m(\bm\xi)$.}
%new{, $p_{\Q}(\x^*)=[\x^*_q]_{q\in\Q}$}. 
%We will use square brackets to denote the multiset and may then write $p_{\Q'}(\x^*)=[\x^*_q]_{q\in\Q'}$, where $\Q'$ is the index set that is marginalized over. 
\end{definition}

The individual features in a sequence may be scalar, sequences of scalars, or  other more general functions. 
This means that multiple types of distributions may be defined from a sequence of features $\x$, depending on the type of sequence.

The distribution constructed from a sequence of scalar features, $\x=(x_q)_{q\in\Q}$, is simply the multiset containing all individual feature values, $p(\x)=[x_q]_{q\in\Q}$.

\begin{example}
The sequence of static degrees, $\k$, is a sequence of scalar features, and we can construct the simple distribution $p(\k)=[k_i]_{i\in\N}$ from it.
\end{example}

From a sequence of vector valued features, $\x = (\x_q)_{q\in\Q}$ with 
$\x_q=(x_q^r)_{\R_q\in\Q}$, we may construct several different types of distributions by marginalizing over the inner or outer indices, or both.
We show the different types of distributions that can be obtained in the following example (see Supplementary Note 2 for formal definitions of each type of distribution).

% (1) by marginalizing over the outer index $\Q$ we obtain a distribution of {\sl local} sequences, $p_\Q(\x)=[\x_q]_{q\in\Q}$;
% (2) by marginalizing over the inner indices $\R_q$ we get a sequence of {\sl local} distributions, $\piq(\x)=(p(\x_q))_{q\in\Q}$ with $p(\x_q)=[x_q^r]_{r\in\R_q}$;
% finally, (3) by marginalizing over all indices we obtain the {\sl global} distribution, $p(\x^*)=\cup_{q\in\Q}p(\x_q) = [x_q^r]_{r\in\R_q, q\in\Q}$.

%\new{
%\begin{example}
%For the sequence of static degrees, we can construct the distribution  $p(\k)=[k_i]_{i\in\N}$.
%The sequence of instantaneous degrees, $\d=((\dit)_{t\in\T})_{i\in\N}$, defined as a sequence of sequences, can be used to construct different marginals:
%the global distribution $p(\d)=[\dit]_{t\in\T, i\in\N}$;
%the distribution of the sequences of each individual node's degree over time, $\pxi(\d)=[\d_i]_{i\in\N}$, 
%and the distribution of the degree sequences of each snapshot graph, $\pxt(\d)=[\d^t]_{t\in\T}$, 
%where $\d_i=(\dit)_{t\in\T}$ and $\d^t=(\dit)_{i\in\N}$;
%the sequence of the distributions of each node's instantaneous degrees, $\pii(\d)=(p(\d^t))_{i\in\N}$,
%the sequence of the degree distribution in each snapshot, $\pit(\d)=(p(\d_i))_{t\in\T}$, 
%where $p(\d^t)=[\dit]_{i\in\N}$ and $p(\d_i)=[\dit]_{t\in\T}$.
%\end{example}
%}

\new{
\begin{example}
The sequence of instantaneous degrees, $\d=((\dit)_{i\in\N})_{t\in\T}$, is a sequence of sequences of scalar features. 
We can thus marginalize over the two indices $i$ and $t$ in different ways to construct different types of distributions:
\begin{itemize}
  \item By marginalizing over both $i$ and $t$, we obtain the {\sl global distribution} $p(\d) = \cup_{t\in\T}[\dit]_{i\in\N}p(\x_q) = [\dit]_{t\in\T, i\in\N}$.
  \item  By marginalizing over the outer index $t$, we get the distribution of the degree sequences of each snapshot graph, $\pxt(\d)=[\d^t]_{t\in\T}$, where $\d^t=(\dit)_{i\in\N}$.
  \item Marginalizing over the inner index $i$ gives the temporal sequence of the nodes' instantaneous  degree distributions, $\pit(\d)=(p(\d^t))_{t\in\T}$, where $p(\d^t)=[\dit]_{i\in\N}$ is the instantaneous degree distribution in the $t$th snapshot.
\end{itemize}
While inverting the order of the indices in the definition of the sequence $\d$ (i.e.\ letting $\d=((\dit)_{t\in\T})_{i\in\N}$) leads to exactly the same feature, it will lead to different distributions when marginalizing over the indices:
\begin{itemize}
  \item By marginalizing over $t$ with $t$ as the inner index in $\d=((\dit)_{t\in\T})_{i\in\N}$ we get the sequence of the temporal distribution of each node's instantaneous degree, $\pii(\d)=(p(\d^i))_{i\in\N}$, with $p(\d^i)=[\dit]_{t\in\T}$.
  \item Finally, marginalizing over $i$ with $i$ as the outer index leads to the distribution of the temporal sequences of the individual node's degree, $\pxi(\d)=[\d_i]_{i\in\N}$, with $\d_i=(\dit)_{t\in\T}$.
\end{itemize}
\end{example}
}

\new{Finally, many MRRMs conserve mean values of network features.}

\new{
\begin{definition}
{\sl Mean of a sequence of features.}
\label{def:mean}
The mean $\mu(\x)$ of a sequence of features is defined as the average over all individual scalar elements in $\x$. 
\end{definition}
}

\new{From a sequence of scalars, the mean is naturally given as their average value: $\mu(\x)=\sum_{q\in\Q} x_q/Q$, where Q is the number of elements in $\x$.
For a sequence of sequences of scalars, 
we may construct either the {\sl global} mean, providing an average over all scalar elements: $\mu(\x)=\sum_{q\in\Q}\sum_{r\in\R_q}x_q^r/(\sum_{q\in\Q}R_q)$, where $R_q$ is the number of elements in $\x_q$;
or the sequence of local means, providing the averages over each local sequence $\x_q\in\x$: $\muq(\x)=(\mu(\x_q))_{q\in\Q}$ with $\mu(\x_q)=\sum_{r\in\R_q} x_q^r/R_q$.}

%\note{CLV: add example?}

\new{The distributions and moments are all functions of a sequence of features, so they are coarser than the sequence itself (Proposition~\ref{prop:comparability-functional_relation}). 
This means that a MRRM that constrains a distribution or the mean of a collection of feature values randomizes more than a MRRM that constrains their ordered sequence.}

\new{
%The sequences and distributions of features and are described in more detail in Appendix~\ref{sec:features} \note{CLV: move to supplement / remove?} and 
A detailed list of the general form that different distributions and moments take for features of nodes, links, and snapshots is given in Supplementary Table~\ref{tab:constraint_levels}. %(Appendix~\ref{app:tables}). 
%Other types of features are also defined there, notably moments (e.g.\ the mean) of a sequence of features.
Supplementary Note 2 additionally provides a detailed description of how the all the distributions and moments listed in Table~\ref{tab:constraint_levels} are constructed and discusses how to order them using the finer/coarser relation of Def.~\ref{def:comparability}.
}
 
%\new{The distributions are all functions of a sequence of features, so they are coarser than the sequence itself (Proposition~\ref{prop:comparability-functional_relation}). 
%%The mean $\mu(\x)$ and the distributions of local feature values, $p(\x)$ or $\pxq(\x)$, are functions of the sequence of features, $\x$, so as features they are coarser than the sequence itself (Proposition~\ref{prop:comparability-functional_relation}). 
%This means that a MRRM that constrains a distribution of feature values randomizes more than a MRRM that constrains their ordered sequence.
%Similarly, the mean value of a sequence of features is coarser than the distribution. 
%%In the same manner the mean $\mu(\x)$ is a function of the distribution over $\x$. 
%%We can in general establish the following hierarchy: $\mu(\x) \geq p(\x) \geq \pxq(\x) \geq \x$.
%Appendix~\ref{sec:features} uses this to build hierarchies of the different types of distributions and moments constructed from a sequence of features.} 

We are now ready to build a taxonomy of MRRMs which rigorously characterizes and orders them.
We provide for each MRRM informal definitions of the principal features that it constrains and refer to Table~\ref{tab:features} for detailed definitions of the features. 
The effects of each shuffling on a wide selection of other temporal network features are shown in Table~\ref{tab:effects-contact}.
We additionally provide details and a graphical illustration of the typical algorithmic implementation of each shuffling method.

%%------------------------------------------------------------------------------------------------------------------
%\subsubsWection{\new{How to read the description of the shufflings methods}}
%%------------------------------------------------------------------------------------------------------------------

%------------------------------------------------------------------------------------------------------------------
\subsubsection{The basic instant-event and event shufflings}
\label{sec:contact_shuffling}
%------------------------------------------------------------------------------------------------------------------
\noindent
We first present the coarsest (i.e.\ the most random) instant-event and event shufflings  possible. 
% \note{CLV: add features conserved indirectly by all shufflings?}
% \new{These both conserve the average node strength, $\mu(\s)$, the average number of events per time-lapse, $\mu(\Ebf)$, and the mean instantaneous node degree, $\mu(\d)$ (see Tables~\ref{tab:features} and \ref{tab:constraint_levels} in Appendix~\ref{app:tables} for detailed definitions of  features). 
%\new{They randomize most temporal network features %(see Supplementary Table~\ref{tab:effects-contact} %in Appendix~\ref{app:tables} 
%for a comprehensive overview ). 
%Table~\ref{tab:features} provides detailed definitions of elementary features and Table~\ref{tab:constraint_levels} of distributions and moments constructed from these.}

\paragraph{Instant-event shuffling}

\begin{itemize}
  \item[{P[$E$]}.] 
  \new{{\bf Common name:} {\sl Instant-event shuffling}.
  {\bf Features constrained:} The number of instantaneous events, $E$.
%   {\bf Other features conserved:} $\mu(\s)$, $\mu(\Ebf)$, $\mu(\d)$.
  {\bf Reference:} \cite{Holme2016T}.}
  \vspace{0.2cm}\\
%    \begin{center} 
%      \note{add illustration?}
%      %\includegraphics[height=0.04\textheight]{P__pTheta.pdf}
%    \end{center}
  \new{P[$E$] draws $i$, $j$, and $t$ at random without replacement for each instantaneous event $(i,j,t)\in\E$.}
  %shuffles the instantaneous events at random without any constraints\new{, i.e.\ it randomizes $i$, $j$, and $t$ in each instantaneous event $(i,j,t)\in\E$.}
%  P[$1$] is thus the coarsest instant-event shuffling possible, and more generally the coarsest MRRM in the space of all MRRMs that conserve the nodes $\N$, the recording interval $[\tmin,\tmax]$ and the number of events $E$. 
%P[$1$] was employed in Ref.~\cite{Holme2016T}.
\end{itemize}

\paragraph{Event shuffling}

\begin{itemize}
  \item[{P[$p(\tauf)$]}.] 
  \new{{\bf Common name:} {\sl Event shuffling}.
   {\bf Features constrained:} 
     the distribution (Def.~\ref{def:unordered_set}) of event durations, $p(\tau)=[\tau_q]_{q=1}^C$. 
%   {\bf Other features conserved:} $\mu(\s)$, $\mu(\Ebf)$, $\mu(\d)$, $\mu(\a)$.
   {\bf Reference:} %Section~\ref{sec:walktrough_example}.
   Supplementary Note~3.
  \vspace{0.2cm}\\
%    \begin{center} 
%      \note{add illustration?}
%      %\includegraphics[height=0.04\textheight]{P__pTheta.pdf}
%    \end{center}
  %P[$p(\tauf)$] constrains only the set of event durations but randomizes everything else in the network. 
  P[$p(\tauf)$] draws $i$, $j$, and $t$ at random without replacement for each  event $(i,j,t,\tau)\in\C$.
  It conserves the events durations but not their order, which is equivalent to constraining their distribution $p(\tauf)$.}
  %P[$p(\tauf)$] randomizes $i$, $j$, and $t$ in each event $(i,j,t,\tau)\in\C$.}
%  It is the coarsest event shuffling, and we thus refer to it as {\sl event shuffling}. }
\end{itemize}

%------------------------------------------------------------------------------------------------------------------
\subsubsection{Link shufflings}
\label{sec:link-shufflings}
%------------------------------------------------------------------------------------------------------------------
\noindent
Link shufflings %(Def.~\ref{def:link_shuffling}) 
alter the aggregated network topology but conserve temporal structure locally on each link. 
%Link shufflings always conserve $\pxl(\Thf)$ and all coarser features (see Fig.~\ref{fig:dependency_diagram}), 
%\note{Add features conserved indirectly?} 
% \new{namely: 
% the mean node strength, static degree, and instantaneous degree, $\mu(\s)$, $\mu(\d)$, and $\mu(\k)$; 
% the number of links in the static graph, $L$;
% the distributions of link weights (number of events and cumulative duration), $p(\n)$ and $p(\w)$; 
% number of events per time-lapse, $\Ebf$; 
% the multisets of the first and last events and the sequences of event and inter-event durations in the timelines, $\pxl(\t^1)$, $\pxl(\t^w)$, $\pxl(\tauf)$, and $\pxl(\Dtauf)$ 
% (see Tables~\ref{tab:features} and \ref{tab:constraint_levels} in Appendix~\ref{app:tables} for detailed definitions of  features).} 
Different time-aggregated features may be constrained or randomized depending on the model. 
% \new{We provide informal definitions of the principal features that are constrained by each shuffling method in the description 
% (see Table~\ref{tab:features} for detailed definitions). 
% The effects of each shuffling on a wide selection of other temporal network features are shown in Table~\ref{tab:effects-contact}.}
%\new{as described below (see also Table~\ref{tab:effects-contact}).
% We provide informal definitions of the principal features that are constrained by each shuffling method, and we list other features that are conserved as a consequence.
% Table~\ref{tab:features} provides detailed definitions of elementary features and Table~\ref{tab:constraint_levels} of distributions and moments constructed from these.} 
%Link shufflings are compatible with timeline shufflings (Subsec.~\ref{sec:timeline-shufflings}) and with sequence shufflings (Subsec.~\ref{sec:sequence_shufflings}) (Propositions~\ref{prop:link-timeline_independent} and \ref{prop:link-sequence_independent}).
%They may thus be combined in composition with these to generate new MRRMs from the ones listed here.

%All link shufflings \new{are} event shufflings since they automatically constrain the contact durations (i.e.\ since $p(\tauf)\geq\pxl(\Thf)$). 
Link shufflings are defined and implemented exactly the same way for temporal networks with and without event durations. We thus do not need to distinguish between instant-event and event shuffling versions of these.
They are ordered hierarchically in the Hasse diagram shown in Fig.~\ref{fig:RRMs}(a).

%\paragraph{Event shufflings}

\begin{itemize}
  \item[{P[$\pxl(\Thf)$]}.] 
    \new{{\bf Common name:} {\sl Link shuffling.}
    {\bf Features constrained:} 
      the distribution (Def.~\ref{def:unordered_set}) of timelines, $\pxl(\Thf)=[\Thij]_{\ij\in\L}$.
    % {\bf Other features conserved:} 
    %   $\mu(\s)$, $\mu(\d)$, $\mu(\k)$, $L$, $p(\n)$, $p(\w)$, $\Ebf$, $\pxl(\t^1)$, $\pxl(\t^w)$, $\pxl(\tauf)$, $\pxl(\Dtauf)$.
    {\bf References:} \cite{Karimi2013T} ({\sl Erd\H{o}s-R\'{e}nyi model}); Supplementary Note~3.}
    \begin{figure}[H]
        \begin{center} 
          \includegraphics[height=0.04\textheight]{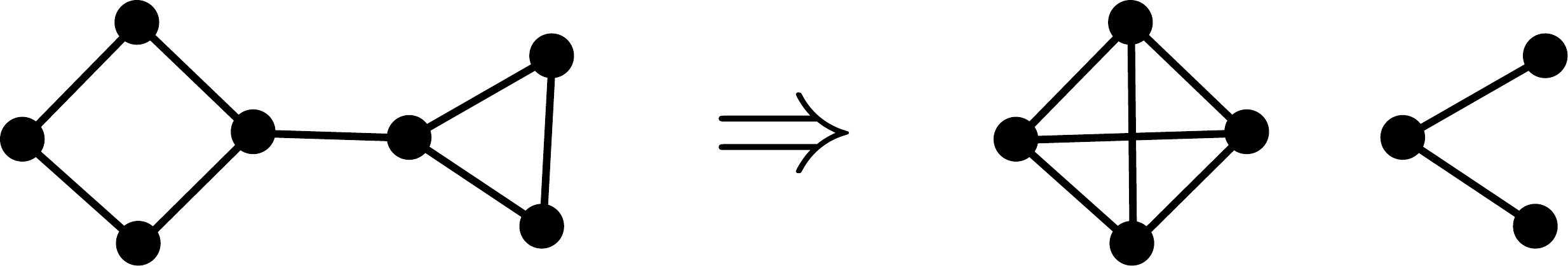}
        \end{center}
      \vspace{-8pt}
      \caption{
        P[$\pxl(\Thf)$] shuffles the links and associated timelines $\Thij\in\Thf$ between all node pairs $(i,j)$ without any constraints on the static network structure (i.e.\ corresponding to the Erd\H{o}s-R\'{e}nyi (ER)~\cite{Erdos1960O} model).  
        All nodes and links in the static network are equivalent and are shown in the same color. }
    \end{figure}
%    (ii) redistributes the timelines $\Thij\in\Thf$ on the new links at random.
    % in the illustration.
        % shuffles the links and associated timelines between all node pairs $(i,j)$ without any constraints on the static network structure. 
    % \new{This corresponds to drawing $\Gstat$ uniformly from the set of all graphs with the same nodes $\N$ and number of links $\Estat$ as the original network (equivalent to Erd\H{o}s-R\'{e}nyi (ER)~\cite{Erdos1960O} random graphs)} 
    % %This corresponds to drawing $\Gstat$ uniformly from the ensemble of all Erd\H{o}s-R\'{e}nyi (ER)~\cite{Erdos1960O} random graphs with the same nodes $\N$ and number of links $\Estat$ as the original network 
    % and redistributing the timelines $\Thij\in\pxl(\Thf)$ on the new links at random. 
%  {P[$\pxl(\Thf)$]} was employed in Ref.~\cite{Karimi2013T} where it was referred to as the {\sl Erd\H{o}s-R\'{e}nyi model}. 
%  \new{P[$\pxl(\Thf)$]} is the coarsest possible link shuffling, %(i.e.\ the most random). We 
%  so we may simply refer to it as {\sl link shuffling}.
  %
  \item[{P[$\lc,\pxl(\Thf)$]}.]  
    \new{{\bf Common name:} {\sl Connected link shuffling.}
     {\bf Features constrained:} 
       the connectedness $\lc(\Gstat)$ of the static graph $\Gstat=(\N,\L)$; 
       the distribution (Def.~\ref{def:unordered_set}) of timelines, $\pxl(\Thf)=[\Thij]_{\ij\in\L}$.
    % {\bf Other features conserved:} 
    %   $\mu(\s)$, $\mu(\d)$, $\mu(\k)$, $L$, $p(\n)$, $p(\w)$, $\Ebf$, $\pxl(\t^1)$, $\pxl(\t^w)$, $\pxl(\tauf)$, $\pxl(\Dtauf)$.
    {\bf References:} \cite{Takaguchi2012I} ({\sl rewiring}); \cite{Kivela2012M} ({\sl random network}).}
    \begin{figure}[H]
        \begin{center} 
          \includegraphics[height=0.04\textheight]{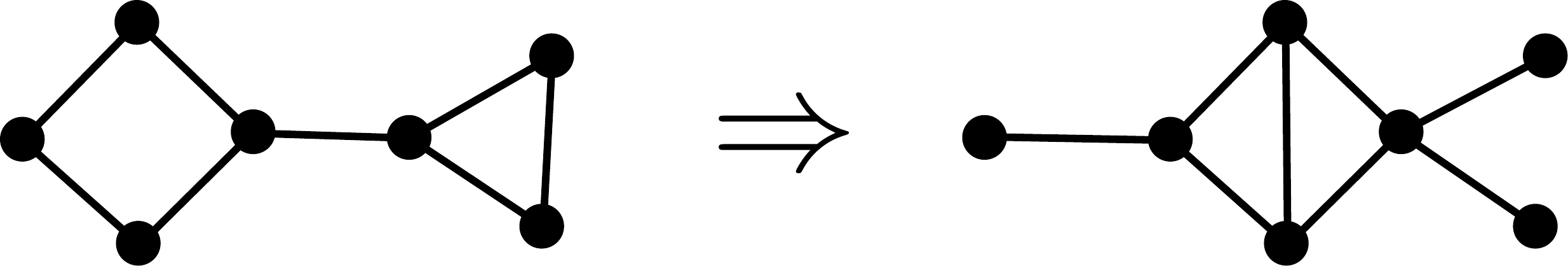}
        \end{center}
        \vspace{-8pt}
        \caption{
            P[$\lc,\pxl(\Thf)$] generates randomized networks in the same manner as P[$\pxl(\Thf)$], with the additional constraint that the static graph $\Gstat$ of the sampled networks must be connected if it was in the input network.
            All nodes and links in the static network are equivalent and are shown in the same color. }
        \end{figure}
        % in the illustration.
    
    % adds the additional constraint to {P[$\pxl(\Thf)$]} that the static network of the sampled networks must be connected if it was in the input network and disconnected if it was not. 
 % P[$\lc,\pxl(\Thf)$] was called {\sl rewiring} in Ref.~\cite{Takaguchi2012I} and {\sl random network} in Ref.~\cite{Kivela2012M}. 
  %
  \item[{P[$\kstat,\pxl(\Thf)$]}.] 
    \new{{\bf Common name:} {\sl Degree-constrained link shuffling.}
     {\bf Features constrained:} 
       the static degree sequence, $\k=(k_i)_{i\in\N}$; 
       the distribution (Def.~\ref{def:unordered_set}) of timelines, $\pxl(\Thf)=[\Thij]_{\ij\in\L}$.
    % {\bf Other features conserved:} 
    %   $\mu(\s)$, $\mu(\d)$, $\mu(\k)$, $L$, $p(\n)$, $p(\w)$, $\Ebf$, $\pxl(\t^1)$, $\pxl(\t^w)$, $\pxl(\tauf)$, $\pxl(\Dtauf)$.
    {\bf References:} \cite{Holme2005N,Holme2012T,Li2017T} ({\sl randomized edges}); \cite{Holme2015M} ({\sl random link shuffling}); \cite{Delvenne2015D} ({\sl randomized structure}); Supplementary Note~3.}
    \begin{figure}[H]
        \begin{center} 
          \includegraphics[height=0.04\textheight]{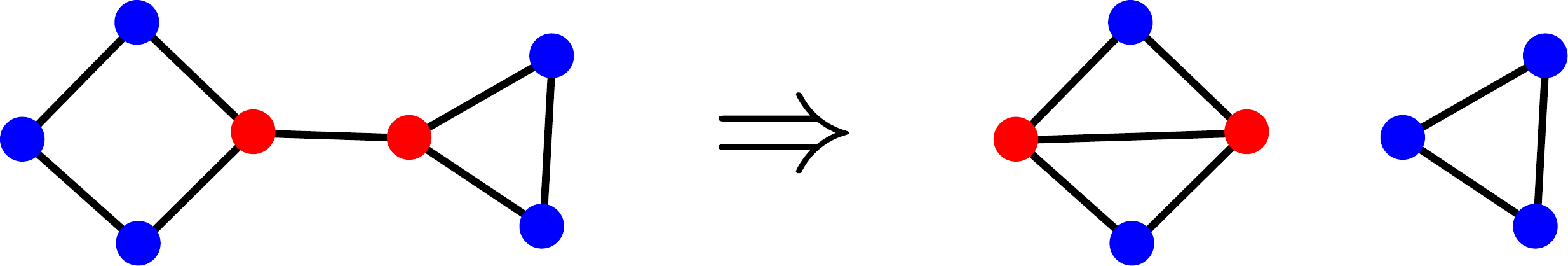}
        \end{center}
      \vspace{-8pt}
      \caption{
        P[$\kstat,\pxl(\Thf)$] shuffles the links and associated timelines $\Thij\in\Thf$ between all node pairs $(i,j)$ while \new{constraining the sequence of degrees of the nodes in the static network, $\kstat$} 
        (typically implemented using the algorithm of Maslov and Sneppen~\cite{Maslov2002S} or by using a stub matching algorithm~\cite{Newman2003S, Fosdick2018C}).
        Nodes are colored by their degree in the static graph $\Gstat$, which is conserved by the shuffling.}
    \end{figure}
%    (ii) redistributes the timelines $\Thij\in\Thf$ on the new links at random.
    % shuffles the links and associated timelines between all node pairs $(i,j)$ while \new{constraining the sequence of degrees of the nodes in the static network, $\kstat$}. The procedure is typically implemented using the algorithm of Maslov and Sneppen~\cite{Maslov2002S} or by using the configuration model~\cite{Newman2003S, Fosdick2018C}. 
%      P[$\kstat,\pxl(\Thf)$] was called {\sl randomized edges} in Refs.~\cite{Holme2005N,Holme2012T,Li2017T} {\sl random link shuffling} in Ref.~\cite{Holme2015M}, and {\sl randomized structure} in Ref.~\cite{Delvenne2015D}. 
      %
  \item[{P[$\k,\lc,\pxl(\Thf)$]}.] 
    \new{{\bf Common name:} {\sl Connected degree-constrained link shuffling.}
     {\bf Features constrained:} 
       the connectedness $\lc(\Gstat)$ of the static graph $\Gstat=(\N,\L)$; 
       the static degree sequence, $\k=(k_i)_{i\in\N}$; 
       the distribution (Def.~\ref{def:unordered_set}) of timelines, $\pxl(\Thf)=[\Thij]_{\ij\in\L}$.
    % {\bf Other features conserved:} 
    %   $\mu(\s)$, $\mu(\d)$, $\mu(\k)$, $L$, $p(\n)$, $p(\w)$, $\Ebf$, $\pxl(\t^1)$, $\pxl(\t^w)$, $\pxl(\tauf)$, $\pxl(\Dtauf)$.
    {\bf References:} \cite{Kivela2012M} ({\sl configuration model}).}
    \begin{figure}[H]
        \begin{center} 
          \includegraphics[height=0.045\textheight]{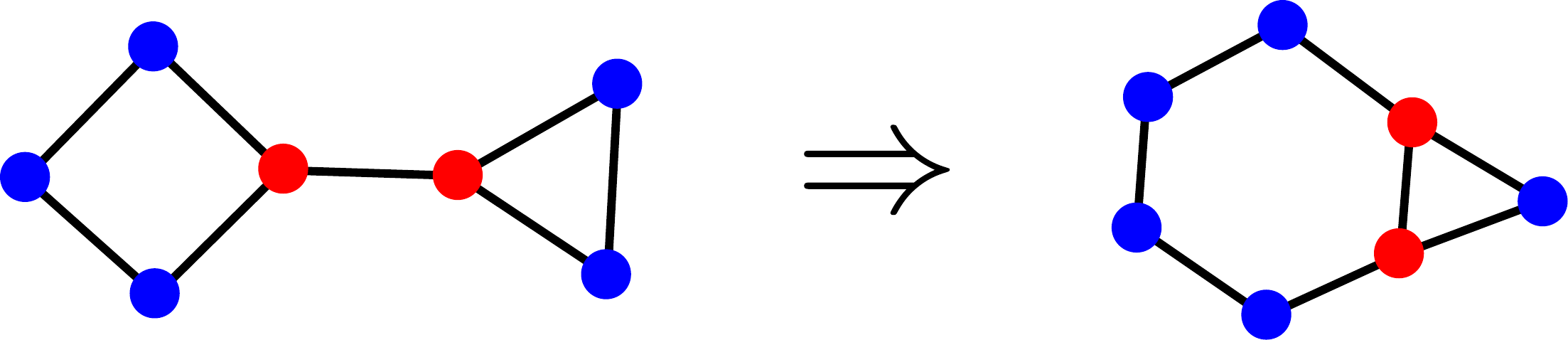}
        \end{center}
      \vspace{-8pt}
      \caption{
        P[$\k,\lc,\pxl(\Thf)$] generates randomized networks in the same manner as P[$\k,\pxl(\Thf)$], with the additional constraint that the static network of the sampled networks must be connected if it was in the input network.
        Nodes are colored by their degree in the static graph $\Gstat$, which is conserved by the shuffling.}
    \end{figure}
%    adds the additional constraint to {P[$\k,\pxl(\Thf)$]} that the static network must be connected. 
    %P[$\k,\lc,\pxl(\Thf)$] was called {\sl configuration model} in Ref.~\cite{Kivela2012M}. 
\end{itemize}

%------------------------------------------------------------------------------------------------------------------
\subsubsection{Timeline shufflings}
\label{sec:timeline-shufflings}
%------------------------------------------------------------------------------------------------------------------
\noindent
Timeline shufflings %(Def.~\ref{def:timeline_shuffling}) 
randomize the individual timelines $\Thij$ without changing the topology of the aggregated network. 
% Thus, they always constrain $\Gstat$ and all coarser features (see Fig.~\ref{fig:dependency_diagram}), 
% \note{Add features conserved indirectly?} 
% \new{notably: 
% the static degree sequence, $\k$; 
% the number of links in the static graph, $L$;
% the mean number of events at each time, $\mu(\Ebf)$;
% they also conserve the mean link weight, node strength, and instantaneous degree, $\mu(\w)$, $\mu(\s)$, $\mu(\d)$, and $\mu(\k)$;
% (see Tables~\ref{tab:features} and \ref{tab:constraint_levels} in Appendix~\ref{app:tables} for detailed definitions of  features).}
They typically randomize temporal features of both links and nodes to different extents as described below. 
%(see also Table~\ref{tab:effects-contact}). 
% We provide informal definitions of the principal features that are constrained by each shuffling method in the description 
% (see Table~\ref{tab:features} for detailed definitions). 
% The effects of each shuffling on a wide selection of other temporal network features are shown in Table~\ref{tab:effects-contact}.
% We provide informal definitions of the principal features that are constrained by each shuffling method, and we list other features that are conserved as a consequence.
% Table~\ref{tab:features} provides detailed definitions of elementary features and Table~\ref{tab:constraint_levels} of distributions and moments constructed from these.} 
%Timeline shufflings are compatible with link shufflings (Subsec.~\ref{sec:link-shufflings}) (Proposition~\ref{prop:link-timeline_independent}), and they may thus be applied together in composition to generate new MRRMs. 

The timeline shufflings listed below are ordered hierarchically in the Hasse diagram shown in Fig.~\ref{fig:RRMs}(b).

\paragraph{Instant-event shufflings}

\begin{itemize}
  \item[{P[$\L,E$]}.] 
    \new{{\bf Common name:} {\sl Timeline shuffling.}
     {\bf Features constrained:} 
       the static graph $\Gstat=(\N,\L)$;
       the number of instantaneous events, $E$.
    % {\bf Other features conserved:} 
    %   $\k$, $L$, $\mu(\w)$, $\mu(\s)$, $\mu(\Ebf)$, $\mu(\d)$.
    {\bf References:} \cite{Holme2005N,Holme2012T} ({\sl random(ized) contacts}).} 
    \begin{figure}[H]
        \begin{center} 
          \includegraphics[height=0.04\textheight]{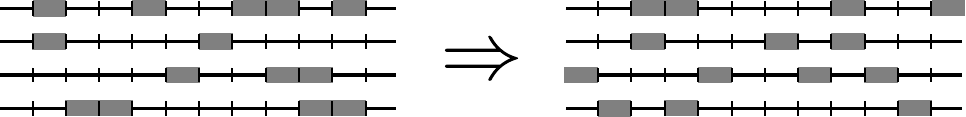}
        \end{center}
          \vspace{-8pt}
        \caption{
          P[$\L,E$] redistributes the instantaneous events completely at random between the existing timelines.
          Since all events are equivalent, they are marked in the same color (grey). 
          %in the illustration.
          }
    \end{figure}
    %, while conserving {the static graph, i.e. $\L$}. 
    %It was called {\sl random(ized) contacts} in Refs.~\cite{Holme2005N,Holme2012T}. 
%    It is the coarsest possible timeline shuffling and we may thus refer to it simply as the {\sl timeline shuffling}.
  %
  \item[{P[$\w$]}.] 
    \new{{\bf Common name:} {\sl Weight-constrained timeline shuffling.}
    {\bf Features constrained:} 
       the sequence of link weights, $\w=(\wij)_{\ij\in\L}$ (numbers of instantaneous events per link). 
       %, which constrains their order and thus also $\L$ and $\Gstat=(\N,\L)$.
    % {\bf Other features conserved:} 
    %   $\k$, $L$, $\s$, $\mu(\Ebf)$ $\mu(\d)$.
    {\bf References:} \cite{Holme2005N, Pan2011P, Holme2012T, Posfai2014S, Holme2015M} ({\sl random time(s)}); \cite{Kivela2012M} ({\sl uniformly random times}); \cite{Bajardi2011D} ({\sl temporal mixed edges}); \cite{Takaguchi2012I} ({\sl poissonized inter-event intervals}); \cite{Starnini2012R} ({\sl SRan}); \cite{Perotti2014T, Holme2016T}; Section~\ref{sec:spreading_example}.}
    \begin{figure}[H]
        \begin{center} 
          \includegraphics[height=0.04\textheight]{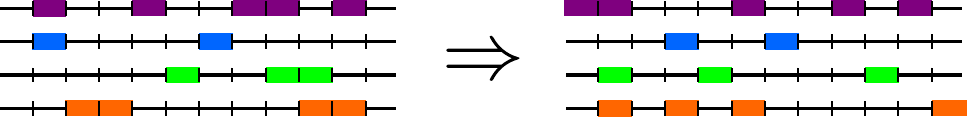}
        \end{center}
        \caption{
        P[$\w$] randomizes the timestamps of the instantaneous events inside each individual timeline.
    	Events of the same color stay on the same timeline after shuffling.}
    \end{figure}
  \item[{P[$\piij(\Dtauf)$,$\t^1$]}.] 
    \new{{\bf Common name:} {\sl Inter-event shuffling.}
    {\bf Features constrained:}
      the sequence of local distributions of inter-event durations on each link, $\piij(\Dtauf) = ([\Dtau_\ij^m]_{m\in\Ml})_{\ij\in\L}$; %, where $\Ml=1,2,\ldots,M_\ij$ indexes the inter-event durations on the link $\ij$;
	  the sequence of times of the first event on each link,
	  $\t^1=(t_\ij^1)_{\ij\in\L}$.
%      since $\piij(\Dtauf)\leq\w$, the static graph and link weights, $\w=(\wij)_{\ij\in\L}$.
    % {\bf Other features conserved:} 
    %   $\k$, $L$, $\a$, $\s$, $\n$, $\w$, $\mu(\Ebf)$, $\mu(\d)$, $\t^w$, $\muij(\tauf)$.
    {\bf References:} \cite{Takaguchi2012I,Takaguchi2013I} ({\sl shuffled inter-event intervals}); Section~\ref{sec:spreading_example}. }
    \begin{figure}[H]
        \begin{center} 
          \includegraphics[height=0.04\textheight]{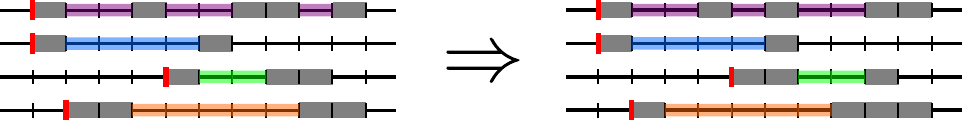}
        \end{center}
      \vspace{-8pt}
      \caption{
    	P[$\piij(\Dtauf)$,$\t^1$] shuffles the inter-event durations between the instantaneous events on each link while keeping the times of the first event on each link fixed. 
    	%In the illustration, 
    	Inter-event durations on the same link are marked in the same color and red vertical lines mark the start times of the first event on each link. Both are conserved by the shuffling.}
    \end{figure}
	%P[$\pit(\Dtauf)$,$\t^1$] was named {\sl shuffled IEIs} in Refs.~\cite{Takaguchi2012I,Takaguchi2013I}. 
 \end{itemize}

\paragraph{Event shufflings}

\begin{itemize}
  \item[{P[$\L$,$p(\tauf)$]}] 
  \new{{\bf Common name:} {\sl Timeline  shuffling.}
    {\bf Features constrained:}
      the static graph $\Gstat=(\N,\L)$;
      the distribution (Def.~\ref{def:unordered_set}) of event durations, $p(\tau)=[\tau_q]_{q=1}^C$. 
    % {\bf Other features conserved:}  
    %   $\k$, $L$, $\mu(\w)$, $\mu(\n)$, $\mu(\s)$, $\mu(\Ebf)$, $\mu(\d)$.
    {\bf Reference:} Supplementary Note~3.}
    \begin{figure}[H]
        \begin{center} 
          \includegraphics[height=0.04\textheight]{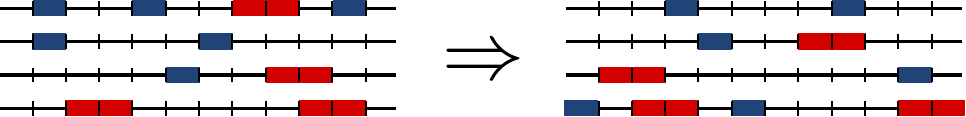}
        \end{center}      
      \vspace{-8pt}
      \caption{
        P[$\L$,$p(\tauf)$] constrains the static network structure $\Gstat$ and otherwise shuffles the events completely at random between all timelines. 
        %In the illustration, 
        Colors mark the events' durations, which are conserved by the shuffling. }
    \end{figure}
  %, while . %The intercontact durations are randomized and asymptotically obeys exponential distributions. 
%  P[$\L$,$p(\tauf)$] is the coarsest timeline event shuffling, % which constrains event durations, 
%  and is the event shuffling equivalent of the instant-event shuffling P[$\L,E$]. We may consequently refer to P[$\L,p(\tauf)$] as the {\sl timeline (event) shuffling}.
  %
  \item[{P[$\piij(\tauf)$]}] 
    \new{{\bf Common name:} {\sl Local timeline shuffling.}
    {\bf Features constrained:}
       the sequence of local distributions of event durations on each link, $\piij(\tauf) = ([\tau_\ij^m]_{m\in\Ml})_{\ij\in\L}$.
    % {\bf Other features conserved:} 
    %   $\k$, $L$, $\a$, $\s$, $\n$, $\w$, $\mu(\Ebf)$, $\mu(\d)$.
    {\bf Reference:} Supplementary Note~3.}
    \begin{figure}[H]
        \begin{center} 
          \includegraphics[height=0.04\textheight]{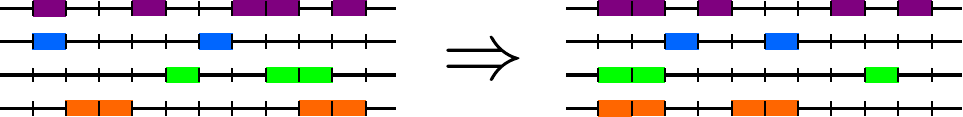}
        \end{center}
      \vspace{-8pt}
      \caption{
        P[$\piij(\tauf)$] redistributes the events uniformly inside each timeline, but not in-between them. 
        %In the illustration, t
        The events are colored by the timeline they belong to.}
    \end{figure}
%The inter-event durations are randomized and asymptotically follow exponential distributions. 
%  P[$\piij(\tauf)$] is an event-shuffling version of the instant-event shuffling P[$\w$].
  %
  \item[{P[$\piij(\tauf)$,$\t^1$,$\t^w$]}] 
    \new{{\bf Common name:} {\sl Activity-constrained timeline shuffling.}
    {\bf Features constrained:}
       the sequence of local distributions of event durations on each link, $\piij(\tauf) = ([\tau_\ij^m]_{m\in\Ml})_{\ij\in\L}$.
	  the sequences of times of the first and last events on each link,
	  $\t^1=(t_\ij^1)_{\ij\in\L}$ and $\t^w=(t_\ij^w)_{\ij\in\L}$, respectively.
    % {\bf Other features conserved:} 
    %   $\k$, $L$, $\a$, $\s$, $\n$, $\w$, $\mu(\Ebf)$, $\mu(\d)$, $\muij(\Dtauf)$.
    {\bf Reference:}  Supplementary Note~3.}
    \begin{figure}[H]
        \begin{center} 
          \includegraphics[height=0.04\textheight]{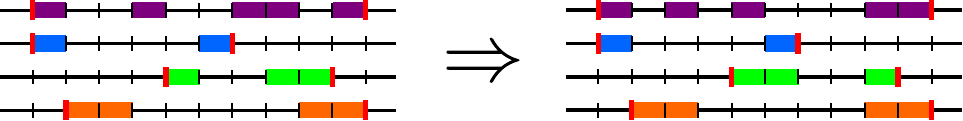}
        \end{center}
      \vspace{-8pt}
      \caption{
        P[$\piij(\tauf)$,$\t^1$,$\t^w$] redistributes the events \new{at random inside each timeline, while constraining the start time of the first event and the end time of the last event in each timeline (i.e.\ the timelines' {\sl activity} intervals).}
        %In the illustration, 
        The colors of events mark the timeline they belong to and vertical red lines mark the start time of the first event and end time of last event on each timeline, 
        %which are all conserved by the shuffling.
        }
    \end{figure}
    %P[$\piij(\tauf)$, $\t^1$, $\t^w$] is a natural refinement of P[$\w$, $\t^1$, $\t^w$] to make it an event shuffling.
  %
  \item[P{[$\piij(\tauf)$, $\piij(\Dtauf)$]}]
    {\bf Common name:} {\sl Interval shuffling.}
    {\bf Features constrained:}
       the sequences of local distributions of event and inter-event durations on each link, $\piij(\tauf) = ([\tau_\ij^m]_{m\in\Ml})_{\ij\in\L}$ and $\piij(\Dtauf) = ([\Dtau_\ij^m]_{m\in\Ml})_{\ij\in\L}$, respectively.
    % {\bf Other features conserved:} 
    %   $\k$, $L$, $\a$, $\s$, $\n$, $\w$, $\mu(\Ebf)$, $\mu(\d)$.
    {\bf References:} \cite{Gauvin2013A} ({\sl interval shuffling});  Supplementary Note~3.
    \begin{figure}[H]
        \begin{center} 
          \includegraphics[height=0.04\textheight]{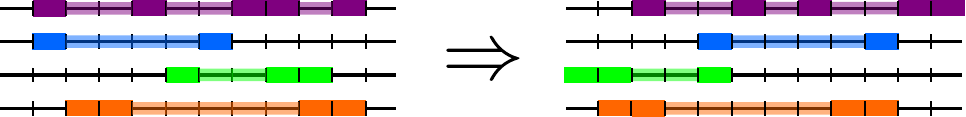}
        \end{center}
      \vspace{-8pt}
      \caption{
         P{[$\piij(\tauf)$, $\piij{\Dtauf}$]} shuffles \new{the start time of the first event as well as the order of} the event and inter-event durations on each link. 
         The events and inter-event intervals are colored by the timeline they belong to.}
    \end{figure}
     %P{[$\piij(\tauf)$, $\piij{\Dtauf}$]} is referred to as {\sl interval shuffling} in Ref.~\cite{Gauvin2013A}.
  %
  \item[{P[$\piij(\tauf)$, $\piij(\Dtauf)$,$\t^1$]}] 
    \new{{\bf Common name:} {\sl Inter-event shuffling.}
    {\bf Features constrained:}
      the sequences of local distributions of event and inter-event durations on each link, $\piij(\tauf) = ([\tau_\ij^m]_{m\in\Ml})_{\ij\in\L}$ and $\piij(\Dtauf) = ([\Dtau_\ij^m]_{m\in\Ml})_{\ij\in\L}$, respectively;
	  the sequence of times of the first  event on each link,
	  $\t^1=(t_\ij^1)_{\ij\in\L}$.
    % {\bf Other features conserved:} 
    %   $\k$, $L$, $\a$, $\s$, $\n$, $\w$, $\mu(\Ebf)$, $\mu(\d)$, $\t^w$.
    {\bf Reference:}  Supplementary Note~3.}
    \begin{figure}[H]
        \begin{center} 
          \includegraphics[height=0.04\textheight]{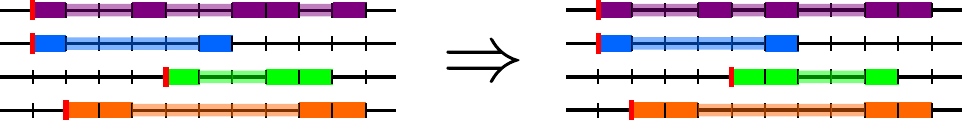}
        \end{center}
      \vspace{-8pt}
      \caption{
        P[$\piij(\Dtauf)$,$\t^1$] adds another constraint to P[$\piij(\tauf)$, $\piij(\Dtauf)$] so that it conserves the time of the first event each link. 
        The events and inter-event intervals are colored by the timeline they belong to, and red vertical lines mark the start time of the first events in each timeline.}
    \end{figure}
    % (it is an event shuffling variant of the instant-event shuffling P[$\piij(\Dtauf)$,$\t^1$]).
 %
  \item[{P[$\perl(\Thf)$]}] 
    \new{{\bf Common name:} {\sl Timeline shifting.}
    {\bf Features constrained:}
      The sequence of sets of all possible translations of each timeline with periodic boundary conditions, $\perl(\Thf)=({\rm per}(\Thij))_{\ij\in\L}$.
    % {\bf Other features conserved:}
    %   $\k$, $L$, $\a$, $\s$, $\n$, $\w$, $\mu(\Ebf)$, $\mu(\d)$, $\tauf$, $\Dtauf$.  
    {\bf Reference:} \cite{Backlund2014E} ({\sl random offset}).}
    \begin{figure}[H]
        \begin{center} 
          \includegraphics[height=0.04\textheight]{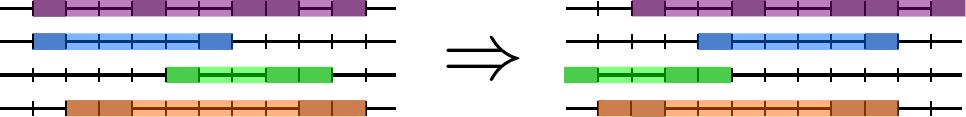}
        \end{center}
      \vspace{-8pt}
      \caption{
        P[$\perl(\Thf)$] randomly translates the %whole sequence of event and inter-event durations on each timeline, i.e\ it randomizes $\t^1$ for each 
          timelines on each link individually using periodic boundary conditions, randomizing the start time of the first event in each timeline but otherwise conserving their temporal order and placement.
          Colors highlight the intervals between the first and last events in each timeline before and after shuffling.}
    \end{figure}
      %P[$\perl(\Thf)$] was named {\sl random offset} in  Ref.~\cite{Backlund2014E}.
  %
%   \item[{P[$\tauf$,$\Dtauf$]}] \note{CLV: removed (not event-shuffling version of instant-event shuffling, not used in literature or in walkthrough example).}
%    \new{{\bf Common name:} {\sl .}
%    {\bf Features constrained:}
%    {\bf References:} .}
%    \begin{center} 
%%      \includegraphics[height=0.04\textheight]{P__.pdf}
%    \end{center}
%    is a refinement of P[$\perl(\Thf)$] that imposes hard boundary conditions instead of periodic ones. %which randomly translates the whole sequence of event and inter-event durations on each timeline using hard boundary conditions.
\end{itemize}

%------------------------------------------------------------------------------------------------------------------
\subsubsection{Sequence shufflings}
\label{sec:sequence_shufflings}
%------------------------------------------------------------------------------------------------------------------
\noindent
Sequence shufflings %(Def.~\ref{def:sequence_shuffling}) 
%are are a particular kind of instant-event timeline shufflings. 
%They 
randomize the sequence of snapshots while leaving the individual snapshots unchanged, 
% \new{i.e., they constrain the multiset of snapshot graphs $\pxt(\Gs)$. 
% \note{indirect:} They indirectly conserve the weighted aggregated network, $\w$, the static graph $\Gstat=(\N,\L)$, degree sequence $\k$ and number of links, $L$, as well as all instantaneous topological correlations inside snapshots, notably the multisets of instantaneous degree sequences, $\pxt(\d)$, and number of events per snapshot, $p(\Ebf)$ (see Tables~\ref{tab:features} and \ref{tab:constraint_levels} in Appendix~\ref{app:tables} for detailed definitions of  features).}
%(see Fig.~\ref{fig:dependency_diagram}). 
% in the snapshot-sequence representation. % (Def.~\ref{def:snapshot-sequence}).
They generally destroy temporal correlations inside timelines and in node activities.
% We provide informal definitions of the principal features that are constrained by each shuffling method in the description 
% (see Table~\ref{tab:features} for detailed definitions). 
% The effects of each shuffling on a wide selection of other temporal network features are shown in Table~\ref{tab:effects-contact}.
%(see Table~\ref{tab:effects-contact}). 
% We provide informal definitions of the principal features that are constrained by each shuffling method, and we list other features that are conserved as a consequence.
% Table~\ref{tab:features} provides detailed definitions of elementary features and Table~\ref{tab:constraint_levels} of distributions and moments constructed from these.} 
%, including $\tauf$, $\Dtauf$, $\n$, $\alphaf$, $\Dalphaf$, and $\a$.
%This in particular means that sequence shufflings are also timeline shufflings (see Def.~\ref{def:sequence_shuffling}). 
%They are thus compatible with both snapshot shufflings (Subsec.~\ref{sec:snapshot_shufflings}) and timeline shufflings (Subsec.~\ref{def:timeline_shuffling}) (Propositions~\ref{prop:sequence-snapshot_independent} and \ref{prop:link-sequence_independent}) and may be combined with these to generate new MRRMs.

We have identified the following two sequence shufflings in the literature. These are included in the Hasse diagram shown in Fig.~\ref{fig:RRMs}(c).

%\paragraph{Instant-event shufflings}
\begin{itemize}
  \item[{P[$\pxt(\Gs)$]}] 
    \new{{\bf Common name:} {\sl Sequence shuffling.}
    {\bf Features constrained:} 
      the distribution (Def.~\ref{def:unordered_set})  of snapshot graphs $\pxt(\Gs)=[\Gt]_{t\in\T}$.
    % {\bf Other features conserved:}
    %   $\Gstat$, $\k$, $L$, $\w$, $\s$ $p(\Ebf)$, $\pxt(\d)$.
    {\bf References:} \cite{Tang2010S} ({\sl reshuffled sequences}); \cite{Cardillo2014E,Bajardi2011D} ({\sl random ordered}); \cite{Posfai2014S} ({\sl shuffled times}); \cite{Valdano2015I} ({\sl reshuffle}).}
    \begin{figure}[H]
        \begin{center} 
          \includegraphics[height=0.05\textheight]{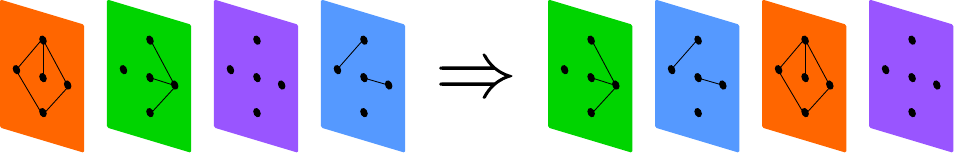}
        \end{center}
        \vspace{-8pt}
        \caption{
        P[$\pxt(\Gs)$] randomly shuffles the order of the snapshots. 
        Colors mark each individual snapshot graph, which the shuffling conserves. 
        %temporal placement of each snapshot before and after shuffling.
        }
    \end{figure}
    %It is both conceptually and algorithmically simple, and represents a quick way to gauge the impact of temporal structures and correlations.
       %P[$\pxt(\Gs)$] was named {\sl reshuffled sequences} in Ref.~\cite{Tang2010S}; it appears in Refs.~\cite{Cardillo2014E}, in Ref.~\cite{Bajardi2011D} as {\sl random ordered}, in Ref.~\cite{Posfai2014S} as {\sl shuffled times}, and in Ref.~\cite{Valdano2015I} as {\sl reshuffle}. 
%       Since it is the coarsest possible sequence shuffling, we simply name it {\sl sequence shuffling}.
  %
  \item[{P[$\pxt(\Gs)$,$\sgnf(\Ebf)$]}] 
    \new{{\bf Common name:} {\sl Activity-constrained sequence shuffling.}
    {\bf Features constrained:}
      the distribution (Def.~\ref{def:unordered_set}) of snapshot graphs $\pxt(\Gs)=[\Gt]_{t\in\T}$;
      the times during which events take place, formally defined as the indicator function for $A^t\in\mathbb{N}^+$ (i.e.\ $A^t \leq1$) at each time, $\sgnf(\Ebf)=(\sgn(\Et))_{t\in\T}$. 
    % {\bf Other features conserved:}
    %   $\Gstat$, $\k$, $L$, $\w$, $\s$ $p(\Ebf)$, $\pxt(\d)$.
    {\bf References:} \cite{Posfai2014S} ({\sl shuffled times}).}
    \begin{figure}[H]        \begin{center} 
          \includegraphics[height=0.05\textheight]{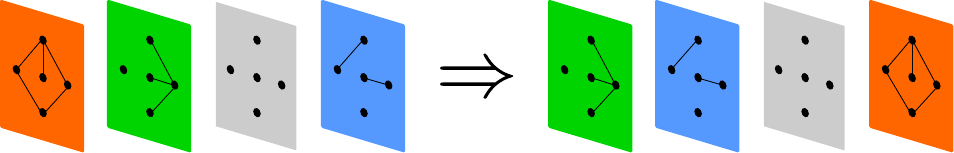}
        \end{center}
        \vspace{-8pt}
        \caption{
        P[$\pxt(\Gs)$,$\sgnf(\Ebf)$] shuffles the timestamps \new{between} snapshots where at least one event takes place.
        Colors mark each individual snapshot graph, which the shuffling conserves. 
        % Colors mark the temporal placement of each snapshot before and after shuffling.
        The grey snapshot contains no events and its placement is not shuffled.}
    \end{figure}
    %It was employed in Ref.~\cite{Posfai2014S} with the name of {\sl shuffled times}.
\end{itemize}

%------------------------------------------------------------------------------------------------------------------
\subsubsection{Snapshot shufflings}
\label{sec:snapshot_shufflings}
%------------------------------------------------------------------------------------------------------------------
\noindent
Snapshot shufflings %(Def.~\ref{def:snapshot_shuffling}) 
conserve the start times $t$ of all events. %, as sequence shufflings, are all instant-event shufflings. 
They are typically implemented by randomizing the instantaneous snapshot graphs $\Gt$ corresponding to each time $t\in\T$. %temporal snapshot individually 
As a consequence, all snapshot shufflings found in the literature are instant-event shufflings, but they may also be implemented as event shufflings---we give one example in this section and two others in Section~\ref{sec:intersections-TS-Snaps}.
%while c. % at which it takes place.

%Snapshot shufflings all constrain \new{the start times of all events} $\t$.
%\new{They consequently conserve the number of events at each time, $\Ebf$  (see Tables~\ref{tab:features} and \ref{tab:constraint_levels} in Appendix~\ref{app:tables} for detailed definitions of  features). 
Snapshot shufflings generally destroy temporal features of the links, but may conserve some temporal node features, such as the sequence of instantaneous degrees, $\d$.
% We provide informal definitions of the principal features that are constrained by each shuffling method in the description 
% (see Table~\ref{tab:features} for detailed definitions). 
% The effects of each shuffling on a wide selection of other temporal network features are shown in Table~\ref{tab:effects-contact}.
%(see Table~\ref{tab:effects-contact}).
% We provide informal definitions of the principal features that are constrained by each shuffling method, and we list other features that are conserved as a consequence.
% Table~\ref{tab:features} provides detailed definitions of elementary features and Table~\ref{tab:constraint_levels} of distributions and moments constructed from these.} 
%Snapshot shufflings are compatible with sequence shufflings (Subsec.~\ref{sec:sequence_shufflings}) (Proposition~\ref{prop:sequence-snapshot_independent}) and can thus be combined with them in a composition to form new MRRMs.

The snapshot shufflings listed below are ordered hierarchically in the Hasse diagram shown in Fig.~\ref{fig:RRMs}(d).
%We list below snapshot shufflings found in the literature as well as some new ones.

\paragraph{Instant-event shufflings}
\begin{itemize}
  \item[{P[$\t$]}] 
    \new{{\bf Common name:} {\sl Snapshot shuffling.}
    {\bf Features constrained:}
      the times of all instantaneous events, $\t=(t_q)_{q\in\{1,2,\ldots,E\}}$.
    % {\bf Other features conserved:}
    %   $\mu(\s)$, $\Ebf$, $\mut(\d)$.
    {\bf Reference:} \cite{Posfai2014S} ({\sl random network}).}
    \begin{figure}[H]
        \begin{center} 
          \includegraphics[height=0.05\textheight]{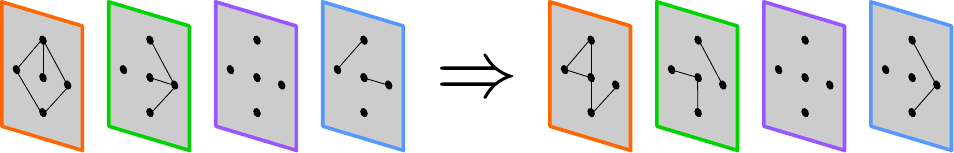}
        \end{center}
       \vspace{-8pt}
       \caption{
        P[$\t$] randomly shuffles the instantaneous events inside each snapshot. % while constraining only the number of events at each time $t\in\T$. 
        This is equivalent to generating each snapshot $\Gt$ as an instance of an Erd\H{o}s-R\'enyi graph with $N$ nodes and $\Et=|\E^t|$ edges. 
        Colored outlines mark the temporal placement of each snapshot, which is conserved.}
    \end{figure}
  %P[$\t$] is the coarsest possible snapshot shuffling, and we thus also refer to it simply as {\sl snapshot shuffling}. 
  %It was called {\sl random network} in Ref.~\cite{Posfai2014S}.
  %  
%  \item[{P[$\t$,$\Phif$]}] \note{CLV: removed}
%    \new{{\bf Common name:} {\sl .}
%    {\bf Features constrained:}
%    {\bf References:} .}
%    \begin{center} 
%%      \includegraphics[height=0.04\textheight]{P__.pdf}
%    \end{center}
%    shuffles the events inside each snapshot while additionally constraining the set of nodes that are active at each $t$, i.e.\ it constrains $\Phif$. %We have here included 
%  P[$\t$,$\Phif$] provides a rare MRRM that conserves the nodes' activity and inactivity durations (besides the finer P[$\d$] introduced below).
  %  
  \item[{P[$\d$]}] 
    \new{{\bf Common name:} {\sl Degree-constrained snapshot shuffling.}
    {\bf Features constrained:}
      sequence of instantaneous degrees, $\d=((\dit)_{i\in\N})_{t\in\T}$.
    % {\bf Other features conserved:} 
    %   $\mu(\s)$, $\alphaf$, $\Dalphaf$.
    {\bf References:} \cite{Bajardi2011D} ({\sl time ordered and reshuffled networks});\cite{Posfai2014S} ({\sl degree preserved network}); \cite{Sun2015C, Galimberti2018M}.}
    \begin{figure}[H]
        \begin{center} 
          \includegraphics[height=0.05\textheight]{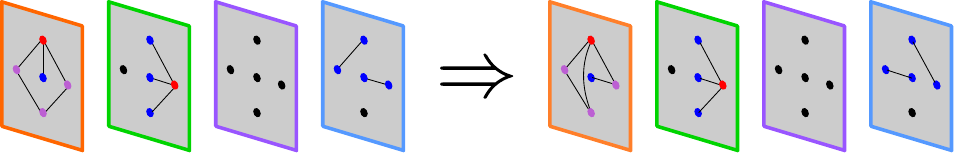}
        \end{center}
        \vspace{-8pt}
        \caption{
        P[$\d$] shuffles the events inside each snapshot while constraining the instantaneous degree sequence $\d^t=(\dit)_{i\in\N}$, using e.g.\ the Maslov-Sneppen model. 
        Outline colors mark the temporal placement of each snapshot and the nodes' colors mark their instantaneous degrees.}
    \end{figure}
    % of each node. 
    %P[$\d$] was called {\sl time ordered and reshuffled networks} in Ref.~\cite{Bajardi2011D} and {\sl degree preserved network} in Ref.~\cite{Posfai2014S}, and was also applied in Ref.~\cite{Sun2015C}.
  %  
  \item[{P[$\isof(\Gs)$]}] 
    \new{{\bf Common name:} {\sl Isomorphic snapshot shuffling.}
    {\bf Features constrained:}
      the isomorphism class of each snapshot graph, $\isof(\Gs)=(\iso(\Gt))_{t\in\T}$.
    % {\bf Other features conserved:} 
    %   $\mu(\s)$, $\Ebf$, $\pit(\d)$.
    {\bf Reference:} \cite{Valdano2015I} ({\sl anonymize}).}
    \begin{figure}[H]
        \begin{center} 
          \includegraphics[height=0.05\textheight]{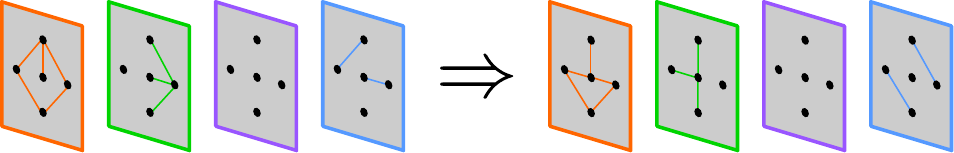}
        \end{center}
        \vspace{-8pt}
        \caption{
        P[$\isof(\Gs)$] consists in randomizing the identity of the nodes in each time snapshot. 
        This produces snapshot graphs, $(\Gt)'$, that are 
        %Each snapshot graph in the randomized network $(\Gt)'$ is thus isomorphic to 
        isomorphic to those of the original network, $(\Gt)'\simeq\Gt$. 
        Outline colors mark the temporal placement of the snapshots and link colors mark the isomorphism class of the snapshot graphs.}
    \end{figure}
    %, i.e.\ the shuffling constrains .
      %Nodes that are not active in that particular time step, however, may become active when their identity is swapped with active nodes. As a result, $\Phi$ is no longer preserved. This model preserves only the instantaneous topological structure of $\G^{(t)}$, breaking all temporal and aggregated correlations.
      %{P[$\isof(\Gs)$]} was named {\sl anonymize} in Ref.~\cite{Valdano2015I}.
  %  
%  \item[{P[$\isof(\Gs)$,$\Phif$]}]  \note{CLV: removed}
%    \new{{\bf Common name:} {\sl .}
%    {\bf Features constrained:}
%    {\bf References:} .}
%    \begin{center} 
%%      \includegraphics[height=0.04\textheight]{P__.pdf}
%    \end{center}
%    consists in randomizing the identity of nodes at each time step, but only nodes that are active are shuffled. It thus combines P[$\isof(\Gs)$] and P[$\Phif$] by intersection.
\end{itemize}

\paragraph{Event shufflings}

\begin{itemize}
  \item[{P[$p(\t,\tauf)$]}] 
    \new{{\bf Common name:} {\sl Snapshot shuffling.}
    {\bf Features constrained:}
      the timestamps and durations of the events, $p(\t,\tauf)=[(t,\tau)]_{(i,j,t,\tau)\in\C}$.
    % {\bf Other features conserved:} 
    %   $\mu(\s)$, $\Ebf$, $\mut(\d)$, $\pit(\tauf)$.
    {\bf References:} Supplementary Note~3.}
      \vspace{0.2cm}\\
%    \begin{center} 
%      \includegraphics[height=0.04\textheight]{P__pttau.pdf}
%    \end{center}
%     \note{illustration?}
%     \vspace{0.2cm}\\
    P[$p(\t,\tauf)$] \new{randomizes the values of $i$ and $j$ for each event $(i,j,t,\tau)\in\C$ while constraining the time $t$ at which it occurs as well as its duration $\tau$}. 
    %P[$p(\t,\tauf)$] is an event-shuffling version of the instant-event shuffling P[$\t$] defined above. % in order to preserve event durations and make . It conserves the distribution of event durations and the times at which they occur.
\end{itemize}

%------------------------------------------------------------------------------------------------------------------
\subsubsection{Intersections of link and timeline shufflings}
\label{sec:intersections-LS-TS}
%------------------------------------------------------------------------------------------------------------------
\noindent
\new{Several shuffling methods constrain both the static graph $\Gstat=(\N,\L)$ and the multiset of timelines, $p(\Thf)=[\Thl]_{\ij\in\L}$,
and are thus intersections of link and timeline shufflings.}

%The most random of these, $\model{\L,\pxl(\Thf)}$, is 
%the intersection of any pair of link and timeline shufflings and 
%both a link and a timeline shuffling, but other, more constrained, shufflings are neither since they constrain features that depend on the configuration of the timelines in the static graph and not only on their content.} 

% \note{indirect:} 
% \new{They conserve
% the static degree sequence, $\k$; 
% the number of links in the static graph, $L$;
% the number of events per time-lapse, $\Ebf$; 
% the mean node activity, strength, and instantaneous degree, $\mu(\a)$, $\mu(\s)$, and $\mu(\d)$; 
% the distributions of link weights (number of events and cumulative duration), $p(\n)$ and $p(\w)$; 
% and the multisets of the first and last events and the sequences of event and inter-event durations in the timelines, $\pxl(\t^1)$, $\pxl(\t^w)$, $\pxl(\tauf)$, and $\pxl(\Dtauf)$
% (see Tables~\ref{tab:features} and \ref{tab:constraint_levels} in Appendix~\ref{app:tables} for detailed definitions of  features).

The shufflings randomize temporal-topological correlations.
%(see Table~\ref{tab:effects-contact}).
% We provide informal definitions of the principal features that are constrained by each shuffling method in the description 
% (see Table~\ref{tab:features} for detailed definitions). 
% The effects of each shuffling on a wide selection of other temporal network features are shown in Table~\ref{tab:effects-contact}.
% We provide informal definitions of the principal features that are constrained by each shuffling method, and we list other features that are conserved as a consequence.
% Table~\ref{tab:features} provides detailed definitions of elementary features and Supplementary Table~\ref{tab:constraint_levels} of distributions and moments constructed from these.} 

%\new{All of these shufflings are event shufflings as they conserve the multiset of event durations.}
%The three following event shufflings are intersections between link shufflings and timeline shufflings. 
%They are thus compatible with link shufflings (Subsec.~\ref{sec:link-shufflings}), as well as timeline (Subsec.~\ref{sec:timeline-shufflings}) and sequence shufflings (Subsec.~\ref{sec:sequence_shufflings}) and may be combined with these to form new MRRMs (Propositions~\ref{prop:link-timeline_independent} and \ref{prop:sequence-snapshot_independent}).
The shufflings are included in the Hasse diagrams in Figs.~\ref{fig:RRMs}(a) and \ref{fig:RRMs}(b).

%\paragraph{Event shufflings.}

\begin{itemize}
  \item[{P[$\L$,$\pxl(\Thf)$]}] 
    \new{{\bf Common name:} {\sl Topology-constrained link shuffling.}
    {\bf Features constrained:}
       the static graph $\Gstat=(\N,\L)$;
       the distribution (Def.~\ref{def:unordered_set}) of timelines, $\pxl(\Thf)=[\Thij]_{\ij\in\L}$.
    % {\bf Other features conserved:} 
    %   $\k$, $L$, $\mu(\a)$, $\mu(\s)$, $p(\n)$, $p(\w)$, $\Ebf$, $\mut(\d)$, $\pxl(\tauf)$, $\pxl(\Dtauf)$, $\pxl(\t^1)$, $\pxl(\t^w)$.
    {\bf References:} 
      \cite{Karsai2011S,Kivela2012M,Thomas2015D} ({\sl link-sequence shuffled});  
      \cite{Holme2012T} ({\sl edge randomization});
      \cite{Gauvin2013A,Genois2015C} ({\sl link shuffling}); 
    Supplementary Note~3.}
    \begin{figure}[H]
        \begin{center} 
          \includegraphics[height=0.04\textheight]{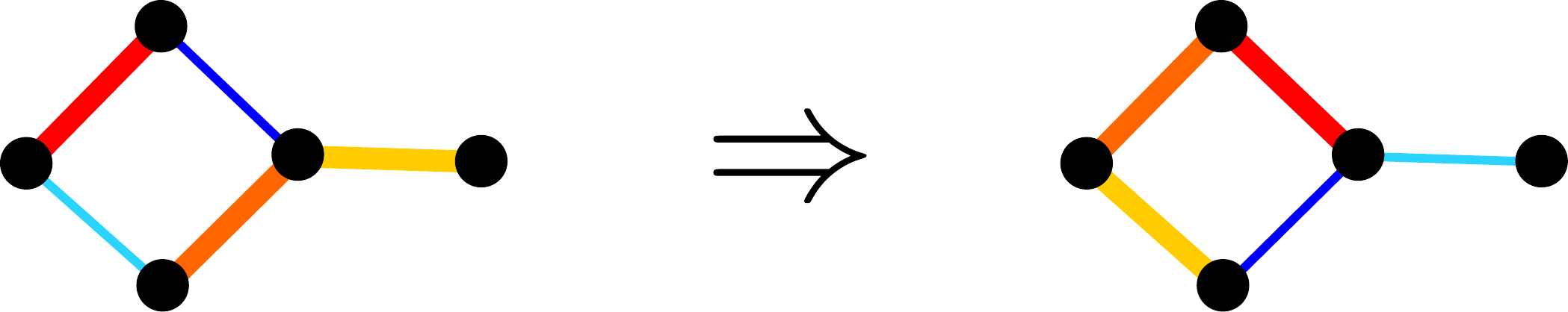}
        \end{center}
        \vspace{-8pt}
        \caption{
        P[$\L$,$\pxl(\Thf)$] randomly shuffles the timelines between all links while keeping the static graph $\Gstat$ fixed.
        Colors mark links corresponding to different timelines and are randomized by the shuffling.
        P[$\L$,$\pxl(\Thf)$] is the most random intersection between a link and a timeline shuffling.
        }
    \end{figure}
%    It is also both a link shuffling and a timeline shuffling (and the finest possible one).}
    %P[$\L$,$\pxl(\Thf)$] was named {\sl link-sequence shuffled} in Refs.~\cite{Karsai2011S,Kivela2012M,Thomas2015D}, {\sl edge randomization} in Ref.~\cite{Holme2012T}, and {\sl link shuffling} in Refs.~\cite{Gauvin2013A,Genois2015C}.
  %  
  \item[{P[$\w$,$\pxl(\Thf)$]}] 
    \new{{\bf Common name:} {\sl Weight-constrained link shuffling.}
    {\bf Features constrained:}
       the sequence of link weights, $\w=(\wij)_{\ij\in\L}$ (cumulative duration of events on each link);
       the distribution (Def.~\ref{def:unordered_set}) of timelines, $\pxl(\Thf)=[\Thij]_{\ij\in\L}$.
%   {\bf Other features conserved:} 
%       $\k$, $L$, $\s$, $\mu(\a)$, $p(\n)$, $\Ebf$, $\mut(\d)$, $\pxl(\tauf)$, $\pxl(\Dtauf)$, $\pxl(\t^1)$, $\pxl(\t^w)$.
    {\bf References:} 
      \cite{Karsai2011S,Pan2011P,Kivela2012M,Thomas2015D} ({\sl equal-weight link-sequence shuffled}); 
      \cite{Holme2012T} ({\sl equal-weight edge randomization (EWER)}).}
      \begin{figure}[H]
        \begin{center} 
          \includegraphics[height=0.04\textheight]{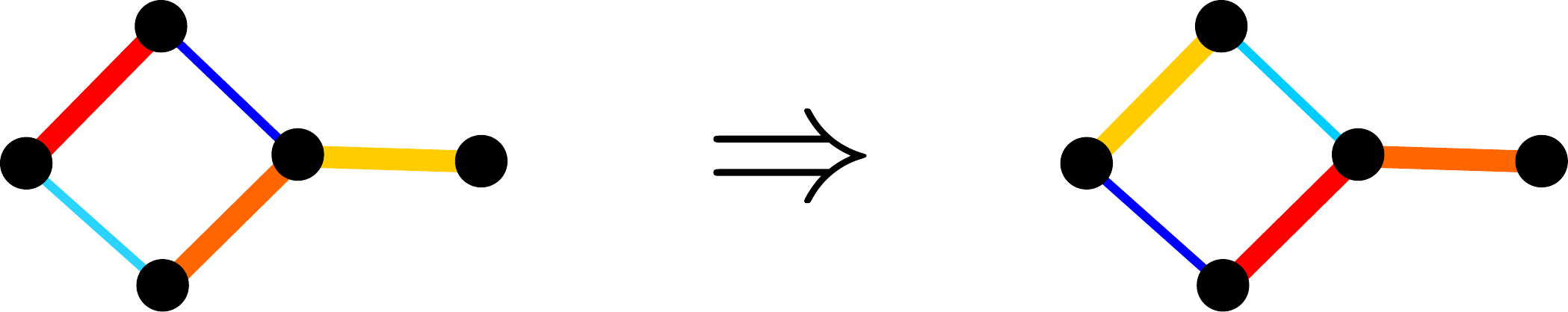}
        \end{center}
        \vspace{-8pt}
        \caption{
        P[$\w$,$\pxl(\Thf)$] shuffles timelines $\Thij$ between links with the same cumulative event duration $\wij$ (for instant-event networks defined as the number of instantaneous events).
        Colors mark links corresponding to different timelines, while link thickness marks their weights $\wij$. The former are randomized under the constraint that the latter are conserved.}
    \end{figure}
      %P[$\w$,$\pxl(\Thf)$] was named {\sl equal-weight link-sequence shuffled} in Refs.~\cite{Karsai2011S,Kivela2012M,Thomas2015D} and was also called {\sl equal-weight edge randomization (EWER)} in Ref.~\cite{Holme2012T}.
  %  
  \item[{P[$\n$,$\pxl(\Thf)$]}] 
    \new{{\bf Common name:} {\sl Weight-constrained link shuffling.}
    {\bf Features constrained:}
       the sequence of link weights, $\n=(\nij)_{\ij\in\L}$ (number of events on each link);
       the distribution (Def.~\ref{def:unordered_set}) of timelines, $\pxl(\Thf)=[\Thij]_{\ij\in\L}$.
    % {\bf Other features conserved:} 
    %   $\k$, $L$, $\a$, $\mu(\s)$, $p(\w)$, $\Ebf$, $\mut(\d)$, $\pxl(\tauf)$, $\pxl(\Dtauf)$, $\pxl(\t^1)$, $\pxl(\t^w)$.
    {\bf Reference:} 
      Supplementary Note~3.}
      \begin{figure}[H]
        \begin{center} 
          \includegraphics[height=0.04\textheight]{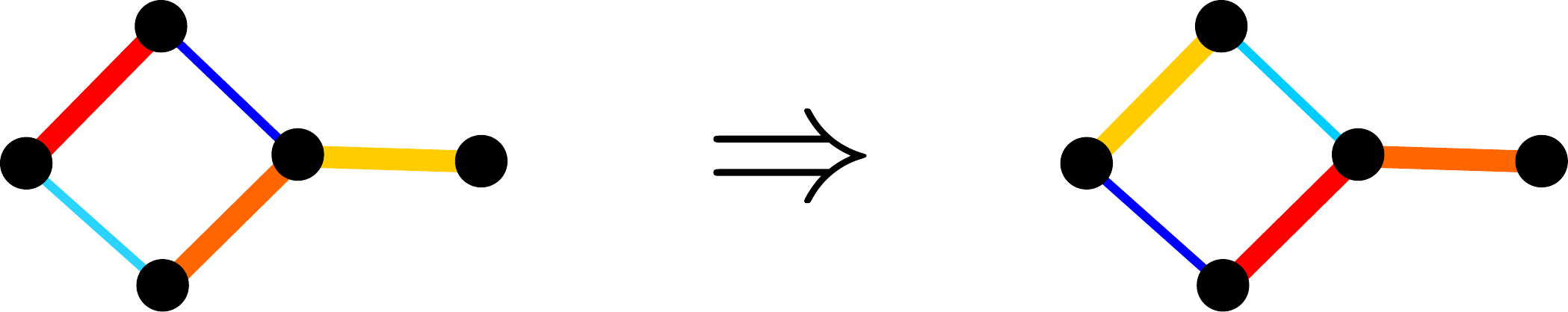}
        \end{center}
        \vspace{-8pt}
        \caption{
        P[$\n$,$\pxl(\Thf)$] shuffles timelines between links with the same number of events, $\nl$.
        Colors mark links corresponding to different timelines, while link thickness marks their weights $\nij$. The former are randomized under the constraint that the latter are conserved.}
    \end{figure}
    %P[$\n$,$\pxl(\Thf)$] is a natural alternative to P[$\w$,$\pxl(\Thf)$] for temporal networks with event durations, where $\n$ has a similar role to the one $\w$ has in networks with instantaneous events.
\end{itemize}

%------------------------------------------------------------------------------------------------------------------
\subsubsection{Intersections of timeline and snapshot shufflings}
\label{sec:intersections-TS-Snaps}
%------------------------------------------------------------------------------------------------------------------
\noindent
\new{Yet other shuffling methods constrain both the static graph $\Gstat=(\N,\L)$ and the timestamps of each event $\t$ and are thus intersections of timeline and snapshot shufflings.}
%\note{indirect: $\mut(\d)$.}
% \new{They conserve
% the static degree sequence, $\k$; 
% the number of links in the static graph, $L$;
% the number of events per time-lapse, $\Ebf$; 
% and the mean node strength $\mu(\s)$; 
% and the mean link weight $\mu(\w)$
% (see Tables~\ref{tab:features} and \ref{tab:constraint_levels} in Appendix~\ref{app:tables} for detailed definitions of  features).
%The shufflings generally randomize other features of the timelines.
%(see Table~\ref{tab:effects-contact}).
% We provide informal definitions of the principal features that are constrained by each shuffling method in the description 
% (see Table~\ref{tab:features} for detailed definitions). 
% The effects of each shuffling on a wide selection of other temporal network features are shown in Table~\ref{tab:effects-contact}.
% We provide informal definitions of the principal features that are constrained by each shuffling method, and we list other features that are conserved as a consequence.
% Table~\ref{tab:features} provides detailed definitions of elementary features and Table~\ref{tab:constraint_levels} of distributions and moments constructed from these.}

\begin{figure*}
  \centerline{
  \begin{tikzpicture}
    \node at(-4.2,0.7){ \includegraphics[height=0.275\textheight]{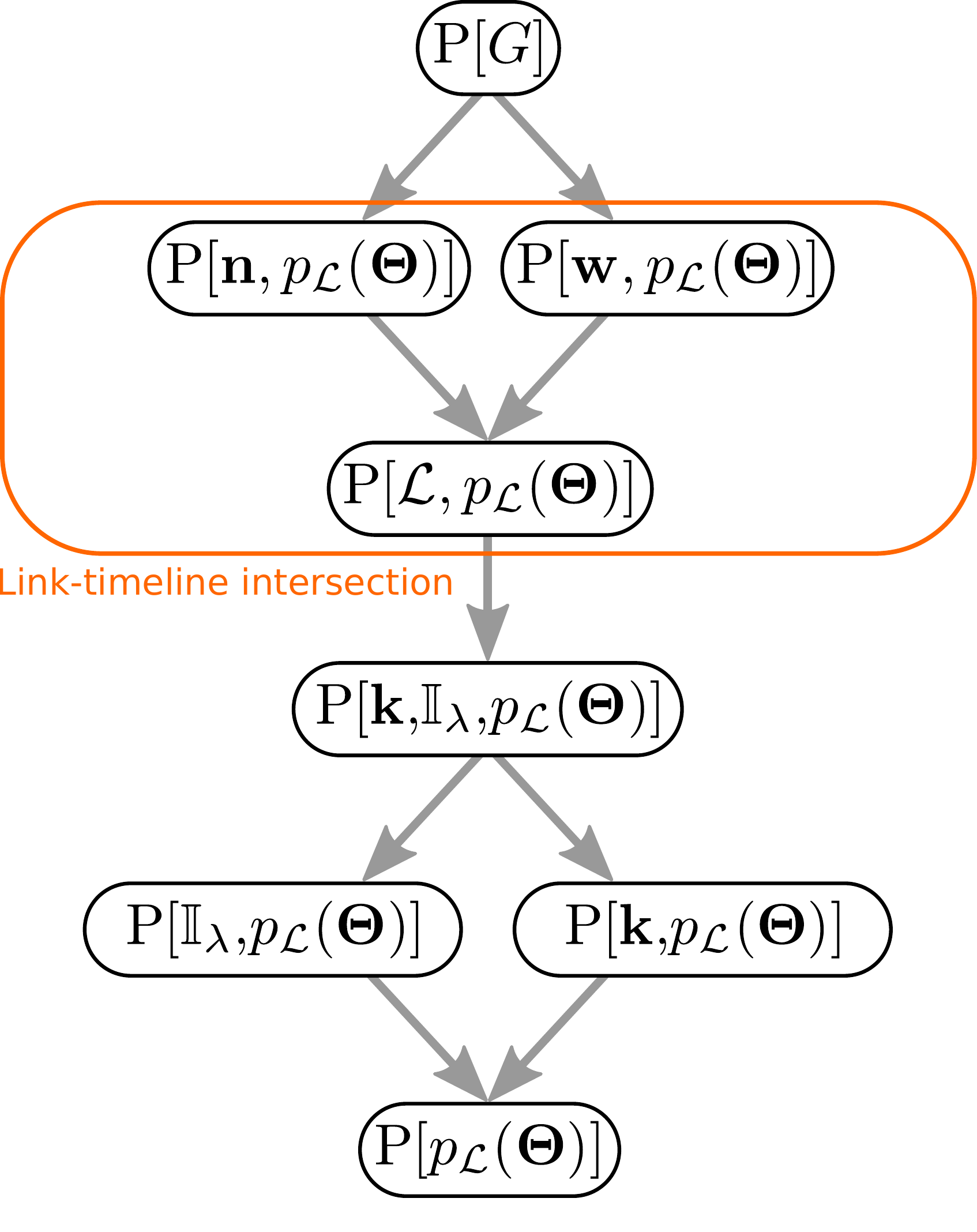} }; 
    \node at(-6.5,3.8){ (a) };
    %---------------
    \node at(5,0){ \includegraphics[height=0.325\textheight]{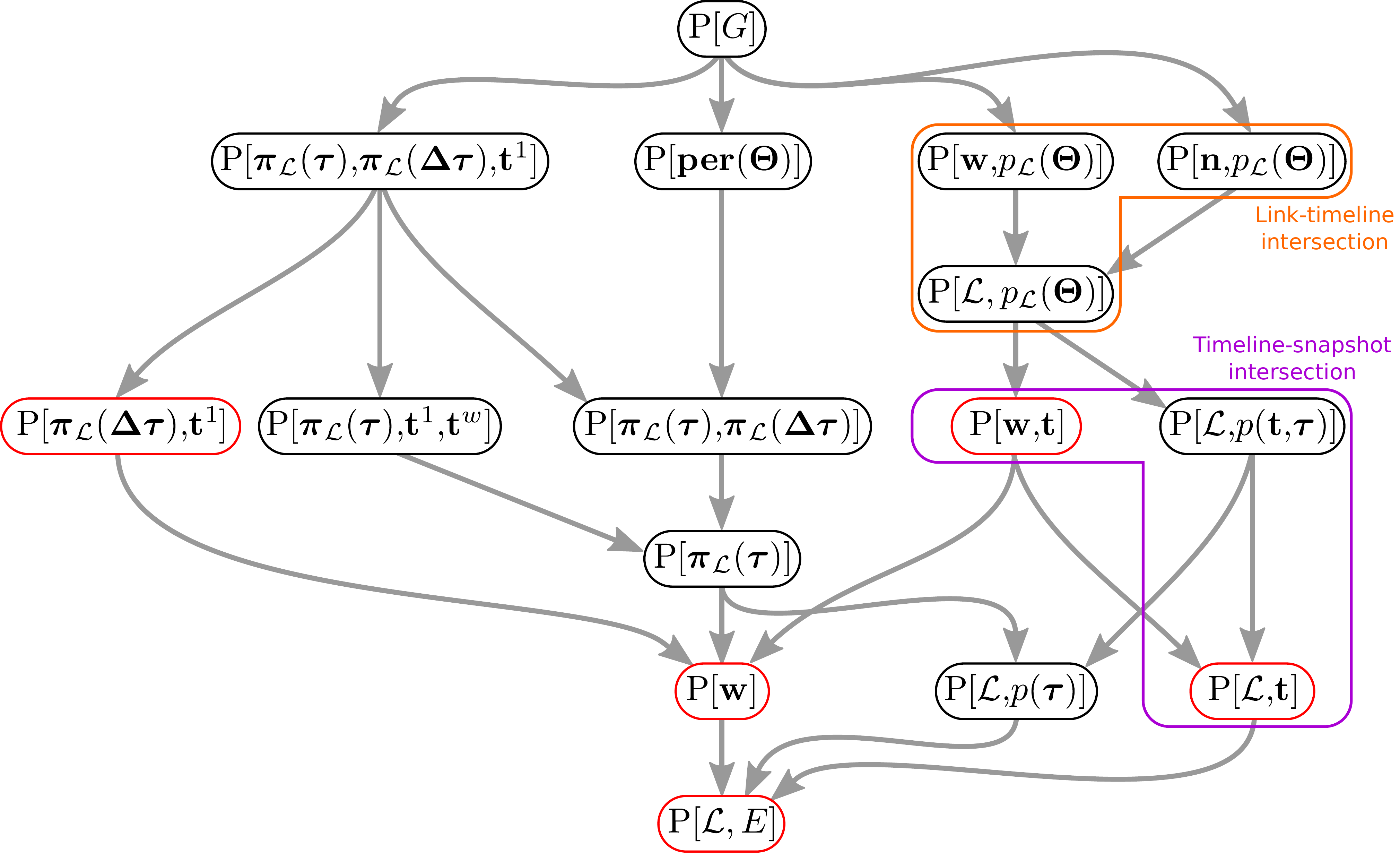} };
    \node at(0.,3.8){ (b) };
    %---------------
    \node at(-4.2,-5.8){ \includegraphics[height=0.125\textheight]{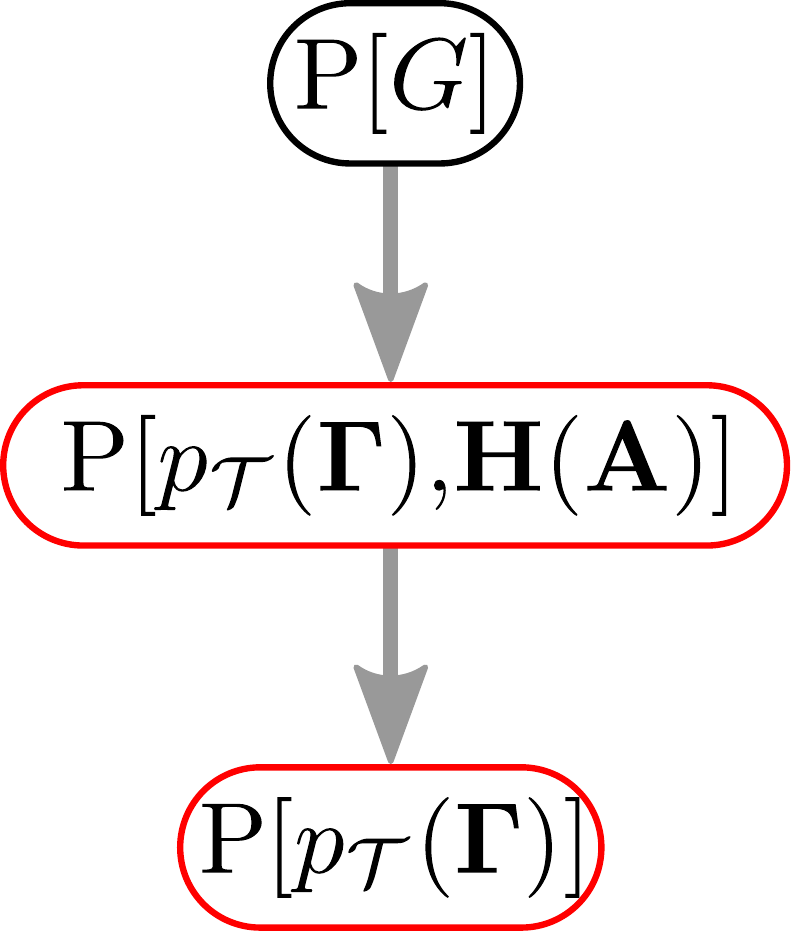} };
    \node at(-5.5,-4.4){ (c) };
    %---------------
    \node at(3.,-7.){ \includegraphics[height=0.225\textheight]{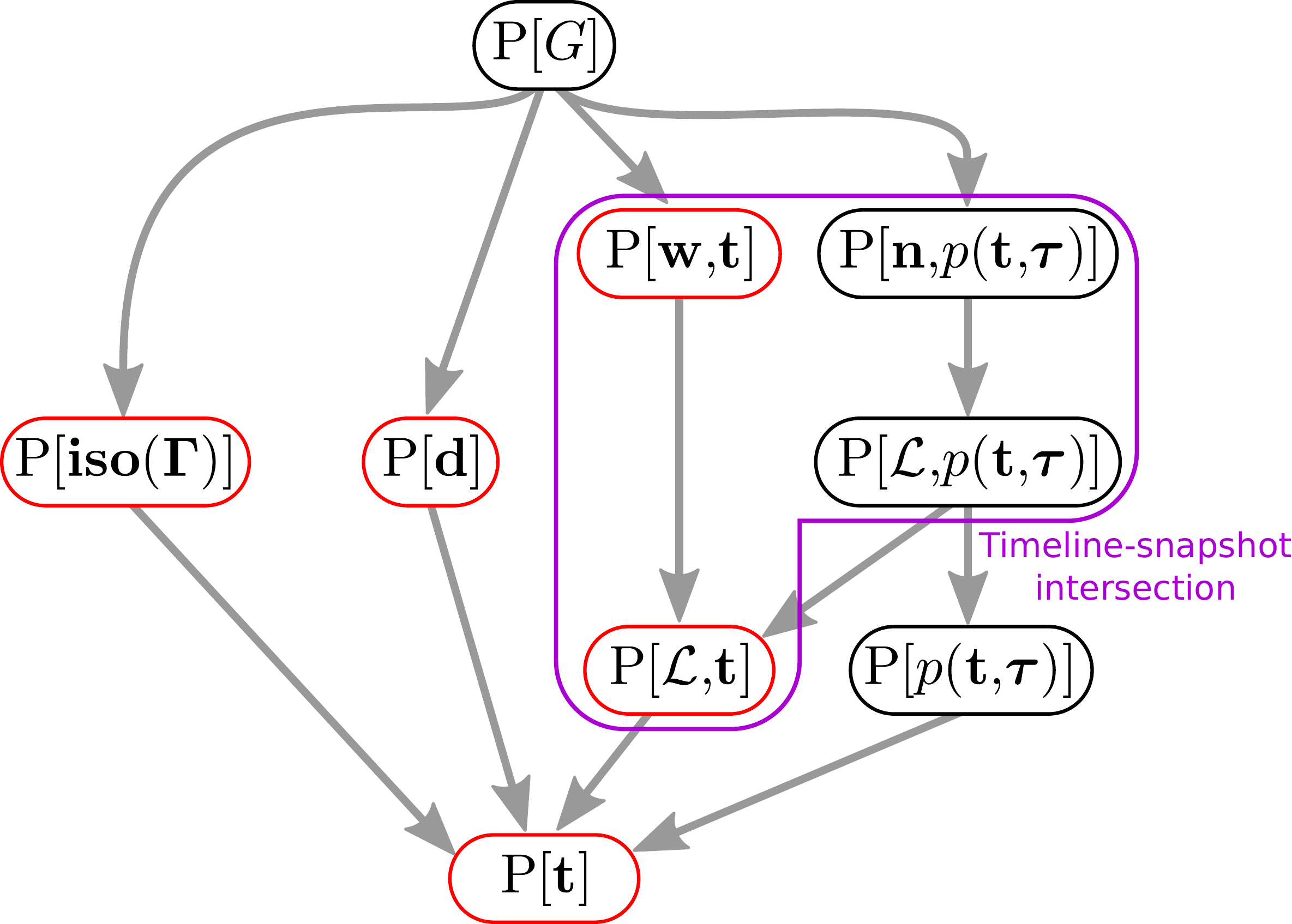} };
    \node at(0,-4.4){ (d) };
  \end{tikzpicture}
  }
  \caption{{\bf Hierarchies of shuffling methods.}
  (a) Link shufflings (Sec.~\ref{sec:link-shufflings}).
  (b) Timeline shufflings (Sec.~\ref{sec:timeline-shufflings}).
  (c) Sequence shufflings (Sec.~\ref{sec:sequence_shufflings}).
  (d) Snapshot shufflings (Sec.~\ref{sec:snapshot_shufflings}).
  An arrow from a higher MRRM to a lower one indicates that the former MRRM is finer than the latter and thus randomizes less. 
  Nodes with red outlines represent instant-event shufflings, black outlines mark event shufflings.
  The link-timeline intersections are defined in Section~\ref{sec:intersections-LS-TS} and the timeline-snapshot intersections in Section~\ref{sec:intersections-TS-Snaps}.
  }
  \label{fig:RRMs}
\end{figure*}

The shufflings are included in the Hasse diagrams in Figs.~\ref{fig:RRMs}(b) and \ref{fig:RRMs}(d).

%features corresponding to two of the classes above %of the restricted shufflings, 
%and can thus be classified as intersections of these. 
%We have found shuffling methods in the literature that are intersections of timeline and snapshot shufflings and of link and timeline shufflings.
%%Intersections of timeline and snapshot shufflings are compatible with both link and sequence shufflings, while intersections of link and timeline shufflings are compatible with  link shufflings , timeline shufflings, and sequence shufflings.

%We list first intersections of instant-event shufflings, followed by event shufflings.

\paragraph{Instant-event shufflings.}
%\paragraph{Intersections of timeline and snapshot shufflings.}
%The two following instant-event shufflings are intersections of timeline and snapshot shufflings.
% They are thus compatible with both link shufflings (Subsec.~\ref{sec:link-shufflings}) and sequence shufflings (Subsec.~\ref{sec:sequence_shufflings}) and may be combined with these to form new MRRMs (Propositions~\ref{prop:link-timeline_independent} and \ref{prop:sequence-snapshot_independent}).

\begin{itemize}
%   \item[{P[$\L$,$\t$]}] \note{CLV: keep (coarsest intersection between a timeline and snapshot shuffling) or remove (not used anywhere)?}
%     \new{{\bf Common name:} {\sl Topology-constrained snapshot shuffling.}
%     {\bf Features constrained:}
%       the static graph $\Gstat=(\N,\L)$;
%       the times of all instantaneous events, $\t=(t_q)_{q\in\{1,2,\ldots,E\}}$.
% %    {\bf References:} .
% }    
%     \begin{center} 
%       \includegraphics[height=0.04\textheight]{P__L_t.pdf}
%     \end{center}
%     P[$\L$,$\t$] resamples the events inside each snapshot while constraining $\Gstat$, i.e.\ assigning the resampled events only to node pairs with at least one event in $G$. Each snapshot is thus a subgraph of $\Gstat$ in which $\Et$ links are chosen at random.
  %  
  \item[{P[$\w$,$\t$]}] 
    \new{{\bf Common name:} {\sl Timestamp shuffling.}
    {\bf Features constrained:} 
       the sequence of link weights, $\w=(\wij)_{\ij\in\L}$ (number instantaneous events on each link);
       the times of all instantaneous events, $\t=(t_q)_{q\in\{1,2,\ldots,E\}}$.
    % {\bf Other features conserved:} 
    %   $\k$, $L$, $\s$, $\Ebf$, $\mut(\d)$.
    {\bf References:} \cite{Saramaki2015E} ({\sl time-stamp shuffling}); \cite{Holme2005N} ({\sl permuted times}); \cite{Karsai2011S, Kovanen2011T, Pan2011P, Kivela2012M, Kovanen2013T, Thomas2015D} ({\sl time-shuffled} or {\sl time-shuffling}); \cite{Karimi2013T} ({\sl time reshuffle}); \cite{Holme2012T, Takaguchi2013B, Li2017T, Takaguchi2016C} ({\sl randomly permuted times}); \cite{Rocha2011S} ({\sl random dynamic}); \cite{Backlund2014E} ({\sl random time shuffle}); \cite{Valdano2015I} ({\sl reconfigure}); \cite{Holme2015M} ({\sl shuffled time stamps}); \cite{Miritello2011D, Karimi2013T, Takaguchi2013I, Redmond2014I, Zhang2017C}.}
    \begin{figure}[H]
        \begin{center} 
          \includegraphics[height=0.04\textheight]{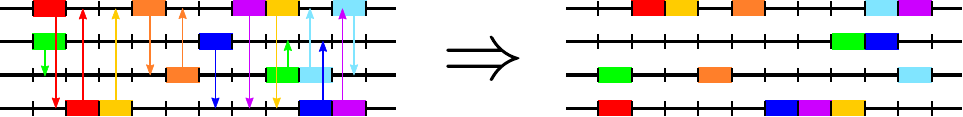}
        \end{center}
        \vspace{-8pt}
        \caption{
        P[$\w$,$\t$] randomly shuffles the timestamps $t$ of all instantaneous events $(i,j,t)\in\E$, while keeping $i$ and $j$ fixed.
    %, i.e.\ it constrains $\ijf=(i_q,j_q)_{q=1}^Q$ and $p(\t)=\{t_q\}_{q=1}^Q$.  
      In a completely equivalent manner, we may define the shuffling by constraining the timestamps $\t$ and permuting the pairs $(i,j)$.
      The figure illustrates the procedure, where pairs of instantaneous events (of the same color) are swapped between timelines (red arrows) while conserving their timestamps.
      %, i.e.\ constraining $\w$. 
      Due to the indistinguishability of networks obtained through permutation of event indices, both are equivalent to conserving $\w$ and $\Ebf$.
      For convenience, we choose the canonical name P[$\w$,$\t$] which conveys that it is  both a timeline shuffling and a snapshot shuffling. %and it is thus compatible with link shufflings and sequence shufflings.
      }
    \end{figure}
%      P[$\w$,$\t$] is a very popular MRRM. It was named {\sl permuted times} in Ref.~\cite{Holme2005N}, and called {\sl time-shuffled} or {\sl time-shuffling} in Refs.~\cite{Karsai2011S,Kovanen2011T,Kivela2012M,Kovanen2013T,Thomas2015D}, {\sl randomly permuted times} in Refs.~\cite{Holme2012T,Takaguchi2013B,Li2017T,Takaguchi2016C}, {\sl random dynamic} in Ref.~\cite{Rocha2011S}, {\sl random time shuffle} in Ref.~\cite{Backlund2014E}, {\sl reconfigure} in Ref.~\cite{Valdano2015I}, and {\sl shuffled time stamps} in Ref.~\cite{Holme2015M}.
%      It was also employed in Refs.~\cite{Miritello2011D,Karimi2013T,Takaguchi2013I,Redmond2014I}.
\end{itemize}

\paragraph{Event shufflings}
%The two following event shufflings are intersections of timeline and snapshot shufflings. %They are thus compatible with both link shufflings (Subsec.~\ref{sec:link-shufflings}) and sequence shufflings (Subsec.~\ref{sec:sequence_shufflings}) and may be combined with these to form new MRRMs (Propositions~\ref{prop:link-timeline_independent} and \ref{prop:sequence-snapshot_independent}). 
%The shufflings are included in the Hasse diagrams in Figs.~\ref{fig:RRMs-TS} and \ref{fig:RRMs-SnapS}.

\begin{itemize}
  \item[{P[$\L$,$p(\t,\tauf)$]}] 
    \new{{\bf Common name:} {\sl Topology-constrained snapshot shuffling.}
    {\bf Features constrained:}
       the static graph $\Gstat=(\N,\L)$.
    % {\bf Other features conserved:} 
    %       $\k$, $L$, $\mu(\a)$, $\mu(\s)$, $\mu(\n)$, $\mu(\w)$, $\Ebf$, $\mut(\d)$, $\pit(\tauf)$.
    {\bf References:} Supplementary Note~3.}
    \begin{figure}[H]
        \begin{center} 
          \includegraphics[height=0.04\textheight]{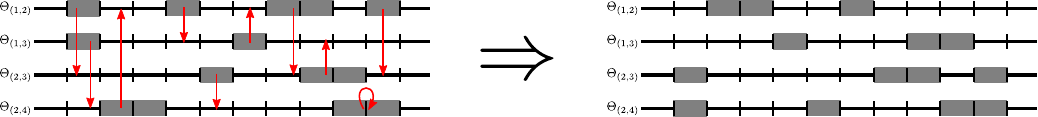}
        \end{center}
        \vspace{-8pt}
        \caption{
        P[$\L$,$p(\t,\tauf)$] \new{shuffles the events between existing links while constraining their starting times and their durations}. 
        Each event is moved a random timeline (red arrows---note that the new placement may be the same as the old).
        %, respecting the constraint the events do not overlap. 
        }
    \end{figure}
    %It can be seen as a refinement of P[$\L$,$\t$] defined above to networks with event durations. %the contacts are shuffled only among the node pairs that have at least one contact in $G$.
  %
%  \item[{P[$\n$,$p(\t,\tauf)$]}] \note{CLV: removed.} 
  %or add to walkthrough \ref{sec:walktrough_example}?}
%    \new{{\bf Common name:} {\sl Weight-constrained snapshot shuffling.}
%    {\bf Features constrained:}
%       the static graph $\Gstat=(\N,\L)$;
%       the sequence of link weights, $\n=(\nij)_{\ij\in\L}$ (number of events on link);
%    {\bf Other features conserved:} $\a$, $\pit(\tauf)$.
%    {\bf References:} .}
%    \begin{center} 
%%      \includegraphics[height=0.04\textheight]{P__.pdf}
%    \end{center}
%    P[$\n$,$p(\t,\tauf)$] 
%    %is an event shuffling variant of P[$\w$,$\t$] defined above. It 
%    conserves the distribution of event durations, their starting times, and the links and their event frequencies $\n$, but not necessarily their weights $\w$.
\end{itemize}

%%------------------------------------------------------------------------------------------------------------------
%\subsubsection{Intersections of two shufflings}
%\label{sec:intersections}
%%------------------------------------------------------------------------------------------------------------------

%------------------------------------------------------------------------------------------------------------------
\subsubsection{Randomization based on metadata}
\label{sec:metadata}
%------------------------------------------------------------------------------------------------------------------
\noindent
The availability of metadata offers the possibility to impose additional constraints in the MRRMs. 
This allows to study effects that are not purely due to network structure and dynamics.
For instance, in Ref.~\cite{Kovanen2013T}, the age, gender, and type of subscription of mobile phone users were known; in Ref.~\cite{Rocha2011S}, the authors used shuffling methods respecting the bipartite structure of a sex worker-buyer interaction network, and Ref.~\cite{Genois2015C} used a shuffling that rewired links between each pair of predefined node groups in face-to-face networks.

These metadata MRRMs are all a type of stochastic blockmodel\footnote{These node-grouped MRRMs can be seen as microcanonical variants of the stochastic block model~\cite{Holland1983S}. However, typical stochastic block models found in the literature assign either the links at random inside each block (i.e.\ equivalent to P[$\gf,\Sigma_\L$]) or while constraining the degree sequence (equivalent to P[$\k$,$\gf,\Sigma_\L$])~\cite{Karrer2011S}, while the metadata MRRMs we consider here may impose any structural constraints.}. 
They may be defined by assigning a {\sl color} to each node, i.e.\ to which group it belongs among a set of $R$ predefined groups. The node colors are fixed by the vector $\gf=(\sigma_1, \sigma_2, \ldots, \sigma_N)$, where $\sigma_i\in\{1, 2, \ldots, R\}$.  An $R\times R$ {\sl group contact matrix}, $\Sigma_\L$ [with elements given by the number of links between groups, $(\Sigma_\L)_{\sigma\sigma'}=\sum_{(i,j)\in\L}(\delta_{\sigma_i,\sigma}\delta_{\sigma_j,\sigma'}+\delta_{\sigma_j,\sigma}\delta_{\sigma_i,\sigma'})$], typically fixes the number of links between members of each group [we may alternatively fix the number of events instead using a matrix $\Sigma_\C$, with elements given by $(\Sigma_\C)_{\sigma\sigma'}=\sum_{(i,j,t,\tau)\in\C}(\delta_{\sigma_i,\sigma}\delta_{\sigma_j,\sigma'}+\delta_{\sigma_j,\sigma}\delta_{\sigma_i,\sigma'})$].
%($\Sigma_\L$ is necessarily symmetric for undirected networks).
These two additional constraints enables us to define MRRMs that impose structure or dynamics determined by the metadata.

We may also directly use this blockmodel MRRM construction to conserve the bipartite structure of a network as in~\cite{Rocha2011S} by imposing two groups and a perfectly antidiagonal $\Sigma_\L$, with $(\Sigma_\L)_{11}=(\Sigma_\L)_{22}=0$ and $(\Sigma_\L)_{12}=(\Sigma_\L)_{21}=L$.
We may finally allow both $\gf$ and $\Sigma$ to vary over time  in order to capture temporal changes in the group structure.
%For example, we may introduce a contact matrix $\Sigmaf=(\Sigma_{\E^t})_{t\in\T}$ that fixes the number of events taking place in each snapshot in order to capture temporal changes in the group structure. %, as was e.g.\ done by the {\sl inducement} shufflings introduced in Ref.~\cite{Thomas2015D}. 
%Then $\gf=(\gf^m)_{m=1}^T$, with $\gf=(\sigma_1^m, \sigma_2^m, \ldots, \sigma_N^m)$, and $\Sigmaf_\E=(\Sigma^m_\E)_{m=1}^T$, where each $\Sigma^m_\E$ gives the number of events taking place between each group in snapshot $m$.

We describe below %and in Table~\ref{tab:effects-metadata} 
MRRMs relying on node metadata.
These are all link shufflings, so they conserve the same network features as those. 
%, namely: 
% the mean node strength, static degree, and instantaneous degree, $\mu(\s)$, $\mu(\d)$, and $\mu(\k)$; 
% the number of links in the static graph, $L$;
% the distributions of link weights (number of events and cumulative duration), $p(\n)$ and $p(\w)$; 
% number of events per time-lapse, $\Ebf$; 
% the multisets of the first and last events and the sequences of event and inter-event durations in the timelines, $\pxl(\t^1)$, $\pxl(\t^w)$, $\pxl(\tauf)$, and $\pxl(\Dtauf)$ 
% (see Tables~\ref{tab:features} and \ref{tab:constraint_levels} in Appendix~\ref{app:tables} for detailed definitions of features).
% We provide informal definitions of the principal features that are constrained by each shuffling method in the description 
% (see Table~\ref{tab:features} for detailed definitions). 
% The effects of each shuffling on a wide selection of other temporal network features are shown in Table~\ref{tab:effects-contact}.
% Table~\ref{tab:effects-metadata} lists the effects of the MRRMs on selected temporal network features. 
% We provide informal definitions of the principal features that are constrained by each shuffling method, and we list other features that are conserved as a consequence.
% Table~\ref{tab:features} provides detailed definitions of elementary features and Table~\ref{tab:constraint_levels} of distributions and moments constructed from these.}

\begin{itemize}
  \item[{P[$\pxl(\Thf)$,$\gf$,$\Sigma_\L$]}] 
    \new{{\bf Common name:} {\sl block-constrained link shuffling.}
    {\bf Features constrained:}
      the distribution (Def.~\ref{def:unordered_set}) of timelines, $\pxl(\Thf)=[\Thij]_{\ij\in\L}$;
      the sequence of node colors, $\gf=(\sigma_i)_{i\in\N}$;
      the group contact matrix, $\Sigma_\L$.
    % {\bf Other features conserved:} 
    %   $\mu(\k)$, $L$, $\mu(\s)$, $\mu(\d)$, $p(\n)$, $p(\w)$, $\Ebf$, $\pxl(\t^1)$, $\pxl(\t^w)$, $\pxl(\tauf)$, $\pxl(\Dtauf)$.
    {\bf References:} \cite{Genois2015C} ({\sl CM-shuffling}).}
      \vspace{0.2cm}\\
%    \begin{center} 
%      \includegraphics[height=0.04\textheight]{P__.pdf}
%    \end{center}
    P[$\pxl(\Thf)$,$\gf$,$\Sigma_\L$ shuffles the links in the static graph while constraining the group membership of each node, $\gf$, and the number of links between each group, $\Sigma_\L$. 
    %It was employed in Ref.~\cite{Genois2015C}, where it was called {\sl CM-shuffling}.
  %  
  \item[{P[$\kstat$,$\pxl(\Thf)$,$\gf$,$\Sigma_\L$]}] 
    \new{{\bf Common name:} {\sl degree- and block-constrained link shuffling.}
    {\bf Features constrained:}
      the degree sequence $\k=(k_i)_{i\in\N}$;
      the distribution (Def.~\ref{def:unordered_set}) of timelines, $\pxl(\Thf)=[\Thij]_{\ij\in\L}$;
      the sequence of node colors, $\gf=(\sigma_i)_{i\in\N}$;
      the group contact matrix, $\Sigma_\L$.
    % {\bf Other features conserved:} 
    %   $L$, $\mu(\s)$, $\mu(\d)$, $p(\n)$, $p(\w)$, $\Ebf$, $\pxl(\t^1)$, $\pxl(\t^w)$, $\pxl(\tauf)$, $\pxl(\Dtauf)$.
    {\bf References:} \cite{Rocha2011S} ({\sl random topological}).}
      \vspace{0.2cm}\\
%    \begin{center} 
%      \includegraphics[height=0.04\textheight]{P__.pdf}
%    \end{center}
    P[$\kstat$,$\pxl(\Thf)$,$\gf$,$\Sigma_\L$] randomizes $\Gstat$ while constraining the group structure, as P[$\pxl(\Thf)$,$\gf$,$\Sigma_\L$] does, while additionally constraining the node degrees $\kstat$. 
%    \new{It was used to constrain the bipartite structure of the network in} Ref.~\cite{Rocha2011S}.
  %  
  \item[{P[$G$,$p(\gf$)]}] 
    \new{{\bf Common name:} {\sl color shuffling.}
    {\bf Features constrained:}
      the complete temporal network $G$;
      the distribution (Def.~\ref{def:unordered_set}) of node colors, $p(\gf)=[\sigma_i]_{i\in\N}$.
    % {\bf Other features constrained:} all network features.
    {\bf References:} \cite{Kovanen2013T} ({\sl node type shuffled data}).}
      \vspace{0.2cm}\\
%    \begin{center} 
%      \includegraphics[height=0.04\textheight]{P__.pdf}
%    \end{center}
    P[$G$,$p(\gf$)] shuffles the group affiliations (colors) of the nodes at random.
    % (i.e. it randomizes the order of $\gf$). 
    It thus destroys all correlations between node color and network structure and dynamics, but conserves the network structure and dynamics completely. 
    %It is equivalent to permuting the links while constraining the static graph to be isometric to the original static graph, and thus could also be named {P[$\iso(\Gstat)$,$\pxl(\Thf)$,$\gf$,$\Sigma_\L$]}, but we use the above name for conciseness.
%      It was employed in Ref.~\cite{Kovanen2013T}, where it was called {\sl node type shuffled data}.
\end{itemize}

%------------------------------------------------------------------------

%------------------------------------------------------------------------
\begin{table*}
  \caption{{\bf Effects of MRRMs on features of temporal networks.}
  \new{See Table~\ref{tab:features} for definitions of features.
  Colored symbols show to what extent each feature is conserved. 
  Informal definitions are found in the tablenotes (detailed definitions are found in Supplementary Table~\ref{tab:constraint_levels}).}}
  \label{tab:effects-contact}
\centerline{
  \begin{threeparttable}
  \begin{tabular}{ll|ccc|cccc|cc|cccc|ccccc}
  \hline
  \hline
    Canonical name & Common name & \multicolumn{3}{l|}{topological} & \multicolumn{4}{l|}{weighted} & \multicolumn{2}{c|}{temp.} & \multicolumn{4}{l|}{node} & \multicolumn{5}{l}{link} \\
    && $\Gstat$ & $k_i$ & $\Estat$ & ${a_i}^{\dagger}$ & $s_i$ & ${n_\ij}^{\dagger}$ & $w_\ij$ & $\Et$ & &  & ${\alpha_i^m}^{\dagger}$ & $\Dalpha_i^m$ & $\dit$ &  & ${\tau_\ij^m}^{\dagger}$ & $\Dtau_\ij^m$ & $t^1_\ij$ & $t^w_\ij$   \\
  \hline
  \hline
    P[$E$] & {\sl Instant-event shuffling} & \xmark & \xmark & \xmark & \xmark & \mumark & \xmark & \xmark & \mumark & &  & \xmark & \xmark & \mumark &  & \xmark & \xmark & \xmark & \xmark  \\
    P[$p(\tauf)$] & {\sl Event shuffling} & \xmark & \xmark & \xmark & \xmark & \mumark & \xmark & \xmark & \mumark &   &  & \xmark & \xmark & \mumark &   & \pmark & \xmark & \xmark & \xmark  \\
  \hline
  \hline
    \multicolumn{2}{l|}{\textsl{\textbf{Link shufflings (LS):}}} &&&&&&&&&&&&&&&&&\\
    P[$\pxl(\Thf)$] & {\sl LS} & \xmark & \mumark & \cmark & \mumark & \mumark & \pmark & \pmark & \cmark & & & \xmark & \xmark & \mutmark &  & \pxlmark & \pxlmark & \pmark & \pmark \\
    P[$\lc,\pxl(\Thf)$] & {\sl Connected LS} & \lcmark & \mumark & \cmark & \mumark & \mumark & \pmark & \pmark & \cmark & & & \xmark & \xmark & \mutmark &  & \pxlmark & \pxlmark & \pmark & \pmark \\
    P[$\k,\pxl(\Thf)$] & {\sl Degree-constrained LS} & \xmark & \cmark & \cmark & \mumark & \mumark & \pmark & \pmark & \cmark & & & \xmark & \xmark & \mutmark &  & \pxlmark & \pxlmark & \pmark & \pmark \\
    P[$\k,\lc,\pxl(\Thf)$] & {\sl Connected, degree-constr.\ LS} & \lcmark & \cmark & \cmark & \mumark & \mumark & \pmark & \pmark & \cmark & & & \xmark & \xmark & \mutmark &  & \pxlmark & \pxlmark & \pmark & \pmark  \\
  \hline
  \hline
    \multicolumn{2}{l|}{\textsl{\textbf{Timeline shufflings (TS):}}} &&&&&&&&&&&&&&&&&\\
    P[$\L,E$] & {\sl TS} & \cmark & \cmark & \cmark & \xmark & \mumark & \xmark & \mumark & \mumark &   &   & \xmark & \xmark & \mumark &   & \xmark & \xmark & \xmark & \xmark  \\
    P[$\w$] & {\sl Weight-constrained TS} & \cmark & \cmark & \cmark & \xmark & \cmark & \xmark & \cmark & \mumark &  &  & \xmark & \xmark & \mumark &  & \xmark & \xmark & \xmark & \xmark \\
    P[$\piij(\Dtauf)$,$\t^1$] & {\sl Inter-event shuffling} & \cmark & \cmark & \cmark & \cmark & \cmark & \cmark & \cmark & \mumark &  &  & \xmark & \xmark & \mumark &  & \muijmark & \piijmark & \cmark & \cmark \\
    P[$\L$,$p(\tauf)$] &  {\sl TS} & \cmark & \cmark & \cmark & \mumark & \mumark & \mumark & \mumark & \mumark & & & \xmark & \xmark & \mumark &  & \pmark & \xmark & \xmark & \xmark  \\
    P[$\piij(\tauf)$] & {\sl Local TS} & \cmark & \cmark & \cmark & \cmark & \cmark & \cmark & \cmark & \mumark &   &   & \xmark & \xmark & \mumark &  & \piijmark & \xmark & \xmark & \xmark    \\
    P[$\piij(\tauf)$,$\t^1$,$\t^w$] & {\sl Activity-constrained TS} & \cmark & \cmark & \cmark & \cmark & \cmark & \cmark & \cmark & \mumark &   &   & \xmark & \xmark & \mumark &   & \piijmark & \muijmark & \cmark & \cmark \\
    P[$\piij(\tauf$),$\piij(\Dtauf$)] & {\sl Interval shuffling} & \cmark & \cmark & \cmark & \cmark & \cmark & \cmark & \cmark & \mumark && & \xmark & \xmark & \mumark & & \piijmark & \piijmark & \xmark & \xmark    \\
    P[$\piij(\tauf$),$\piij(\Dtauf)$,$\t^1$] & {\sl Inter-event shuffling} & \cmark & \cmark & \cmark & \cmark & \cmark & \cmark & \cmark & \mumark & & & \xmark & \xmark & \mumark &  & \piijmark & \piijmark & \cmark & \cmark  \\
    P[$\perl(\Thf)$] & {\sl Timeline shifting} & \cmark & \cmark & \cmark & \cmark & \cmark & \cmark & \cmark & \mumark & & & \xmark & \xmark & \mumark &   & \cmark & \cmark & \xmark & \xmark \\
  \hline
  \hline
    \multicolumn{2}{l|}{\textsl{\textbf{Sequence shufflings (SeqS):}}} &&&&&&&&&&&&&\\
    P[$\pxt(\Gs)$] & {\sl SeqS} & \cmark & \cmark & \cmark & \xmark & \cmark & \xmark & \cmark & \pmark &   &   & \xmark & \xmark & \pxtmark &   & \xmark & \xmark & \xmark & \xmark \\
    P[$\pxt(\Gs)$,$\sgnf(\Ebf)$] & {\sl Activity-constrained SeqS} & \cmark & \cmark & \cmark & \xmark & \cmark & \xmark & \cmark & \pmark, \sgnmark  &  &  & \xmark & \xmark & \pxtmark &  & \xmark & \xmark & \xmark & \xmark  \\
  \hline
  \hline
    \multicolumn{2}{l|}{\textsl{\textbf{Snapshot shufflings (SnapS):}}} &&&&&&&&&&&&&&&&&\\
    P[$\t$] & {\sl SnapS} & \xmark & \xmark & \xmark & \xmark & \mumark & \xmark & \xmark & \cmark &  &   & \xmark & \xmark & \mutmark &   & \xmark & \xmark & \xmark & \xmark \\
    P[$\d$] & {\sl Degree-constrained SnapS} & \xmark & \xmark & \xmark & \xmark & \mumark & \xmark & \xmark & \cmark &  &   & \cmark & \cmark & \cmark &  & \xmark & \xmark & \xmark & \xmark \\
    P[$\isof(\Gs)$] & {\sl Isomorphic SnapS} & \xmark & \xmark & \xmark & \xmark & \mumark & \xmark & \xmark & \cmark &   &   & \xmark & \xmark & \pitmark &   & \xmark & \xmark & \xmark & \xmark \\
    P[$p(\t,\tauf)$] & {\sl SnapS} & \xmark & \xmark & \xmark & \xmark & \mumark & \xmark & \xmark & \cmark & & & \xmark & \xmark & \mutmark &   & \pitmark & \xmark & \xmark & \xmark  \\
  \hline
  \hline
    \multicolumn{2}{l|} {\textsl{\textbf{Link-timeline intersections:}}}  &&&&&&&&&&&&&&&&&\\
    P[$\L$,$\pxl(\Thf)$] &  {\sl Topology-constrained LS} & \cmark & \cmark  & \cmark & \mumark & \mumark & \pmark & \pmark & \cmark & & & \xmark & \xmark & \mutmark &   & \pxlmark & \pxlmark & \pmark & \pmark \\
    P[$\w$,$\pxl(\Thf)$] &  {\sl Weight-constrained LS} & \cmark & \cmark & \cmark & \mumark & \cmark & \pmark & \cmark & \cmark & & & \xmark & \xmark & \mutmark &  & \pxlmark & \pxlmark & \pmark & \pmark   \\
    P[$\n$,$\pxl(\Thf)$] & {\sl Weight-constrained LS}  & \cmark & \cmark  & \cmark & \cmark & \mumark & \cmark & \pmark & \cmark & & & \xmark & \xmark & \mutmark &   & \pxlmark & \pxlmark & \pmark & \pmark  \\
  \hline
  \hline
    \multicolumn{2}{l|}{\textsl{\textbf{Link-snapshot intersections:}}} &&&&&&&&&&&&&&&&&\\
    P[$\w$,$\t$] & {\sl Timestamp shuffling} & \cmark & \cmark & \cmark & \xmark & \cmark & \xmark & \cmark & \cmark &  &  & \xmark & \xmark &\mutmark &  & \xmark  &\xmark & \xmark & \xmark \\
    P[$\L$,$p(\t,\tauf)$] & {\sl Topology-constrained SnapS} & \cmark & \cmark & \cmark & \mumark & \mumark & \mumark & \mumark & \cmark & & & \xmark & \xmark & \mutmark &  & \pitmark & \xmark & \xmark & \xmark     \\
  \hline
  \hline
  \end{tabular}
  \begin{tablenotes}
    \item[$\dagger$] Feature only defined for temporal networks with event durations.
    \item[\cmark] Feature completely conserved, typically the ordered sequence of individual features, e.g.\ $\x=(x_i)_{i\in\N}$ or $\x=((\x_i^t)_{t\in\T})_{i\in\N}$.
    \item[\piijmark] Sequence of local distributions on links, $\piij(\x) = \left([x_\ij^m]_{m\in\Ml}\right)_{\ij\in\L}$.
    \item[\pitmark] Sequence of local distributions in snapshots, $\pit(\x) = \left([x_i^t]_{i\in\N}\right)_{t\in\T}$.
    \item[\pxlmark] Distribution of local sequences on links, $\pxl(\x) = \left[(x_\ij^m)_{m\in\Ml}\right]_{\ij\in\L}$.
    \item[\pxtmark] Distribution of local sequences in snapshots, $\pxt(\x) = \left[(x_i^t)_{i\in\N}\right]_{t\in\T}$.
    \item[\pmark] Distribution (i.e.\ the multiset) of individual scalar values in sequence.
    \item[\muijmark] Sequence of local means on links, $\muij(\x) = \left(\sum_{m\in\Ml} x_\ij^m/M_\ij\right)_{\ij\in\L}$.
    \item[\mutmark] Sequence of local means in snapshots links, $\mut(\x) = \left(\sum_{i\in\N} x_i^t/N\right)_{t\in\T}$.
    \item[\mumark] Mean value of the individual scalar features in sequence.
    \item[\xmark] Feature not conserved.
%    \item[$\lc$] Connectedness of static graph $\Gstat$ conserved.
  \end{tablenotes}
  \end{threeparttable}
}
\end{table*}
%------------------------------------------------------------------------

%------------------------------------------------------------------------
\begin{table*}
  \caption{{\bf Effects of metadata-dependent shufflings on features of temporal networks.}
  Special metadata symbols are the color (group affiliation) of a node, $\sigma_i$, and the group contact matrices $\Sigma_\L$ and $\Sigma_\E$ (see main text for their definition and Table~\ref{tab:features} for other \new{features).
  Colored symbols show to what extent each feature is conserved. 
  Informal definitions are found in the tablenotes (detailed definitions are found in Supplementary Table~\ref{tab:constraint_levels}).}
%   Note that a feature that is not constrained (\xmark) by a MRRM is not necessarily completely randomized \new{as the feature may be correlated with the features constrained  by the MRRM (for a more detailed discussion see Supplementary Note 2).}
  }
  \label{tab:effects-metadata}
\centerline{
  \begin{threeparttable}
  \begin{tabular}{ll|ccc|ccc|cccc|cc|cccc|ccccccc}
    \hline
    \hline
    Canonical name & Common name & \multicolumn{3}{l|}{meta} & \multicolumn{3}{l|}{topological} & \multicolumn{4}{l|}{weighted} & \multicolumn{2}{c|}{temp.} & \multicolumn{4}{l}{node} & \multicolumn{5}{l}{link} \\
    &&  $\sigma_i$ & $\Sigma_\L$ &  $\Sigma_\E$ & $\Gstat$ & $k_i$ & $\Estat$ & $a_i^{\dagger}$ & $s_i$ & $\nij^{\dagger}$ & $\wij$ & $\Et$ & & & ${\alpha_i^m}^{\dagger}$ & $\Dalpha_i^m$ & $\dit$ & & ${\tau_\ij^m}^{\dagger}$ & $\Dtau_\ij^m$ & $t^1_\ij$ & $t^w_\ij$ \\
    \hline
    P[$\pxl(\Thf)$,$\gf$,$\Sigma_\L$] & {\sl Block LS} & \cmark & \cmark & \xmark  &\xmark & \mumark & \cmark  & \mumark & \mumark & \pmark & \pmark & \cmark &  &  & \xmark & \xmark & \mutmark &  & \pxlmark & \pxlmark & \pmark & \pmark \\
    P[$\k$,$\pxl(\Thf)$,$\gf$,$\Sigma_\L$] & {\sl Deg. + block LS} & \cmark & \cmark & \xmark & \xmark & \cmark & \cmark  & \mumark & \mumark & \pmark & \pmark & \cmark &  & & \xmark & \xmark & \mutmark &  & \pxlmark & \pxlmark & \pmark & \pmark \\
    P[$G$,$p(\gf)$] & {\sl Color shuffling} & \pmark & \cmark & \xmark & \cmark & \cmark & \cmark  & \cmark & \cmark & \cmark & \cmark & \cmark & &  & \cmark & \cmark & \cmark &  & \cmark & \cmark & \cmark & \cmark \\
    \hline
    \hline
  \end{tabular}
  \begin{tablenotes}
    \item[$\dagger$] Feature only defined for temporal networks with event durations.
    \item[\cmark] Feature completely conserved, typically the ordered sequence of individual features, e.g.\ $\x=(x_i)_{i\in\N}$ or $\x=((\x_i^t)_{t\in\T})_{i\in\N}$.
    \item[\pxlmark] Distribution of local sequences on links, $\pxl(\x) = \left[(x_\ij^m)_{m\in\Ml}\right]_{\ij\in\L}$.
    \item[\pmark] Distribution (i.e.\ the multiset) of individual scalar values in sequence.
    \item[\mutmark] Sequence of local means in snapshots links, $\mut(\x) = \left(\sum_{i\in\N} x_i^t/N\right)_{t\in\T}$.
    \item[\mumark] Mean value of the individual scalar features in sequence.
    \item[\xmark] Feature not conserved.
  \end{tablenotes}
  \end{threeparttable}
  }
\end{table*}
%------------------------------------------------------------------------

%========================================================================
\section{Generating new microcanonical randomized reference models from existing ones}
\label{sec:composition}
%========================================================================
%, i.e.\ it is {\sl coarser} (Proposition~\ref{the:partition_refinements}) than both. 
%Not all compositions lead to a microcanonical RRM, however. We refer to pairs of MRRMs whose composition does result in a microcanonical RRM as {\sl compatible} (Def.~\ref{def:compatibility}). 
%As shown in Section~\ref{sec:implementation}, link shufflings and timeline shufflings are compatible (Proposition~\ref{prop:link-timeline_independent}), while sequence shufflings are compatible with both snapshot shufflings and link shufflings (Propositions~\ref{prop:sequence-snapshot_independent} and \ref{prop:link-sequence_independent}). 

%As discussed above, we may combine two compatible MRRMs by composition (i.e.\ applying the second MRRM to the networks generated by the first) to randomize at different levels at the same time (see Section~\ref{sec:combination}). For example, complete randomization of an instant-event temporal network, while keeping the number of links fixed, may be obtained by randomly permuting the links between all pairs of nodes using the link shuffling P[$\pxl(\Thf)$] and randomly permuting the instantaneous events on and between the links using the timeline shuffling P[$\L$]. 
%The resulting model is P[$L$] (Proposition~\ref{prop:link-timeline_independent}). %(Theorem~\ref{the:independence}).
%Since compatible MRRMs commute, it does not matter in which order we apply them (Proposition~\ref{prop:compat-commute}). 

\noindent
\new{It is possible to combine two different MRRMs to form a new MRRM by applying the second MRRM to the state (i.e., a network) returned by the first. %each graph in the ensemble generated by the first.
This defines a {\sl composition} of the MRRMs and results in a model that \new{randomizes} more than either of the two original MRRMs. 
%In this section we consider how we may combine a pair of MRRMs to form a new MRRM. 
}
%{\sl Composition} of two reference models by applying one shuffling method after another is a practical way of creating new reference models. 
However, not all MRRMs are {\sl compatible} in a way that their composition would produce another MRRM, and we here develop the theory needed to show that two MRRMs are compatible and to identify the MRRMs resulting from their composition. 
\new{These theorems  are instrumental for showing that several of the important classes of shuffling methods defined in Section~\ref{sec:implementation} are compatible, and thus that new MRRMs can be created by composing pairs of these shufflings.}

\new{In Subsection~\ref{sec:theory-composition}, we start  by developing the theory needed  to formally define the composition of two MRRMs, and we explore its properties. % and to show which MRRMs are compatible. 
We next show that compatibility is equivalent to a certain form of conditional independence between the features constrained by the two MRRMs. 
We use this in the following subsection (Subsection~\ref{sec:shufflings-composition}) to show that certain of the classes of shuffling methods described in Section~\ref{sec:implementation} are compatible and to describe the MRRM that results from their composition.
In Subsection~\ref{sec:compositions}, we finally list and characterize MRRMs found in the literature that are compositions of two other MRRMs. }
% listed in Section~\ref{sec:classification} above.}
%Some MRRMs generated by composition of two other MRRMs are listed in Subsection~\ref{sec:compositions}, but 

%------------------------------------------------------------------------
\subsection{Theory: Composition of MRRMs}
\label{sec:theory-composition}
%------------------------------------------------------------------------
\noindent
In this section we explore how two different MRRMs may be combined \new{in composition to generate another MRRM.
This} generates a RRM that is not necessarily microcanonical, but if it is, it shuffles more than either of the two.
Especially the type of compositions that produce MRRMs are of practical interest as they provide a way of producing new MRRMs by \new{combining existing shuffling algorithms}.

The latter part of this section will be devoted to exploring under which conditions the composition of two MRRMs is microcanonical (we then say that the two MRRMs are {\sl compatible}).
We develop a concept called {\sl conditional independence given a common coarsening} and show that it characterizes compatibility. 
Finally, we show that specific types of refinements of compatible MRRMs are also compatible and identify the way the resulting MRRM inherits the features of the two input models. 

% We use these results in Section~\ref{sec:implementation} to show that 
% \new{link shufflings (Def.~\ref{def:link_shuffling}) are compatible with timeline shufflings (Def.~\ref{def:timeline_shuffling}) and that sequence shufflings (Def.~\ref{def:sequence_shuffling}) are compatible with snapshot shufflings (Def.~\ref{def:snapshot_shuffling}).}
%MRRMs which shuffle the links are compatible with MRRMs which randomize the timelines in the link-timeline representation of a temporal network (Def.~\ref{def:link-timeline_graph}), and that MRRMs shuffling the order of snapshots are compatible with MRRMs randomizing the individual snapshots in the snapshot graph sequence representation (Def.~\ref{def:snapshot-sequence}).

\new{%To keep the presentation as simple as possible, w
We here present propositions and theorems without proofs. They can be found in Appendix~\ref{app:proofs}.}

\subsubsection{\new{Composition of two MRRMs}}

\begin{figure}
  \begin{tikzpicture}
    \node at(0.2, 0.){\includegraphics[width=3.325in]{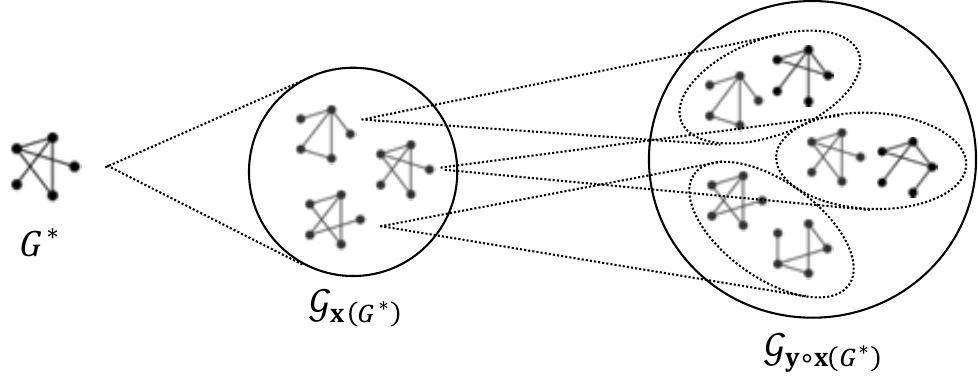}};
      \node at(-4.1,1.3){\bf(a)};
    \node at(-0.5,-3){\includegraphics[height=0.68in]{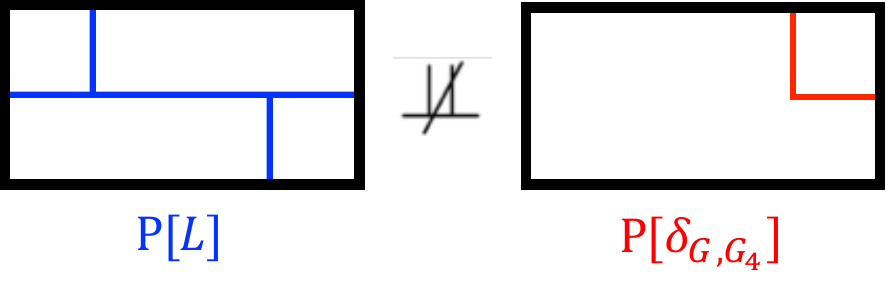}};
      \node at(-4.1,-2){\bf(b)};
    \node at(0.3,-5.2){\includegraphics[height=0.68in]{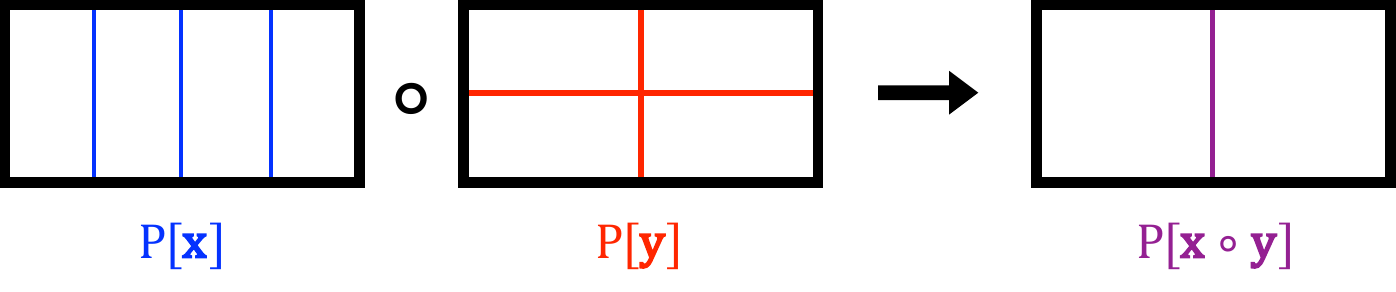}};
      \node at(-4.1,-4.2){\bf(c)};
  \end{tikzpicture}
  \caption{{\bf Composition of MRRMs.} 
%  \note{CLV: add illustration of algorithmic implementation of composition?}  
  \new{(a) The composition $\model{\comp{\x}{\y}}$ of two shuffling methods is implemented by applying $\model{\x}$ to the outputs of $\model{\y}$.
  (b) Example of incompatible partitions of a state space of 8 states (e.g.\ all simple graphs with 3 nodes).
  (c)}  
%  \note{CLV: add example of partitions from Ex.~\ref{ex:incompatible} that is not compatible with the two above?}  
  The composition P[$\x\circ\y$] of two compatible MRRMs, P[$\x$] and P[$\y$], shuffles more than (or as much as) either of the two. In terms of partitions, P[$\x\circ\y$] produces the least upper bound of P[$\x$] and P[$\y$].}
  \label{fig:composition}
\end{figure}

\noindent
In practice, the composition of two MRRMs involves first applying one shuffling method to the input network $G^*$, and then applying the second shuffling method to the outputs of the first \new{[Fig.~\ref{fig:composition}(a)]}. 
This thus defines a composition of the two shuffling methods: 

\begin{definition}
{\sl Composition of MRRMs.}
\label{def:composition}
Consider two MRRMs P[$\x$] and P[$\y$] and an input network $G^*\in\G$. The composition of P[$\y$] on P[$\x$], denoted $\model{\comp{\y}{\x}}$ %P[$\y\circ\x$], 
is defined by the conditional probability:
\beqa
  \Prob{\comp{\y}{\x}}(G|G^*) &=& \sum_{G'\in\G} \Prob{\y}(G|G')\Prob{\x}(G'|G^*) \nonumber\\
  &=& \sum_{G'\in\G} \frac{\delta_{\y(G),\y(G')}}{\Omega_{\y}(G')} \frac{\delta_{\x(G'),\x(G^*)}}{\Omega_{\x(G^*)}} \es.
  \label{eq:composition}
\eeqa
\end{definition}

For a given indexing of the state space $\G$,  
%s.~(\ref{eq:transition_matrix}) and 
\new{Eq.~(\ref{eq:composition}) shows that the transition matrix (Def.~\ref{def:mrrm_representations})} for the composition of P[$\y$] on P[$\x$] is simply the matrix product of the individual transition matrices, $\P^{\y\circ\x}=\P^\y\P^\x$.

\begin{definition}
{\sl Compatibility.}
\label{def:compatibility}
We say that two MRRMs P[$\x$] and P[$\y$] are {\sl compatible} if their composition $\model{\comp{\y}{\x}}$ is also a MRRM. 
\end{definition}

The notion of compatibility is central as it defines which MRRMs we may combine through composition to define a new MRRM.

\begin{proposition}
\label{prop:compat-commute}
{\sl Compatible randomized reference models commute.}
If two MRRMs, P[$\x$] and P[$\y$], are compatible then $\model{\comp{\y}{\x}}=\model{\comp{\x}{\y}}$.
\end{proposition}

Proposition~\ref{prop:compat-commute} means that it does not matter in which order we apply two compatible MRRMs in the composition, and consequently that $\model{\comp{\y}{\x}}$ and $\model{\comp{\x}{\y}}$ define the same MRRM if P[$\x$] and P[$\y$] are compatible.
It also means that in order to show that two MRRMs are not compatible, it suffices to show that they do not commute.

\begin{example}
\label{ex:incompatible}
Let the state space $\G$ be all static graphs with 3 nodes. 
As in Example~\ref{ex:MRRM_representations}, we number the 8 graphs such that $G_1$ is the graph with 0 links, $G_2$, $G_3$, and $G_4$ are the graphs with 1
link, $G_5$, $G_6$, and $G_7$ are the graphs with 2 links and $G_8$ is the graph with 3 links.
Let us now define two MRRMs for this state space \new{[Fig.~\ref{fig:composition}(b)]}:
\begin{enumerate}
  \item The MRRM P[$L$], defined by the number of links $L$ in the graph,
  %, corresponding to the Erd\H{o}s-R\'enyi random graph model ${\rm ER}(3,L)$, 
%is defined by the feature $L$ which returns the number of edges in the network, and 
partitions the state space in 4 sets $\G_0 = \{G_1\}$, $\G_1 = \{G_2, G_3, G_4\}$, $\G_2 = \{G_5, G_6, G_7\}$, and $\G_3=\{G_8\}$. %\note{Add figure of partition [Fig.~\ref{fig:incomparable}(a)]?}
  \item The MRRM P[${\delta_{G,G_4}}$]  keeps the graph $G_4$ unchanged and shuffles all the others. It is defined by the feature $\delta_{G,G_4}$ which returns 1 when applied to $G_4$ and 0 otherwise. 
  This MRRM partitions $\G$ into two partitions $\G_1=\{G_4\}$ and $\G_0 = \{G_1, G_2, G_3, G_5, G_6, G_7, G_8\}$. %\note{Add figure of partition [Fig.~\ref{fig:incomparable}(b)]?}
\end{enumerate}
With these definitions $\model{\comp{L}{\delta_{G,G_4}}}\neq\model{\comp{\delta_{G,G_4}}{L}}$ %${\rm P}[L]{\rm P}[{\delta_{G,G_4}}]\neq{\rm P}[{\delta_{G,G_4}}]{\rm P}[L]$
\new{. 
For example, $\model{\comp{L}{\delta_{G,G_4}}}$ applied to $G_4$ can return the states $\{G_2, G_3, G_4\}$, while the application of $\model{\comp{\delta_{G,G_4}}{L}}$ can return the entire $\G$.}
So the two MRRMs do not commute and are thus not compatible. Consequently, the \new{RRM} %ensembles 
obtained by composition of P[${\delta_{G,G_4}}$] and P[$L$] is not microcanonical.
It is in the above example also easy to verify \new{e.g.\ }that the states generated by $\model{\comp{\delta_{G,G_4}}{L}}$ are not equiprobable (e.g.\ $P_{\comp{\delta_{G,G_4}}{L}}(G_4)=1/3$ while $P_{\comp{\delta_{G,G_4}}{L}}(G_i)=2/21$ for all other graphs).
\end{example}

\subsubsection{Comparability and compatibility}
\noindent
\new{The composition of two compatible MRRMs produces a MRRM that is comparable to the two and, in particular, one that randomizes more than each of them individually (i.e.\ one that is {\sl coarser}).}
%We can use Proposition~\ref{prop:compat-commute} to show that the composition of two compatible MRRMs randomizes more than either of the individual MRRMs.

\begin{proposition}
\label{prop:composition_is_coarser}
{\sl Composition of two compatible MRRMs always results in a MRRM which does not shuffle less.}
Consider two compatible MRRMs, P[$\x$] and P[$\y$]. Their composition, $\model{\comp{\y}{\x}}$, is coarser \new{than (or equal to)} both $\model{\y}$ and $\model{\x}$, i.e.\ $\model{\comp{\y}{\x}}\geq\model{\x}$ and $\model{\comp{\y}{\x}}\geq\model{\y}$, even if $\model{\x}$ and $\model{\y}$ are not comparable.
\end{proposition}

In order for the concept of compatibility to be practically useful we need to be able to find out which MRRMs are compatible and what the result of their compositions is. Comparable MRRMs are an easy special case in this regard, as all comparable MRRMs turn out to be compatible and their composition simply yields the MRRM that shuffles more.

\begin{proposition}
\label{prop:comparable-compatible}
{\sl Comparable microcanonical randomized reference models are compatible.}
Let P[$\x$] and P[$\y$] be two MRRMs \new{satisfying} $\model{\x}\leq \model{\y}$. Then they are compatible and their composition gives $\model{\comp{\y}{\x}} = {\rm P}[\y]$.
\end{proposition}

\begin{example}
Consider again the MRRMs P[$L$] and P[$\kstat$] from Example~\ref{ex:partial_order}.
Since they are comparable, they are compatible according to Proposition~\ref{prop:comparable-compatible}. Consequently they also commute (Proposition~\ref{prop:compat-commute}) and $\model{\comp{\k}{L}} = \model{\comp{L}{\k}} = {\rm P}[L]$.
\end{example}

The effect of the composition operation  works in the opposite manner to the intersection \new{of MRRMs} \new{(Def.~\ref{def:intersection})}. %operation. 
This is also seen in the effect of composition with the zero and unity elements (which are compatible with all MRRMs by Proposition \ref{prop:comparable-compatible})\new{: 
here} zero is the neutral element $\model{\comp{0}{\x}} = \model{\x}$ and unity is the absorbing element $\model{\comp{1}{\x}} = \model{1}$, whereas $\model{0}$ is the neutral element and $\model{1}$ the absorbing element for intersection (Sec.~\ref{sec:MRRM_space}). 
Furthermore, by Proposition \ref{prop:composition_is_coarser}, the composition gives an upper bound for the two MRRMs {(Fig.~\ref{fig:composition})}. 
In fact, the bound is the least upper bound, and any set of compatible MRRMs forms a {\sl lattice} \cite{Stanley2011E}, but this connection to the theory of partially ordered sets is not pursued further here.

\subsubsection{Conditional independence and compatibility}
\noindent
Our aim in this section is to be able to compose MRRMs to produce new ones, and even though comparable MRRMs are always compatible they are not useful for this purpose as they do not produce a new MRRM. There are more interesting compositions, but in order to be able to access these we need a way of characterizing which pairs of MRRMs are compatible outside of comparable ones. We will next define the concept of conditional independence between two features given a common coarsening of these 
and show in Theorem~\ref{the:independence} that it is equivalent to compatibility. 
We next show in Theorem~\ref{the:independent_refinement} that certain refinements of compatible MRRMs (termed {\sl adapted} refinements) are themselves compatible.
Theorem~\ref{the:independent_refinement} furthermore shows which features their composition inherits from the original MRRMs. 

%The theorems will be used in Section~\ref{sec:implementation} to show that certain important classes of MRRMs are compatible. 
%In particular, we will show \new{first that link shufflings} %, which shuffle the links in the link-timeline representation of a temporal network 
%(Def.~\ref{def:link-timeline_graph}) are compatible with \new{timeline shufflings}. %, which randomize the individual timelines. 
%This will be done by first showing that the coarsest possible link and timeline shufflings are compatible using Theorem~\ref{the:independence}. 
%Second, noting that all link and timeline shufflings are adapted refinements of these, Theorem~\ref{the:independent_refinement} then gives us that they are compatible, as well as telling us which features their composition constrains.
%The same reasoning will be applied to show that \new{sequence shufflings} %, which shuffle the order of snapshots in the snapshot-graph sequence 
%(Def.~\ref{def:snapshot-sequence}) and \new{snapshot shufflings} %, which randomize events inside snapshots, 
%are compatible and to derive which features their compositions constrain.

Before we can define the concept of conditional independence given a common coarsening, we first need to define the concepts of conditional probability \new{of a feature} and conditional independence between features. 

\begin{definition}
{\sl Conditional probability of a feature.}
\label{def:conditional_probability}
The conditional probability of a feature $\y$ given another feature $\x$ is the probability $P_{\y|\x}(\y^\dagger|\x^*)$ that the feature $\y$ takes the value $\y^\dagger$ conditioned on the value $\x^*$ of the feature $\x$. It is given by 
\begin{eqnarray}
  \CProb{\y}{\x}(\y^\dagger|\x^*) &=& \sum_{G'\in\G} \delta_{\y^\dagger,\y(G')} \Prob{\x}(G'|\x^*)  \nonumber\\
  &=& \frac{\Omega_{(\y^\dagger,\x^*)}}{\Omega_{\x^*}} \es.
  \label{eq:conditional_measure}
\end{eqnarray}
\end{definition}

The conditional probability of a feature satisfies all properties of usual conditional probabilities. 
We may notably relate the composition of two MRRMs to the conditional probability of their features using the law of total probability as $P_{\comp{\y}{\x}}(G|\x^*)=\sum_{\y^\dagger}P_{\y}(G|\y^\dagger)P_{\y|\x}(\y^\dagger|\x^*)$.
%(\new{we provide a proof of this in the proof of Theorem~\ref{the:independence} in Appendix~\ref{app:proofs}}). 
It also allows us to define the conditional independence in the usual sense as when $P_{\y|\x,\z}(\y^\dagger|\x^\ddagger,\z^*)=P_{\y|\z}(\y^\dagger|\z^*)$ for a given a third feature $\z$. % and for all $\y^\dagger$, $\x^*$.
We shall here be concerned with a stricter version of conditional independence which is satisfied when the feature $\z$ is coarser than both $\y$ and $\x$. %As we show below, 
This {\sl conditional independence given a common coarsening} is equivalent to $\model{\x}$ and $\model{\y}$ being compatible.

\begin{definition}
{\sl Conditional independence given a common coarsening.}
\label{def:independence}
If there exist a feature $\z$ that is a common coarsening of both $\x$ and $\y$ (i.e.\ $\z\geq\x$ and $\z\geq\y$) such that $P_{\y|\x}(\y^\dagger|\x(G^*))=P_{\y|\z}(\y^\dagger|\z(G^*))$ for all $G^*\in\G$, we will say that $\y$ is conditionally independent of $\x$ given their common coarsening $\z$.
\end{definition}

%Note that 

As for the usual conditional independence, the conditional independence given a common coarsening defined above is symmetric in $\x$ and $\y$. 
%, we show this in the following proposition.

\begin{proposition}
\label{prop:symmetry}
{\sl Symmetry of the conditional independence given a common coarsening.}
If $\x$ is conditionally independent of $\y$ given a common coarsening $\z$ then $\y$ is conditionally independent of $\x$ given $\z$
\end{proposition}

Because of the symmetry, we can simply say that $\x$ and $\y$ are conditionally independent given the common coarsening $\z$.

As we stated above, the concept of conditional independence given a common coarsening is important because it is a characterization of compatibility. 
The following theorem proves this.

\begin{theorem}
{\sl Conditional independence given a common coarsening is equivalent to compatibility.}
\label{the:independence}
$\model{\x}$ and $\model{\y}$ are compatible if and only if they are conditionally independent given the common coarsening $\z = \comp{\x}{\y}$. 
\end{theorem}

Conditional independence is important in practice for designing MRRMs that randomize both the topology and the time-domain of a temporal network by implementing them as compositions of MRRMs that \new{individually randomize either the topological or temporal aspects of a network (see Section~\ref{sec:shufflings-composition} below).} 
%different levels in the two-level temporal network representations introduced in Section~\ref{sec:nested}. 
The \new{following example illustrates the concepts of conditional independence and compatibility in terms of the abstract state space}. 

%\begin{example}
%\label{ex:orthogonality}
%Consider a state space with 9 states, $\G=\{G_1,\ldots,G_9\}$, which are placed into a square formation such that the states 1 to 3 are in the first row, 4 to 6 in the second row and 7 to 9 in the third row.
%Now we can define two features: $f_r$ that returns the row number, and $f_c$ that returns the column number. The partitions these two features induce are $\G_{f_r}=\{\{G_1, G_2, G_3\}, \{G_4, G_5, G_6\}, \{G_7, G_8, G_9\}\}$ for $f_r$ and $\G_{f_c}=\{\{G_1, G_4, G_7\}, \{G_2, G_5, G_8\}, \{G_3, G_6, G_9\}\}$ for $f_c$.
%The two features $f_r$ and $f_c$ are conditionally independent given $1$ (the unity element, i.e., a constant function).
%Thus, the corresponding MRRMs, $\model{f_r}$ and $\model{f_c}$, are compatible and their composition is the unity element $\model{f_r}\model{f_c}=\model{f_c}\model{f_r}=\model{1}$, which shuffles everything.
%Indeed, direct computation shows that their composition is the unity element $\P^{f_r}\P^{f_c} = \P^{1}$.
%%\note{CLV: should we change this example to illustrate conditional independence instead of complete independence? We could also add an example of complete independence at the end of the section.}
%%\note{CLV: Figure?}
%\end{example}

\new{
\begin{example}
\label{ex:orthogonality}
Consider a state space with 8 states, $\G=\{G_1,\ldots,G_8\}$, which are placed into a square formation such that the states 1 to 4 are in the first row and 5 to 8 are in the third row.
Now we can define two features: $f_c$ that returns the column number, and $f_q$ that returns the quadrant that the state is in. 
The partitions these two features induce are $\G_{f_c}=\{\{G_1, G_5\}, \{G_2, G_6\}, \{G_3, G_7\}, \{G_4, G_8\}\}$ for $f_c$ and $\G_{f_c}=\{\{G_1, G_2\}, \{G_3, G_4\}, \{G_5, G_6\}, \{G_7, G_8\}\}$ for $f_q$ [Fig.~\ref{fig:composition}(c)].
The two features $f_c$ and $f_q$ are conditionally independent given the function $f_h$ that indicates which half of the state space the state is in.
Thus, the corresponding MRRMs, $\model{f_c}$ and $\model{f_q}$, are compatible and their composition is the MRRM $\model{f_h}$ [Fig.~\ref{fig:composition}(c)].
\end{example}
}

%\note{CLV: replaced the example above to 8 states in a $4\times2$ rectangle to match illustration in Fig.~\ref{fig:composition}. This way we furthermore show  a conditionally independent example instead of a fully independent one.} 
%\note{Functions would be column number and quadrant if we want a conditionally independent example, or column and row number for a fully independent one.
%The column and quadrant are independent conditioned on the function returning zero for states in the left half and one for states in the right.
%%This way we show something that is not fully independent...
%Depending on whether we keep the discussion of full independence, we may even want to show first a conditionally independent example and second a fully independent one (in relation to Def.~\ref{def:independence}).}

%\note{CLV: removed the  paragraph on relation to temporal net MRRMs.}
%The above example illustrates the abstract nature of the problem of combining MRRMs. In terms of temporal networks, $\G$ may be thought of as the space of all networks consisting of 3 nodes and a single event that takes place during one of three possible snapshots. 
%The two features may then be identified with the features $\Gstat$ returning the static network and $\pxl(\Thf)$ returning the number events on each link and their timings. If we let $f_r=\Gstat$ and $f_c=\pxl(\Thf)$, then the row number determines the placement of the link in the static network and the column number the snapshot during which the event takes place. 

Our main aim when defining compositions has been to be able to produce new useful MRRMs. With the help of the concept of conditional independence we are now ready to write down a theorem 
%, and a proof, 
that will allow us to compose non-comparable MRRMs and know the features of the resulting model.
To do so we define a special type of {\sl adapted} refinements of compatible MRRMs which we can show are also compatible.

\begin{definition}
\label{def:adapted_refinement}
{\sl Adapted refinement.}
Consider two compatible MRRMs, P[$\x$] and P[$\y$]. 
Any refinement of P[$\x$] of the form P[$\x,\f(\y)$], where $\f$ is any function of $\y$, is said to be {\sl adapted} to P[$\y$]. We will refer to P[$\x,\f(\y)$] as an {\sl adapted refinement of {\rm P}$[\x]$ with respect to {\rm P}$[\y]$}.
\end{definition}

In the following theorem we will now demonstrate that all adapted refinements of compatible MRRMs are themselves compatible, as well as showing which features the composition of such MRRMs inherits from the individual MRRMs. 
This theorem will be very useful in practice: if we can show that a given pair of MRRMs are compatible (i.e.\ using Theorem~\ref{the:independence}), we get for free that a whole class of MRRMs, consisting of all adapted refinements of the original MRRMs, are also compatible. 

\begin{theorem}
{\sl Adapted refinements of compatible MRRMs are compatible.}
\label{the:independent_refinement}
Consider two compatible MRRMs $\model{\x}$ and $\model{\y}$, and any adapted refinements of these, $\model{\y, \f(\x)}$ and $\model{\x, \g(\y)}$. 
%, that shuffle less. %$\f(\x) \leq \x$ and $\g(\y) \leq \y$.
%two functions $\f$ and $\g$ that take $\x$ and $\y$ as input, respectively. 
Then $\model{\y, \f(\x)}$ and $\model{\x, \g(\y)}$ are compatible, and their composition is given by $\model{\comp{\x}{\y},\f(\x),\g(\y)}$.
\end{theorem}

\subsection{Compositions of shuffling methods}
\label{sec:shufflings-composition}
%------------------------------------------------------------------------
\noindent
\new{Theorems~\ref{the:independence} and \ref{the:independent_refinement} let us show with the two following propositions that all link shufflings are compatible with any timeline shuffling, and likewise that all sequence shufflings are compatible with any snapshot shuffling.
We may thus combine pairs of such shufflings to generate new MRRMs that randomize both the topological and temporal domains of a temporal network in different ways.}

\new{Formal proofs for the propositions are given in Appendix~\ref{app:proofs}. 
In essence they boil down to first using Theorem~\ref{the:independence} to show that the coarsest (i.e.\ most random) link and timeline  (sequence and snapshot) shufflings are compatible.
Second, by noting that all link and timeline  (sequence and snapshot) shufflings are adapted refinements (Def.~\ref{def:adapted_refinement}) of these, Theorem~\ref{the:independent_refinement} gives that these are also compatible.}
%one class of MRRM randomizing features in a manner that is independent of the features randomized by the other class.

\begin{proposition}
\label{prop:link-timeline_independent}
{\sl Link shufflings and timeline shufflings are compatible.}
Any link shuffling P[$\f(\L),\Thf$] and timeline shuffling P[$\L,\g(\Thf)$] are compatible and their composition is given by P[$L,\f(\L),\g(\Thf)$].
\end{proposition}

\new{
\begin{example}
The composition of the coarsest link shuffling $\model{\pxl(\Thf)}$ with the coarsest timeline shuffling $\model{\L,E}$ results in the MRRM $\model{L,E}$ which randomizes both the static topology and the temporal order of events while conserving the number of links $L$ in the static graph 
%\note{CLV: add figure?}
%(Fig.~\ref{fig:link-timeline-shuffling}).
\end{example}
}

%\begin{figure}
%  \caption{\new{{\bf The coarsest link-timeline shuffling.} $\model{L,E}$ is obtained as the composition of $\model{\pxl(\Thf)}$ and $\model{\L,E}$.}}
%  \label{fig:link-timeline-shuffling}
%\end{figure}

\begin{proposition}
\label{prop:sequence-snapshot_independent}
{\sl Sequence shufflings and snapshot shufflings are compatible.}
Any sequence shuffling ${\rm P}[\f(\t),\pxt(\Gf)]$ and snapshot shuffling ${\rm P}[\t,\g(\pxt(\Gf))]$ are compatible and their composition is given by ${\rm P}[p(\Ebf),\f(\t),\g(\pxt(\Gf))]$.
\end{proposition}

\new{
\begin{example}
The composition of the coarsest sequence shuffling $\model{\pxt(\Gs)}$ with the coarsest snapshot shuffling $\model{\t}$ results in the MRRM $\model{p(\Ebf)}$ which randomizes both the topology of snapshots and their temporal order while conserving the multiset of the numbers of events per snapshot, $p(\Ebf)=[|\E^t|]_{t\in\T}$. 
%\note{CLV: add figure?}
%(Fig.~\ref{fig:sequence-snapshot-shuffling}).
\end{example}
}

%\begin{figure}
%  \caption{\new{{\bf The coarsest sequence-snapshot shuffling.} $\model{p(\Ebf)}$ is obtained as the composition of $\model{\pxt(\Gs)}$ and $\model{\t}$.}}
%  \label{fig:sequence-snapshot-shuffling}
%\end{figure}

%\new{All sequence shufflings are timeline shufflings since $\pxt(\Gf)\leq\L$. 
%It thus follows directly from Proposition~\ref{prop:link-timeline_independent} that they are also compatible with link shufflings. }
%
%\begin{corollary}
%\label{prop:link-sequence_independent}
%{\sl Link shufflings and sequence shufflings are compatible.}
%Any link shuffling P[$\f(\L),\pxl(\Thf)$] and any sequence shuffling ${\rm P}[\f(\t),\pxt(\Gf)]$ are compatible.
%\end{corollary}

%------------------------------------------------------------------------
\subsection{Examples of compositions of temporal network MRRMs}
\label{sec:compositions}
%------------------------------------------------------------------------
\noindent
\new{As we showed in the previous subsection, link shufflings are compatible with  timeline shufflings (Subsec.~\ref{sec:timeline-shufflings}), 
and sequence shufflings (Subsec.~\ref{sec:sequence_shufflings})  % (Proposition~\ref{prop:link-timeline_independent} and Corollary~\ref{prop:link-sequence_independent}).
 are compatible with snapshot shufflings (Subsec.~\ref{sec:snapshot_shufflings}).}

We list here examples we have found in the literature of MRRMs that are compositions of two compatible MRRMs. 
%We add two new MRRMs that we will use in the walkthrough example in Section~\ref{sec:spreading_example}.
%\new{Appendix~\ref{app:tables} provides a table 
\new{Table~\ref{tab:effects-compositions}) lists the effects of these MRRMs on a selection of important temporal network features (see Table~\ref{tab:features} %and \ref{tab:constraint_levels}
%Appedices~\ref{sec:features} and \ref{sec:feature-tables} 
for formal definitions of the features).}
%All of these are compositions of link and and instant-event timeline shufflings. 
%The number of different MRRMs that may generated by composition is given by the number of combinations of compatible shufflings, and is thus much larger than this (see discussion at the start of this section).

\new{Many more MRRMs than those listed here may be generated directly from the MRRMs surveyed in Section~\ref{sec:classification}
as any composition of a pair of a link and a timeline shuffling or of a sequence and a snapshot shuffling forms a new MRRM.}
% by applying any pair of compatible ones in composition. 
%The number of new MRRMs we can generate this way is given by the number of pairs of compatible MRRMs that are not comparable.}
% (for the shufflings listed in Section~\ref{sec:classification}, %7 link shufflings, 21 timeline shufflings---including 2 sequence shufflings, and 7 snapshot shufflings described here, this number is $21\cdot7+2\cdot7=161$
%this number is 156 \note{CLV: revise this number after adding/removing MRRMs}).}

\subsubsection{Instant-event shufflings}

\begin{itemize}
  \item[{P[$L,E$]}]     
    \new{{\bf Composition of:}  P[$\pxl(\Thf)$] (Sec.~\ref{sec:link-shufflings}) and P[$\L,E$] (Sec.~\ref{sec:timeline-shufflings}).
    {\bf Features constrained:}
      the number of links in static graph, $L$;
      the number of instantaneous events, $E$.
    {\bf Reference:} Ref.~\cite{Holme2005N} ({\sl all random}).} %\vspace{0.2cm}\\
%    is generated by the composition of P[$\pxl(\Thf)$] and P[$\L$]. %It completely randomizes both the topological and temporal structure of the network while constraining $\Estat$. 
%  It was called {\sl all random} in Ref.~\cite{Holme2005N}.
  %
  \item[{P[$\k,E$]}]     
    \new{{\bf Composition of:}  P[$\k,\pxl(\Thf)$] (Sec.~\ref{sec:link-shufflings}) and P[$\L,E$] (Sec.~\ref{sec:timeline-shufflings}).
    {\bf Features constrained:}
      the static degree sequence $\k = (k_i)_{i\in\N}$;
      the number of instantaneous events, $E$.
    {\bf Reference:} Ref.~\cite{Sun2019D}.} 
  \item[{P[$\kstat$,$p(\w)$,$\t$]}] 
    \new{{\bf Cothe mposition of:} P[$\kstat$,$\pxl(\Thf)$] (Sec.~\ref{sec:link-shufflings}) and P[$\w$,$\t$] (Sec.~\ref{sec:intersections-TS-Snaps}).
    {\bf Features constrained:}
      the static degree sequence $\k = (k_i)_{i\in\N}$;
      the multiset of link weights, $p(\w) = [\wij]_{\ij\in\L}$;
      the sequence of event timestamps, $\t=[t]_{(i,j,t)\in\E}$.
    {\bf References:} \cite{Holme2012T,Li2017T} ({\sl randomized edges with randomly permuted times}).} %\vspace{0.2cm}\\
   %applies P[$\pxl(\Thf)$,$\kstat$] and P[$\w$,$\t$] in composition. It was called {\sl randomized edges with randomly permuted times} in Refs.~\cite{Holme2012T,Li2017T}.
  %
  \item[{P[$\kstat$,$\lc$,$p(\w)$,$\t$]}] 
    \new{{\bf Composition of:} P[$\kstat$,$\lc$,$\pxl(\Thf)$] (Sec.~\ref{sec:link-shufflings}) and P[$\w$,$\t$]  (Sec.~\ref{sec:intersections-TS-Snaps}).
    {\bf Features constrained:}
      the static degree sequence $\k = (k_i)_{i\in\N}$;
      the connectedness of static graph, $\lc$;
      the multiset of link weights, $p(\w) = [\wij]_{\ij\in\L}$;
      the sequence of event timestamps, $\t=[t]_{(i,j,t)\in\E}$.
    {\bf References:} \cite{Karsai2011S,Thomas2015D} ({\sl configuration model}).} %\vspace{0.2cm}\\
%    adds the additional constraint to P[$\kstat$,$p(\w)$,$\t$] that the static graph must be connected. It is the composition of P[$\pxl(\Thf)$,$\lc$,$\kstat$] and P[$\w$,$\t$] and was called {\sl configuration model} in Refs.~\cite{Karsai2011S,Thomas2015D}.
%
  \item[\new{P[$\L,p(\w)$]}] 
      \new{{\bf Composition of:} P[$\L$,$\pxl(\Thf)$] (Sec.~\ref{sec:intersections-LS-TS}) and P[$\w$]  (Sec.~\ref{sec:timeline-shufflings}).
    {\bf Features constrained:}
      the static topology $\Gstat=(\N,\L)$;
      the multiset of link weights, $p(\w) = [\wij]_{\ij\in\L}$;      
    {\bf Reference:} Section~\ref{sec:spreading_example}.}
  \item[\new{P[$\L,p(\w),\t$]}] 
      \new{{\bf Composition of:} P[$\L$,$\pxl(\Thf)$] (Sec.~\ref{sec:intersections-LS-TS}) and P[$\w,\t$]  (Sec.~\ref{sec:intersections-TS-Snaps}).
    {\bf Features constrained:}
      the static topology $\Gstat=(\N,\L)$;
      the multiset of link weights, $p(\w) = [\wij]_{\ij\in\L}$;     
      the sequence of event timestamps, $\t=[t]_{(i,j,t)\in\E}$.
    {\bf Reference:} Section~\ref{sec:spreading_example}.}
\end{itemize}

\subsubsection{Metadata-dependent shufflings}
\begin{itemize}
  \item[{P[$\kstat$,$p(\w)$,$\t$,$\gf$,$\Sigma_\L$]}] 
    \new{{\bf Composition of:} P[$\pxl(\Thf)$,$\kstat$,$\gf$,$\Sigma_\L$] with P[$\w$,$\t$].
    {\bf Features constrained:}
      the static degree sequence $\k = (k_i)_{i\in\N}$;
      the multiset of link weights, $p(\w) = [\wij]_{\ij\in\L}$;
      the sequence of event timestamps, $\t=[t]_{(i,j,t)\in\E}$;
      the sequence of node colors, $\gf=(\sigma_i)_{i\in\N}$;
      the the group contact matrix, $\Sigma_\L$.
    {\bf Reference:} Ref.~\cite{Rocha2011S} ({\sl random dynamic topological}).} %\vspace{0.2cm}\\
%    is generated by composition of P[$\pxl(\Thf)$,$\kstat$,$\gf$,$\Sigma_\L$] with P[$\w$,$\t$]. It was named {\sl random dynamic topological} in Ref.~\cite{Rocha2011S}.
\end{itemize}

%------------------------------------------------------------------------
\begin{table*}
  \caption{{\bf Effects of compositions of two MRRMs on the features of a temporal network.}
  \new{See Table~\ref{tab:features} for definitions of features.
  Colored symbols show to what extent each feature is conserved. 
  Informal definitions are found in the tablenotes (detailed definitions are found in Supplementary Table~\ref{tab:constraint_levels}).}}
  \label{tab:effects-compositions}
\centerline{
  \begin{threeparttable}
  \begin{tabular}{ll|ccc|ccc|cccc|cc|cccc|ccccc}
  \hline
  \hline
    Canonical name & Composition of & \multicolumn{3}{l|}{meta} & \multicolumn{3}{l|}{topological} & \multicolumn{4}{l|}{weighted} & \multicolumn{2}{c|}{temp.} & \multicolumn{4}{l|}{node} & \multicolumn{5}{l}{link} \\
    && $\sigma_i$ & $\Sigma_\L$ &  $\Sigma_\E$ & $\Gstat$ & $k_i$ & $\Estat$ & ${a_i}^{\dagger}$ & $s_i$ & ${n_\ij}^{\dagger}$ & $w_\ij$ & $\Et$ & &  & ${\alpha_i^m}^{\dagger}$ & $\Dalpha_i^m$ & $\dit$ &  & ${\tau_\ij^m}^{\dagger}$ & $\Dtau_\ij^m$ & $t^1_\ij$ & $t^w_\ij$   \\
  \hline
  \hline
    %\multicolumn{1}{l}{${\rm P}[\pxl(\Thf)]{\rm P}[\L,E]$} %=
    $\model{L,E}$  & $\model{\pxl(\Thf)} \circ \model{\L,E}$ &  \xmark  & \xmark & \xmark & \xmark & \mumark & \cmark & \xmark & \mumark & \xmark & \mumark & \mumark &   &   & \xmark & \xmark & \mumark &  & \xmark & \xmark & \xmark & \xmark  %& \cite{Holme2005N}
    \\
    %\multicolumn{1}{l}{${\rm P}[\pxl(\Thf),\k]{\rm P}[\w,\t]$} %=
    $\model{\k,p(\w),\t}$ &  $\model{\k,\pxl(\Thf)} \circ \model{\w,\t}$ &  \xmark & \xmark & \xmark & \xmark & \cmark & \cmark & \xmark & \mumark & \xmark & \pmark & \cmark &   &   & \xmark & \xmark & \mutmark &   & \xmark & \xmark & \xmark & \xmark  %& \cite{Holme2012T,Takaguchi2016C,Li2017T}
    \\
    %\multicolumn{1}{l}{${\rm P}[\pxl(\Thf),\lc,\k]{\rm P}[\w,\t]$} %=
    $\model{\k,\lc,p(\w),\t}$  & $\model{\k,\lc,\pxl(\Thf)} \circ \model{\w,\t}$ &   \xmark & \xmark & \xmark & $\lc$ & \cmark & \cmark & \xmark & \mumark & \xmark & \pmark & \cmark &   & & \xmark & \xmark & \mutmark & & \xmark & \xmark & \xmark & \xmark  %& \cite{Karsai2011S,Thomas2015D}
    \\
    $\model{\L,p(\w)}$  & $\model{\L,\pxl(\Thf)} \circ \model{\w}$ &   \xmark & \xmark & \xmark & \cmark & \cmark & \cmark & \xmark & \mumark & \xmark & \pmark & \mumark &   & & \xmark & \xmark & \mumark & & \xmark & \xmark & \xmark & \xmark  %& \cite{Karsai2011S,Thomas2015D}
    \\
    $\model{\L,p(\w),\t}$  & $\model{\L,\pxl(\Thf)} \circ \model{\w,\t}$ &   \xmark & \xmark & \xmark & \cmark & \cmark & \cmark & \xmark & \mumark & \xmark & \pmark & \cmark &   & & \xmark & \xmark & \mutmark & & \xmark & \xmark & \xmark & \xmark  %& \cite{Karsai2011S,Thomas2015D}
    \\
  \hline
  \multicolumn{2}{l|}{\underline{\textsl{\textbf{Metadata dependent:}}}} &&&&&&&&&&&&&&&&&\\
%  \multicolumn{2}{ll}{\underline{\textsl{\textbf{Metadata dependent:}}}} &&&&&&&&&&&&&\\
    $\model{\k,p(\w),\t,\gf,\Sigma_\L}$ & $\model{\k,\pxl(\Thf),\gf,\Sigma_\L} \circ \model{\w,\t}$ & \cmark & \cmark & \xmark & \xmark & \cmark & \cmark  & \xmark & \mumark & \xmark & \pmark & \cmark &  &  & \xmark & \xmark & \mutmark &  & \xmark & \xmark & \xmark & \xmark  %& \cite{Rocha2011S}
    \\
  \hline
  \hline
  \end{tabular}
  \begin{tablenotes}
    \item[$\dagger$] Feature only defined for temporal networks with event durations.
    \item[\cmark] Feature completely conserved, typically the ordered sequence of individual features, e.g.\ $\x=(x_i)_{i\in\N}$ or $\x=((\x_i^t)_{t\in\T})_{i\in\N}$.
    \item[\pxlmark] Distribution of local sequences on links, $\pxl(\x) = \left[(x_\ij^m)_{m\in\Ml}\right]_{\ij\in\L}$.
    \item[\pmark] Distribution (i.e.\ the multiset) of individual scalar values in sequence.
    \item[\mutmark] Sequence of local means in snapshots links, $\mut(\x) = \left(\sum_{i\in\N} x_i^t/N\right)_{t\in\T}$.
    \item[\mumark] Mean value of the individual scalar features in sequence.
    \item[\xmark] Feature not conserved.
%    \item[$\lc$] Connectedness of static graph $\Gstat$ conserved.
  \end{tablenotes}
  \end{threeparttable}
}
\end{table*}
%------------------------------------------------------------------------

%========================================================================
\section{Other reference models}
\label{sec:other_models}
%========================================================================
\noindent
We have restricted this review to microcanonical RRMs as they are the only maximum entropy reference models that can be generated by shuffling elements of an empirical temporal network and they constitute the largest part of RRMs for temporal networks found in the literature. 

In this section, we briefly discuss other types of reference models for temporal networks. These models can be divided into three general classes:
(1) {\sl canonical} RRMs, which correspond to generalized canonical ensembles of random networks; % defined by a generative model;
(2) data-driven reference models that do not maximize entropy;
(3) bootstrap methods, which are a particular type of reference models that do not maximize entropy. %;
%(4) models that do not randomize the entire temporal network but only considers and randomizes summary statistics (i.e.\ features) directly.

\subsection{Canonical randomized reference models}
\label{sec:generative}

\noindent
Canonical RRMs present alternatives that are very close in spirit to the microcanonical RRMs considered here.
They permit to sample canonical ensembles of networks, i.e.\ ensembles where selected features are constrained only on average, $\langle \x(G)\rangle=\x(G^*)$, instead of exactly, $\x(G)=\x(G^*)$, as is the case for MRRMs. (One often talks of {\sl soft} constraints for the canonical ensemble and {\sl hard} constraints for the microcanonical ensemble). Such canonical generative models are also known as {\sl exponential random graph models} (ERGMs)~\cite{Newman2003S,Zhang2017R} and allow to model the expected variability between samples (see discussion in \cite[Section 4]{Squartini2015U}). %They are thus expected to have a lower generalization error than microcanonical RRMs. 
%Additionally, since they are parametric, %generative, they may be incorporated in a Bayesian or information theoretic model selection framework~\cite{Peixoto2014H,Peixoto2015I}.
Additionally, due their soft constraints, canonical models are typically more amenable to analytical treatment than their microcanonical counterparts~\cite{Squartini2011A}. %\footnote{A counterexample to this rule is given in~\cite{Peixoto...}}.

Conversely, the main advantage of MRRMs is that they are usually defined as data shuffling methods, which are often easier to construct than methods that generate networks from scratch.
They are thus generally the only type of models that realistically capture many of the temporal and topological correlations present in empirical networks, which explains their popularity for analyzing temporal networks.
In particular, it is easy to generate microcanonical RRMs that impose features such as the global activity timeline $\Ebf$ or temporal correlations in individual timelines. 
%, which is difficult and often currently impossible to do using a generative model.
Perhaps due to the difficulty in defining generative reference models that capture empirical temporal correlations, these are currently almost exclusively defined for static networks or to model either memoryless dynamics~\cite{Karsai2011S,Perra2012A,Delvenne2015D,Takaguchi2012I} or dynamics with limited temporal correlations~\cite{Pfitzner2013B,Rosvall2014M,Scholtes2014C,Peixoto2015I,Peixoto2017M} 
A notable exception is a recent study combining Markov chains with change point detection to model multiscale temporal dynamics~\cite{Peixoto2017C}. 
We shall not discuss canonical RRMs in more detail here, but refer to \cite{Zhang2017R} for a recent review of ERGMs for temporal networks and to \cite{Squartini2015U,Casiraghi2016G} for recent developments in such models for static networks.

\subsection{Reference models that do not maximize entropy}
\label{sec:non-MaxEnt}

\noindent
Several reference models exist that impose a constraint that is not justified solely by the data (the empirical temporal network) in conjunction with the maximum entropy principle~\cite{Presse2013P}. Such reference models thus introduce new order that is not found in the original network. Here we discuss different types of such reference models and give examples.

\paragraph*{Delta function constraints.}
Some studies have considered reference models where what we may call a {\sl delta function constraint} was imposed on a set of features of the temporal network. Specifically they constrained all instances of this feature to have the same value, i.e.\ to follow a delta distribution. 
This is different from (and more constrained than) the maximum entropy distribution.
The {\sl SStat} method introduced in Ref.~\cite{Starnini2012R} imposes a fixed number of events in each snapshot (equal to the mean number of events per snapshot in the empirical network).
Holme~\cite{Holme2014B} introduced three reference models that all three impose a delta-function constraint (referred to as {\sl poor man's reference models} since they do not satisfy the maximum entropy principle and provide only a single reference network instead of an ensemble~\cite{Holme2015M}):  equalizing the inter-event durations $\Dtau_\ij^m$ while constraining $t^1_\ij$, $t^w_\ij$ and $w_\ij$ for each link $\ij\in\L$;  %(a non-MaxEnt version of P[$\w$, $\t^1$, $\t^w$]); 
shifting the whole sequence of events (sequences of event and inter-event times) on each link in order to make $t_\ij^1=\tmin$ or to make $t_\ij^w=\tmax$ for all $\ij\in\L$.

\paragraph*{Biased sampling.}
Kovanen et al.~\cite{Kovanen2011T} proposed a biased version of P[$\w$,$\t$], where instead of swapping timestamps of events at random, for each instantaneous event $\et$ they drew $m$ other events at random from the set of instantaneous events $\E$ and swapped the timestamps of $\et$ and the other event $(i',j',t')$ among the $m$ drawn for which $t'$ was closest to $t$.
This reference model thus retains some temporal correlations due to the biased sampling, where the parameter $m$ controls the force of this bias and thus of temporal correlations (for $m=1$ the reference model is equal to P[$\w$,$\t$]).
The same method was also employed in Refs.~\cite{Jurgens2012T,Kovanen2013T}.
%Another biased variant of P[$\w$,$\Ebf$] was employed in Refs.~\cite{Karsai2012C,Karsai2014T,Aledavood2015D} focusing on ego-centric networks. For each node, this variant shuffles the partner nodes of events involving the focal node while retaining time stamps. The result of this variant converges to that of P[$\w$,$\Ebf$] if the static graph $\Gstat$ of the input network is connected and one additionally applies the method to all nodes and all times repeatedly until convergence. However, in general it is not a maximum entropy randomization method. 
Valdano et al.~\cite{Valdano2015I} considered a heuristic variant of P[$\pxt(\g)$] (called {\sl reshuffle-social}, where they only permuted snapshots inside intervals where nodes showed approximately the same median {\sl social strategy}~\cite{Miritello2013L}, where the social strategy of a node $i$ is defined as the ratio $\gamma_i^t=k_i^{\delta,t}/s_i^{\delta,t}$ of its degree $k_i^{\delta,t}$ and its strength $s_i^{\delta,t}$ in a network aggregated over $\delta=20$ consecutive snapshots from $t-\delta\Dt$ to $t$.  The empirical temporal network that they investigated showed very clear spikes in $\gamma_i^t$ separated by low-$\gamma_i^t$ intervals, referred to as $\gamma$-slices, which allowed them to permute snapshots within each $\gamma$-slice only.

\paragraph*{Time reversal.}
A quick but informal %and dirty 
way to gain insight into the role of causality in the interaction dynamics is to reverse  the order of the snapshots~\cite{Kovanen2011T,Holme2012T,Holme2015M,Li2017T}. This method obviously does not increase entropy as the time-reversed network is unique, but it may be used as a simple way to study the importance of causality in the temporal network.
(Note that a {\sl time-reversal} MRRM may in principle be defined as one that returns an input temporal network and its time-reversed version with equal probability.)

\subsection{Bootstrap methods}
\label{sec:bootstrapping}

\noindent
Bootstrap methods are based on sampling with replacement, whereas MRRMs are based on sampling without replacement (i.e.\ shuffling).
Resampling with replacement means that network features are not constrained exactly as for shuffling methods.
This means that bootstrapping algorithms may be easier to implement than shuffling methods when the exact constraints are hard to satisfy.
The hope when using bootstrapping can additionally be to capture some of the expected out-of-sample variability. 
The set of states that may be generated is strongly constrained by the particular dataset however, so bootstrapping does not generate a maximum entropy model. 
Though it may be seen as a means to approximate one, it does not come with the same statistical guarantees as microcanonical  and canonical RRMs do. 
So the theoretical results and guarantees that exist for microcanonical RRMs %(see Sec.~\ref{sec:theory}) 
do not hold for bootstrapping, and additional care is advised when analyzing results obtained using bootstrapping.

Two bootstrap methods used in the literature are described below. %, can be regarded as variants of existing microcanonical randomized methods.
The method called {\sl time shuffling} in Ref.~\cite{Gauvin2013A} constrains the number of events per link $\n$ exactly and resamples the event durations $\tau$ from the global distribution $p(\tauf)$ with replacement.
The method called {\sl time shuffling} in Ref.~\cite{Genois2015C} constrains the static network $\Gstat$ and bootstraps $n_\ij$, $t^1_\ij$ for all links from the global distributions $p(\n)$ and $p(\t^1)$, respectively, and then bootstraps the $n_\ij$ event durations $\tau_\ij^m$ and $(n_\ij-1)$ of inter-event durations $\Dtau_\ij^m$ for each link $\ij\in\L$ from the global distributions $p(\tauf)$ and $p(\Dtauf)$, respectively.

\section{Analyzing temporal distances in a communication network using a series of MRRMs}
\label{sec:spreading_example}

\begin{figure}
  \includegraphics[width=\columnwidth]{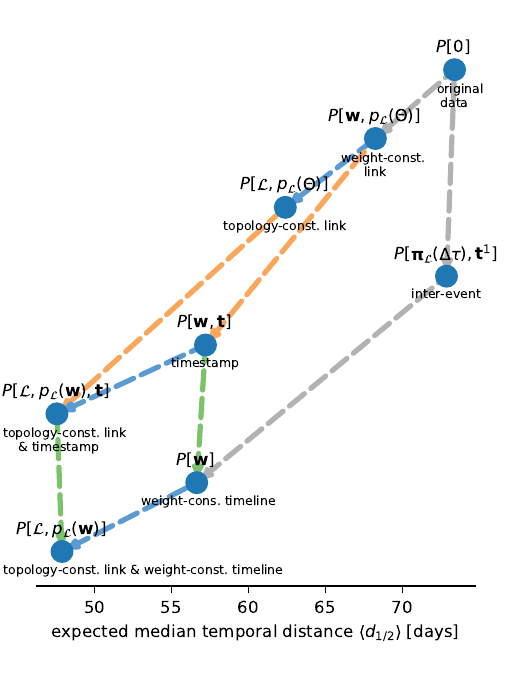}
  \caption{{\bf Expected median temporal distance for a hierarchy of MRRMs.} 
  Each circle in the figure represents a single MRRM. The horizontal location of the circle reports the expected median temporal distance $\langle d_{1/2}(G) \rangle$ of the MRRM applied to the empirical network. An arrow from a MRRM %(P[$\x$]) 
  at a higher location to a lower one %(P[$\y$]) 
  means that the former shuffles less than the latter. %(${\rm P}[\x] \leq {\rm P}[\y]$). 
  (Note that the absolute positions of the nodes along the vertical axis are arbitrary and are thus not indicative of how random the MRRMs are. Only MRRMs linked by are directed path in the diagram can be formally compared.)  
  A canonical name of each MRRM is given above each circle and a common name below it (see Section~\ref{sec:classification}).
  In the common names the word \emph{shuffling} is always removed for brevity, and the $\&$-sign denotes that both MRRMs are applied to the data in composition (Def.~\ref{def:composition}).
  Colored links indicate that the same features were removed: ${\bf t}$ for green links, ${\bf w} \hookrightarrow \mathcal{L}, p_{\mathcal{L}}({\bf w})$  for blue links, and $p_{\mathcal{L}}(\Theta) \hookrightarrow {\bf t}$ for orange links. } 
  \label{fig:spreading_example}
\end{figure}

\noindent
 \new{In this section we go through a walk-through example in which we}  use the hierarchy of MRRMs to investigate how different features of a temporal communication network affects the temporal distances between nodes in the network (defined as the minimal times required for any contagion process to spread between the nodes). 
This example additionally serves to showcase a graphical representation that incorporates both the hierarchy of the MRRMs and their effects on a scalar feature, and which provides an intuitive way to interpret the results (Fig.~\ref{fig:spreading_example}).
As discussed in Section \ref{sec:applications}, understanding how different features affect spreading was the starting point of some of the early studies employing MRRMs in temporal networks, and here we reproduce some of those results with a different data set. However, the analysis pipeline introduced here does not only work for temporal distances, but can be used for any other scalar-valued feature.

The dataset is a publicly available temporal mobile phone communication network published by Wu et al. \cite{Wu2010Evidence}.
Here we focus on the first company with $44431$ nodes and around $5.5 \times 10^5$ instantaneous events taking place over $30$ days. Distances in temporal networks is a multifaceted topic \cite{Pan2011P}, but here we quantify the distances in a network by a single number describing the typical temporal distance in the network. More specifically, we calculate the expected temporal distance to reach half of the nodes in the network, i.e.\ the {\sl expected median temporal distance} $\langle d_{1/2}(G) \rangle$, where the expectation is evaluated over all nodes and all times as source points. 
The temporal distance from one node $i$ to another node $j$ is defined as the time required for the fastest possible spreading process starting at a given time $t$ to spread from $i$ to $j$\footnote{Note that the temporal distance is not a metric distance as the temporal distance from $i$ to $j$ generally differs from the temporal distance from $j$ to $i$.}. 
Formally, this fastest possible spreading is modeled by a deterministic susceptible-infectious (SI) process where susceptible nodes are infected immediately by contact with an infectious node.
When evaluating the distances we use periodic boundary conditions in time to remove boundary effects \cite{Karsai2011S}.

\new{The MRRMs explored are listed in Table~\ref{tab:effects-walkthrough2}. 
%together with their effects on a selection of temporal network features relevant for the following discussion.
The table also lists to which extent the MRRMs conserve selected network features.
It is constructed formally by showing for feature and each MRRM whether or not the feature can be defined as function of the features constrained by the MRRM.
For example, because the activity timeline $\Ebf=(|\E^t|)_{t\in\T}$ can be calculated from the multiset of timelines $\pxl(\Thf)=[\Thij]_{\ij\in\L}$, the MRRM $\model{\L,\pxl(\Thf)}$ conserves $\Ebf$.
Similarly, $p(\w)=[\wij]_{\ij\in\L}$ is known from $\pxl(\Thf)$, but $\w=(\wij)_{\ij\in\L}$ is not, so $\model{\L,\pxl(\Thf)}$ conserves $p(\w)$.}

\begin{table}
  \caption{\new{{\bf Effects of selected MRRMs on temporal network features.}
  The features considered are: 
  the static graph $\Gstat=(\N,\L)$;
  the link weights and their configuration in $\Gstat$, $\w=(\wij)_{\ij\in\L}$;
  the activity timeline $\Ebf=(|\E^t|)_{t\in\T}$;
  higher order temporal correlations in and between timelines, i.e.\ $\Thf=(\Thij)_{\ij\in\L}$;
  the inter-event durations on the links, $\Dtauf=((\Dtau_\ij^m)_{m\in\Ml})_{\ij\in\L}$, where $\Ml$ is a temporal index;
  and the timing of the first event on each link $\t^1=(t^1)_{\ij\in\L}$}
  (Table~\ref{tab:features} provides detailed definitions each feature).}
  \label{tab:effects-walkthrough2}
\centerline{
  \begin{threeparttable}
  \begin{tabular}{ll|cccccc}
  \hline
  \hline
%    Canonical name &  &\multicolumn{9}{l|}{One-level} & \multicolumn{9}{l}{Two-level} %& References
%    \\
%    && \multicolumn{3}{l}{topological} & \multicolumn{4}{l}{weighted} & \multicolumn{2}{c|}{temp.} & \multicolumn{4}{l}{node} & \multicolumn{5}{l}{link} \\
    Model && \multicolumn{6}{l}{Features}      \\
     Name & Common name & $\Gstat$ &  $\w$ & $\Ebf$ & $\Thf$ & $\Dtauf$ & $\t^1$   \\
  \hline
    P[$\w$,$\pxl(\Thf)$] &  {\sl Weight-constrained LS} & \checkmark & \checkmark & \checkmark & \pmark & \pmark & \pmark       \\
    P[$\L$,$\pxl(\Thf)$] &  {\sl Topology-constrained LS} & \checkmark & \pmark & \checkmark & \pmark & \pmark & \pmark   \\
    P[$\piij(\Dtauf)$,$\t^1$] & {\sl Inter-event shuffling} & \checkmark & \checkmark & \mumark & \crossmark & \pmark & \checkmark   \\
    P[$\w$,$\t$] & {\sl Timestamp shuffling} & \checkmark & \checkmark & \checkmark & \crossmark & \crossmark & \crossmark    \\
    P[$\w$] & {\sl Weight-constrained TS} & \checkmark & \checkmark & \mumark & \crossmark & \crossmark & \crossmark     \\
    $\model{\L,p(\w),\t}$  & $\model{\L,\pxl(\Thf)} \circ \model{\w,\t}$ &    \checkmark & \pmark & \checkmark & \crossmark & \crossmark & \crossmark    \\
    $\model{\L,p(\w)}$  & $\model{\L,\pxl(\Thf)} \circ \model{\w}$ & \checkmark & \pmark & \mumark & \crossmark & \crossmark & \crossmark    \\
  \hline
  \hline
  \end{tabular}
  \begin{tablenotes}
    \item[\checkmark] Feature completely conserved.
    \item[\pmark] Distribution (i.e.\ the multiset) of individual values in sequence conserved.
    \item[\mumark] Mean value of the individual features in sequence conserved.
    \item[\crossmark] Feature not conserved.
  \end{tablenotes}
  \end{threeparttable}
}
\end{table}

Figure \ref{fig:spreading_example} displays $\langle d_{1/2}(G) \rangle$ for the original data and for several MRRMs. The figure is organized in a way that the hierarchies (see Section
\ref{sec:hierarchies})
%\ref{sec:classification})
are visible similar to Figure \ref{fig:RRMs}. % \ref{fig:RRMs-LS}-\ref{fig:RRMs-SnapS}.
Reading the figure from top to bottom now yields a picture of what happens when the original data is shuffled more and more, i.e., when the temporal features present in the data are destroyed one by one by the MRRMs. 
All of the arrows are pointing either almost directly downwards or down and left, which means that, for this network and this set of MRRMs, randomizing more never leads to longer temporal distances.  

The overall activity sequence ${\bf t}$, including the daily and weekly changes in the activity, does not have a noticeable effect on the temporal distances on these MRRMs: Removing the constraint on ${\bf t}$ when going from $P[{\bf w},{\bf t}]$ to $P[{\bf w}]$ and from $P[\mathcal{L},p_{\mathcal{L}}({\bf w}),{\bf t}]$ to $P[\mathcal{L},p_{\mathcal{L}}({\bf w})]$ almost does not change the temporal distances. Similarly, shuffling the inter-event times, \new{$\Dtauf$,} while keeping the first activation time of each link, \new{${\bf t}^1$}, with $P[\piij(\Dtauf), {\bf t}^1]$ barely changes $\langle d_{1/2}(G) \rangle$, showing that higher-order temporal correlations between events over the same link has a very small effect on the temporal distances of the original data.

Adding the shuffling of the weights of the network -- i.e. replacing the feature that keeps the weights of the links, ${\bf w}$, with the one only keeping the links and the weight distribution, $\mathcal{L}$ and $p({\bf w})$ -- makes the temporal paths around 7--9 days faster. The pairs of MRRMs corresponding to this replacement are $P[{\bf w},{\bf t}]$ to $P[\mathcal{L},p_{\mathcal{L}}({\bf w}),{\bf t}]$, $P[{\bf w}]$ to $P[\mathcal{L},p_{\mathcal{L}}({\bf w})]$, and $P[{\bf w},p_{\mathcal{L}}(\Theta)]$ to $P[\mathcal{L}, p_{\mathcal{L}}(\Theta)]$. 
Note that in the MRRM $P[\mathcal{L}, p_{\mathcal{L}}(\Theta)]$ the weight distribution $p_{\mathcal{L}}({\bf w})$ is kept implicitly by the link sequence distribution $p_{\mathcal{L}}(\Theta)$, because $p_{\mathcal{L}}(\Theta) \le p_{\mathcal{L}}({\bf w})$.

Finally the largest change in the temporal distance are seen when the times of events in the timelines $\Thij$ are shuffled such that they simply follow the overall activity sequence ${\bf t}$. In these transitions, from $P[{\bf w},p_{\mathcal{L}}(\Theta)]$ to $P[{\bf w},{\bf t}]$ and from $P[\mathcal{L}, p_{\mathcal{L}}(\Theta)]$ to $P[\mathcal{L},p_{\mathcal{L}}({\bf w}),{\bf t}]$, the temporal distances are reduced on average by around 12--14 days.  

%\note{CLV: add discussion of the importance of $\t^1$ (as seen from $\model{\piij(\Dtauf),\t^1}$)? -- Add additional models to analysis to support this ? Namely: $\model{\w,\t^1}$ and/or $\model{\L,p(\w),\t^1}$.}

Almost no combination effects were observed for these data: removing each feature had a very similar effect -- with variations of around 2 days -- independently of the other features that were kept.
This allows a very simple summarization of the results: The typical temporal distance in the data is around 73 days and in the most random MRRM applied here around 48 days. Out of that difference, around 12--14 days is explained by link activation sequence features (such as bursts \cite{Karsai2011S} \new{and the timings of the links' first activations~\cite{Holme2014B}}), 7--9 days are by weight-topology correlations (such as weak links located in bridge positions \cite{Granovetter1973S,Onnela2007S}), and 4 days by link-timeline-topology correlations (such as correlations in times at which two neighbors of a node are communicated with \cite{Karsai2011S,Kivela2012M}).

This analysis can be made more detailed by adding more fine-scaled features related to timings of events or link weights. Alternatively, the analysis could be expanded by including topological MRRMs such as the configuration model \new{$\model{\k,\pxl(\Thf)}$}.

%========================================================================
\section{Conclusion}
%========================================================================
\noindent
%Many MRRMs are easily implemented by numerical shuffling methods.
Microcanonical randomized reference models (MRRMs) provide a \new{versatile} and generally applicable toolbox for the analysis of dynamical networked systems. 
Their main advantages are their wide applicability and relative ease of implementation: they only require the definition of a corresponding unbiased shuffling method\new{, which is often easily implemented as a purely numerical randomization scheme}.
This means that they can be used to test the importance of any given feature provided a corresponding shuffling method, and may in principle be used to generate model networks that are arbitrarily close to empirical ones.

Shuffling methods provide an interesting alternative to more elaborate generative models, and can be seen as a {\sl top-down} approach to modeling by progressively randomizing features of an empirical network, as opposed to the {\sl bottom-up} approach of generative models.
Each approach has its strengths and weaknesses (as discussed in Sec.~\ref{sec:generative}). We believe that shuffling methods are best used as exploratory tools to identify important qualitative features and effects. Generative models can then be used to explore them quantitatively and to perform model selection in order to identify potential underlying generative mechanisms.

We here introduced a fundamental framework for MRRMs. 
This enabled consistent naming, analysis, and classification of MRRMs for temporal networks.
We have used this framework to describe numerical shuffling procedures found in the literature rigorously in terms of microcanonical RRMs, built a taxonomy of these RRMs, and surveyed their applications to the study of temporal networks.
This framework also allowed us to define conditions for when we may combine two  MRRMs in a composition to generate a new MRRM and to derive which features it inherits from them. 
Such compositions of compatible MRRMs make it possible to easily generate %hundreds of 
new MRRMs from the %dozens  
existing ones.

We have focused on undirected and unweighted temporal networks, but the extension of the MRRM framework introduced in Section~\ref{sec:definitions} to any other types of network is trivial. 
Such extensions may require defining new ways of representing the structure and defining appropriate features. 
This is straightforward for \new{temporal networks with directed (see Supplementary Table~\ref{tab:features-directed}) or weighted events as well as networks with bipartite or even multipartite structure (Sec.~\ref{sec:metadata})}. 
%Temporally varying multilayer networks~\cite{Kivela2014M} will require the definition of three-level nested features. 
Furthermore, a MRRM-based framework can be developed for any other types of multilayer networks~\cite{Kivela2014M}, such as multiplex networks or networks of networks, and even for structures beyond networks such as hypergraphs. 
Finally, it should be helpful to define a similar framework for canonical reference models (see Section~\ref{sec:generative}) as more of such models are emerging.

It is our intention that this framework and collection of MRRMs will serve as a reference for researchers who want to employ MRRMs to analyze the dynamics of networks and if processes take place on them.
\new{It is straightforward to incorporate many more temporal network MRRMs into the framework than those presented here, e.g.\ models that constrain correlations between features instead of only their marginal values.}
With the foundations for MRRMs laid here we are \new{thus} ready to repeat the success stories of RRMs for static networks, and may even go much further.

Notable important challenges remain which may now be addressed using the formalism defined here. 
For example, how to automatize the definition and classification of new MRRMs, which would allow a user to simply state the set of features she wants to constrain to generate a corresponding \new{set} of networks.
How to automatize the choice of MRRMs in order to most efficiently infer which features of an empirical temporal network control a given dynamical phenomenon, i.e.\ identifying which models best divide the space of network features.
%(for single features this corresponds to cuts in the dependency diagram, Fig.~\ref{fig:dependency_diagram}).
\new{How to compare MRRMs (e.g. in terms of sizes or overlap of partitions) that are neither comparable nor compatible. 
Being able to do this would notably make it possible to characterize automatically the effects} of a MRRM on temporal network features that are not comparable to nor independent of the features constrained by the MRRM.

We note finally that it is a difficult problem to design unbiased shuffling methods for MRRMs that take higher order topological correlations into account~\cite{Jerrum1986R, Orsini2015Q}. This may put natural barriers on the possible resolution of exact MRRMs.
Instead, approximate procedures for generating such MRRMs would have to be considered, and their accuracy may be gauged by how closely they reproduce features which we know they should constrain.

%========================================================================
\section*{Acknowledgments}
\noindent
L.G.  acknowledges support from the Lagrange Laboratory of the ISI Foundation
funded by the CRT Foundation. 
M.G. was supported by the French ANR HARMS-flu, ANR-12-MONU-0018.
M. Karsai acknowledges support from the DylNet (ANR-16-CE28-0013) and SoSweet (ANR-15-CE38-0011) ANR projects and the MOTIf STIC-AmSud project.
T.T. was supported by the JST ERATO Grant Number JPMJER1201, Japan.
C.L.V. was supported by the EU FET project MULTIPLEX 317532, the french ANR SiNCoBe, ANR-20-CE45-0021, and by the French government under management of Agence Nationale de la Recherche as part of the ``Investissements d’avenir'' program, reference ANR-19-P3IA-0001 (PRAIRIE 3IA Institute).
M.G, T.T., and C.L.V. acknowledge financial support through the Bilateral Joint Research Program between MAEDI, France, and JSPS, Japan (SAKURA Program).

\section*{Author contributions}
\noindent
C.L.V. conceived and directed the study.
All authors contributed to the definition of the classification system, to classifying existing MRRMs found in the literature, and to writing.
M. Kivel\"a, T.T., and C.L.V. performed theoretical calculations.
M.G. and M. Kivel\"a wrote the Python/C++ software packages.
M.G., M. Karsai, M. Kivel\"a, T.T., and C.L.V. drafted the final manuscript.

%========================================================================
\appendix
%========================================================================
\renewcommand{\thetable}{\Alph{section}.\arabic{table}}
\renewcommand{\thefigure}{\Alph{section}.\arabic{figure}}

%------------------------------------------------------------------------
\section*{Appendix: Formal definitions and proofs}
\label{app:proofs}
\addtocounter{section}{1}
%------------------------------------------------------------------------
\noindent
This appendix provides \new{formal definitions of the different classes of shuffling methods as well as} proofs for the propositions and and theorems given in Sections~\ref{sec:theory-comparison} and \ref{sec:theory-composition}.

%------------------------------------------------------------------------
\tocless\subsection{Formal definitions of classes of shuffling methods}
\label{app:definitions-shufflings}
%------------------------------------------------------------------------
\noindent
\new{To formalize the set of possible constraints imposed by a class of shuffling methods, we shall use functions that take network features as input.}

\new{
\begin{definition}
{\sl Function of a network feature.}
\label{def:function_feature}
A function of a temporal network feature $\x$ (Def.~\ref{def:feature}) is any function $\f$ that has as domain the entire co-domain of $\x$, i.e.\ a function that takes $\x(G)$ as input and returns a value $\f(\x(G))$ for all $G\in\G$.
\end{definition}
}

\new{
\begin{example}
The sequence of static degrees $\k=(k_i)_{i\in\N}$ of a temporal network $G$ can be written as a function of the static graph $\Gstat(G)=(\N(G),\L(G))$ (Def.~\ref{def:static_graph}). In particular, it is a function of the configuration of links, $\L$, with each $k_i(G)\in\k(G)$ given by $k_i(G) = |\{i : \ij\in\L(G)\}|$.
\end{example}
}

Definition~\ref{def:function_feature} lets us formally define classes of shuffling %shuffling
methods using the naming convention developed in Def.~\ref{def:naming-MRRMs}.
%We are in general interested in methods that shuffle the events in a temporal network ({\sl event shufflings}) or shuffle the instantaneous events in an instant-event temporal network ({\sl instant-event shufflings}).}
%We shall call the former type of shuffling method an {\sl event shuffling} and the latter an {\sl instant-event shuffling}.}

\begin{definition}
{\sl Event shuffling \new{{\rm P}$[p(\tauf),\f(\C)]$}.}
\label{def:event_shuffling}
We define an {\sl event shuffling} as a shuffling method that generates networks from an input temporal network by randomizing one or multiple of the indices $i$, $j$, $t$ in all of the events $(i,j,t,\tau)\in\C$. 
%while conserving their durations. 
\new{Formally, any event shuffling is of the form $\model{p(\tauf),\f(\C)}$.
It thus constrains the multiset of the event durations of the network, $p(\tauf)=[\tau_q]_{q=1}^C$, as well as any additional constraint that can be written as a function $\f$ of the set of events $\C$.}
\end{definition}

\begin{definition}
{\sl Instant-event shuffling \new{{\rm P}$[E,\f(\E)]$}.}
\label{def:instant-event_shuffling}
We define an {\sl instant-event shuffling} as a shuffling method that generates networks by randomizing one or multiple of the indices $i$, $j$, \new{$t$} in all of the instantaneous events $(i,j,t)\in\E$ of an instant-event temporal network. \new{Formally, any instant-event shuffling is of the form $\model{E,\f(\E)}$. 
It thus constrains the number of events, $E$, as well as any additional constraint that can be written as a function $\f$ of the set of instantaneous events $\E$.}
\end{definition}

% \new{
% \begin{example}
% The most random event shuffling possible, $\model{p(\tauf)}$, is the one that conserves only the events' durations and otherwise redistributes them completely at random.
% The most random instant-event shuffling is $\model{E}$, which redistributes the instantaneous events at random in an instant-event temporal network and thus conserves only the number of instantaneous events.
% $\model{p(\tauf)}$ and $\model{E}$ correspond to setting $\f$ equal to a constant in Defs.~\ref{def:event_shuffling} and \ref{def:instant-event_shuffling}, respectively.
% %\note{Add figures?}
% %\note{MKi: cst. ? is it a constant? Should we just spell it out?}
% \end{example}
% }

% \new{We furthermore define several more restricted classes of shuffling methods that randomize specific temporal or topological aspects of a network using the two level representations introduced in Section~\ref{sec:nested} above.}

%We will now define four further constrained event shufflings, namely , {\sl link shufflings}, , and  These can be implemented directly using the nested network representations introduced in Section~\ref{sec:nested} (timeline and link shufflings using the link-timeline representation, and snapshot and sequence shufflings using the snapshot-sequence representation).  
%Timeline and link shufflings, as well as snapshot and sequence shufflings, are compatible. This lets us generate new microcanonical RRMs as compositions of these.

%-----------------------------------------------------------------------
\subsubsection{Link and timeline shufflings}
%------------------------------------------------------------------------
\noindent
% \new{Based on the link-timeline representation (Def.~\ref{def:link-timeline_graph}), we define {\sl link shufflings}, which randomize the static graph of a network but not the individual timelines, and {\sl timeline shufflings}, which randomize the timelines but not the static topology.}

\begin{definition}
\label{def:link_shuffling}
{\sl Link shuffling} \new{P[$\f(\L),\Thf$].}
A {\sl link shuffling} %is an event or instant-event shuffling that
constrains all the individual timelines, i.e.\ the multiset $\pxl(\Thf)=[\Thij]_{\ij\in\L}$. 
It randomizes the \new{links in the static graph, i.e.\ the} 
values of $i$ and $j$ for each link $\ij\in\L$, 
while respecting a constraint given by any function $\f$ of \new{the configuration of links} $\L$. 
\end{definition}

% \begin{example}
% In practice a link shuffling is implemented by randomizing the links $\L$ in the static graph and redistributing the timelines $\Thij\in\pxl(\Thf)$ at random on the new links without replacement. 
% \new{Using the Erd\H{o}s-R\'enyi  model for randomizing the static graph leads to the most random link shuffling possible, P[$\pxl(\Thf)$] (Fig.~\ref{fig:ex-link_shuffling}), corresponding to setting $\f = {\rm cst.}$ in Def.~\ref{def:link_shuffling} above.}
% %\note{CLV: use instead the Maslov-Sneppen model (P[$\kstat,\pxl(\Thf)$]) as example? -- this way we apply the concept of a function of a network feature to a shuffling.} 
% \end{example}

% \begin{figure}
%   \includegraphics[width=0.3\textwidth]{P__pTheta.pdf}
%   \caption{\new{{\bf Illustration of the link shuffling P[$\pxl(\Thf)$].}
%   P[$\pxl(\Thf)$] completely randomizes the static graph but conserves the content of the individual timelines.}}
%   \label{fig:ex-link_shuffling}
% \end{figure}

%The coarsest link shuffling is P[$\pxl(\Thf)$], which shuffles the static graph using P[$L$]. 

\begin{definition}
{\sl Timeline shuffling} \new{P[$\L,\f(\Thf)$].}
\label{def:timeline_shuffling}
A {\sl timeline shuffling} %is an event or instant-event shuffling that 
constrains \new{the network's static topology, i.e. $\Gstat=(\N,\L)$}. 
It shuffles the events in the timelines % either at random or 
while respecting a constraint given by any function $\f$ of \new{the timelines $\Thf=(\Thij)_{\ij\in\L}$}.
\end{definition}

% \new{
% \begin{example}
% The most random link shuffling, P[$\L,E$], redistributes the instantaneous events in an instant-event network $\G=(\N,\E)$ at random between the timelines while conserving the static structure (Fig.~\ref{fig:ex-timeline_shuffling}).
% It corresponds to setting $\f(G) = E(G) = |\E|$ in Def.~\ref{def:timeline_shuffling}.
% \end{example}
% }

% \begin{figure}
%   \includegraphics[width=0.45\textwidth]{P__L_E.pdf}
%   \caption{\new{{\bf Illustration of the timeline shuffling P[$\L,E$].
%   P[$\L,E$] completely redistributes the instantaneous events between all timelines at random while conserving the static topology.}}}
%   \label{fig:ex-timeline_shuffling}
% \end{figure}

%Timeline shufflings conserve the static graph $\Gstat$ by construction.
%The coarsest timeline instant-event shuffling is P[$\L$], while the coarsest timeline  event shuffling is P[$\L,p(\deltaf)$]. 

%-----------------------------------------------------------------------
\subsubsection{Sequence and snapshot shufflings}
%-----------------------------------------------------------------------
\noindent
% \new{Based on the snapshot-sequence representation (Def.~\ref{def:snapshot-sequence}), 
% we define {\sl sequence shufflings}, which randomize the order of the snapshots but not the individual snapshot graphs, 
% and {\sl  snapshot shufflings}, which randomize individual snapshot graphs but not their order.}

\begin{definition}
\label{def:sequence_shuffling}
{\sl Sequence shuffling.} \new{P[$\f(\t),\pxt(\Gf)$].}
A {\sl sequence shuffling} %is a timeline shuffling that \new{additionally 
\new{constrains the multiset of instantaneous snapshot graphs, $\pxt(\Gf)=[\Gt]_{t\in\T}$}. 
It randomizes the order of snapshots
%, i.e.\ the set $\T$, 
in a manner that may depend on any function $\f$ of the times of the events, \new{$\t=(t)_{(i,j,t)\in\E}$, but otherwise conserves the individual snapshot graphs}.
\end{definition}

%The coarsest sequence shuffling is P[$\pxt(\Gf)$]. 

% \new{
% \begin{example}
% Shuffling the temporal order of the individual snapshots completely at random leads to the sequence shuffling P[$\pxt(\Gs)$]. 
% It corresponds to setting $\f = {\rm cst.}$ in Def.~\ref{def:sequence_shuffling}. 
% (Fig.~\ref{fig:ex-sequence_shuffling}).
% \end{example}
% }

% \begin{figure}
%   \includegraphics[height=0.055\textheight]{P__pGamma.pdf}  
%   \caption{\new{{\bf Illustration of the sequence shuffling P[$\pxt(\Gs)$].}
%   P[$\pxt(\Gs)$] conserves all the individual snapshot graphs but randomizes their temporal order.}}
%   \label{fig:ex-sequence_shuffling}
% \end{figure}

\begin{definition}
\label{def:snapshot_shuffling}
{\sl Snapshot shuffling} \new{P[$\t,\f(\pxt(\Gf))$].}
A {\sl snapshot shuffling} %is an event or instantaneous-event shuffling that 
constrains the time of each event, i.e.\ \new{$\t=(t)_{(i,j,t)\in\E}$}.
 It randomizes each snapshot graph $\Gt$ individually in a manner that may be constrained by any function $\f$ of $\pxt(\Gf)$. %, i.e. of the individual $\Gt$.
\end{definition}

\tocless\subsection{Proofs of propositions in Section~\ref{sec:theory-comparison}}
\label{app:proofs-comparison}
%------------------------------------------------------------------------

\proofprop{prop:comparability-functional_relation}
{Two MRRMs $\model{\x}$ and $\model{\y}$ are comparable and $\model{\x} \leq \model{\y}$ if and only if there exists a function $\f$ for which $\y(G)=\f(\x(G))$ for all states $G\in\G$.}
{\new{If there exists} a function $\f$ that allows to calculate $\y$ solely from $\x$ \new{then} all networks in a given $\x$-equivalence class $\G_{\x^*}$ (Def.~\ref{def:mrrm_representations}) correspond to the same value of $\y$.
%Conversely, each $\y$-value may correspond to multiple values for $\x$, so different networks in a given $\y$-equivalence class $\G_{\y^*}$ may correspond to different values of $\x$. 
Thus $\G_{\x(G^*)}\subseteq\G_{\y(G^*)}$ for all $G^*\in\G$.
\new{Conversely, if the $\x$-equivalence class $\G_{\x(G^*)}$ is contained in the $\y$-equivalence class $\G_{\y(G^*)}$ for all $G^*\in\G$, it means that a unique value of $\y$ corresponds to each possible value of $\x\in\X=\{\x(G) : G\in\G\}$, thus defining a functional relation from $\X$ to $\Y=\{\y(G) : G\in\G\}$.}}

\proofprop{the:partition_refinements}
{$\model{\x} \leq \model{\y}$ if and only if the partition $\partition{\x}$ is finer than $\partition{\y}$.}
{The propositions follows directly from the fact that the definition of comparability (Def.~\ref{def:comparability}) is exactly the definition of the refinement relation for partitions~\cite{Hrbacek1999I}, i.e.\ that for all $G\in\G$, and thus for all sets $\G_{\x(G)}$ and $\G_{\y(G)}$ in the partitions generated by P[$\x$] and P[$\y$], respectively, we have $\G_{\x(G)}\subseteq\G_{\y(G)}$.}

%------------------------------------------------------------------------
\tocless\subsection{Proofs of propositions and theorems in Section~\ref{sec:theory-composition}}
\label{app:proofs-composition}
%------------------------------------------------------------------------

\proofprop{prop:compat-commute}
{If two MRRMs, P[$\x$] and P[$\y$], are compatible then $\model{\comp{\x}{\y}} = \model{\comp{\y}{\x}}$.\\}
{This can be shown by direct calculation by noting that the transition matrices of MRRMs are symmetric.
For any two compatible MRRMs, P[$\x$] and P[$\y$], their compositions $\model{\comp{\x}{\y}}$ and $\model{\comp{\y}{\x}}$ both define a MRRM. So the associated transition matrices must be symmetric (Def.~\ref{def:mrrm_representations}), and
\beq
  \P^\y\P^\x = (\P^\y\P^\x)^\intercal = (\P^\x)^\intercal(\P^\y)^\intercal = \P^\x\P^\y \es.
\eeq}

\proofprop{prop:composition_is_coarser}
{Consider two compatible MRRMs, P[$\x$] and P[$\y$]. Their composition, $\model{\comp{\y}{\x}}$, is coarser (or equal) than both $\model{\y}$ and $\model{\x}$, i.e.\ $\model{\comp{\y}{\x}} \geq \model{\x}$ and $\model{\comp{\y}{\x}} \geq \model{\y}$, even if $\model{\x}$ and $\model{\y}$ are not comparable.}
{Since P[$\x$] and P[$\y$] are compatible, $\model{\comp{\y}{\x}}$ is a MRRM by the definition of compatibility (Def.~\ref{def:compatibility}).
Since $G$ is itself by definition in the set obtained by applying any MRRM to $G$, the target set of $\model{\comp{\y}{\x}}$ can never be smaller than the target set of $\model{\x}$. Thus, $\model{\comp{\y}{\x}} \geq \model{\x}$. By the same reasoning we obtain that $\model{\comp{\x}{\y}} \geq \model{\y}$, and since $\model{\comp{\x}{\y}} = \model{\comp{\y}{\x}}$ (Proposition~\ref{prop:compat-commute}), that $\model{\comp{\y}{\x}} \geq \model{\y}$.}

\proofprop{prop:comparable-compatible}
{Let P[$\x$] and P[$\y$] be two MRRMs and $\model{\x}\leq \model{\y}$. Then they are compatible and their composition gives $\model{\comp{\y}{\x}} = {\rm P}[\y]$.}
{Since $\y$ is a function of $\x$, its value is the same for all $G$ which satisfy $\x(G)=\x^*$ and is equal to $\y^*=\y(G^*)$. This means that $\Prob{\y}(G|G')=\Prob{\y}(G|G^*)$ for all $G'\in\G_{\x^*}$, and Eq.~(\ref{eq:composition}) reduces to:
\beqa
  \Prob{\comp{\y}{\x}}(G|G^*) &=& \Prob{\y}(G|G^*)\sum_{G'\in\G_{\x^*}} \Prob{\x}(G'|\x^*) \nonumber\\
   &=& \Prob{\y}(G|G^*) \es,
\eeqa
where the second equality is obtained from the requirement that $\Prob{\x}$ must be normalized on $\G_{\x^*}$.
So $\model{\comp{\y}{\x}} = {\rm P}[\y]$, which is a MRRM, showing that P[$\x$] and P[$\y$] are compatible.}

\proofprop{prop:symmetry}
{If $\x$ is conditionally independent of $\y$ given a common coarsening $\z$ then $\y$ is conditionally independent of $\x$ given $\z$}
{We note that since $\z$ is coarser than $\x$, it follows that $P_{\y|\x}=P_{\y|\x,\z}$ (since $\G_{\x^*}\subseteq\G_{\z^*}$, conditioning only on $\x^*$  is equivalent to conditioning on both $\x^*$ and $\z^*$), and in the same manner that $P_{\x|\y}=P_{\x|\y,\z}$. Thus, the symmetry of the conditional independence given a common coarsening follows directly from the symmetry of the traditional conditional independence.
For completeness we demonstrate the symmetry of conditional independence below.

Consider that $\x$ is independent of $\y$ conditioned on $\z$, i.e.\ $P_{\x|\z}(\x^\dagger|\z^*)=P_{\x|\y,\z}(\x^\dagger|\y^\ddagger,\z^*)$. 
To show that this implies the symmetric relation, we use the definition of the conditional probability [Eq.~(\ref{eq:conditional_measure})]:
\beqa
  \frac{P_{\x|\y,\z}(\x^\dagger|\y^\ddagger,\z^*) }{P_{\x|\z}(\x^\dagger|\z^*)}
  &=& \frac{\Omega_{(\x^\dagger,\y^\ddagger,\z^*)}}{\Omega_{(\y^\ddagger,\z^*)}} \frac{\Omega_{\z^*}}{\Omega_{(\x^\dagger,\z^*)}} \nonumber\\
  &=& \frac{\Omega_{(\x^\dagger,\y^\ddagger,\z^*)}}{\Omega_{(\x^\dagger,\z^*)}} \frac{\Omega_{\z^*}}{\Omega_{(\y^\ddagger,\z^*)}} \nonumber\\
  &=& \frac{P_{\y|\x,\z}(\y^\ddagger|\x^\dagger,\z^*)}{P_{\y|\z}(\y^\ddagger|\z^*)}  \es.
\eeqa
This relation (which must be true) is only satisfied if $P_{\y|\x,\z}(\y^\ddagger|\x^\dagger,\z^*)=P_{\y|\z}(\y^\ddagger|\z^*)$, i.e.\ if $\y$ is independent of $\x$ conditioned on $\z$, thus completing the proof.}

\prooftheorem{the:independence}
{$\model{\x}$ and $\model{\y}$ are compatible if and only if they are conditionally independent given the common coarsening $\z = \comp{\x}{\y}$.}
{To show that conditional independence given a common coarsening is equivalent to compatibility, we first show that the former implies the latter and then that the latter implies the former.
To avoid clutter, we will in the following use the notation $\y'$ as short for $\y(G')$, as well as $\y''$ for $\y(G'')$ and $\x''$ for $\x(G'')$.

{\bf Conditional independence given a common coarsening implies compatibility.}
We will use the law of total probability and Proposition~\ref{prop:comparable-compatible} to show this. We first show that a law of total probability applies to the probabilities of features. Taking the probability distribution defining the composition of P[$\y$] on P[$\x$] [Eq.~(\ref{eq:composition})] and multiplying by the term $\sum_{\y^\dagger}\delta_{\y^\dagger,\y'}=1$, we get:
\beqa
  \Prob{\comp{\y}{\x}}(G|\x^*) &=& \sum_{G'\in\G}\sum_{\y^\dagger} \delta_{\y^\dagger,\y'} \frac{\delta_{\y(G),\y'}}{\Omega_{\y}(G)} \Prob{\x}(G'|\x^*) \nonumber\\
  &=& \sum_{\y^\dagger} \frac{\delta_{\y(G),\y^\dagger}}{\Omega_{\y}(G)}  \sum_{G'\in\G}  \delta_{\y^\dagger,\y'} \Prob{\x}(G'|\x^*) \nonumber\\
  &=& \sum_{\y^\dagger} \Prob{\y}(G|\y^\dagger) \Prob{\y|\x}(\y^\dagger|\x^*) \es.
  \label{eq:total_proba}
\eeqa
To obtain the second equality above we used the property of Kronecker delta functions that $\delta_{a,b}\delta_{b,c}=\delta_{a,c}\delta_{b,c}$, and the last equality was obtained from the definitions of $\Prob{\y}(G|\y^\dagger)$ (Def.~\ref{def:MRRM}) and $\Prob{\y|\x}(\y^\dagger|\x^*)$ (Def.~\ref{def:conditional_probability}).
Using the law of total probability [Eq.~(\ref{eq:total_proba})], we now expand $P_{\comp{\y}{\x}}(G|\x^*)$ to get:
\beqa
  \Prob{\comp{\y}{\x}}(G|\x^*) &=& \sum_{\y^\dagger} \Prob{\y}(G|\y^\dagger) \CProb{\y}{\x}(\y^\dagger|\x^*) \nonumber \\ 
  &=& \sum_{\y^\dagger} \Prob{\y}(G|\y^\dagger) \CProb{\y}{\z}(\y^\dagger|\z^*)  \nonumber \\ 
  &=& \Prob{\comp{\y}{\z}}(G|\z^*) \nonumber\\
  &=& \Prob{\z}(G|\z^*) \es.
\eeqa
Here, the second equality follows from the conditional independence of $\x$ and $\y$ (Def.~\ref{def:independence}), the second-to-last equality follows from the law of total probability, and the last from Proposition~\ref{prop:comparable-compatible} since $\z\geq\y$.

{\bf Compatibility implies conditional independence given a common coarsening.}
Because $\x$ and $\y$ are compatible their composition is a MRRM and we can choose $\z = \comp{\x}{\y}$, which by construction is a common coarsening of $\x$ and $\y$. The conditional independence of $\x$ and $\y$ given this $\z$ can now be shown from its definition via a direct calculation: 
\beqa
  \CProb{\y}{\comp{\x}{\y}}(\y^\dagger | \comp{\x}{\y}(G^*)) 
  &=& \sum_{G^{\prime} \in \G} \delta_{\y^\dagger, \y^\prime} \Prob{\comp{\x}{\y}}(G'|\comp{\x}{\y}(G^*)) \nonumber\\
  &=& \sum_{G^{\prime} \in \G}  \sum_{G^{\prime\prime} \in \G} \frac{\delta_{\x^{\prime\prime},\x^*} \delta_{\y^\dagger, \y^\prime} \delta_{\y^\prime, \y^{\prime\prime}} }{\Omega_{\y^{\prime\prime}} \Omega_{\x^*} }  \nonumber \\
  &=&   \sum_{G^{\prime\prime} \in \G} \frac{\delta_{\x^{\prime\prime},\x^*} \delta_{\y^\dagger, \y^{\prime\prime} }  }{\Omega_{\y^{\prime\prime}} \Omega_{\x^*} } \sum_{G^{\prime} \in \G} \delta_{\y^\prime, \y^{\prime\prime} } \nonumber \\
  &=&   \sum_{G^{\prime\prime} \in \G} \frac{\delta_{\x^{\prime\prime},\x^*} \delta_{\y^\dagger, \y^{\prime\prime} }  }{ \Omega_{\x^*} } \nonumber \\
  &=&   \sum_{G^{\prime\prime} \in \G}  \delta_{\y^\dagger, \y^{\prime\prime} } \Prob{\x}(G''|\x^*)   \nonumber \\
  &=& \CProb{\y}{\x}(\y^\dagger | \x^* )
  \es.
\eeqa
In the third equality we again used the property of Kronecker delta functions $\delta_{a,b}\delta_{b,c}=\delta_{a,c}\delta_{b,c}$, and in the fourth equality we used the definition of the partition function $\Omega_{\y^{\prime\prime}}$.}

\prooftheorem{the:independent_refinement}
{Consider two compatible MRRMs $\model{\x}$ and $\model{\y}$, and any adapted refinements of these, $\model{\y, \f(\x)}$ and $\model{\x, \g(\y)}$. 
Then $\model{\y, \f(\x)}$ and $\model{\x, \g(\y)}$ are compatible, and their composition is given by $\model{\comp{\x}{\y},\f(\x),\g(\y)}$.}
{To prove Theorem~\ref{the:independent_refinement}, it is sufficient to prove that $(\y,\f(\x))$ and $\x$ are independent conditioned on their common coarsening $(\y\circ\x,\f(\x))$. From this it then immediately follows that $(\y,\f(\x))$ and $(\x,\g(\y))$ are independent conditioned on their common coarsening $(\y\circ\x,\f(\x),\g(\y))$ and thus that P[$\y$,$\f(\x)$] and P[$\x$,$\g(\y)$] are compatible with ${\rm P}[(\y,\f(\x))\circ(\x,\g(\y))]={\rm P}[\y\circ\x,\f(\x),\g(\y)]$.
To develop the proof, we consider the conditional probability of $\x$ given $(\y,\f(\x))$:
\beq
  P_{\x|\y,\f(\x)}(\x^\dagger|\y^*,\f(\x^*)) 
  = \frac{\Omega_{(\x^\dagger,\y^*,\f(\x^*))}}{\Omega_{(\y^*,\f(\x^*))}} \es.
\eeq
%from the definition of the conditional probability of a feature (Def.~\ref{def:conditional_probability}).
Since $\x\leq\f(\x)$, we have $\Omega_{(\x^\dagger,\y^*,\f(\x^*))}=\Omega_{(\x^\dagger,\y^*)}$ whenever $\f(\x^\dagger)=\f(\x^*)$. We use this to rewrite the equation above, multiplying by the factor $\Omega_{\y^*}/\Omega_{\y^*}=1$, along the way,
\beqa
  P_{\x|\y,\f(\x)}(\x^\dagger|\y^*,\f(\x^*)) 
  &=& \frac{\Omega_{(\x^\dagger,\y^*)}\delta_{\f(\x^\dagger),\f(\x^*)}}{\Omega_{\y^*}} \frac{\Omega_{\y^*}}{\Omega_{(\y^*,\f(\x^*))}} \nonumber\\
  &=& \frac{P_{\x|\y}(\x^\dagger|\y^*)\delta_{\f(\x^\dagger),\f(\x^*)}}{P_{\f(\x)|\y}(\f(\x^*)|\y^*)} \es.
  \label{eq:proof:refinement1}
\eeqa
Now, since $\x$ and $\y$ are compatible, we have that $P_{\x|\y}(\x^\dagger|\y^*)=P_{\x|\z}(\x^\dagger|\z^*)$, with $\z=\y\circ\x$. Furthermore, it also means that $P_{\f(\x)|\y}(\f(\x^*)|\y^*)=P_{\f(\x)|\z}(\f(\x^*)|\z^*)$. The following calculation shows this:
\beqa
  P_{\f(\x)|\y}(\f(\x^*)|\y^\dagger) 
  &=& \sum_{\x^\ddagger} P_{\f(\x)|\x}(\f(\x^*)|\x^\ddagger) P_{\x|\y}(\x^\ddagger|\y^\dagger) \nonumber\\ 
  &=& \sum_{\x^\ddagger} P_{\f(\x)|\x}(\f(\x^*)|\x^\ddagger) P_{\x|\z}(\x^\ddagger|\z^\dagger) \nonumber\\ 
  &=& P_{\f(\x)|\z}(\f(\x^*)|\z^\dagger) \es.
  \label{eq:proof:refinement2}
\eeqa
(Note that we do not necessarily have $\f(\x)\leq\z$, though.)
Plugging these two identities into Eq.~(\ref{eq:proof:refinement1}) gives:
\beqa
  P_{\x|\y,\f(\x)}(\x^\dagger|\y^*,\f(\x^*)) 
  &=& \frac{P_{\x|\z}(\x^\dagger|\z^*)\delta_{\f(\x^\dagger),\f(\x^*)}}{P_{\f(\x)|\z}(\f(\x^*)|\z^*)} \nonumber\\
    &=& \frac{\Omega_{(\x^\dagger,\z^*)}\delta_{\f(\x^\dagger),\f(\x^*)}}{\Omega_{\z^*}} \frac{\Omega_{\z^*}}{\Omega_{(\z^*,\f(\x^*))}} \nonumber\\
    &=& \frac{\Omega_{(\x^\dagger,\z^*,\f(\x^*))}}{\Omega_{(\z^*,\f(\x^*))}} \nonumber\\
    &=& P_{\x|\z,\f(\x)}(\x^\dagger|\z^*,\f(\x^*))  \es.
  \label{eq:proof:refinement3}
\eeqa
Since both $\z^*\geq\x$ and $\f(\x^*)\geq\x$, we have $(\z,\f(\x))\geq\x$, which together with Eq.~(\ref{eq:proof:refinement3}) shows that $\x$ is independent of $(\y,\f(\x))$ conditionally on their common coarsening $(\z,\f(\x))$, thus completing the proof.}

%\proofcor{cor:combining_independent}
%{Consider two independent MRRMs $\model{\x}$ and $\model{\y}$, and two MRRMs, $\model{\y,\f(\x)}$ and $\model{\x,\g(\y)}$, that shuffle less. 
%Then $\model{\y,\f(\x)} \model{\x,\g(\y)}=\model{\f(\x),\g(\y)}$.}
%{Because independence is a special case of conditional independence the proof is a direct application of Theorem \ref{the:independent_refinement}: $\model{\y,\f(\x)} \model{\x,\g(\y)}= \model{\comp{\x}{\y},\f(\x),\g(\y)}= \model{1,\f(\x),\g(\y)}=\model{\f(\x),\g(\y)}$.}

\proofprop{prop:link-timeline_independent}
{Any link shuffling P[$\f(\L),\Thf$] and timeline shuffling P[$\L,\g(\Thf)$] are compatible and their composition is given by P[$L,\f(\L),\g(\Thf)$].}
{It is clear that the content of the individual timelines $\Thij\in\pxl(\Thf)$ does not in any way constrain what values $\L$ may take, only their number $L$ does. 
Furthermore, the number of ways that we can distribute the $L$ timelines on the links is independent of the particular configuration of $\L$, so $\Omega_{\L',\pxl(\Thf^*)}=\Omega_{\L'',\pxl(\Thf^*)}$ for all $\L'$, $\L''$. 
Similarly, the way we can distribute the 
 $E$ instantaneous events on the timelines depends only on $\L$ through $L$, so also $\Omega_{\L',L^*}=\Omega_{\L'',L^*}$ for all $\L'$, $\L''$.
This means that $\Omega_{\L',L^*} \propto \Omega_{\L',\pxl(\Thf^*)}$ for all $L'$, and since the conditional probabilities must be normed, that $P_{\L|\pxl(\Thf)}(\L^{\dagger}|\pxl(\Thf^*))=P_{\L|L}(\L^{\dagger}|L^*)$, i.e. that $\L$ and $\pxl(\Thf)$ are independent conditioned on $L$. 
Since $(L,E)\geq(\L,E)$ and $(L,E)\geq\pxl(\Thf)$, it then follows from Theorem~\ref{the:independence} that $(\L,E)$ and $\pxl(\Thf)$ are compatible. 
This shows that the coarsest link shuffling, P[$\Thf$] (equivalent to P[$\pxl(\Thf)$]), and the coarsest timeline shuffling, P[$\L,E$], are compatible. 
We next note that any link and timeline shufflings are adapted refinements of P[$\Thf$] and P[$\L,E$], respectively (compare Defs.~\ref{def:link_shuffling} and \ref{def:timeline_shuffling} with Def.~\ref{def:adapted_refinement}).
So applying Theorem~\ref{the:independent_refinement} gives that any link shuffling P[$\f(\L)$,$\Thf$] and any timeline shuffling P[$\L,\g(\Thf)$] are compatible and that their composition is ${\rm P}[L,\f(\L),\g(\Thf)]$.}

\proofprop{prop:sequence-snapshot_independent}
{Any sequence shuffling ${\rm P}[\f(\t),\pxt(\Gf)]$ and snapshot shuffling ${\rm P}[\t,\g(\pxt(\Gf))]$ are compatible and their composition is given by ${\rm P}[p(\Ebf),\f(\t),\g(\pxt(\Gf))]$.}
{Following the same reasoning as in the proof of Proposition~\ref{prop:link-timeline_independent}, we note that $\Omega_{\t',\pxt(\Gf^*)}=\Omega_{\t',p(\Ebf^*)}$ for all $\t'$ %, where $\Ebf=(2|\E^t|)_{t\in\T}$ is the cumulative activity at each time 
and that $p(\Ebf)$ satisfies $p(\Ebf)\geq\t$ and $p(\Ebf)\geq\pxt(\Gf)$.
Thus, $\t$ is independent of $\pxt(\Gf)$ conditioned on $p(\Ebf)$. So the coarsest sequence and snapshot shufflings, P[$\pxt(\Gf)$] and P[$\t$] are compatible by Theorem~\ref{the:independence}. Consequently, since all sequence shufflings P[$\f(\t),\pxt(\Gf)$] and snapshot shufflings P[$\Ebf,\f(\pxt(\Gf))$] are adapted refinements of P[$\pxt(\Gf)$] or P[$\t$], respectively (Def.~\ref{def:adapted_refinement}), Theorem~\ref{the:independent_refinement} tells us that they are compatible and that their composition is ${\rm P}[p(\Ebf),\f(\t),\g(\pxt(\Gf))]$.}

%\proofprop{prop:link-sequence_independent}
%{Any link shuffling P[$\f(\L),\pxl(\Thf)$] and any sequence shuffling ${\rm P}[\f(\t),\pxt(\Gf)]$ are compatible.}
%{Since sequence shufflings are timeline shufflings, they are by virtue of Proposition~\ref{prop:link-timeline_independent} also compatible with link shufflings.}

% %------------------------------------------------------------------------
% \section{Tables}
% \label{app:tables}
% %------------------------------------------------------------------------
% \input{Sec-Tables}

%=========================================================
% Bibliography
%=========================================================
\bibliographystyle{naturemag}
\bibliography{random_refs}

%=========================================================
% Tables
%=========================================================
% \input{Tables}
% \clearpage

%=========================================================
% Supplement
%=========================================================

%\newcounter{lastpage}
%\setcounter{lastpage}{\value{page}}
%\begin{titlepage}
%\end{titlepage}
%\setcounter{page}{\value{lastpage}}
%\addtocounter{page}{1}

\begin{titlepage}
  \vspace{1cm}\.\\
  \centering{\Large{{\bf Supplementary material}}}
 % \clearpage
\end{titlepage}

\counterwithout{table}{section}
\counterwithout{figure}{section}

\renewcommand{\thefigure}{{S\arabic{figure}}}
\renewcommand{\thetable}{{S\arabic{table}}}

\renewcommand{\figurename}{{\bf Supplementary FIG.}}
\setcounter{figure}{0}
\renewcommand{\tablename}{{\bf Supplementary TABLE}}
\setcounter{table}{0}
%------------------------------

%------------------------------------------------------------------------
%------------------------------------------------------------------------
%------------------------------------------------------------------------
\begin{table*}[p]
  \caption{{\bf Sequences, distributions, and moments of features.}
  Below, $(\cdot)$ denotes an ordered sequence and $[\cdot]$ denotes a multiset, equivalent to the empirical distribution.}
  \label{tab:constraint_levels}
  \begin{threeparttable}
  \begin{tabular}{clll}
    \hline
    \hline
    {\sl Symbol} & {\sl Meaning of symbol} & {\sl Definition} %& {\sl Example(s)}
    \\
    \hline
    \cxmark & One-level sequence of link features. & $\x=(x_\ij)_{\ij\in\L}$ & \\
    & One-level sequence of node features. & $\x=(x_i)_{i\in\N}$ %& Fig.~\ref{fig:examples-kstat}(a) 
    \\
    & One-level sequence of snapshot features. & $\x=(x^t)_{t\in\T}$ \\
    & Two-level sequence of link features. & $\x=(\x_\ij)_{\ij\in\L}$~$^{\rm a}$ %& Fig.~\ref{fig:examples-Dtau}(a)
    \\
    & Two-level sequence of node features. & $\x=(\x_i)_{i\in\N}$~$^{\rm b}$ %& Fig.~\ref{fig:examples-d}(a)
    \\
    \hline
    \piijmark & Sequence of local distributions on links. & $\piij(\x)=(\pi_\ij(\x_\ij))_{\ij\in\L}$~$^{\rm d}$ %& Fig.~\ref{fig:examples-Dtau}(b) 
    \\
    \piimark & Sequence of local distributions on nodes. & $\pii(\x)=(\pi_i(\x_i))_{i\in\N}$~$^{\rm e}$ %& Fig.~\ref{fig:examples-d}(b) 
    \\
    \pitmark & Sequence of local distributions in snapshots. & $\pit(\x)=(\pi^t(\x^t))_{t\in\T}$~$^{\rm f}$ %& Fig.~\ref{fig:examples-d}(e) 
    \\
    \hline
    \pxlmark & Distribution of local sequences on links. & $\pxl(\x)=[\x_\ij]_{\ij\in\L}$~$^{\rm a}$ %& Fig.~\ref{fig:examples-Dtau}(c) 
    \\
    \pximark & Distribution of local sequences on nodes. & $\pxi(\x)=[\x_i]_{i\in\N}$~$^{\rm b}$ %& Fig.~\ref{fig:examples-d}(c) 
    \\
    \pxtmark & Distribution of local sequences in snapshots. & $\pxt(\x)=[\x^t]_{t\in\T}$~$^{\rm c}$ %& Fig.~\ref{fig:examples-d}(d) 
    \\
    \hline
    \pmark & Distribution of one-level link features.  & $p(\x)=[x_\ij]_{\ij\in\L}$ \\
    & Distribution of one-level node features & $p(\x)=[x_i]_{i\in\N}$ %& Fig.~\ref{fig:examples-kstat}(b) 
    \\
    & Distribution of one-level snapshot features & $p(\x)=[x^t]_{t\in\T}$ \\
    & Global distribution of two-level link features. & %$p(\x)=\cup_{\ij\in\L}\pi_\ij(\x_\ij)$~$^{\rm d}$ 
    $p(\x)=[x_\ij^m]_{m\in\Ml,\,\ij\in\L}$ %& Fig.~\ref{fig:examples-Dtau}(g) 
    \\
    & Global distribution of two-level node features. & %$p(\x)=\cup_{i\in\N}\pi_i(\x_i)$~$^{\rm e}$ 
    $p(\x)=[x_i^m]_{m\in\Mi,\,i\in\N}$ %& Fig.~\ref{fig:examples-d}(k) 
    \\
    \hline
    \muijmark & Sequence of local means on links. & $\muij(\x)=(\mu_\ij(\x_\ij))_{\ij\in\L}$~$^{\rm g}$ %& Fig.~\ref{fig:examples-Dtau}(d)
    \\
    \muimark & Sequence of local means on nodes.  & $\mui(\x)=(\mu_i(\x_i))_{i\in\N}$~$^{\rm h}$ %& Fig.~\ref{fig:examples-d}(f)
    \\
    \mutmark & Sequence of local means in snapshots.  & $\mut(\x)=(\mu^t(\x^t))_{t\in\T}$~$^{\rm i}$  %& Fig.~\ref{fig:examples-d}(i)
    \\
    \hline
    \mumark & Mean of one-level link features. & $\mu(\x)=\sum_{\ij\in\L} x_\ij/L$ \\
    & Mean of one-level node features. & $\mu(\x)=\sum_{i\in\N} x_i/N$ %& Fig.~\ref{fig:examples-kstat}(c) 
    \\
    & Mean of one-level snapshot features. & $\mu(\x)=\sum_{t\in\T} x^t/T$ \\
    & Global mean of two-level link features. & $\mu(\x)=\sum_{\ij\in\L}\sum_{m\in\Ml}x_\ij^m/(\sum_{\ij\in\L}M_\ij)$ %& Fig.~\ref{fig:examples-Dtau}(h) 
    \\
    & Global mean of two-level node features. & $\mu(\x)=\sum_{i\in\N}\sum_{m\in\Mi}x_i^m/(\sum_{i\in\N}M_i)$ %& Fig.~\ref{fig:examples-d}(m) 
    \\
    \hline
    \xmark & Feature is not conserved.\\
    \hline
    \hline
  \end{tabular}
  \begin{tablenotes}
    \item[{\rm a}]  $\x_\ij$ : Local sequence on link,  $\x_\ij=(x_\ij^m)_{m\in\Ml}$, where $\Ml$ is a temporally ordered index set. \\
    \item[{\rm b}] $\x_i$ : Local sequence on node, $\x_i=(x_i^m)_{m\in\Mi}$, where $\Mi$ is a temporally ordered index set. \\
    \item[{\rm c}] $\x^t$ : Local sequence in snapshot, $\x^t=(x_i^t)_{i\in\N}$. \\
    \item[{\rm d}] $\pi_\ij(\x_\ij)$ : Local distribution on link, $\pi_\ij(\x_\ij)=[x_\ij^m]_{m\in\Ml}$. \\
    \item[{\rm e}] $\pi_i(\x_i)$ : Local distribution on node, $\pi_i(\x_i)=[x_i^m]_{m\in\Mi}$. \\
    \item[{\rm f}] $\pi^t(\x^t)$ : Local distribution in snapshot, $\pi^t(\x^t)=[x_i^t]_{i\in\N}$. \\
    \item[{\rm g}] $\mu_\ij(\x_\ij)$ : Local mean on link, $\mu_\ij(\x_\ij)=\sum_{m\in\Ml} x_\ij^m/M_\ij$. \\
    \item[{\rm h}] $\mu_i(\x_i)$ : Local mean on node, $\mu_i(\x_i)=\sum_{m\in\Mi} x_i^m/M_i$. \\
    \item[{\rm i}] $\mu^t(\x^t)$ : Local mean in snapshot, $\mu^t(\x^t)=\sum_{i\in\N} x_i^t/N$. \\
  \end{tablenotes}
  \end{threeparttable}
\end{table*}
%------------------------------------------------------------------------

%------------------------------------------------------------------------
\begin{table*}[p]
  \caption{{\bf Additional features of directed temporal networks. }
  \new{Several of the studies in the literature survey (Sec.~\ref{sec:applications}) considered MRRMs specifically defined for directed temporal networks, namely P[$\doutf$] and P[$\soutf,p(\Dtauf)$].  
  The \new{definition of MRRMs that take the directionality of events into account} is straightforward as it simply requires defining the appropriate directed features. 
  For features of links, no generalizations are necessary since they all generalize automatically to directed networks by using the convention that $(i,j)$ designates an interaction from $i$ to $j$.
  However, since in directed networks a link from $i$ to $j$ does not imply the presence of the reciprocal link from $j$ to $i$, the interpretation of link features may change. 
For each feature of nodes, three generalizations typically exist: an outgoing version, e.g., the out-strength $\sout{i}$, an ingoing version, e.g., the in-strength $\sin{i}$, and a combined version, e.g., the total strength $s_i=\sout{i}+\sin{i}$. 
  We here list some generalizations of node features to directed temporal networks. }
%  Link features are the same as for undirected networks.
  Below, $(\cdot)$ denotes an ordered sequence, $\{\cdot\}$ denotes an unordered set, $|\cdot|$ denotes the cardinality of a set, $:$ means ``for which'' or ``such that''. 
  }
  \label{tab:features-directed}
  \begin{tabular}{lll}
    \hline
    \hline
    {\sl Symbol} & {\sl Meaning of symbol} & {\sl Definition}\\
    \hline
    \multicolumn{3}{l}{{\bf\em Topological-temporal features}}\\
    $\N_\iout$ & Outgoing neighborhood of node. & $\N_\iout=\{j : (i,j)\in\L\}$\\
    $\N_\iin$ & Incoming neighborhood of node. & $\N_\iin=\{j : (j,i)\in\L\}$\\
    $\N_i$ & Neighborhood of node. & $\N_i=\{j : (i,j)\in\L\ {\rm or}\ (j,i)\in\L\}$\\
    \multicolumn{3}{l}{{\bf\em Topological-temporal features}}\\
    $\dtout{i}{t}$ & Instantaneous out-degree. & $\dtout{i}{t}=|\{j : (i,j)\in\E^t\}|$\\
    $\dtin{i}{t}$ & Instantaneous in-degree. & $\dtin{i}{t}=|\{j : (j,i)\in\E^t\}|$\\
    $\dit$ & Instantaneous (total) degree & $\dit=|\{j : (i,j)\in\E^t\ {\rm or}\ (j,i)\in\E^t\}|$\\
    $\Phi_{\iout}$ & Node activity timeline. & $\Phi_\iout = \left((v_\iout^1,\alpha_\iout^1),(v_\iout^2,\alpha_\iout^2),\ldots,(v_\iout^{a_\iout},\alpha_\iout^{a_\iout})\right)$\\
    {$\alpha_\iout^m$} & Activity duration. & Consecutive interval during which $i$ has at least one outgoing contact.\\
    $\Dalpha_\iout^m$ & Inactivity duration. & $\Dalpha_\iout^m=v_\iout^{m+1}-(v_\iout^m+\alpha_\iout^m)$ \\
    \multicolumn{3}{l}{{\bf\em Aggregated features}}\\
    $\aout{i}$ & Outgoing node activity. & $\aout{i}=\sum_{j\in\N_\iout} \nl$\\
    $\ain{i}$ & Ingoing node activity. & $\ain{i}=\sum_{j\in\N_\iin} \nl$\\
    $\ai$ & (Total) node activity. & $\ai=\sum_{\ij\in\L_i} \nl$\\
    $\sout{i}$ & Node out-strength. & $\sout{i}=\sum_{\ij\in\L_\iout} \wl$\\
    $\sin{i}$ & Node in-strength. & $\sin{i}=\sum_{\ij\in\L_\iin} \wl$\\
    $\si$ & Node (total) strength. & $\si=\sum_{\ij\in\L_i} \wl$\\
    $\kout{i}$ & Node out-degree. & $\kout{i}=|\N_\iout|$\\
    $\kin{i}$ & Node in-degree. & $\kin{i}=|\N_\iin|$\\
    $k_i$ & Node (total) degree. & $k_i=|\N_i|$\\
    \hline
    \hline
  \end{tabular}
\end{table*}
%------------------------------------------------------------------------

%------------------------------------------------------------------------
%------------------------------------------------------------------------
%\input{SM-Figures.tex}

\clearpage
%------------------------------------------------------------------------
\begin{titlepage}
  \centering{\Large{{\bf Supplementary Note 1: Applying instant-event shufflings to temporal networks with event durations}}}
  \vspace{1cm}
\end{titlepage}
%\section{Applying instant-event shufflings to temporal networks with event durations}
%\note{CLV: May be moved to supplement as well?}
%\label{sec:instant-event_shuffling}
%------------------------------------------------------------------------
\noindent
Event shufflings by definition conserve the events' durations in a temporal network, but we may randomize the durations \new{events in a temporal network} by first representing the network as an instant-event network and then applying an instant-event shuffling to it using the following procedure: 

i) Choose an appropriate time-resolution for discretization. A natural choice may be the time-resolution of recordings, but for high resolution measurements a lower time-resolution may be more practical. 

ii) Construct the corresponding instant-event temporal network by defining an instantaneous event between a pair of nodes at the start of each time-interval during which the nodes are in contact (in the temporal network shown Fig.~\ref{fig:link-timeline} this splits the four long events into two instantaneous events  each and creates one instantaneous event for each of the shorter events). 

iii) Randomize this instant-event network using an instant-event shuffling.

iv) Recreate a randomized version of the temporal network by concatenating consecutive instantaneous events between the same nodes into single events. 

Table~\ref{tab:effects-contact} show the effects of both event and instant-event shufflings on the features of a temporal network. To understand the effects of the same shufflings on an instant-event network one should simply ignore the features that are not defined for instant-event temporal networks, i.e.\ $\ai$, $\nij$, $\alpha^m_\ij$ and $\tau^m_\ij$.

From the above procedure, the number of instantaneous events on a link in the generated instant-event network is seen to correspond to the weight $\wij$ of the link in the original temporal network. Using $\wij$ to designate the number of instantaneous events on the links in the instant-event network thus makes it possible to name each event and instant-event shuffling based on the features they conserve in a consistent manner, no matter whether the shuffling is applied to a temporal network or an instant-event temporal network. 
It follows from this alternative definition of $\wij$ that $s_i$ designates the activity of the node $i$ in an instant-event network (see the section ``Notation for features of instant-event networks'' in Supplementary Note~2 %Section~\ref{sec:instant-event_features} 
for further discussion).

%------------------------------------------------------------------------
\setcounter{definition}{0}
\setcounter{example}{0}
\renewcommand{\thesection}{S2}
\begin{titlepage}
  \centering{\Large{{\bf Supplementary Note 2: Features of temporal networks}}}
  \vspace{1cm}
\end{titlepage}
%\section{Features of temporal networks}
%\label{sec:features}	
%\note{CLV: moved to Supplement or perhaps even removed completely if we can do without it.}
%------------------------------------------------------------------------
\noindent
In this supplementary note we provide detailed definitions for a selection of temporal network features that have been shown to play an important role in network dynamics, and which are sufficient to name the MRRMs found in our literature survey presented in Section~\ref{sec:classification}.
Table~\ref{tab:features} lists basic features of temporal networks, often describing single elements of a network such as nodes or events, and Table~\ref{tab:constraint_levels} lists different general ways to construct features describing the whole temporal network by using the basic features as building blocks.

A feature of a temporal network is any function that takes a network as an input (Def.~\ref{def:feature}). Clearly, there is a very large number of such functions which could be defined\footnote{The number of functions leading to different MRRMs is equal to the the number of possible partitions of the state space of temporal networks of a given size. In practice, the number of possible temporal networks is very large and the number of partitions of the state space is a super-exponential function of this number (see Section~\ref{sec:RRM}).}, and as we will see later, a multitude of such functions have been (often implicitly) used in the literature. Here we attempt to organize this set of functions in a way that it is compatible with the different temporal network representations introduced in Section~\ref{sec:temp-net} and the concept of order of features \new{developed in Section~\ref{sec:theory-comparison}}.
% given by Def.~\ref{def:comparability} and Proposition~\ref{the:partial_order}. 

For definiteness we will here consider features of temporal networks with event durations. In order to make our description of MRRMs consistent for both networks with and without event durations, we shall need to modify the definitions of some of these features for instant-event networks. We do this % in Subsection~\ref{sec:instant-event_features} 
at the end of this \new{note}. %section.

Many features are ones returning a sequence of lower-dimensional features \new{(Def.~\ref{def:sequence})}, e.g., the degree sequence of a static graph, $\kstat$, is given by the sequence of the individual node degrees $k_i$.
Temporal network features are often given by a nested sequence where individual features in the sequence themselves are a %function of a 
sequence of scalar features.
Sequences and nested sequences can further be turned into distributions and average values in multiple ways.

The basic building blocks of many features constructed as sequences and used in MRRMs are scalar features describing single elements of the network (Table~\ref{tab:features}). %Due to their importance and for ease of reference, we denote these as {\sl features}.
%
%\begin{definition}
%{\sl Characteristic.} 
%A temporal network {\sl characteristic} is a scalar function $x$ pertaining to a single element of a network, such as a node $i\in\N$, a link $\ij\in\L$, a snapshot in time $t\in\T$, an event $c\in\C$, or both a point in time and a node or a link. 
%\end{definition}
%
%
%Table~\ref{tab:features} lists important temporal features and, in particular, scalar ones.
We mainly consider two types of sequences: {\sl one-level} sequences of scalar features and {\sl two-level} sequences of sequences of scalar features.

\begin{definition}
{\sl One-level and two-level sequences.} 
\begin{enumerate}
  \item{\sl One-level sequence.} We refer to a non-nested sequence of scalar features, $\x=(x_q)_{q\in\Q}$, as a {\sl one-level sequence}. 
  \item{\sl Two-level sequence.} We refer to a nested sequence of features that are themselves one-level sequences, $\x=(\x_q)_{q\in\Q}$ with $\x_q=(x_q^r)_{r\in\R_q}$, as a {\sl two-level sequence}. We refer to the $\x_q$ as {\sl local sequences}. 
\end{enumerate}  
Detailed definitions of particular types of one- and two-level sequences are given in Table~\ref{tab:constraint_levels} (symbol: $\x$).
\end{definition}

One-level sequences are typically used to represent features that are aggregated over the temporal or topological dimension of the temporal network, while two-level sequence are composed of features that depend both on topology and time.
The following examples illustrate this.

%\begin{definition}
%{\sl One-level sequence of characteristics.}
%A {\sl one-level} sequence is a sequence of single characteristics (i.e.\ scalar features): $\x=(x_q)_{q\in\Q}$.
%\end{definition}
%
%A one-level sequence is typically a sequence of aggregated network features: for a set of time-aggregated features of the nodes or links, the one-level sequence is given as $\x=(x_i)_{i\in\N}$ or $\x=(x_\ij)_{\ij\in\L}$, respectively; for a set of aggregated features of snapshots, e.g., the cumulative activity $A^t$ of all nodes  at each time $t$, the one-level sequence is of the form $\x=(x^t)_{t\in\T}$.

\begin{example}
\label{ex:complete_one-level_sequence}
{\sl One-level sequence of static degrees.}
A well known example of an aggregated graph feature is the node degree, $k_i$, giving the number of nodes in $\Gstat$ that are connected to the node $i$.
%\note{CLV: I've removed the figure illustrating this, as I don't think it was very helpful.}
%Figure~\ref{fig:examples-kstat}(a) shows the ({\sl one-level}) sequence of static degrees $\kstat^*=(k_i^*)_{i\in\N}$ of the temporal network shown in Fig.~\ref{fig:link-timeline}.
\end{example}

\begin{example}
\label{ex:dit}
{\sl Two-level sequence of instantaneous degrees.}
A generalization to temporal networks of the static degree $k_i$ of a node is the {\sl instantaneous degree} $\dit$. It is given by the number of nodes that the node $i$ is in contact with at time $t$.
The (two-level) sequence of instantaneous degrees is $\d=(\d^t)_{t\in\T}=((\dit)_{i\in\N})_{t\in\T}$, or alternatively $\d=(\d_i)_{i\in\N}=((\dit)_{t\in\T})_{i\in\N}$ since the order of the indices $i$ and $t$ does not matter here.
%\note{CLV: I've removed the figure illustrating this, as I don't think it was very helpful.}
%Figure~\ref{fig:examples-d}(a) shows the sequence of instantaneous degrees $\d^*$ of the temporal network shown in Figs.~\ref{fig:link-timeline} and \ref{fig:snapshot-sequence}.
\end{example}

\begin{example}
\label{ex:Dtau}
{\sl Two-level sequence of inter-event durations.}
A feature of temporal networks that has been shown to have a profound impact on dynamic processes is the durations between consecutive events in the timelines, termed the {\sl inter-event durations} and defined by $\Dtau_\ij^m=t_\ij^{m+1}-(t_\ij^m+\tau_\ij^m)$. Their (two-level) sequence is
$\Dtauf=(\Dtauf_\ij)_{\ij\in\L}$, where $\Dtauf_\ij=(\Dtau_\ij^m)_{m\in\Ml}$. 
Here $\Ml=\{1, 2, \ldots, \nij-1\}$ indexes the inter-event durations in the timeline $\Thij$ by temporal order, with $\nij$ the number of events in the timeline. %Note that since the timelines are not of equal length, the indices $\ij$ and $m$ cannot be inversed. %This is generally the case for
Note that due to the temporal extent of the inter-event durations, we cannot inverse the order of the indices $m$ and $\ij$ as we could for the instantaneous degrees presented in the previous example.
%\note{CLV: I've removed the figure illustrating this, as I don't think it was very helpful.}
%Figure~\ref{fig:examples-Dtau}(a) shows the two level sequence of inter-event durations $\Dtauf^*$ of the temporal network in Fig.~\ref{fig:link-timeline}.
\end{example}

\begin{example}
\label{ex:snaphot_seq}
{\sl Sequence of snapshot graphs.}
A notable example of a sequence of features that are neither scalar nor sequences of scalars is the sequence of snapshot graphs $\Gs=(\Gt)_{t\in\T}$ (Def.~\ref{def:snapshot-sequence}). 
%The sequence of snapshot graphs for the temporal network of Fig.~\ref{fig:link-timeline} is shown in Fig.~\ref{fig:snapshot-sequence}.
%One such sequence is shown in Fig.~\ref{fig:snapshot-sequence}.
\end{example}

Instead of constraining an ordered sequence itself, many MRRMs constrain marginal distributions (Def.~\ref{def:unordered_set}) or moments of a sequence. %Since any statistic $\f$ of a sequence $\x$ is a measurable function of $\x$, by Theorem~\ref{the:partial_order} it thus corresponds to a MRRM, P[$\f(\x)$], that is coarser than P[$\x$]. 
%Such a statistic is always given by a function of the sequence, so it corresponds to a MRRM that is coarser than the MRRM defined by constraining the sequence itself.
Before we define these marginals and moments in detail for temporal networks, we consider as a simpler example the degree sequence of a static graph.

\begin{example}
\label{ex:marginals-static}
From the sequence of degrees in a static graph, $\kstat$, we may calculate their marginal distribution $p(\kstat)$ (equivalent to the multiset of their values)
%, see Def.~\ref{def:unordered_set} % and \ref{def:unordered_set-1level} below), 
as well as their mean $\mu(\kstat)$. 
%(Fig.~\ref{fig:examples-kstat}).
This leads to three different features, each corresponding to a different MRRM: 
one that constrains the complete sequence of degrees, %(the Maslov-Sneppen model),
P[$\kstat$], one that constrains their distribution, P[$p(\kstat)$], and one that constrains their mean %(an ensemble of Erd\H{o}s-R\'enyi random networks), 
${\rm P}[\mu(\kstat)]$ (which is equivalent to P[$L$] if the number of nodes $N$ is kept constant).
(P[$\kstat$] and P[$p(\kstat)$] are both often referred to as the configuration model or the Maslov-Sneppen model, and ${\rm P}[\mu(\kstat)]={\rm P}[L]$ is the Erd\H{o}s-R\'{e}nyi model with a fixed number of links.)
%Figure~\ref{fig:examples-kstat} shows values of these three features for the temporal network illustrated in Fig.~\ref{fig:link-timeline}. % and \ref{fig:snapshot-sequence}.
The three features (and corresponding MRRMs) satisfy a linear order: $\kstat\leq p(\kstat)\leq\mu(\kstat)$.
%if we constrain the sequence of degrees we necessarily constrain the distribution and mean, while if we constrain the distribution the mean is constrained; the Maslov-Sneppen model is finer than the model that constrains the distribution of degrees, which is again finer than the Erd\H{o}s-R\'enyi random network ensemble.
\end{example}

Since we have both a topological and temporal dimension in temporal networks, a much larger number of different ways to marginalize the sequence of features is possible than for a simple static graph.
We define here those needed to characterize the MRRMs surveyed in this article, but many more could be defined.
%(we provide an extended list in Supplementary Table~1 %\ref{suptab:constraint_levels} 
%as well as hierarchies of this extended list of features in Supplementary Figs.~1 and 2. %\ref{supfig:constraint_hierarchy} and \ref{supfig:dependency_diagram}).
%We define the most important ones below and specify them for features of links, nodes, and snapshots in Table~\ref{tab:constraint_levels}.
%Specifically, we let $\x$ denote the ordered sequence of a set of temporal network features. By ordered we mean that we retain both its value and what it designates in the input network.
%The basic feature function corresponding to a set of features is their ordered sequence.
The different ways to marginalize a sequence of features give rise to the following different three types of distributions.

\begin{definition}
{\sl Distributions over a sequence.}
We here list different ways to construct distributions of feature values by marginalizing over a sequence $\x$. The different types of distributions are defined in more detail in Table~\ref{tab:constraint_levels} (symbols: $p$, $\pxl$, $\pxi$, $\pxt$, $\piij$, $\pii$, and $\pit$).
\begin{enumerate}
  \item{\sl Global distribution $p(\x)$.} The global distribution $p(\x)$ returns the number of times each possible scalar value appears in a measured sequence $\x^*=\x(G^*)$. 
For a one-level sequence $\x^*=(x^*_q)_{q\in\Q}$, it is obtained by marginalizing over the sole index set $\Q$: $p(\x^*)=[x_q^*]_{q\in\Q}$.
For a two-level sequence $\x^*=(((x^r_q)^*)_{r\in\R_q})_{q\in\Q}$,  
it is obtained by marginalizing both over the inner and outer index sets, $\R_q$ and $\Q$: $p(\x^*)=[(x^r_q)^*]_{r\in\R_q, q\in\Q}$.
%For simplicity, we have left out the subscripts in the notation $p(\x)$ since all indices are marginalized over. 
%  Formally, or a one-level sequence, $p(\x)$ is given by the empirical measure $\emp(\xi|\x^*)=\sum_{q\in\Q} \delta_{\xi,x_q^*}$, while for a two-level sequence it is given by $\emp(\xi|\x^*)=\sum_{q\in\Q}\sum_{r\in\R_q} \delta_{\xi,(x^r_q)^*}$.

  \item{\sl Distribution of local features $p_\Q(\x)$.} For a sequence of non-scalar features, $\x=(\x_q)_{q\in\Q}$, the distribution of local features $\pxq(x^*)=[\x_q^*]_{q\in\Q}$ reports the number of times each possible value of the local features $\x_q$ appears in a measured sequence $\x^*$. %Formally, it is most simply defined as the multiset of local sequences, $\pqx=[\x_q]_{q\in\Q}$.
%  Here each $\x_q$ may be a local one-level sequence or a more general feature such as a graph.

  \item{\sl Sequence of local distributions $\piq(\x)$.} For a two-level sequence of features, $\x=(\x_q)_{q\in\Q}$, the sequence of local distributions $\piq(\x)$ is given by the ordered tuple $\piq(\x^*)=(\pi_q(\x_q^*))_{q\in\Q}$, where each local distribution $\pi_q(\x_q^*)=[(x_q^r)^*]_{r\in\R_q}$ is the distribution of the scalar features in the local sequence $\x_q^*$. % as defined in Definition~\ref{def:unordered_set}. 
%To avoid confusion with the distribution of local sequences, $p_\Q(\x)$, we use the different symbol $\piq(\x)$ to designate the sequence of local distributions.
%That is, $\pi_q(\x_q^*)$ %=(\xi, \epsilon(\xi|\x_q^*)_{\xi\in\X}$ %$\pi_q(\x_q^*)=(\xi, \sum_{(x_q^m)^*\in\x_q^*}\delta_{\xi,(x_q^m)^*})_{\xi\in\X}$ 
%returns the number of times each possible value $\xi$ appears in $\x_q^*$.
\end{enumerate}
\end{definition}

%Here %$\delta$ is the Kronecker delta function and 
%the sum runs over all scalar elements in $\x^*$, i.e.\ over the single index for a one-level sequence and over both indices for a two-level sequence.

%\begin{definition}
%\label{def:global_set}
%{\sl Distribution of features.} 
%Consider a one- or two-level sequence of features, $\x$. 
%The distribution of the features, denoted $p(\x)$, returns the number of times each possible scalar value $\xi$ %, corresponding to a single feature in $\x$, 
% appears in a measured sequence $\x^*=\x(G^*)$.
%Formally, the distribution can be defined as %is given as the set of tuples $p(\x)=\{\xi,\emp(\xi|\x^*)\}_{\xi\in\X}$. Here $\X$ is the space of possible values of $\xi$, and $\emp(\xi|\x^*)$ is 
%the unnormed empirical measure that for each value $\xi$ found in $\x^*$ assigns the value given by the function $\emp(\xi|\x^*)=\sum_{x^*\in\x^*} \delta_{\xi,x^*}$. 
%Here %$\delta$ is the Kronecker delta function and 
%the sum runs over all scalar elements in $\x^*$, i.e.\ over the single index for a one-level sequence and over both indices for a two-level sequence.
%\end{definition}

The following examples illustrate the different types of distributions.

\begin{example}
\label{ex:global_sets}
{\sl Global distribution.}
\new{The global distribution of the static degrees is given by $p(\kstat^*)=[k_i^*]_{i\in\N}$.}
%The global distribution $p(\kstat^*)=[k_i^*]_{i\in\N}$ of the one-level static degrees of the temporal network of Figs.~\ref{fig:link-timeline} and \ref{fig:snapshot-sequence} is shown in Fig.~\ref{fig:examples-kstat}(b).
The global distributions of the two-level instantaneous \new{node degrees and the inter-event durations on links are given by $p(\d^*)=[(\dit)^*]_{t\in\T, i\in\N}$ and $p(\Dtauf^*)=[(\Dtau_\ij^m)^*]_{m\in\Ml, \ij\in\L}$,} respectively. 
%, of the same network are shown in Figs.~\ref{fig:examples-d}(g) and \ref{fig:examples-Dtau}(e). 
\end{example}

\begin{example}
\label{ex:set_seqs}
{\sl Distribution of local features.}
Two different types of distributions of local sequences of instantaneous degrees can be constructed from the sequence of instantaneous degrees:
the distribution of local sequences of the instantaneous degrees of each node, \new{ given by $\pxi(\d^*)=[\d_i^*]_{i\in\N}$,}
% shown in Fig.~\ref{fig:examples-d}(c)] for the temporal network illustrated in Figs.~\ref{fig:link-timeline} and \ref{fig:snapshot-sequence}, 
and the distribution of local sequences of instantaneous degrees of nodes in each snapshot, \new{given by $\pxt(\d^*)=[(\d^t)^*]_{t\in\T}$.}
%, shown in Fig.~\ref{fig:examples-d}(d).
From the sequence of inter-event durations, we can construct the distribution of local sequences of inter-event durations on the links, $\pxl(\Dtauf^*)=[\Dtauf_\ij^*]_{\ij\in\L}$. 
%, shown in Fig.~\ref{fig:examples-Dtau}(c).
From the snapshot-graph sequence, we can construct the distribution of snapshot graphs $\pxt(\Gs^*)=[(\Gt)^*]_{t\in\T}$.
\end{example}

\begin{example}
\label{ex:seq_sets}
{\sl Sequence of local distributions.}
We can also construct two different sequences of local distributions from the sequence of instantaneous degrees:
the sequence of local distributions of the instantaneous degrees of each node, $\pii(\d^*)=([(\dit)^*]_{t\in\T})_{i\in\N}$. 
%, shown in Fig.~\ref{fig:examples-d}(b),
and  the sequence of local distributions of instantaneous degrees in each snapshot, $\pit(\d^*)=([(\dit)^*]_{i\in\N})_{t\in\T}$.
%, shown in Fig.~\ref{fig:examples-d}(e).
%respectively, of the temporal network of Figs.~\ref{fig:link-timeline} and \ref{fig:snapshot-sequence}. 
The sequence of local distributions of inter-event durations \new{is given by} $\piij(\Dtauf^*)=([(\Dtau_\ij^m)^*]_{m\in\Ml})_{\ij\in\L}$.
%, is shown in Fig.~\ref{fig:examples-Dtau}(b).
We cannot construct a sequence of local distributions from the sequence of snapshot graphs since they are not sequences.
\end{example}

%\begin{definition}
%{\sl Distribution of local distributions of features.}
%The distribution of local distributions $\ppiq$ is obtained by marginalizing over the sequence of local distributions. 
%We define $\ppiq$ formally as a function that returns the multiset of the local distributions, $\ppiqx{\x}=[\pi_q(\x_q)]_{q\in\Q}$.
%\end{definition}
%
%As the notation indicates, the distribution of local distributions $\ppiq$ is the composition of the function $\pxq$, which takes a sequence of sequences and returns the distribution if the individual (local) sequences, on $\piq$, which returns the sequence of the local distributions of the individual sequences. 
%This composition should not be confused with the compositions of MRRMs, which does not return a function.

%\begin{example}
%\label{ex:set_sets-d}
%Figure~\ref{fig:examples-d}(g) shows the distribution of local distributions of the instantaneous degrees of each node, $\ppiix{\d^*}$, and Fig.~\ref{fig:examples-d}(h) shows the distribution of local distributions of instantaneous degrees of nodes in each snapshot, $\ppitx{\d^*}$, for the temporal network shown  of Figs.~\ref{fig:link-timeline} and \ref{fig:snapshot-sequence}.
%\end{example}
%
% \begin{example}
%\label{ex:set_sets-Dtau}
%Figure~\ref{fig:examples-Dtau}(e) shows the distribution of local distributions of inter-event durations on the links, $\ppilx{\Dtauf^*}$, of the temporal network  of Figs.~\ref{fig:link-timeline} and \ref{fig:snapshot-sequence}.
%\end{example}

After the above definitions of different local and global marginalizations of a sequence of features, we now consider ways to define its moments. We shall here consider only first-order moments, i.e.\ means, but note that one may generally consider also higher order moments such as the variance.
As for the distributions, it is natural to define the mean of a one- or two-level sequence simply as the average over the values of their scalar elements.
For a two-level sequence, we may additionally construct a sequence of means of the local sequences. 
% makes it natural to consider two additional features based on the means of local sequences. These are obtained as the sequence of the means of each local sequence (the  {\sl local means}) and as the distribution of the local means.
%When needed to avoid ambiguity, we will refer to the mean of features as the {\sl global} mean to distinguish it from those features.

\begin{definition}
{\sl Means of a one- or two-level sequence of features.}
We shall consider two different ways to average over a one- or two-level sequence. Detailed definitions for specific kinds of features are given in Table~\ref{tab:constraint_levels} (symbols: $\mu$, $\muij$, $\mui$, and $\mut$). 
\begin{enumerate}
  \item{\sl Global mean $\mu(\x)$.} The global mean $\mu(\x)$ of a sequence of features is defined as the average over all individual scalar features in $\x$. 
  For a one-level sequence, it is given by $\mu(\x^*)=\sum_{q\in\Q} x_q^*/Q$, where Q is the number of elements in $\x^*$.
  For a two-level sequence, it is  $\mu(\x^*)=\sum_{q\in\Q}\sum_{r\in\R_q}(x_q^r)^*/(\sum_{q\in\Q}R_q)$, where $R_q$ is the number of elements in $\x^*_q$.
  \item{\sl Sequence of local means $\muq(\x)$.} For a two-level sequence $\x=(\x_q)_{q\in\Q}$, the sequence of local means $\muq(\x)$ is defined as the sequence of the means of each local sequence $\x_q$. Each of these {\sl local} means is given by $\mu_{q}(\x_q^*)=\sum_{r\in\R_q} (x_q^r)^*/R_q$.
\end{enumerate}
\end{definition}

The different types of means obtained are illustrated in the following examples.

\begin{example}
\label{ex:global_mean}
{\sl Global mean.}
The mean of the one-level sequence of static degrees \new{is given by} $\mu(\kstat^*)=\sum_{i\in\N}k_i^*/N$.
%,  of the temporal network in Fig.~\ref{fig:link-timeline} is shown in Fig.~\ref{fig:examples-kstat}(c).
The global mean of the two-levels instantaneous degrees \new{is given by} $\mu(\d^*)=\sum_{i\in\N}\sum_{t\in\T}(\dit)^*/(NT)$, 
%is shown in Fig.~\ref{fig:examples-d}(j), 
and the global mean of the inter-event durations \new{is} $\mu(\Dtauf^*)=\sum_{\ij\in\L}\sum_{m\in\Ml}(\Dtau_\ij^m)^*/\sum_{\ij\in\L}M_\ij$.
%, is shown in Fig.~\ref{fig:examples-Dtau}(f).
\end{example}

\begin{example}
\label{ex:seq_means}
{\sl Sequence of local means.}
Sequences of local means of the instantaneous degrees can be constructed in two ways: 
as the sequence of local means of the instantaneous degrees of each node, $\mui(\d^*)=(\sum_{t\in\T}(\dit)^*/T)_{i\in\N}$, 
%shown in Fig.~\ref{fig:examples-d}(f), 
and as the sequence of local means of the instantaneous degrees in each snapshot, $\mut(\d^*)=(\sum_{i\in\N}(\dit)^*/N)_{t\in\T}$.
%, shown in Fig.~\ref{fig:examples-d}(h). %  of the temporal network  of Figs.~\ref{fig:link-timeline} and \ref{fig:snapshot-sequence}.
For the inter-event durations, %, a single sequence of local means can be constructed, namely 
the sequence of local means on each link is $\muij(\Dtauf^*)=(\sum_{m\in\Ml}(\Dtau_\ij^m)^*/M_\ij)_{\ij\in\L}$.
% [Fig.~\ref{fig:examples-Dtau}(d)].
\end{example}

%\begin{definition}
%{\sl Distribution of local means of features.}
%The distribution of local means $\pmuq$ is obtained by marginalizing over the sequence of local means. It can be thought of as the composition of  $\pxq$ on $\muq$, and is defined as the multiset of local means, $\pmuqx{\x^*}=[\mu_{q}(\x_q^*)]_{q\in\Q}$.
%\end{definition}
%
%\begin{example}
%\label{ex:set_means-d}
%Figure~\ref{fig:examples-d}(j) shows the distribution of local means of the  instantaneous degrees of each node, $\pmuix{\d^*}$, and Fig.~\ref{fig:examples-d}(l) shows the distribution of local means of the instantaneous degrees in each snapshot, $\pmutx{\d^*}$, of the temporal network  of Figs.~\ref{fig:link-timeline} and \ref{fig:snapshot-sequence}.
%\end{example}
%
%\begin{example}
%\label{ex:set_means}
%Figure~\ref{fig:examples-Dtau}(f) shows the distribution of local means of the  inter-event durations on each link, $\pmulx{\Dtauf^*}$, of the temporal network of Figs.~\ref{fig:link-timeline} and \ref{fig:snapshot-sequence}.
%\end{example}

\begin{figure*}
  \begin{tikzpicture}
     \node at(-2.1,0){ \includegraphics[height=0.11\textheight]{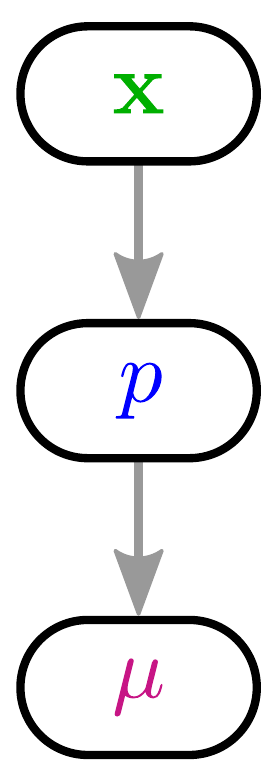} };
    \node at(4.8,0){ \includegraphics[height=0.154\textheight]{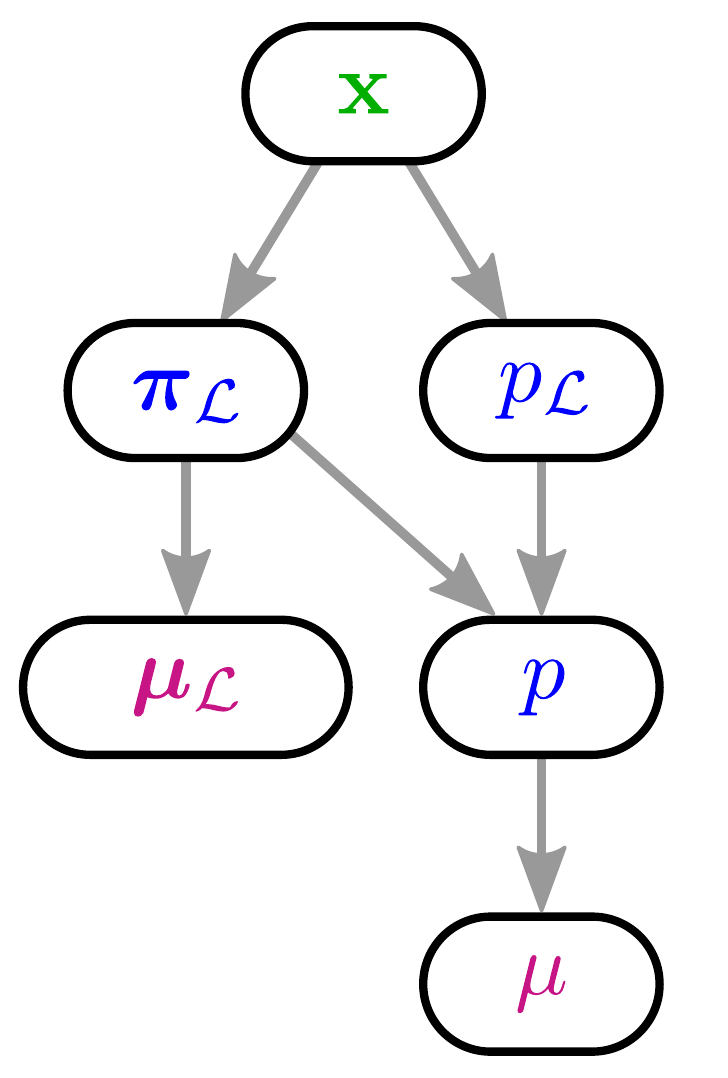} };
   \node at(1,0){ \includegraphics[height=0.154\textheight]{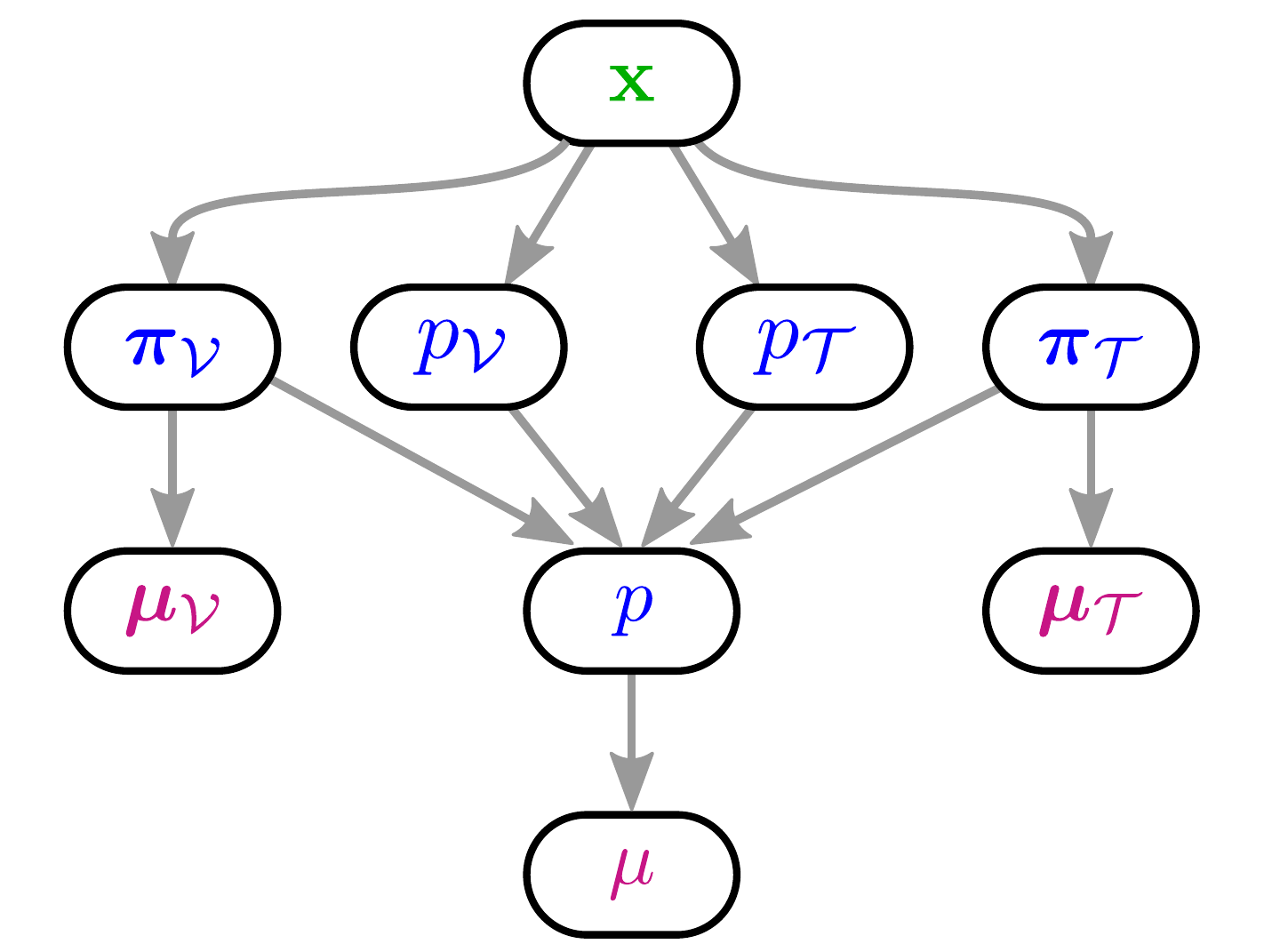} };
    \node at(-2.3,1.6){(a)};
    \node at(-.9,1.6){(b)};
    \node at(4.,1.6){(c)};
  \end{tikzpicture}
  \caption{{\bf Hierarchies of the marginals and moments of a sequence of features.}
  An arrow from a higher node to a lower one indicates that the former feature is finer than the latter. Thus, a MRRM that conserves the former feature necessarily conserves all  downstream features.
  Conversely, a MRRM that randomizes a given feature also randomizes any features upstream of it as well.
  (a) For a sequences of scalar features, e.g.\ the static degrees $\k=(k_i)_{i\in\N}$.
  %namely the aggregated features $\kstat$, $\a$, $\n$, $\s$, and $\w$, and $\Ebf$, and the link-timeline features $\t^1$, and $\t^w$.
  (b) For two-level sequences of features of nodes , namely $\alphaf$, $\Dalphaf$ and $\d$. 
  (c) For two-level sequences of features of link timelines, namely $\tauf$ and $\Dtauf$. 
  }
  \label{fig:constraint_hierarchy}
\end{figure*}

The distributions and means defined above are functions of the sequence of features, so they are all coarser than the sequence. Many of them are also comparable (though not all of them), so we can establish a hierarchy between them using \new{Definition \ref{def:comparability}}.%Proposition~\ref{the:partial_order}. %, albeit not a linear one.
Table~\ref{tab:constraint_levels} lists all the distributions and moments for features of links, nodes, and snapshots, and we establish their hierarchies in Fig.~\ref{fig:constraint_hierarchy}.

%------------------------------------------------------------------------
%------------------------------------------------------------------------

By combining Tables~\ref{tab:features} and \ref{tab:constraint_levels}, as shown in the examples above, we may describe most features constrained by MRRMs found in the literature.

\begin{figure*}
  \includegraphics[width=0.75\textwidth]{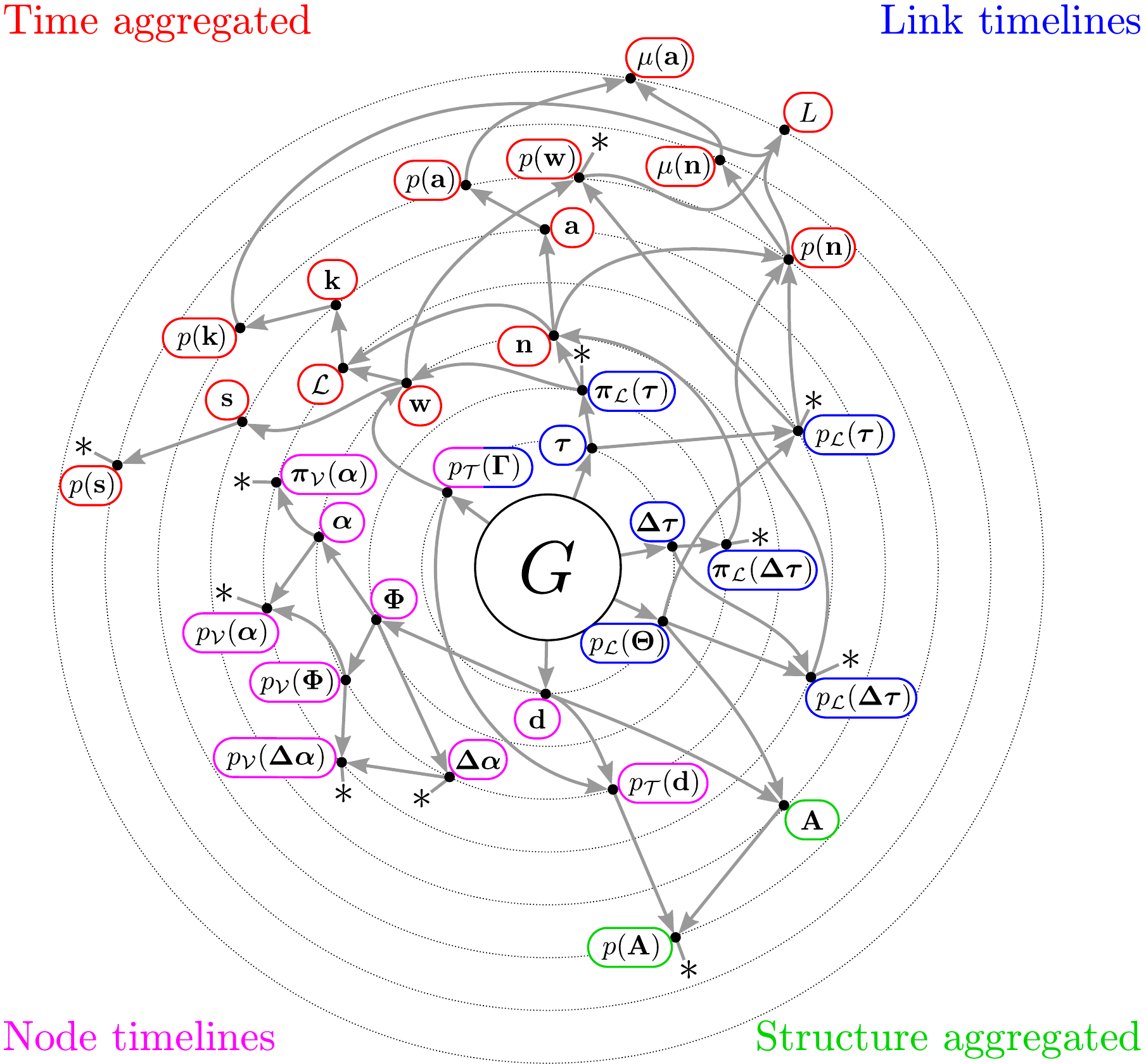}
  \caption{{\bf Feature hierarchy diagram.}
    An arrow from a higher ranking (more central) to a lower ranking feature (vertex in the diagram) indicates that the former feature is finer than the latter. Thus, a MRRM that constrains a given feature also constrains all downstream features.
    Conversely, a MRRM that randomizes (i.e.\ does not constrain) a given feature  does not constrain any of the upstream features either.
    See Tables~\ref{tab:features} and \ref{tab:constraint_levels} for definitions of the features.
   A star ($*$) emanating from a node indicates that lower hierarchical levels follow as shown in Fig.~\ref{fig:constraint_hierarchy}.
    The color coding shows a division of features into different types: time-aggregated features (i.e.\ topological and weighted), link-timeline features, node-timeline features, and structure-aggregated features (i.e.\ purely temporal).}
  \label{fig:dependency_diagram}
\end{figure*}

Some of the different basic features listed in Table~\ref{tab:features} are also pairwise comparable. This enables us to construct a hierarchy of the different features listed in Table~\ref{tab:features} together with their marginals and moments (Table~\ref{tab:constraint_levels}).
Figure~\ref{fig:dependency_diagram} shows such a hierarchy. 
It may be used to derive which features are conserved by a MRRM that constrains a given feature: the MRRM conserves all features that are below the constrained feature in the hierarchy.

%------------------------------------------------------------------------
%------------------------------------------------------------------------

Note  that if two features are not comparable, it does not imply that they are  independent.
So one cannot conclude from the absence of a link between two features in Fig.~\ref{fig:dependency_diagram} that one does not influence the other, only that it does not constrain it completely; the features may be correlated. 
The correlations between features that are neither comparable nor
 independent depend on the input temporal network that is considered. Thus, they can only be investigated on a case-by-case basis. %We illustrate this in the following example.

\subsection*{Notation for features of instant-event networks}
\label{sec:instant-event_features}

\noindent
Whenever possible we use the same symbols and names for features of instant-event temporal networks as for networks with event durations. 
However, we have adopted a different notation for the number of instantaneous events on a link, $\wij$, than for the number of events on a link, $\nij$  (Table~\ref{tab:features}). This is needed to make our description of MRRMs consistent when they are applied to both temporal networks and instant-event networks (see discussion in Supplementary Note~1). This furthermore means that $\si$ denotes the total number of instantaneous events that a node partakes in for instant-event networks, and that $\ai$, and $\nij$ are not defined for these networks. 
Similarly, the event durations $\tau_\ij^m$  and activity durations $\alpha_\ij^m$
are not defined for instant-event networks.

Several other definitions are changed slightly to accommodate the fact that instantaneous events do not have durations. This is namely the case for $\L$, $\Thij$, $\Dtau_\ij^m$, $t^w_\ij$, $\Dalpha_\ij^m$, and $\Phi_i$ (see Table~\ref{tab:features}).  
Conversely, the snapshot-graph sequence is only defined for instant-event networks, and thus so are also the following associated features: $\T$, $\Gt$, and $\E^t$.

%------------------------------------------------------------------------
%\clearpage
\begin{titlepage}
  \centering{\Large{{\bf Supplementary Note 3: Analyzing a face-to-face interaction network using MRRMs}}}
  \vspace{1cm}
\end{titlepage}
%------------------------------------------------------------------------
% \subsection{Analyzing face-to-face a interaction network using MRRMs}
% \label{sec:walktrough_example}
%\note{CLV: Remove panels in Figure and corresponding text: $p(\tauf)$, $p(\s)$, $p(\w)$, and $p(\alphaf)$!}

\noindent
In this supplementary note we apply a selection of MRRMs to 
%illustrate how they can be used to methodically explore different features and 
\new{build a statistical portrait of a temporal network %. 
% We analyze a SocioPatterns dataset 
of face-to-face interactions recorded in a primary school \cite{Stehle2011H,Gemetto2014M}.
The dataset, which is freely available at \url{www.sociopatterns.org/datasets} 
%The data were 
was recorded with a time-resolution of 20\,s and forms} a temporal network consisting of 242 nodes and 77\,521 events of varying duration.

\new{
%Figure~\ref{fig:MRRMs-walktrough} shows the MRRMs applied to study the network and orders them hierarchically.
%\note{CLV: I removed the figure of the hierarchy of the methods. I don't think it contributed much.}
We apply a series of MRRMs to gradually randomize different network features, allowing us to unravel the features' effects on the network's topology and temporal structure.
Supplementary Table~\ref{tab:effects-walkthrough1} lists which temporal network features the different models conserve (a more detailed description is found in Table~\ref{tab:effects-contact}). %in Appendix~\ref{app:tables}).
The MRRMs are described in detail in Section~\ref{sec:classification}.}

%\begin{figure}
%  \includegraphics[height=0.35\textheight]{Methods_example.pdf}
%  \caption{{\bf MRRMs employed in the analysis of face-to-face interactions in a primary school.}
%  \note{CLV: remove box around $\model{p(\tauf)}$}
%  \note{CLV: add instant-event shufflings $\model{\w,\t}$ and $\model{\t}$? (in red?)}
%  P[$p(\tauf)$], described in detail in Section~\ref{sec:contact_shuffling}, is the coarsest (most random) event shuffling possible. 
%  See Section~\ref{sec:link-shufflings} for definitions of link shufflings, 
%  Section~\ref{sec:timeline-shufflings} for timeline shufflings, 
%  Section~\ref{sec:snapshot_shufflings} for the snapshot shuffling 
%  P[$p(t,\tauf)$], 
%  and Section~\ref{sec:intersections} for the intersections P[$\n,\pxl(\Thf)]$, P[$\L,\pxl(\Thf)]$, and P[$\L,p(\t,\tauf)]$.}
%  \label{fig:MRRMs-walktrough}
%\end{figure}

\begin{table*}
  \caption{\new{{\bf Effects of selected MRRMs on temporal network features.}
  See Table~\ref{tab:features} for definitions of features.
  Colored symbols show to what extent each feature is conserved. 
  Informal definitions are found in the tablenotes (detailed definitions are found in Supplementary Table~\ref{tab:constraint_levels}).
  }}
  \label{tab:effects-walkthrough1}
\centerline{
  \begin{threeparttable}
  \begin{tabular}{ll|cccccccccccccccccc}
  \hline
  \hline
%    Canonical name &  &\multicolumn{9}{l|}{One-level} & \multicolumn{9}{l}{Two-level} %& References
%    \\
%    && \multicolumn{3}{l}{topological} & \multicolumn{4}{l}{weighted} & \multicolumn{2}{c|}{temp.} & \multicolumn{4}{l}{node} & \multicolumn{5}{l}{link} \\
    Model && Features &  &  &  &   &   &  &  & &  &  &  &   &  &   &   &  &    \\
     && $\Gstat$ & $\k$ & $\Estat$ & $\a$ & $\s$ & $\n$ & $\w$ & $\Ebf$ & &  & $\alphaf$ & $\Dalphaf$ &   &  & $\tauf$ & $\Dtauf$ & $\t^1$ & $\t^w$   \\
  \hline
    P[$p(\tauf)$] & & \crossmark & \crossmark & \crossmark & \crossmark & \mumark & \crossmark & \crossmark & \mumark &   &  & \crossmark & \crossmark &   &   & \pmark & \crossmark & \crossmark & \crossmark  %& \cite{Holme2016T}
    \\
    P[$p(\t,\tauf)$] & & \crossmark & \crossmark & \crossmark & \crossmark & \mumark & \crossmark & \crossmark & \checkmark & & & \crossmark & \crossmark &   &   & \pmark & \crossmark & \crossmark & \crossmark  %& \cite{Posfai2014S}
    \\
    P[$\pxl(\Thf)$] & & \crossmark & \mumark & \checkmark & \mumark & \mumark & \pmark & \pmark & \checkmark & & & \crossmark & \crossmark &  &  & \pmark & \pmark & \pmark & \pmark  %& \cite{Erdos1960O,Karimi2013T}
    \\
    P[$\k,\pxl(\Thf)$] &  & \crossmark & \checkmark & \checkmark & \mumark & \mumark & \pmark & \pmark & \checkmark & & & \crossmark & \crossmark &   &  & \pmark & \pmark & \pmark & \pmark %& \cite{Maslov2002S,Holme2005N,Holme2012T,Holme2015M,Delvenne2015D,Li2017T}
    \\
    P[$\L$,$p(\tauf)$] &   & \checkmark & \checkmark & \checkmark & \mumark & \mumark & \mumark & \mumark & \mumark & & & \crossmark & \crossmark &   &  & \pmark & \crossmark & \crossmark & \crossmark  %& \cite{Holme2016T}
    \\
    P[$\L$,$p(\t,\tauf)$] & & \checkmark & \checkmark & \checkmark & \mumark & \mumark & \mumark & \mumark & \checkmark & & & \crossmark & \crossmark &  &  & \pmark & \crossmark & \crossmark & \crossmark %&
    \\
    P[$\piij(\tauf)$] & & \checkmark & \checkmark & \checkmark & \checkmark & \checkmark & \checkmark & \checkmark & \mumark &   &   & \crossmark & \crossmark &  &  & \pmark & \crossmark & \crossmark & \crossmark    \\
    P[$\piij(\tauf)$,$\t^1$,$\t^w$] && \checkmark & \checkmark & \checkmark & \checkmark & \checkmark & \checkmark & \checkmark & \mumark &   &   & \crossmark & \crossmark &  &   & \pmark & \mumark & \checkmark & \checkmark \\
    P[$\piij(\tauf$),$\piij(\Dtauf$)] &  & \checkmark & \checkmark & \checkmark & \checkmark & \checkmark & \checkmark & \checkmark & \mumark && & \crossmark & \crossmark &  & & \pmark & \pmark & \crossmark & \crossmark %& \cite{Gauvin2013A}
    \\
    P[$\piij(\tauf$),$\piij(\Dtauf)$,$\t^1$] &  & \checkmark & \checkmark & \checkmark & \checkmark & \checkmark & \checkmark & \checkmark & \mumark & & & \crossmark & \crossmark &  &  & \pmark & \pmark & \checkmark & \checkmark  %&
    \\
%    P[$\n$,$p(\t,\tauf)$] &  & \checkmark & \checkmark & \checkmark & \checkmark & \mumark & \checkmark & \mumark & \checkmark & & & \crossmark & \crossmark & \mutmark &   & \pitmark & \crossmark & \crossmark & \crossmark %&
%    \\
    P[$\L$,$\pxl(\Thf)$] & & \checkmark & \checkmark  & \checkmark & \mumark & \mumark & \pmark & \pmark & \checkmark & & & \crossmark & \crossmark &   &   & \pmark & \pmark & \pmark & \pmark %&  \cite{Karsai2011S,Kivela2012M,Holme2012T,Gauvin2013A,Genois2015C,Thomas2015D}
    \\
    P[$\n$,$\pxl(\Thf)$] & & \checkmark & \checkmark  & \checkmark & \checkmark & \mumark & \checkmark & \pmark & \checkmark & & & \crossmark & \crossmark &  &   & \pmark & \pmark & \pmark & \pmark  %&
    \\
  \hline
  \hline
  \end{tabular}
  \begin{tablenotes}
    \item[\checkmark] Feature completely conserved.
    \item[\pmark] Distribution (i.e.\ the multiset) of individual values in sequence conserved.
    \item[\mumark] Mean value of the individual features in sequence conserved.
    \item[\crossmark] Feature not conserved.
  \end{tablenotes}
  \end{threeparttable}
}
\end{table*}

\begin{figure*}[p]
  \begin{tikzpicture}
    \node at(-4.5,1.5){ \includegraphics{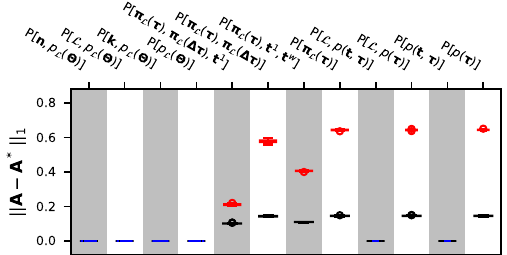} };
    \node at(4.5,1.5){ \includegraphics{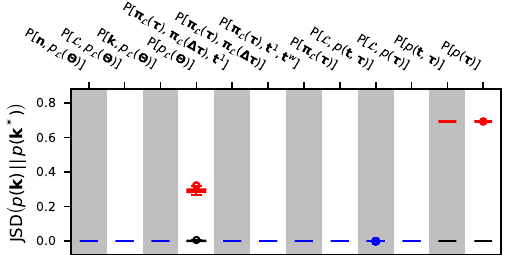} };
    \node at(-4.5,-2.8){ \includegraphics{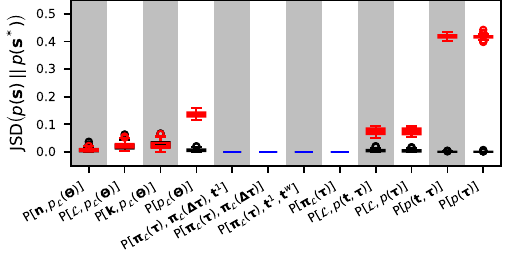} };
    \node at(4.5,-2.8){ \includegraphics{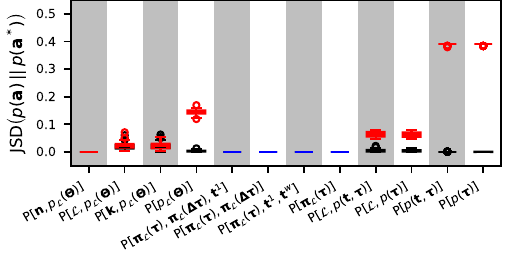} };
    \node at(-4.5,-5.6){ \includegraphics{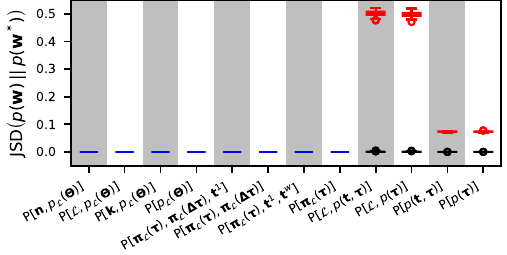} };
    \node at(4.5,-5.6){ \includegraphics{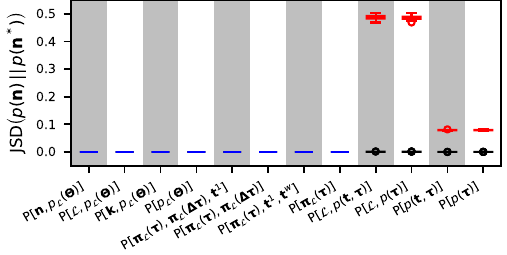} };
    \node at(-4.5,-8.4){ \includegraphics{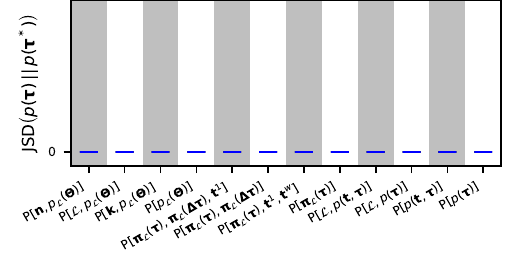} };
    \node at(4.5,-8.4){ \includegraphics{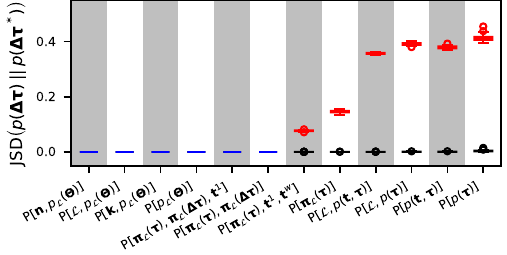} };
    \node at(-4.5,-11.2){ \includegraphics{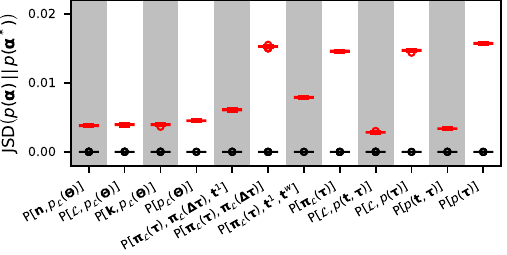} };
    \node at(4.5,-11.2){ \includegraphics{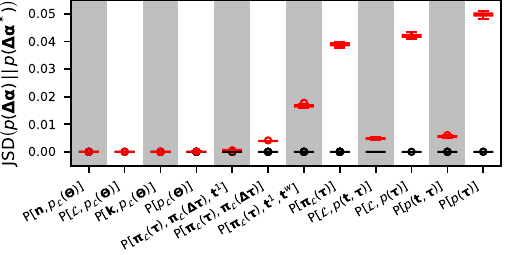} };
  \end{tikzpicture}
  \caption{{\bf Effects of the MRRMs on different features of a temporal network of face-to-face interactions in a primary school.} 
%  \note{Ref 3: Not entirely clear. - Spend more time describing the colors.}
%  \note{CLV: do not show zeroes? i.e. features whose values are conserved by the MRRM.}
%  \note{CLV: add MRRMs that randomize $p(\tauf)$ / replace by other feature, e.g. $p(\d)$?}
  Each panel shows the difference between the value of the feature in the empirical network and its null distribution under each model (red symbols for MRRMs that do not constrain the feature, and blue symbols for MRRMs that do) as well as the differences between its value in different randomized networks in each null \new{model} %ensemble 
  (black). The latter serves as a benchmark that shows the expected difference due to random fluctuations if a null model were true. 
  For the activity timeline $\Ebf$, the difference is quantified as the L1-distance between the activity at each time. For all other features, the difference is quantified as the Jensen-Shannon divergence (JSD) between the global distributions of the values of individual scalar features.  
  Each box-and-whiskers summarizes the distribution of the differences over 100 randomized networks generated by the MRRM in question: boxes show the 1st and 3rd quartiles; whiskers extend to 1.5 times the interquartile range or to the minimum  (maximum) value, whichever is smaller; values beyond the whiskers are marked by open circles.}
  \label{fig:walkthrough-resume}
\end{figure*}

Supplementary Figure~\ref{fig:walkthrough-resume} is our statistical portrait of the temporal network. 
It quantifies how \new{far the values a selection of features of the empirical network are from the networks generated by each MRRM. }
%The features we have chosen to investigate are:} 
%the timeline of cumulative node activity, $\Ebf=(|\E^t|)_{t\in\T}$; 
%the distributions of five time-aggregated (one-level) features: the node degrees, $p(\k)=[k_i]_{i\in\N}$, the link weights and event frequencies, $p(\w)=[\wij]_{\ij\in\L}$, with $\wij=\sum_{m\in\Ml} \tau_\ij^m$, and $p(\n)=[\nij]_{\ij\in\L}$, with $\nij=|\Thij|$, and the node strengths and activities, $p(\s)=[\sum_{j\in\N}\wij]_{i\in\N}$ and $p(\a)=[\sum_{j\in\N}\nij]_{i\in\N}$; 
%as well as global distributions of four temporal-structural features: the event and inter-event durations on links, $p(\tauf)=[\tau]_{(i,j,t,\tau)\in\C}$ and $p(\Dtauf)=\cup_{\ij\in\L} [\Dtau_\ij^m]_{m\in\Ml}$, and the node activity and inactivity durations, $p(\alphaf)=\cup_{i\in\N} [\alpha_i^m]_{m\in\Mi}$ and $p(\Dalphaf)=\cup_{i\in\N} [\Dalpha_i^m]_{m\in\Mi}$.
The differences are quantified by the Jensen-Shannon divergence (JSD)~\cite{Lin1991D} 
between the null distributions and their distribution in the empirical network (for the activity timeline $\Ebf$, the difference is quantified by the L1 distance).
The values of the features in the empirical network and for the networks generated by each MRRM are shown in Supplementary Figs.~\ref{fig:tuto:analyse_link} and \ref{fig:tuto:analyse_time} below. 
We next explore Supplementary Fig.~\ref{fig:walkthrough-resume} panel by panel and discuss what can be learned about the network from it.

We first study the activity timeline $\Ebf=(|\E^t|)_{t\in\T}$.
It is  by construction completely constrained by all link shufflings and snapshot shufflings, while at the opposite end it is essentially completely randomized by P[$\piij(\tauf)$], P[$\L,p(\tauf)$], and P[$p(\tauf)$] (see Supplementary Fig.~\ref{fig:tuto:analyse_time}). %\ref{fig:tuto:analyse_time}).
This shows that $\Ebf$ is not constrained by the static graph of the network. 
Comparison between P[$p(\tauf)$] and P[$\piij(\tauf),\piij(\Dtauf)$] shows that the distribution of inter-event durations does affect $\Ebf$, but not to a large extent. 
Comparing this with P[$\piij(\tauf),\t^1,\t^w$] shows that the timing of the first and last events on each link does on the other hand have a significant effect on $\Ebf$ in the network. 
Furthermore, comparison with P[$\piij(\tauf),\piij(\Dtauf),\t^1$]  shows that constraining both $\t^1$ and $\t^w$ together with $\piij(\Dtauf)$ imposes an even stronger constraint on the activity timeline (see also Supplementary Fig.~\ref{fig:tuto:analyse_time}). %\ref{fig:tuto:analyse_time}).

We next consider time-aggregated features of the temporal network, starting with the distribution of node degrees, $p(\k)=[k_i]_{i\in\N}$. This feature is constrained by most of the MRRMs applied in this example,
with the exception of P[$\pxl(\Thf)$] (which draws $\Gstat$ from an Erd\H{o}s-R\'enyi model), and {{P[$p(\t,\tauf)$]} and P[$p(\tauf)$] (which do not conserve the number of links in $\Gstat$). The high divergence seen for P[$\pxl(\Thf)$] shows that the empirical network's degree distribution is significantly nonrandom (even if it does not seem to follow a broad-tailed distribution, see Supplementary Fig.~\ref{fig:tuto:analyse_link}).} %\ref{fig:tuto:analyse_link}).

Most of the MRRMs also conserve the distributions of link weights and event frequencies, $p(\w)=[\wij]_{\ij\in\L}$, with $\wij=\sum_{m\in\Ml} \tau_\ij^m$ and $p(\n)=[\nij]_{\ij\in\L}$, with $\nij=|\Thij|$, respectively.
The exceptions are P[$\L,p(\tauf)$] (which conserves the static structure, but not the heterogeneity in the number and durations of events in timelines), and {P[$p(\t,\tauf)$]} and P[$p(\tauf)$] (which do not conserve the number of links in $\Gstat$). 
The effects of these shufflings on $p(\w)$ and $p(\n)$ are very similar, highlighting the fact that $\w$ and $\n$ are highly correlated features.
Note that the smaller divergences seen for the more random P[$p(\tauf)$] than for P[$\L,p(\tauf)$] are due to the JSD putting most weight on low values of $\wij$ and $\nij$ since these are most probable. Since P[$p(\tauf)$] produces a much larger number of links than in the original network but conserves the number of events, it also produces a large fraction of links with low $\wij$ and $\nij$, similarly to the original network. Conversely, P[$\L,p(\tauf)$] conserves the number of links and homogenizes $\wij$ and $\nij$, leading to fewer low values of these.  

The majority of the shufflings do not constrain the distributions of node strengths and activities, $p(\s)=[\sum_{j\in\N}\wij]_{i\in\N}$ and $p(\a)=[\sum_{j\in\N}\nij]_{i\in\N}$, respectively.
As for $p(\w)$ and $p(\n)$, their effects on the two features are very similar. Due to this we take $p(\a)$ as example and note that the results are qualitatively the same for $p(\s)$.
The distribution of $a_i$ in the empirical network is indistinguishable from networks generated by P[$\k,\pxl(\Thf)$]. 
This shows that $p(\a)$ is simply determined by the convolution of the individual distributions of $\k$ and $\n$ and that correlations between the two are unimportant. 
Comparison with P[$\L,p(\tauf)$] and P[$\pxl(\Thf)$] shows that both $p(\n)$ (randomized by P[$\L,p(\tauf)$]) and $p(\k)$ (randomized by P[$\pxl(\Thf)$]) are needed to reproduce the non-random shape of $p(\a)$ though. 

We finally investigate temporal-structural features of nodes and links. 
We note first that the distribution of event durations, $p(\tauf)=[\tau]_{(i,j,t,\tau)\in\C}$ is conserved by all the MRRMs employed by construction since these are all event shufflings (Def.~\ref{def:event_shuffling}). 

The distribution of inter-event durations on the links, $p(\Dtauf)=\cup_{\ij\in\L} [\Dtau_\ij^m]_{m\in\Ml}$, is constrained by all link shufflings, but not by most of the other shufflings. 
Comparison of the effects of P[$\piij(\tauf),\t^1,\t^w$] and P[$\piij(\tauf)$] demonstrates that  the timing of the first and last events in the timelines constrain the inter-event durations to some degree in the network (see also Supplementary Fig.~ \ref{fig:tuto:analyse_time}). %\ref{fig:tuto:analyse_time}). 
The much larger divergence found for P[$\L,p(\tauf)$] highlights that the number of events $\nij$ on each link strongly influences the inter-event durations.

None of the MRRMs completely constrain the distributions of the nodes' activity and inactivity durations, $p(\alphaf)=\cup_{i\in\N} [\alpha_i^m]_{m\in\Mi}$ and $p(\Dalphaf)=\cup_{i\in\N} [\Dalpha_i^m]_{m\in\Mi}$, respectively.
However, all link shufflings produce null distributions that are relatively close to the empirical ones, though they are still statistically significantly different. This indicates that the temporal correlations of the individual links' activity strongly constrain the nodes' activity.
More surprisingly, the small divergence observed in $p(\alphaf)$ for P[$\L,p(\t,\tauf)$] and  P[$p(\t,\tauf)$] as compared to the other MRRMs points to the global timing of the events as the most important of the features in determining the node activity durations in the network. It is more important than the number of events and the distributions of inter-event durations on the links. %, and even though {P[$p(\t,\tauf)$]} does not even conserve the number of links in the static graph of the network.
%\note{CLV: discussion of $p(\alphaf)$ could be enriched by included instant-event shufflings.}
Conversely, we see that the distribution of inter-event durations, $\piij(\Dtauf)$, is the most important temporal feature in determining the nodes' inactivity durations, $p(\Dalphaf)$, while the timing of the events is a close second (compare P[$\piij(\tauf),\piij(\Dtauf),\t^1$] and P[$\piij(\tauf),\piij(\Dtauf)$] to {P[$\L,p(\t,\tauf)$]}, and these three to  P[$\piij(\tauf),\t^1,\t^w$] and P[$\piij(\tauf)$]).

As seen in Supplementary Figs.~\ref{fig:tuto:analyse_link} and \ref{fig:tuto:analyse_time}, %\ref{fig:tuto:analyse_link} and \ref{fig:tuto:analyse_time}, 
the distributions of the different features obtained from a single randomized network generally vary little around their median, even though the empirical network studied here is of relatively modest size. 

\begin{figure*}
  \centering
  \begin{tikzpicture}
  	\node at(0,0){\includegraphics[width=0.95\textheight,angle=270]{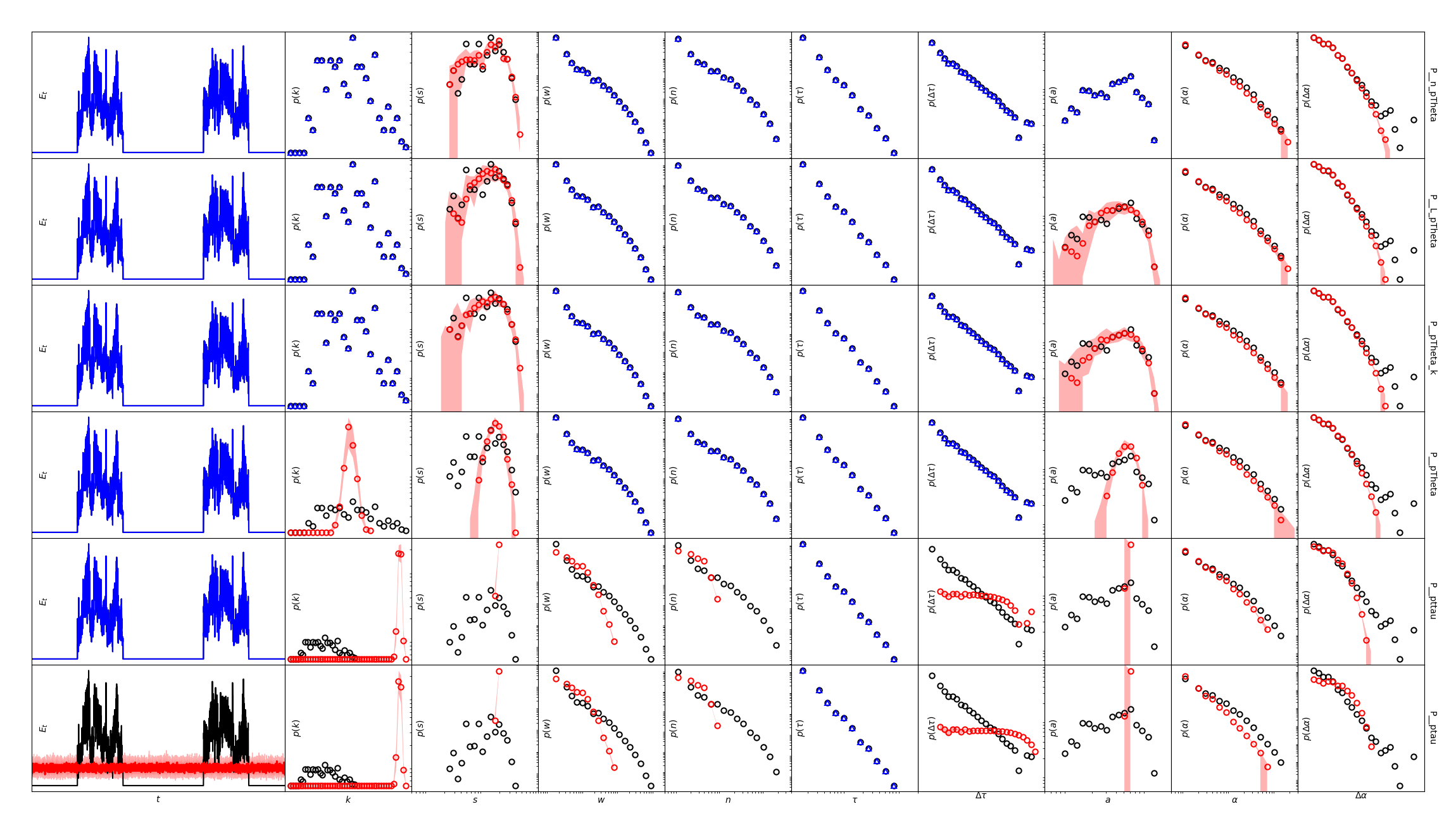}};
  	\node at(-4.3,11.2)[rotate=-20]{P[$p(\t,\tauf)$]};
  	\node at(-2.3,11.2)[rotate=-20]{P[$\pxl(\Thf)$]};
  	\node at(-0.5,11.25)[rotate=-20]{P[$\k,\pxl(\Thf)$]};
  	\node at(1.45,11.25)[rotate=-20]{P[$\L,\pxl(\Thf)$]};
  	\node at(3.4,11.25)[rotate=-20]{P[$\n,\pxl(\Thf)$]};
  	\draw [fill=white,white] (-5,-10.78) rectangle (5,-10.98); 
  \end{tikzpicture}
  \caption{\textbf{Effects of different link and event shufflings on temporal network features.}
  Values of a selection of features in the empirical face-to-face interaction network  %considered in Section~\ref{sec:walktrough_example} of the manuscript 
  and in randomized networks generated from it.
  Original data is in black. Randomized data is in blue if constrained, in red if not. Red lines are medians over 100 randomizations, red areas show 90\,\% confidence intervals.}
  \label{fig:tuto:analyse_link}
\end{figure*}

\begin{figure*}
  \centering
  \begin{tikzpicture}
  	\node at(0,0){\includegraphics[width=0.95\textheight,angle=270]{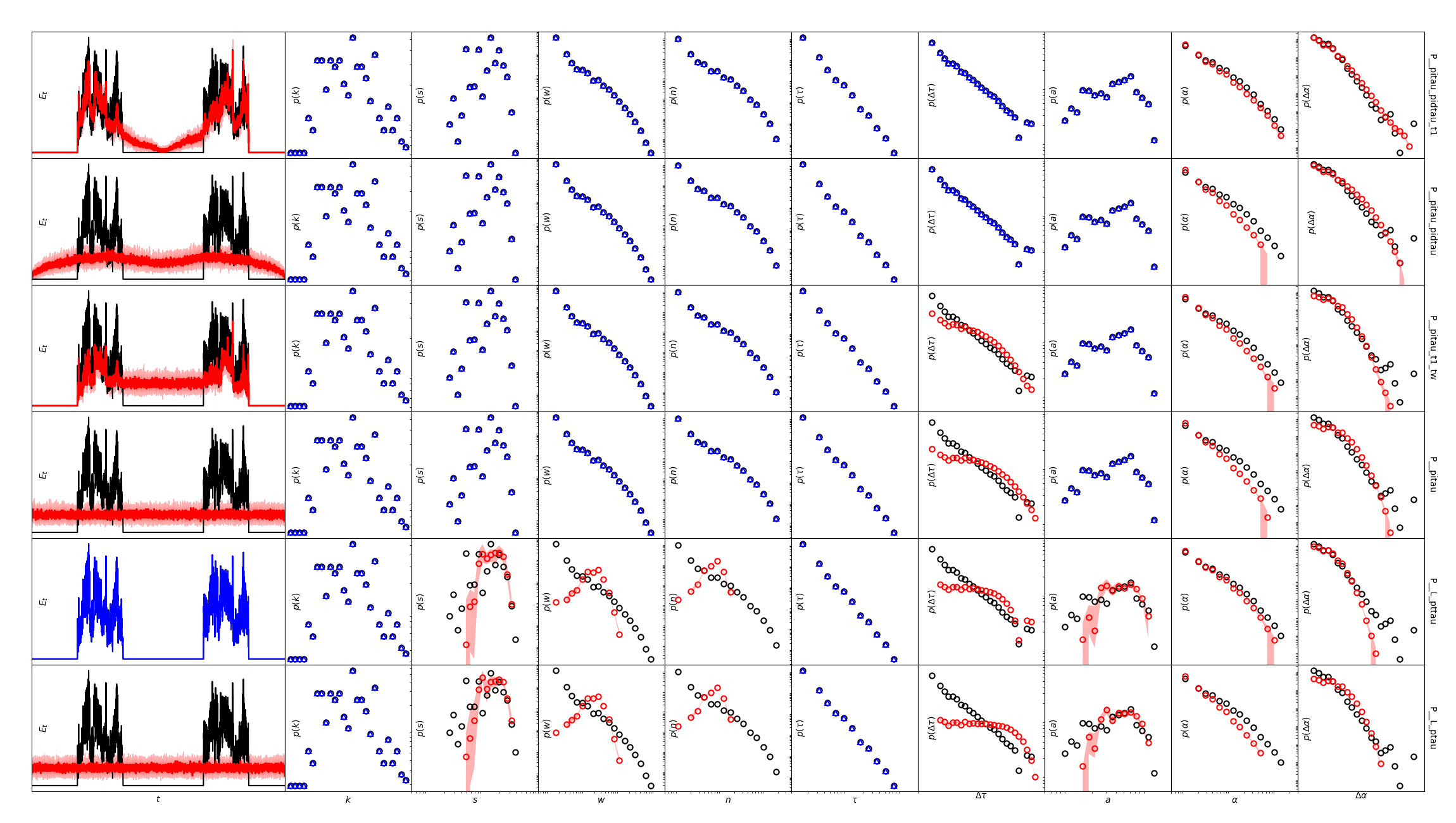}};
  	\node at(-4.3,11.2)[rotate=-20]{P[$\L,p(\tauf)$]};
  	\node at(-2.3,11.2)[rotate=-20]{P[$\piij(\tauf)$]};
  	\node at(-0.7,11.35)[rotate=-20]{P[$\piij(\tauf),\t^1,\t^w$]};
  	\node at(1.1,11.4)[rotate=-20]{P[$\piij(\tauf),\piij(\Dtauf)$]};
  	\node at(2.8,11.5)[rotate=-20]{P[$\piij(\tauf),\piij(\Dtauf),\t^1$]};
  	\draw [fill=white,white] (-5,-10.78) rectangle (5,-10.98); 
  \end{tikzpicture}
  \caption{\textbf{Effects of different timeline shufflings on temporal network features.}
  Values of a selection of features in the empirical face-to-face interaction network % considered in Section~\ref{sec:walktrough_example} of the manuscript 
  and in randomized networks generated from it.
  Original data is in black. Randomized data is in blue if constrained, in red if not. Red lines are medians over 100 randomizations, red areas show 90\,\% confidence intervals.}
  \label{fig:tuto:analyse_time}
\end{figure*}

%------------------------------------------------------------------------
%\clearpage
\begin{titlepage}
  \centering{\Large{{\bf Supplementary Note 4: Names of MRRM algorithms in the Python library}}}
  \vspace{1cm}
\end{titlepage}
%------------------------------------------------------------------------
\subsection*{Instant-event shufflings}

\begin{itemize}
\item P[$E$]: \verb,P__1,
\end{itemize}

\paragraph*{Timeline shufflings.}
\begin{itemize}
\item P[$\L,E$]: \verb,P__L,
\item P[$\w$]: \verb,P__w,
\item P[$\w,\t^1\t^w$]: \verb,P__w_t1_tw,
\item P[$\piij(\Dtauf)$]: \verb,P__pidtau,
\item P[$\piij(\Dtauf),\t^1\t^w$]: \verb,P__pidtau_t1_tw,
\end{itemize}

\paragraph*{Sequence shufflings.}
\begin{itemize}
\item P[$\pxt(\Gs)$]: \verb,P__pGamma,
\item P[$\pxt(\Gs),\sgn(\Ebf)$]: \verb,P__pGamma_sgnA,
\end{itemize}

\paragraph*{Snapshot shufflings.}
\begin{itemize}
\item P[$\t$]: \verb,P__t,
\item P[$\t,\Phif$]: \verb,P__t_Phi,
\item P[$\d$]: \verb,P__d,
\item P[$\isof(\Gs)$]: \verb,P__isoGamma,
\item P[$\isof(\Gs),\Phif$]: \verb,P__isoGamma_Phi,
\end{itemize}

\paragraph*{Intersections.}
\begin{itemize}
\item P[$\L,\t$]: \verb,P__L_t,
\item P[$\w,\t$]: \verb,P__w_t,
\end{itemize}

\paragraph*{Compositions.}
\begin{itemize}
\item P[$L$]: \verb,P__pTheta, with \verb,P__L_E,
\item P[$\k,p(\w),\t$]: \verb,P__pTheta, with \verb,P__w_t, 
\item P[$\k,\lc,p(\w),\t$]: with \verb,P__w_t, 
\end{itemize}

\subsection*{Event shufflings}
\begin{itemize}
\item P[$p(\tauf)$]: \verb,P__ptau,
\end{itemize}

\paragraph*{Link shufflings.}
\begin{itemize}
\item P[$\pxl(\Thf)$]: \verb,P__pTheta,
\item P[$\lc,\pxl(\Thf)$]: \verb,P__I_pTheta,
\item P[$\k,\pxl(\Thf)$]: \verb,P__k_pTheta,
\item P[$\k,\lc,\pxl(\Thf)$]:  \verb,P__k_I_pTheta,
\end{itemize}

\paragraph*{Timeline shufflings.}
\begin{itemize}
\item P[$\L,p(\tauf)$]: \verb,P__L_ptau,
\item P[$\piij(\tauf)$]: \verb,P__pitau,
\item P[$\piij(\tauf),\t^1\t^w$]: \verb,P__pitau_t1_tw,
\item P[$\piij(\tauf),\piij(\Dtauf)$]: \verb,P__pitau_pidtau,
\item P[$\piij(\tauf),\piij(\Dtauf),\t^1$]: \verb,P__pitau_pidtau_t1,
\item P[$\perl(\Thf)$]: \verb,P__perTheta,
\item P[$\tauf,\Dtauf$]: \verb,P__tau_dtau,
\end{itemize}

\paragraph*{Snapshot shufflings.}
\begin{itemize}
\item P[$p(\t,\tauf)$]: \verb,P__pttau,
\end{itemize}

\paragraph*{Intersections.}
\begin{itemize}
\item P[$\L,p(\t,\tauf)$]: \verb,P__L_pttau,
\item P[$\n,p(\t,\tauf)$]: \verb,P__n_pttau,
\item P[$\L,\pxl(\Thf)$]: \verb,P__L_pTheta,
\item P[$\w,\pxl(\Thf)$]: \verb,P__w_pTheta,
\item P[$\n,\pxl(\Thf)$]: \verb,P__n_pTheta,
\end{itemize}

\subsection*{Metadata shufflings}
\paragraph*{Link shufflings.}
\begin{itemize}
\item P[$\pxl(\Thf),\gf,\Sigmaf_\L$]: \verb,P__pTheta_sigma_SigmaL,
\item P[$\k,\pxl(\Thf),\gf,\Sigmaf_\L$]: \verb,P__k_pTheta_sigma_SigmaL,
\item P[$G,p(\gf)$]: \verb,P__G_psigma,
\end{itemize}

\paragraph*{Compositions.}
\begin{itemize}
\item P[$\k,p(\w),\t,\gf,\Sigmaf_\L$]: \verb,P__k_LCM, with \verb,P__w_t, 
\end{itemize}

\end{document}